# Long-Baseline Neutrino Facility (LBNF) and Deep Underground Neutrino Experiment (DUNE)

# Conceptual Design Report

## Volume 3: Long-Baseline Neutrino Facility for DUNE

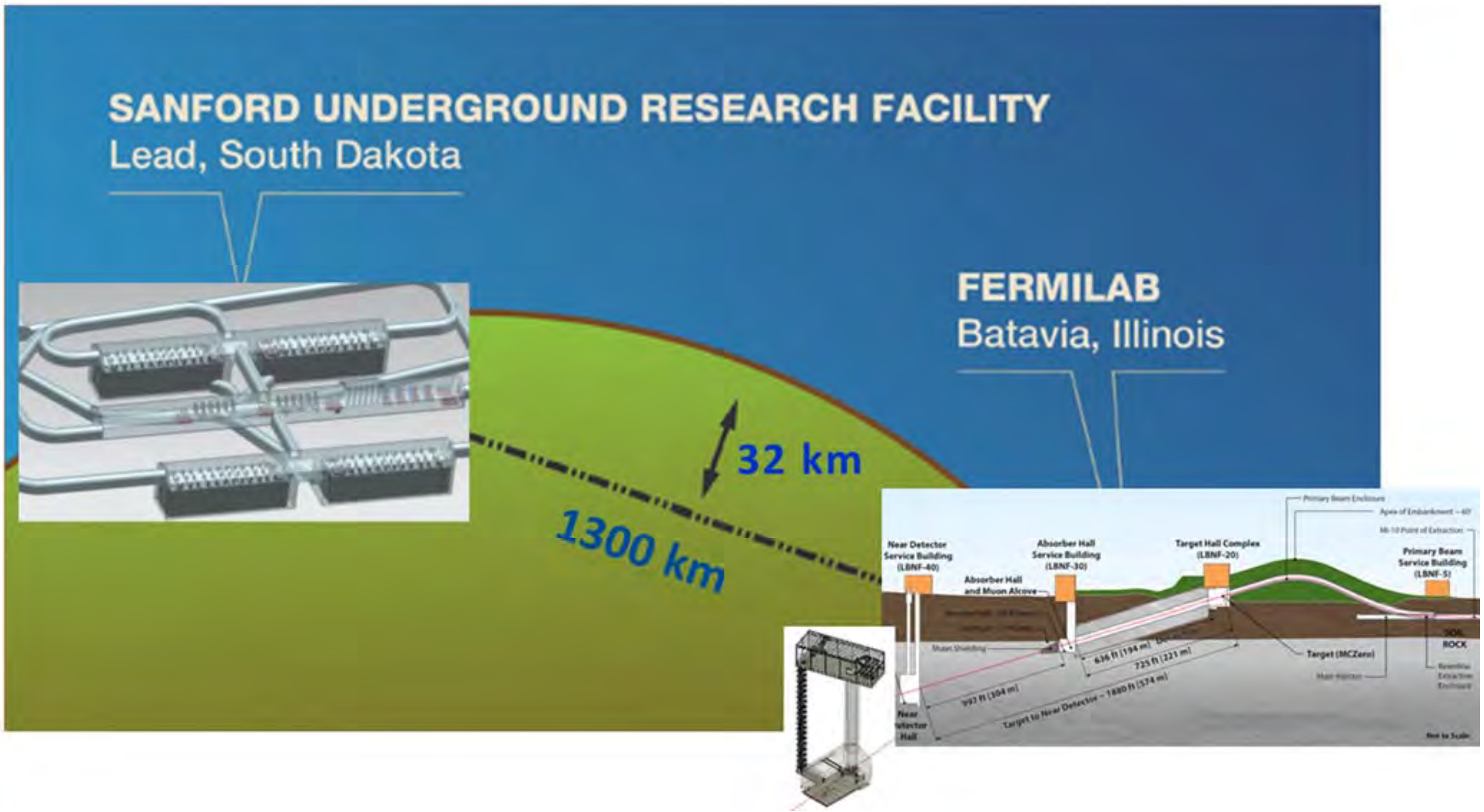

June 24, 2015

This page is intentionally left blank.

# Table of Contents

# List Of Figures

# List of Tables

# Acronyms, Abbreviations and Terms

| | |
|---|---|
| AASHTO | American Association of State Highway and Transportation Officials |
| ACNET | Accelerator Controls NETwork |
| AHJ | authority having jurisdiction |
| AHU(s) | air handling unit(s) |
| APA | anode plane assembly |
| ASME | American Society of Mechanical Engineers |
| ASP | advanced site preparation / air separation plant |
| BLM | beam loss monitor |
| BNL | Brookhaven National Laboratory |
| BPM | beam position monitor |
| CERN | The European Organization for Nuclear Research |
| CDR | Conceptual Design Report |
| CF | conventional facilities |
| CFD | computational fluid dynamics |
| CHW | chilled water |
| CP | charge parity |
| CPW | cooling pond water |
| DC | direct current |
| DOAS | dedicated outside air system |
| DOE | Department of Energy |
| DUNE | Deep Underground Neutrino Experiment |
| DUSEL | Deep Underground Science and Engineering Laboratory |
| ES&H | environment, safety and health |
| FEA | finite element analysis |
| FLS | fire protection-life safety |
| FEMA | Federal Emergency Management Agency |
| FODO | A repetitive configuration of focusing (F) and defocusing (D) magnetic elements separated by a fixed spacing of non-focusing elements (O) |
| FRCM | Fermilab Radiological Control Manual |
| FSCF | LBNF Far Site Conventional Facilities |
| HVAC | heating, ventilation and air conditioning |
| Hx | heat exchanger |
| kt | 1000 metric tons |
| L | used here to indicate the “level” underground at SURF, in feet |
| LArTPC | liquid argon time-projection chamber |
| LBNE | Long-Baseline Neutrino Experiment |
| LBNF | Long-Baseline Neutrino Facility |
| LC | inductive/capacitive |
| LCW | low-conductivity water |
| LEED | Leadership in Energy & Environmental Design, a green building certification program |
| LNG | liquefied natural gas |
| MARS | a monte carlo code |
| MEP | mechanical/electrical/plumbing |
| MER | mechanical electrical room |
| MI | main injector |
| MI-10 (60, 64 etc) | surface buildings in the Main Injector comples |
| MINOS | Main Injector Neutrino Oscillation Search experiment |
| NEPA | National Environmental Policy Act |
| NFPA | National Fire Protection Association |
| NOvA | NuMI Off-axis $\nu_e$Appearance experiment |
| NSCF | Near Site Conventional Facilities |
| NuMI | Neutrinos at the Main Injector (Beamline facility at Fermilab) |
| ODH | oxygen deficiency hazard |
| POT | protons-on-target |
| PIP-II | Proton Improvement Plan-II |
| PMB | Project Management Board |
| PRV | pressure-relief valves |
| RAW | radioactive water |
| RF | radio frequency |
| RMS | root mean square |
| SURF | Sanford Underground Research Facility (in Lead, S.D., the LBNF Far Site) |
| TPC | time-projection chamber |
| WBS | Work Breakdown Structure |

# 1 Introduction to LBNF and DUNE

## 1.1 An International Physics Program

The global neutrino physics community is developing a multi-decade physics program to measure unknown parameters of the Standard Model of particle physics and search for new phenomena. The program will be carried out as an international, leading-edge, dual-site experiment for neutrino science and proton decay studies, which is known as the Deep Underground Neutrino Experiment (DUNE). The detectors for this experiment will be designed, built, commissioned and operated by the international DUNE Collaboration. The facility required to support this experiment, the Long-Baseline Neutrino Facility (LBNF), is hosted by Fermilab and its design and construction is organized as a DOE/Fermilab project incorporating international partners. Together LBNF and DUNE will comprise the world's highest-intensity neutrino beam at Fermilab, in Batavia, IL, a high-precision near detector on the Fermilab site, a massive liquid argon time-projection chamber (LArTPC) far detector installed deep underground at the Sanford Underground Research Facility (SURF) 1300 km away in Lead, SD, and all of the conventional and technical facilities necessary to support the beamline and detector systems.

The strategy for executing the experimental program presented in this Conceptual Design Report (CDR) has been developed to meet the requirements set out in the P5 report [1] and takes into account the recommendations of the European Strategy for Particle Physics [2]. It adopts a model where U.S. and international funding agencies share costs on the DUNE detectors, and CERN and other participants provide in-kind contributions to the supporting infrastructure of LBNF. LBNF and DUNE will be tightly coordinated as DUNE collaborators design the detectors and infrastructure that will carry out the scientific program.

The scope of LBNF is:

- an intense neutrino beam aimed at the far site
- conventional facilities at both the near and far sites
- cryogenics infrastructure to support the DUNE detector at the far site

The DUNE detectors include

- a high-performance neutrino detector and beamline measurement system located a few hundred meters downstream of the neutrino source

- a massive liquid argon time-projection chamber (LArTPC) neutrino detector located deep underground at the far site

With the facilities provided by LBNF and the detectors provided by DUNE, the DUNE Collaboration proposes to mount a focused attack on the puzzle of neutrinos with broad sensitivity to neutrino oscillation parameters in a single experiment. The focus of the scientific program is the determination of the neutrino mass hierarchy and the explicit demonstration of leptonic CP violation, if it exists, by precisely measuring differences between the oscillations of muon-type neutrinos and antineutrinos into electron-type neutrinos and antineutrinos, respectively. Siting the far detector deep underground will provide exciting additional research opportunities in nucleon decay, studies utilizing atmospheric neutrinos, and neutrino astrophysics, including measurements of neutrinos from a core-collapse supernova should such an event occur in our galaxy during the experiment's lifetime.

## 1.2 The LBNF/DUNE Conceptual Design Report Volumes

### 1.2.1 A Roadmap of the CDR

The LBNF/DUNE CDR describes the proposed physics program and technical designs at the conceptual design stage. At this stage, the design is still undergoing development and the CDR therefore presents a *reference design* for each element as well as *alternative designs* that are under consideration.

The CDR is composed of four volumes and is supplemented by several annexes that provide details about the physics program and technical designs. The volumes are as follows:

- Volume 1 [1]**:** *The LBNF and DUNE Projects* provide an executive summary and strategy for the experimental program, and of the CDR as a whole.
- Volume 2 [2]: *The Physics Program for DUNE at LBNF* outlines the scientific objectives and describes the physics studies that the DUNE Collaboration will undertake to address them.
- Volume 3[This Volume]: *The Long-Baseline Neutrino Facility for DUNE* describes the LBNF Project, which includes design and construction of the beamline at Fermilab, the conventional facilities at both Fermilab and SURF, and the cryostat and cryogenics infrastructure required for the DUNE far detector.
- Volume 4 [3]: *The DUNE Detectors at LBNF* describes the DUNE Project, which includes the design, construction and commissioning of the near and far detectors.

More detailed information for each of these volumes is provided in a set of annexes listed on the *CD-1-R Reports and Documents* [4] page.

### 1.2.2 About this Volume

CDR Volume 3 *The Long-Baseline Neutrino Facility for DUNE* describes the facilities at the far and near sites that support the DUNE far and near detectors and the neutrino beamline. The remainder of this chapter provides a short introduction to the Long-Baseline Neutrino Facility. The management structure for LBNF is summarized in Chapter 2. Chapter 3 presents the far site conventional facilities, including evaluation of existing site conditions, the design of surface facilities, underground excavation, and underground infrastructure. The cryogenics infrastructure, including the cryostats which house the DUNE LArTPC detectors, the LAr purification system and the refrigeration system, are presented in Chapter 1. The near site conventional facilities are presented in Chapter 1, including description of the existing site conditions, the overall facility layout, and the designs of surface and underground structures that house LBNF beamline systems and the DUNE near detector. Chapter 6 describes the LBNF beamline, including the primary proton beam, the neutrino beam systems, radiological considerations, and supporting systems. Chapter 6 also includes a short discussion of alternative beamline options that are still under consideration.

## 1.3 Introduction to the Long-Baseline Neutrino Facility

LBNF provides facilities at Fermilab and at SURF to enable the scientific program of DUNE. These facilities are geographically separated into the Far Site Facilities, those to be constructed at SURF, and the Near Site Facilities, those to be constructed at Fermilab. These are summarized as such in sections below.

### 1.3.1 Far Site Facilities

The scope of LBNF at SURF includes both conventional facilities (CF) and cryogenics infrastructure to support the DUNE far detector. Figure 3-1 shows the layout of the underground caverns that house the detector modules with a separate cavern to house utilities and cryogenics systems. The requirements derive from DUNE Collaboration science requirements [5], which drive the space and functions necessary to construct and operate the far detector. Environment, Safety and Health (ES&H) and facility operations (programmatic) requirements also provide input to the design. The far detector is modularized into four 10-kt fiducial mass detectors. The caverns and the services to the caverns will be similar to one another as much as possible to enable repetitive designs for efficiency in design and construction as well as operation.

The scope of the conventional facilities includes design and construction for facilities on the surface and underground. The underground conventional facilities includes new excavated spaces at the 4850L for the detector, utility spaces for experimental equipment, utility spaces for facility equipment, drifts for access, as well as construction-required spaces. Underground infrastructure provided by CF for the experiment includes power to experimental equipment, cooling systems, and cyberinfrastructure. Underground infrastructure necessary for the facility includes domestic (potable) water, industrial water for process and fire suppression, fire detection and alarm, normal and standby power systems, a sump pump drainage system for native and leak water around the detector, water drainage to the facility-wide pump discharge system, and cyberinfrastructure for communications and security. In addition to providing new spaces and infrastructure underground, CF enlarges and provides infrastructure in some existing spaces for use, such as the access drifts from the Ross Shaft to the new caverns. New piping is provided in the shaft for cryogens (gas argon transfer line and the compressor suction and discharge lines) and domestic water as well as power conduits for normal and standby power and cyberinfrastructure.

The existing Sanford Laboratory has many surface buildings and utilities, some of which will be utilized for LBNF. The scope of the above ground CF includes only that work necessary for LBNF, and not for the general rehabilitation of buildings on the site, which remains the responsibility of the Sanford Laboratory. Electrical substations and distribution will be upgraded to increase power and provide standby capability for life safety. Additional surface scope includes a small control room in an existing building and a new building to support cryogen transfer from the surface to the underground near the existing Ross Shaft.

To reduce risk of failure of essential but aging support equipment during the construction and installation period, several SURF infrastructure operations/maintenance activities are included as early activities in LBNF. These include completion of the Ross Shaft rehabilitation, rebuilding of hoist motors, and replacement of the Oro Hondo fan; if not addressed, this aging infrastructure could limit or stop access to the underground if equipment failed.

The scope of the LBNF Cryogenics Infrastructure includes the design, fabrication, and installation of four cryostats to contain the liquid argon (LAr) and the detector components and a comprehensive cryogenics system that meets the performance requirements for purging, cooling down and filling the cryostats, achieving and maintaining the LAr temperature, and purifying the LAr outside the cryostats.

Each cryostat is composed of a free-standing steel-framed structure with a membrane cryostat vessel installed inside, to be constructed in one of the four excavated detector pits. The cryostat is designed to hold a total LAr mass capacity of 17.1 kt. Each membrane tank cryostat has a stainless-steel liner as part of a full membrane system to provide full containment of the liquid cryogen. The hydrostatic pressure loading of the liquid cryogen is transmitted through rigid foam insulation to the surrounding structural steel frame which provides external support for the membrane. All penetrations into the interior of the cryostat will be made through the top plate to minimize the potential for leaks with the exception of the sidewall penetration for connection to the LAr recirculation system.

Cryogenics system components are located both on the surface and within the cavern. The cryogen receiving station is located on the surface near the Ross Shaft to allow for receipt of LAr deliveries for the initial filling period, as well as a buffer volume to accept liquid argon during the extended fill period. A large vaporizer for the nitrogen circuit feeds gas to one of four compressors located in the Cryogenics Compressor Building; the compressor discharges high pressure nitrogen gas to pipes in the Ross shaft. The compressors are located on the surface because the electrical power requirement and cooling requirement is much less than for similar equipment at the 4850L.

Equipment at the 4850L includes the nitrogen refrigerator, liquid nitrogen vessels, argon condensers, external liquid argon recirculation pumps, and filtration equipment. Filling each cryostat with LAr in a reasonable period of time is a driving factor for the refrigerator and condenser sizing. Each cryostat will have its own overpressure protection system, argon recondensers, and argon-purifying equipment, located in the Central Utility Cavern. Recirculation pumps will be placed outside and adjacent to each cryostat to circulate liquid from the bottom of the tank through the purifier.

## 1.3.2 Near Site Facilities

The scope of LBNF at Fermilab is to provide the Beamline plus the CF for the beamline as well as for the DUNE near detector. The layout of these facilities is shown in Figure 5-1. The science requirements as determined by the DUNE Collaboration drive the performance of the Beamline and near detector, which then provide requirements for the components, space, and functions necessary to construct, install, and operate the Beamline and near detector. ES&H and facility operations (programmatic) requirements also provide input to the design.

The beamline is designed to provide a neutrino beam of sufficient intensity and appropriate energy range to meet the goals of DUNE for long-baseline neutrino oscillation physics. The design is a conventional, horn-focused neutrino beamline. The components of the beamline are designed to extract a proton beam from the Fermilab Main Injector (MI) and transport it to a target area where the collisions generate a beam of charged particles which decay into neutrinos to create the neutrino beam aimed at the near and far detectors.

The facility is designed for initial operation at proton-beam power of 1.2 MW, with the capability to support an upgrade to 2.4 MW. The plan is for twenty years of operation, while the lifetime of the Beamline Facility, including the shielding, is for thirty years. The conservative assumption is that for the first five years will be at 1.2 MW operations and the remaining fifteen years at 2.4 MW. The experience gained from the various neutrino projects has been employed extensively in the reference design. In particular, the Neutrinos at the Main Injector (NuMI) beamline serves as the prototype design. Most of the subsystem designs and their integration follow, to a large degree, from previous projects.

The proton beam is extracted at a new extraction point at MI-10. After extraction, this primary beam establishes a horizontally straight compass heading west-northwest toward the far detector, but will be bent upward to an apex before being bent downward at the appropriate angle. The primary beam will be above grade to minimize expensive underground construction and significantly enhances ground-water radiological protection. The design requires, however, construction of an earthen embankment, or hill, whose dimensions are commensurate with the bending strength of the dipole magnets required for the beamline.

The target marks the transition from the intense, narrowly-directed proton beam to the more diffuse, secondary beam of particles that in turn decay to produce the neutrino beam. After collection and focusing, the pions and kaons need a long, unobstructed volume in which to decay. This decay volume in the reference design is a pipe of circular cross section with its diameter and length optimized such that decays of the pions and kaons result in neutrinos in the energy range useful for the experiment. The decay volume is followed immediately by the absorber, which removes the remaining beam hadrons.

Radiological protection is integrated into the LBNF beamline reference design in two important ways. First, shielding is optimized to reduce exposure of personnel to radiation dose and to minimize radioisotope production in ground water within the surrounding rock. Secondly, the handling and control of tritiated ground water produced in or near the beamline drives many aspects of the design.

Beamline CF includes an enclosure connecting to the existing Main Injector at MI-10, concrete underground enclosures for the primary beam, targetry, horns, absorber, and related technical support systems. The service building LBNF 5 provides support utilities for the primary proton beam and LBNF 30 provides support for the absorber. The Target Hall Complex at LBNF 20 houses the targetry and focusing systems. Utilities will be extended from nearby existing services, including power, domestic and industrial water, sewer, and communications. Near Detector CF includes a small muon alcove area in the Beamline Absorber Hall and a separate underground Near Detector Hall that houses the near detector. A service building called LBNF 40 with two shafts to the underground supports the near detector. The underground hall is sized for the reference near detector.

# 2 Project Management

LBNF is organized as a DOE/Fermilab project incorporating in-kind contributions from international partners. At this time, the major international partner is CERN, the European Organization for Nuclear Research. The LBNF Project organization is charged by Fermilab and DOE and through them the international partners to design and construct the conventional and technical facilities needed to support the DUNE Collaboration. LBNF works in close coordination with the DUNE project to ensure that the scientific requirements of the program are satisfied through the mechanisms described in CDR Volume 1. LBNF also works closely with SURF management to coordinate the design and construction of the underground facilities required for the DUNE far detector.

LBNF consists of two major L2 subprojects coordinated through a central Project Office located at Fermilab: Far Site Facilities and Near Site Facilities. Each L2 Project consists of two large L3 subprojects corresponding to the conventional and technical facilities respectively at each site. The project organizational structure, which includes leadership from major partners, is shown in Figure 2-1.

## 2.1 Overview

The LBNF Project will design and construct conventional and technical facilities to support DUNE. LBNF works closely with DUNE through several coordinating groups to ensure scientific direction and coordination to execute the LBNF Project. CERN is providing cryogenics equipment and engineering as part of the Cryogenics Infrastructure at SURF. LBNF also works closely with SURF management to coordinate design and construction for the conventional facilities for the DUNE far detector. In addition, the design and construction of LBNF is supported by consultants and contractors. A full description of LBNF Project Management is contained in the LBNF/DUNE Project Management Plan [6].

The LBNF Project Director reports to the Fermilab Director and manages two Fermilab Divisions that support the Project: Far Site Facilities and Near Site Facilities. The Project Team consists of people from Fermilab, CERN, South Dakota Science and Technology Authority (who manages SURF), and Brookhaven National Laboratory (BNL). The team includes Project Office members and L2 and L3 Project leaders who manage the Project. The team is assembled by the Project Director. The Project team to Level 3 of the Work Breakdown Structure (WBS) is shown in Figure 2-1.

The LBNF Project Management Board (PMB) provides formal advice to the Project Director on matters of importance for the LBNF Project as a whole. Such matters include (but are not limited to) those that

- have significant technical, cost or schedule impact on the Project

- have impacts on more than one L2 Project
- affect the management systems for the Project
- have impacts on or result from impact from other Projects on which the LBNF is dependent
- result from external reviews or reviews called by the PD

The Management Board serves as the

- LBNF Change Control Board, as described in the Configuration Management Plan [7].
- Risk Management Board, as described in the Fermilab Risk Management Procedure for Projects [8].

The PMB comprises the members of the Project Office as well as L2 and L3 Project Managers.
LBNF coordinates with DUNE through regular technical team interactions between the two Projects as well as more formally through the Experiment-Facility Interface Group, where major decisions regarding interfaces and items affecting both Projects are made. In addition, the Projects use an integrated and coordinated project resource-loaded schedule and use a common configuration management system.

## 2.2 Work Breakdown Structure

WBS is a means of organizing the work scope of the Project. A WBS decomposes the Project's tasks and deliverables into smaller, manageable components. The LBNF WBS and Dictionary is maintained by the Project as a separate document under change control that defines the scope of the work.

At the time of CD-1-Refresh, the LBNF WBS is in transition. Both the current and the post CD-1-R WBS is shown in Figure 2-2 to demonstrate how the scope will map from one WBS to the other.

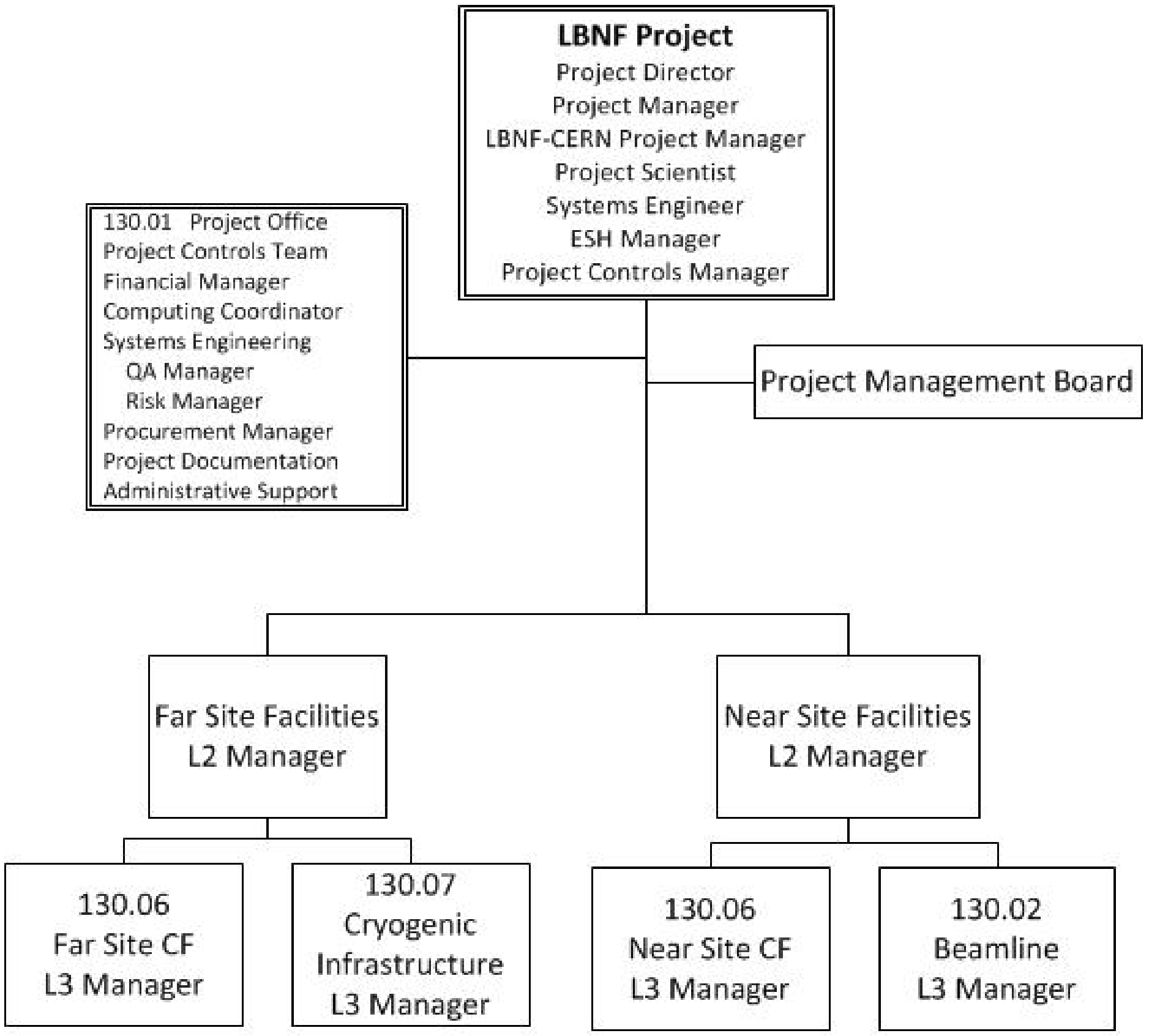

Figure 2-1 LBNF Project Management Organization to WBS L3

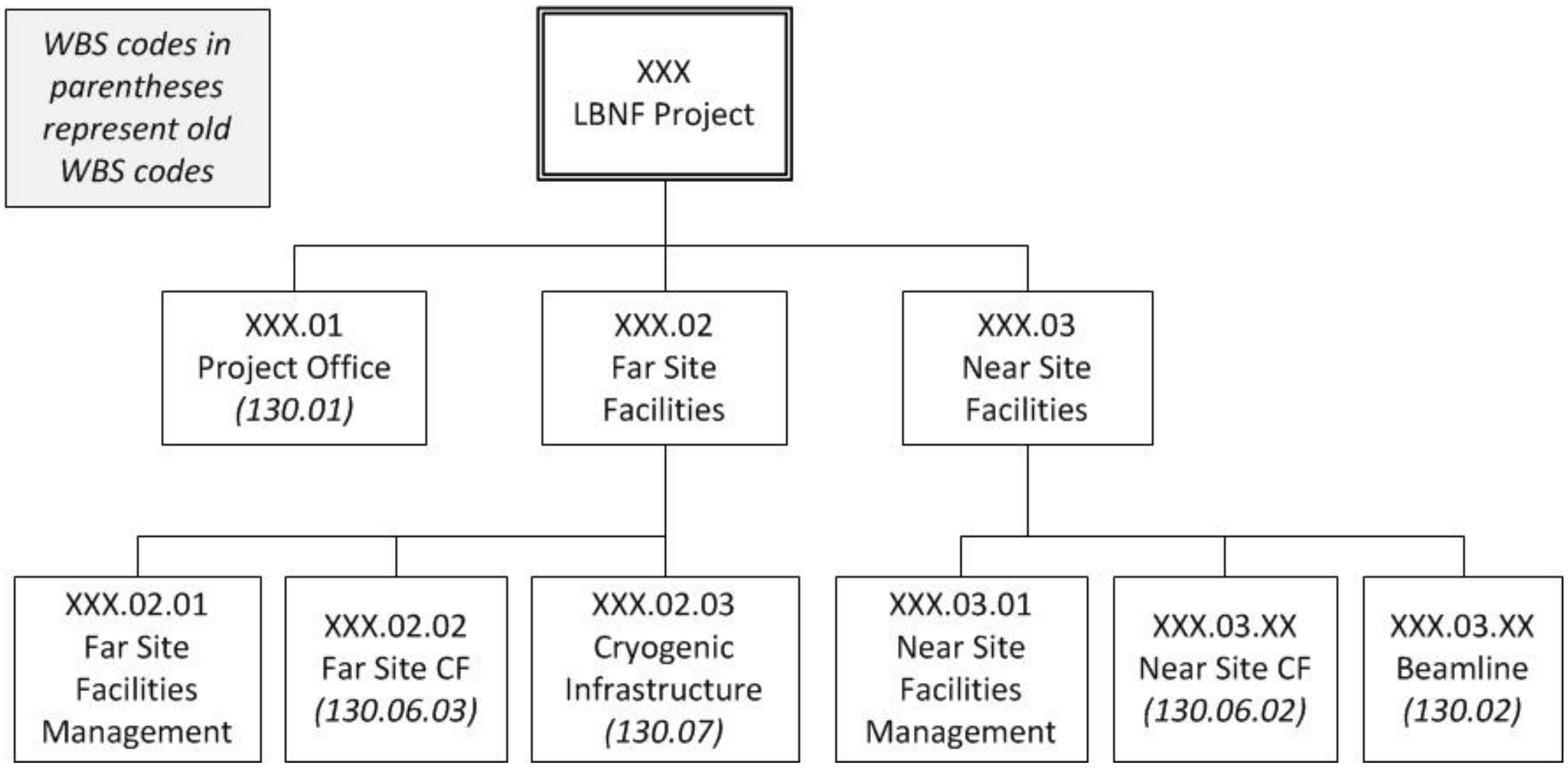

Figure 2-2: LBNF Work Breakdown Structure to WBS Level 3

# 3 Far Site Facilities: Far Site Conventional Facilities

## 3.1 Overview

This chapter presents the scope and necessary steps required to develop the LBNF Far Site Conventional Facilities (FSCF) at the Sanford Underground Research Facility at Lead, South Dakota, the far site location identified for the LBNF project. The key element of the FSCF is the underground space required to install and support the operations of the DUNE far detector. An overview of the 4850L at SURF where the underground facilities will be developed is shown in Figure 3-1.

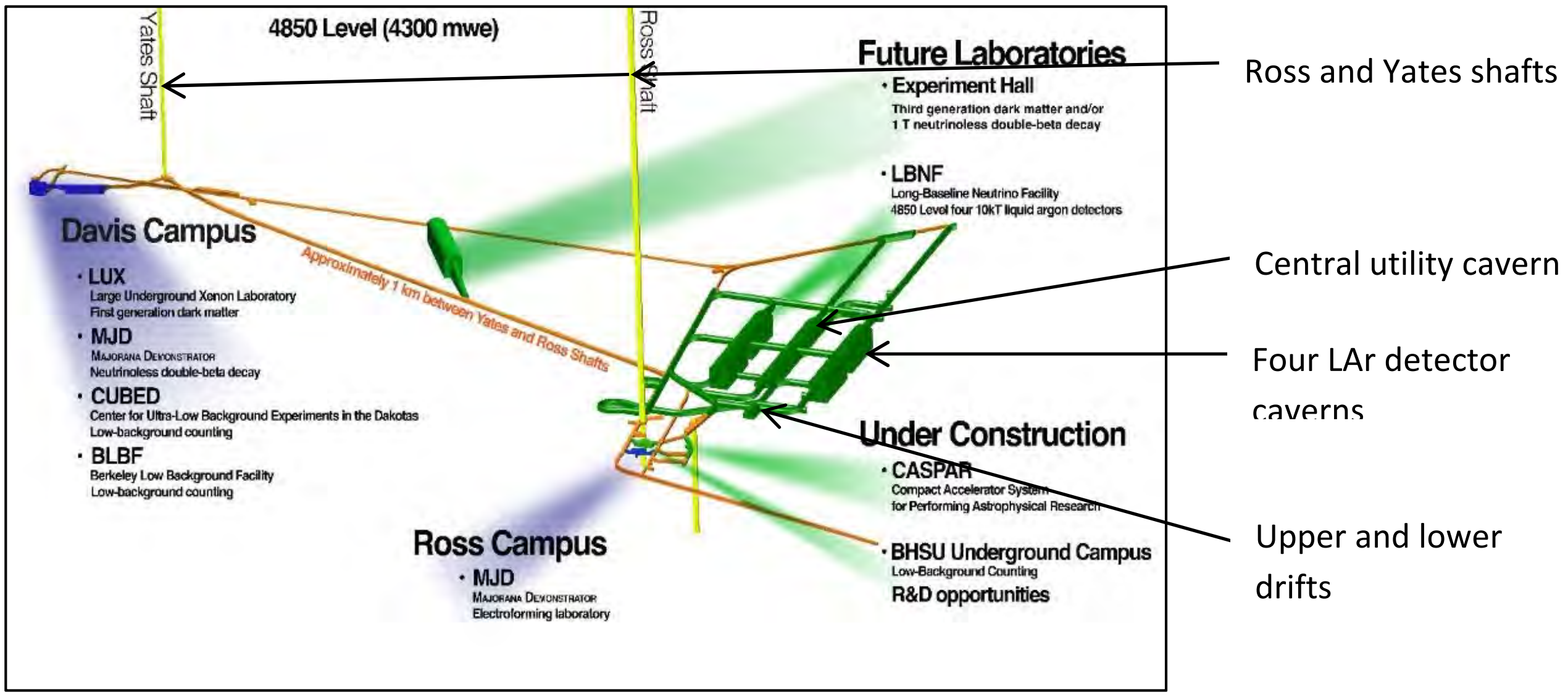

Figure 3-1: Far Site: Main components at the 4850 level (underground)

While the SURF site meets many requirements from the geological, scientific and engineering standpoint, significant work is required to provide adequate space and infrastructure support needed for the experiment's installation and operation. The present and future state of the site, evaluation and assessments of the facilities and the associated provisioning of infrastructure such as power, water, plumbing, ventilation etc., safety measures and planned steps to develop the surface and underground structures are described at a high level.

For more information on FSCF, refer to *Annex 3C: CF at the Far Site* [9]

This chapter contains the following high level topics:

- Existing Site Conditions
- Evaluation of Geology and Existing Excavations
- Surface
  - ✓ Existing Surface Facility
  - ✓ Surface Buildings
- Underground Excavation
- Underground Infrastructure

### 3.1.1 Surface Level Structures

- Cryogenics Compressor Building: Refer to Figure 3-8.
- Control Room: Refer to Figure 3-10.

### 3.1.2 Underground: Main Components

- Four detector pits within two main caverns at the 4850L
- A Central Utility Cavern to house cryogenics system supporting the detector operation and conventional facilities infrastructure (electrical power equipment, air-handling units, etc.)
- Drifts and ramps required for access, egress, and excavation

See Figure 3-11 and Figure 3-12.

## 3.2 Existing Site Conditions

The SDSTA currently operates and maintains the Sanford Underground Research Facilityat Homestake in Lead, South Dakota. The Sanford Laboratory property comprises 186 acres on the surface and 7,700 acres underground. The Sanford Laboratory Surface Campus includes approximately 253,000 gross square feet (gsf) of existing structures. Using a combination of private funds through T. Denny Sanford, South Dakota Legislature-appropriated funding, and a federal Department of Housing and Urban Development (HUD) Grant, the SDSTA has made significant progress in stabilizing and rehabilitating the Sanford Laboratory facility to provide for safe access and prepare the site for new laboratory construction. These efforts have included dewatering of the underground facility and mitigating and reducing risks independent of the former Deep Underground Science and Engineering Laboratory (DUSEL) efforts and funding.

Figure 3-2 shows Sanford Laboratory's location within the region as a part of the northern Black Hills of South Dakota. Figure 3-3 outlines the Sanford Laboratory site in relationship to the city of Lead, South Dakota, and points out various significant features of Lead including the surrounding property that still remains under the ownership of Barrick Gold Corporation.

## 3.2.1 Existing Site Conditions Evaluation

The existing facility conditions were assessed as part of the DUSEL Preliminary Design and documented in the DUSEL PDR, Section 5.2.4, [10] which is excerpted below. The portions of DUSEL's assessment included here have been edited to reflect current activities and to reference only that portion of the assessment that are pertinent to the LBNF Project. References to the DUSEL Project are from that time, and are now considered historic.

Site and facility assessments were performed during DUSEL's Preliminary Design phase by HDR to evaluate the condition of existing facilities and structures on the Yates, and Ross Campuses. The assessments reviewed the condition of buildings proposed for continuing present use, new use, or potential demolition. Building assessments were performed in the categories of architectural, structural, mechanical/electrical/plumbing (MEP), civil, environmental, and historic. Site assessments looked at the categories that included civil, landscape, environmental, and historic. Facility-wide utilities such as electrical, steam distribution lines, water, and sewer systems were also assessed. The assessment evaluation was completed in three phases. The detailed reports are included in the appendices of the DUSEL PDR as noted and are titled:

- Phase I Report, Site Assessment for Surface Facilities and Campus Infrastructure to Support Laboratory Construction and Operations (DUSEL PDR Appendix 5.E)
- Phase II Site and Surface Facility Assessment Project Report (DUSEL PDR Appendix 5.F)
- Phase II Roof Framing Assessment (DUSEL PDR Appendix 5.G)
- The site and facility assessments outlined above were performed during DUSEL's Preliminary Design as listed above and includes a review of the following:
- Buildings proposed for reuse were evaluated for preliminary architectural and full structural, environmental, and historic assessments.
- Buildings proposed for demolition were evaluated for preliminary historic assessments.
- Preliminary MEP assessments were performed on the Ross Substation, #5 Shaft fan, Oro Hondo fan, Oro Hondo substation, and general site utilities for the Ross, Yates, and Ellison Campuses.

- The waste water treatment plant (WWTP) received preliminary architectural and structural assessments and a full MEP assessment.
- Preliminary civil assessments of the Kirk Portal site and Kirk to Ross access road were also completed.

## 3.2.2 Evaluation of Geology and Existing Excavations

LBNF Far Site facilities are planned to be constructed at Sanford Laboratory which is being developed within the footprint of the former Homestake Gold Mine, located in Lead, South Dakota. The accessible underground mine workings are extensive. Over the life of the former gold mine over 360 miles of drifts (tunnels) were mined and shafts and winzes sunk to gain access to depths in excess of 8,000 feet. A number of underground workings are being refurbished by Sanford Laboratory and new experiments are being developed at the 4850L, the same level as proposed for LBNF facilities. Geotechnical investigations and initial geotechnical analyses were completed for the DUSEL Preliminary Design [10] and are described in detail in the DUSEL PDR. Additional geotechnical investigation and analysis was performed in 2014 specific to the LBNF project. Below are summaries these two effort, including work completed for DUSEL that is applicable to LBNF as excerpted from the DUSEL Preliminary Design Report, Chapter 5.3. Much of the work completed for the alternative detector technology considered during DUSEL [water Cherenkov detector (WCD)] is also applicable to the current design at the 4850L.

### 3.2.2.1 Geologic Setting

The Sanford Laboratory is sited within a metamorphic complex containing the Poorman, Homestake, Ellison and Northwestern Formations (oldest to youngest), which are sedimentary and volcanic in origin. An amphibolite unit (Yates Member) is present within the lower known portions of the Poorman Formation. While the Yates Member is the preferred host rock for the LBNF excavations at 4850L, the LBNF cavity has been located in the Poorman formation to isolate it from the remainder of the level. The layout adopted on the 4850L attempts to optimize the needs for ventilation isolation, access control, and orientation relative to the beam line.

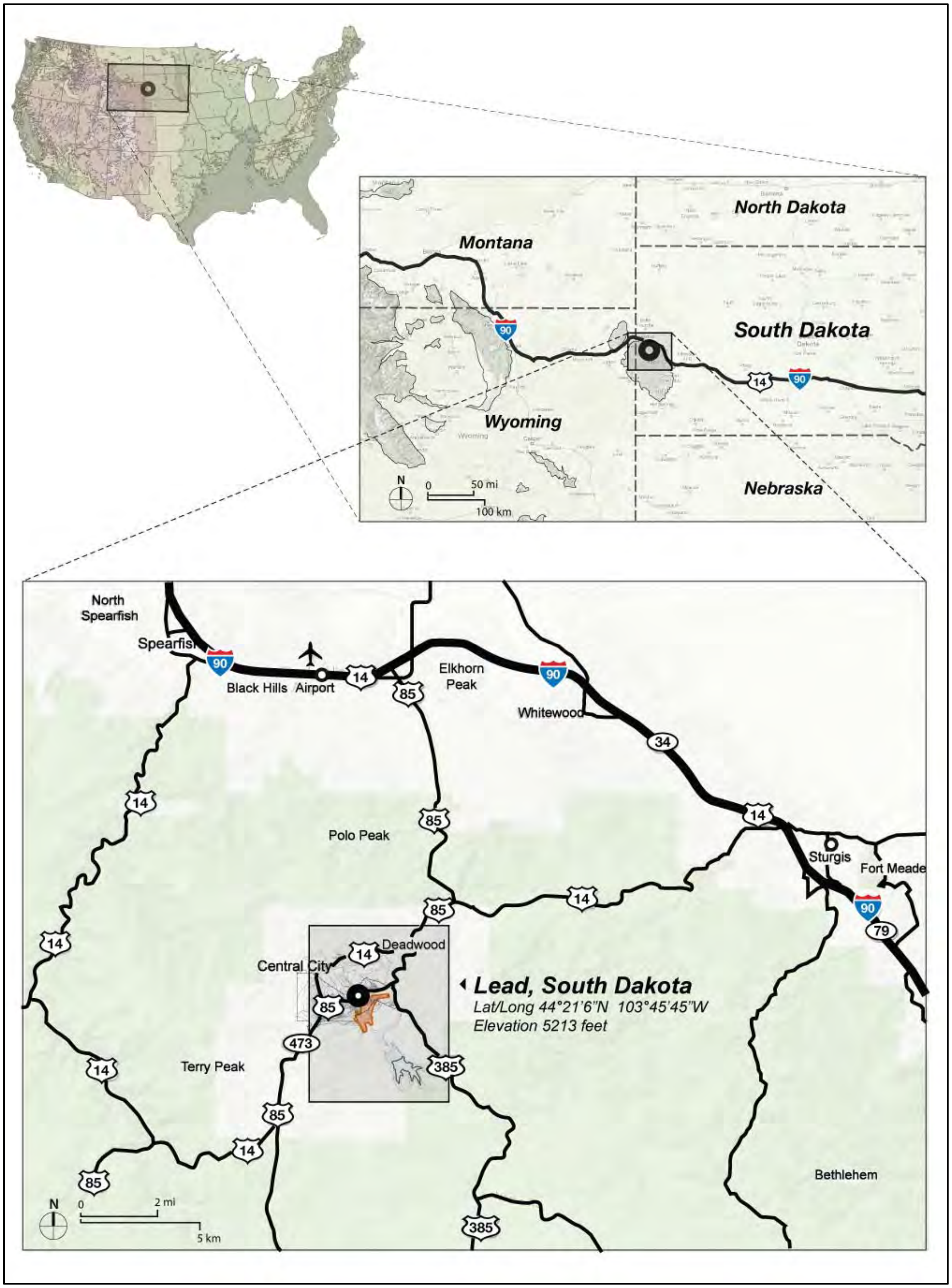

Figure 3-2: Regional context showing the city of Lead, South Dakota. (Dangermond Keane Architecture, Courtesy Sanford Laboratory)

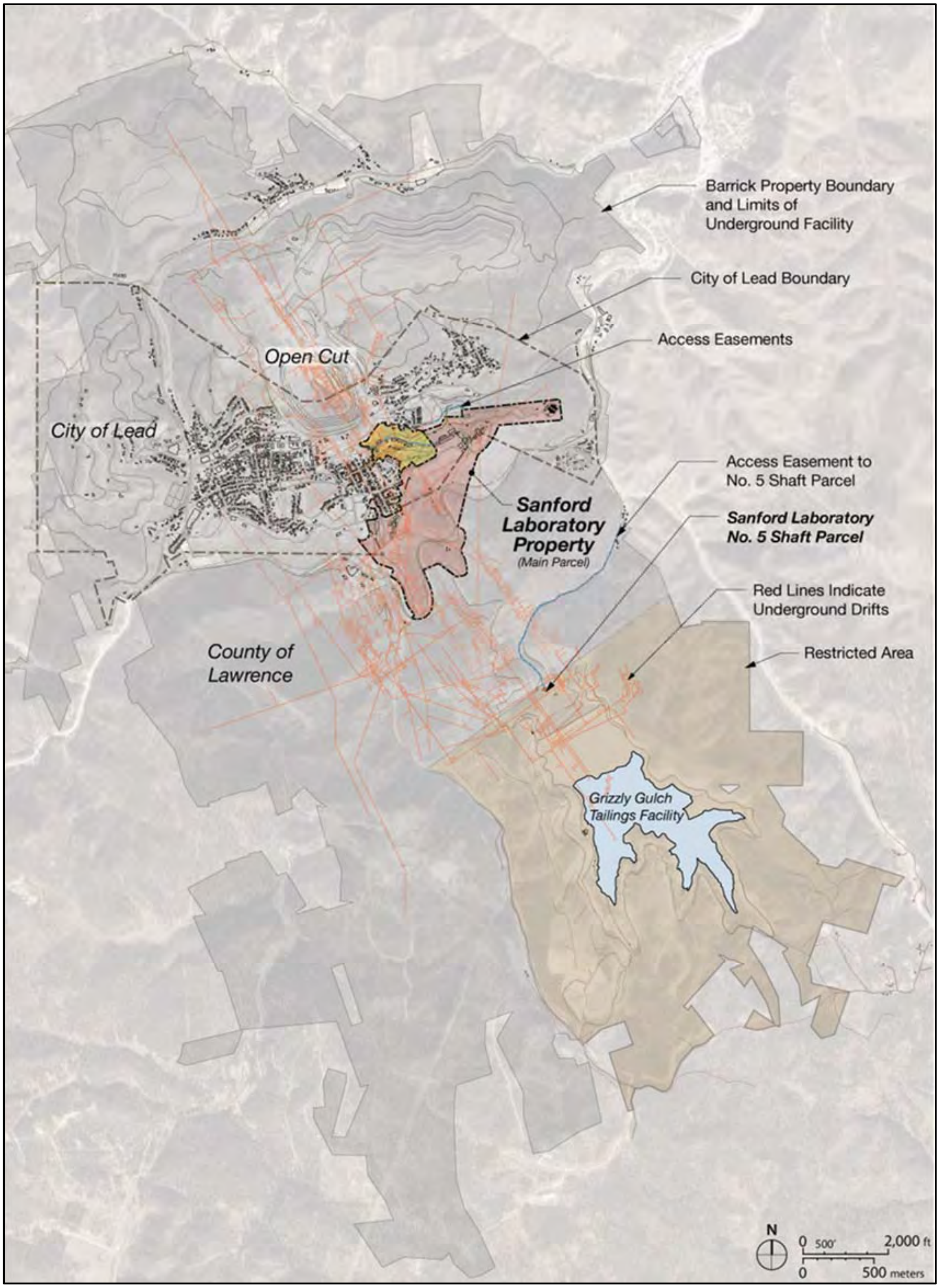

Figure 3-3: Sanford Laboratory Complex shown in the context of the city of Lead, South Dakota, and the property remaining under ownership of Barrick. Area shown in yellow is a potential future expansion of the SDSTA property. [Dangermond Keane Architecture, Courtesy of Sanford Laboratory]

### 3.2.2.2 Rock Mass Characteristics: LBNF

Following a similar strategy as DUSEL, the LBNF project initiated a second geotechnical program in 2013 to evaluate the specific location under consideration and evaluate its appropriateness for the proposed design. This was undertaken in two phases. The first phase was a mapping of the existing spaces surrounding the proposed rock mass using both visual techniques and laser scanning to understand the rock mass and inform the scope of the second phase. The second phase included drilling of four HQ (2.5” diameter) core holes ranging in length from 477 to 801 feet as well as two 6” diameter core holes ~30’ each. The smaller diameter cores were then evaluated for the following characteristics:

- core recovery percent
- rock quality designation (RQD) percent
- rock type, including color, texture, degree of weathering, and strength
- mineralogy and presence of magnetic sulfides
- character of discontinuities, joint spacing, orientation, aperture
- roughness, alteration, and infill (if applicable)

Representative samples were selected from the overall core to test material strength and chemical characteristics. The geotechnical site investigations area on the 4850L, showing boreholes is presented in Figure 3-4.

The holes from which the smaller diameter core was removed were studied in several ways. An absolute survey was conducted to allow the core holes to be plotted relative to cavern designs. An optical televiewer was passed through each small hole to visualize the rock mass. This technique allows visualization of foliation, joint openings, healed joints, and geological contact between rock types. An acoustical imaging device was also used in one hole to complement the optical information. The permeability of the rock was tested by pressurizing the small holes at various intervals to determine if joints allowed for the flow of water outside of the holes (hydraulic conductivity). In all cases, the hydraulic conductivity was well below what can be accomplished using manmade techniques such as grouting. Two of the small holes were plugged and instrumented to determine if water would flow into the holes over time. This test found very low flow rates (.0013-.0087 gpm). Ongoing evaluation of pressure build in these holes was inconclusive, as blast induce fracturing near the existing drifts allow the holes to depressurize outside of the test instruments.

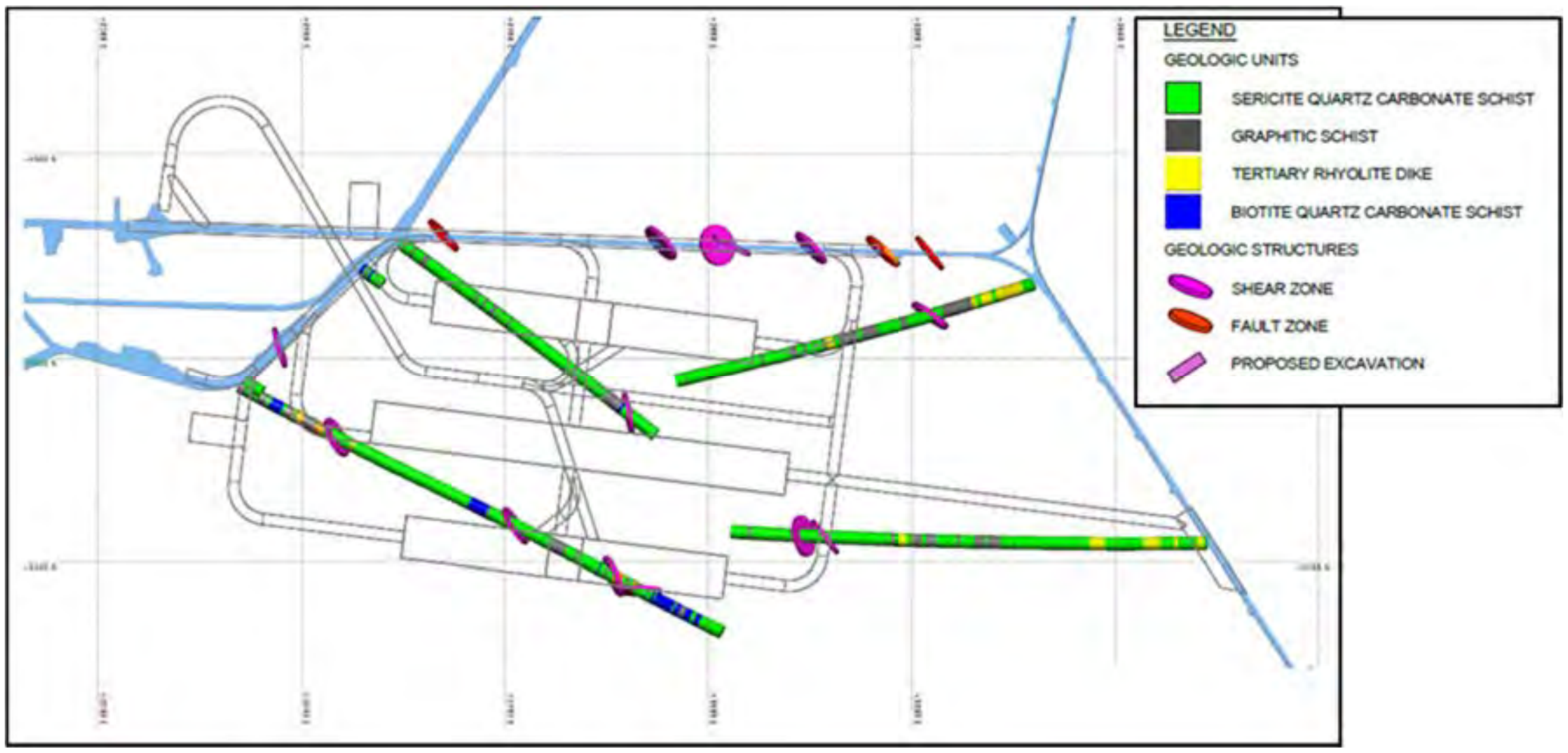

Figure 3-4: LBNF core locations and geological features

The larger (6”) diameter cores and holes were used for strength and stress testing. In-Situ stress was tested by drilling a smaller diameter hole first, then gluing a strain gage at 30-36 feet within the depth. As the larger diameter core was removed, this strain gage recorded the relaxation of the rock. The removed core was re-drilled to provide smaller diameter samples at specific orientations for strength testing, as the strength of the material varies based on applied force direction relative to the foliation of the rock. These samples were also tested for time dependent movement.

LBNF followed a review approach for the analysis performed by Arup by enlisting industry leaders as part of a Neutrino Cavity Advisory Board (NCAB). This board reviewed the philosophy and results of the geotechnical investigation program as well as the preliminary excavation design. Their conclusions indicated that no additional drilling would be required to provide design information for the project and the overall design approach was appropriate. They provided many recommendations that will benefit the advancement of design.

For further details, see Arup’s Geotechnical Interpretive Report [11].

### 3.2.2.3 Geologic Conclusions

The recovery of rock cores, plus geologic mapping, was performed to determine if discontinuities in the rock mass exist that would cause difficulties in the construction and maintenance of planned excavations. In general, the proposed locations of the excavations do not appear to be complicated by geologic structures that cause undue difficulties for construction. This information, along with measurement of in situ stresses, allowed initial numerical modeling of the stresses associated with the anticipated excavations. A sample of some of the modelling done is provided in Figure 3-5.

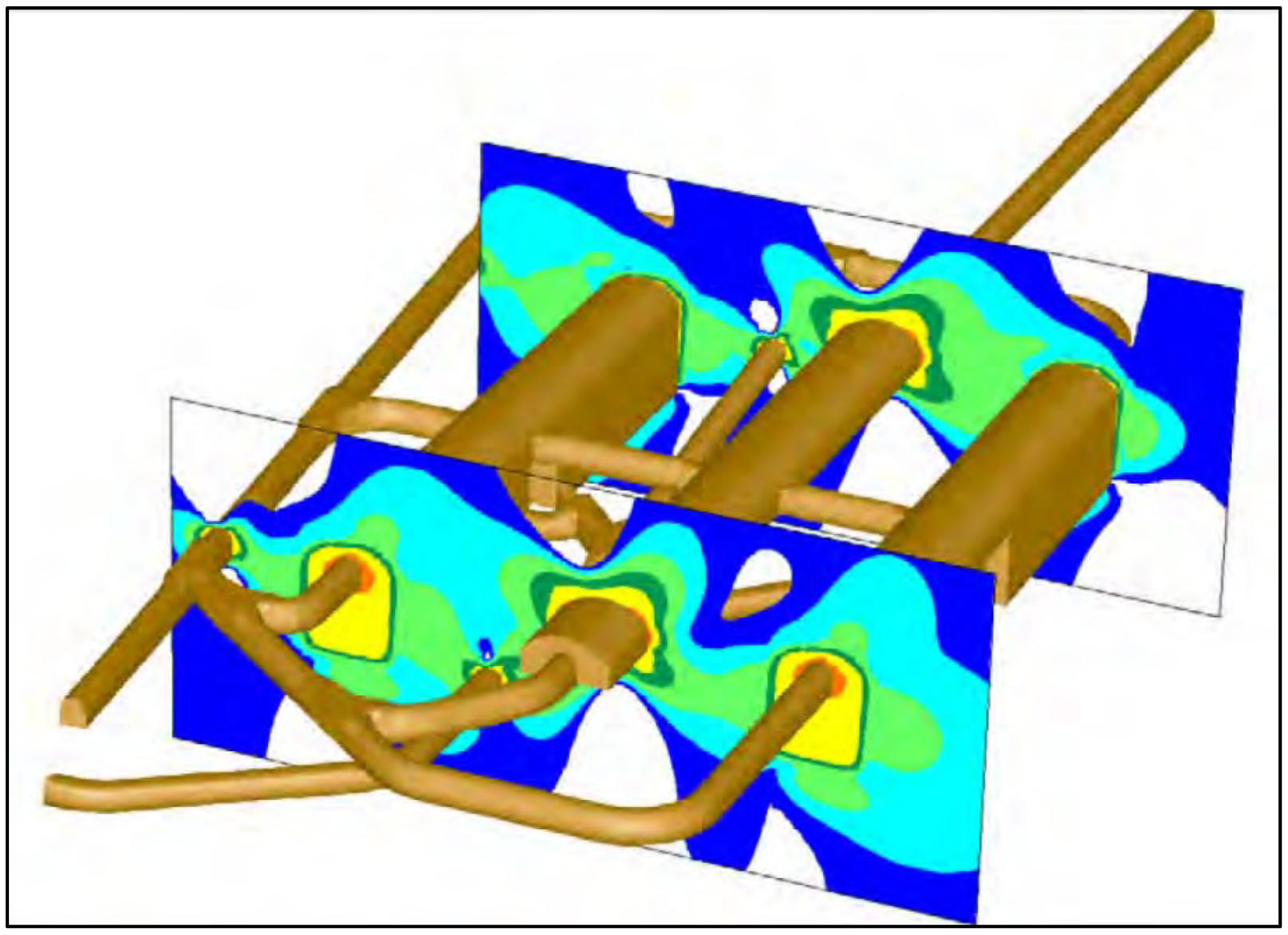

Figure 3-5: Contour of stress safety factor Indicating influences between caverns

The overall analysis of the work indicates that the rock in the proposed location of the LBNF caverns is of good quality for the purposes of the LBNF Project, that preliminary numerical modeling shows that a large cavern of the size envisioned can be constructed, and that a workable excavation design has been developed.

## 3.3 Surface Facility

### 3.3.1 Existing Surface Facility

The Sanford Laboratory property of 186 acres consists of steep terrain and man-made cuts dating from its mining history. There are approximately 50 buildings and associated site infrastructure in various states of repair. A select few of these buildings at the Ross Complex and the main utilities are needed by the LBNF experiment and will be upgraded and rehabilitated as necessary. A layout of the overall Sanford Laboratory architectural site plan for the LBNF Project is found in Figure 3-6.

The Ross Complex will house the facility construction operations, command and control center for the experiment and facility, new cryogenics compressor building, as well as continue to house the Sanford Laboratory maintenance and operations functions. Layout of surface facilities in the vicinity of the Ross Shaft is shown in Figure 3-7.

## 3.3.2 Surface Buildings

Surface facilities utilized for the LBNF include those necessary for safe access and egress to the underground through the Ross Shaft, as well as spaces for temporary offices (by Sanford Laboratory). Existing buildings necessary for LBNF will be rehabilitated to code-compliance and to provide for the needs of the experiment. The only new building will be to provide space for compressors use to transfer cryogens from new receiving tank on surface to the detectors underground. The existing Ross Dry building will be modified to provide space for a surface control room and associated equipment. Much of the text below is excerpted from the 30% Preliminary Design Report [12] provided by Arup, USA.

A new building is planned to provide space for equipment to allow conversion of liquid argon and liquid nitrogen to gaseous form and compression of these gasses for delivery through the shaft to the underground where they are returned to liquid form as described later in this CDR in Chapter 4. The location of this building was selected based on proximity to the shaft and truck accessibility, as thousands of truckloads of argon are required to fill the detectors underground.

In addition to housing nitrogen compressors inside the building, concrete slabs are provided around the building to allow for installation of argon and nitrogen receiving dewars for truck unloading, vaporizers to boil the liquids into gas, and electrical transformer to supply power to the (4) 1,500 Hp compressors, a standby generator, and cooling towers to reject heat generated through compression. All equipment except the cooling towers is provided by the Cryogenics Infrastructure Project. The architectural layout of this building and surrounding equipment is provided in Figure 3-8.

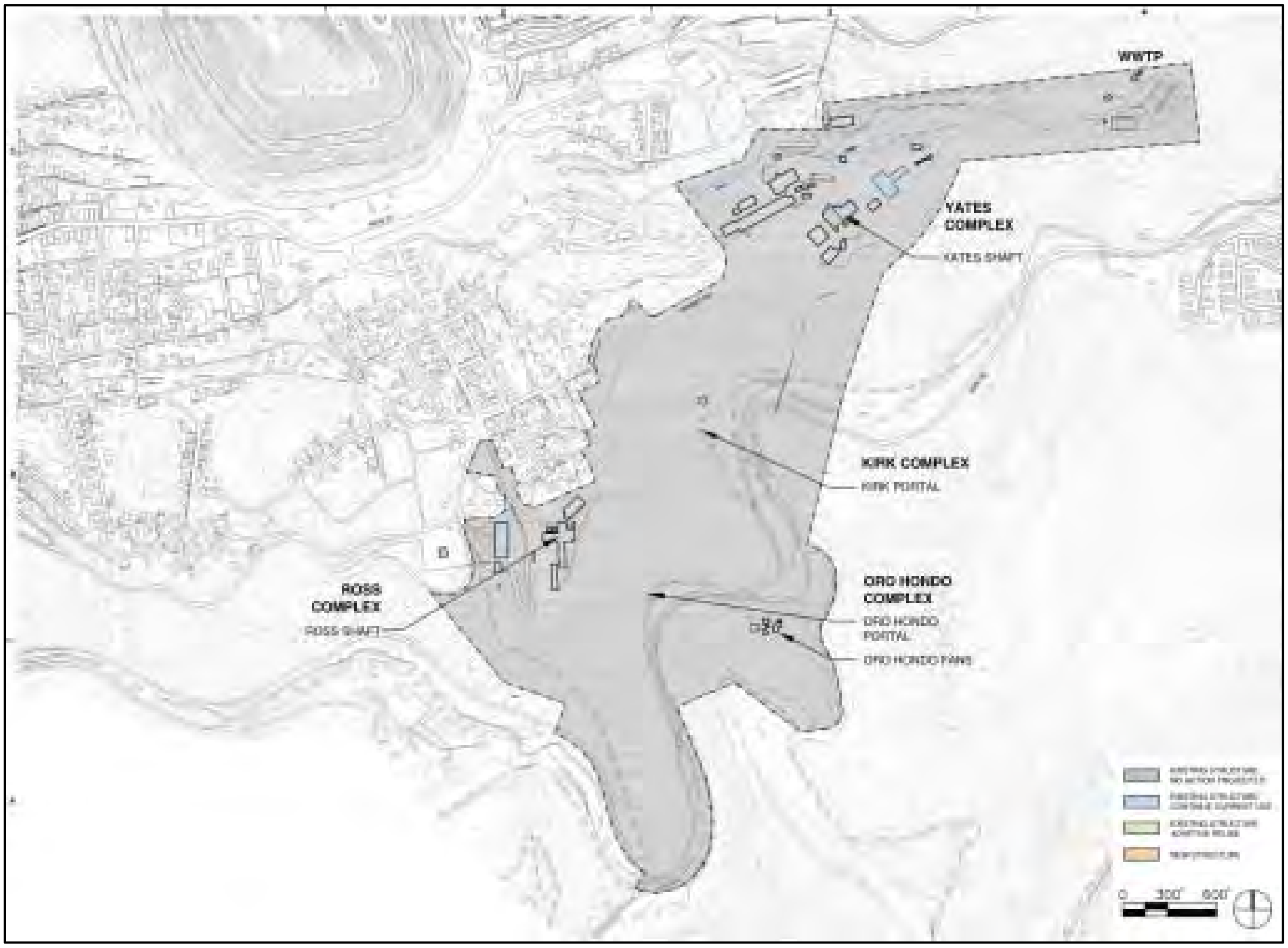

Figure 3-6: Architectural site plan (HDR)

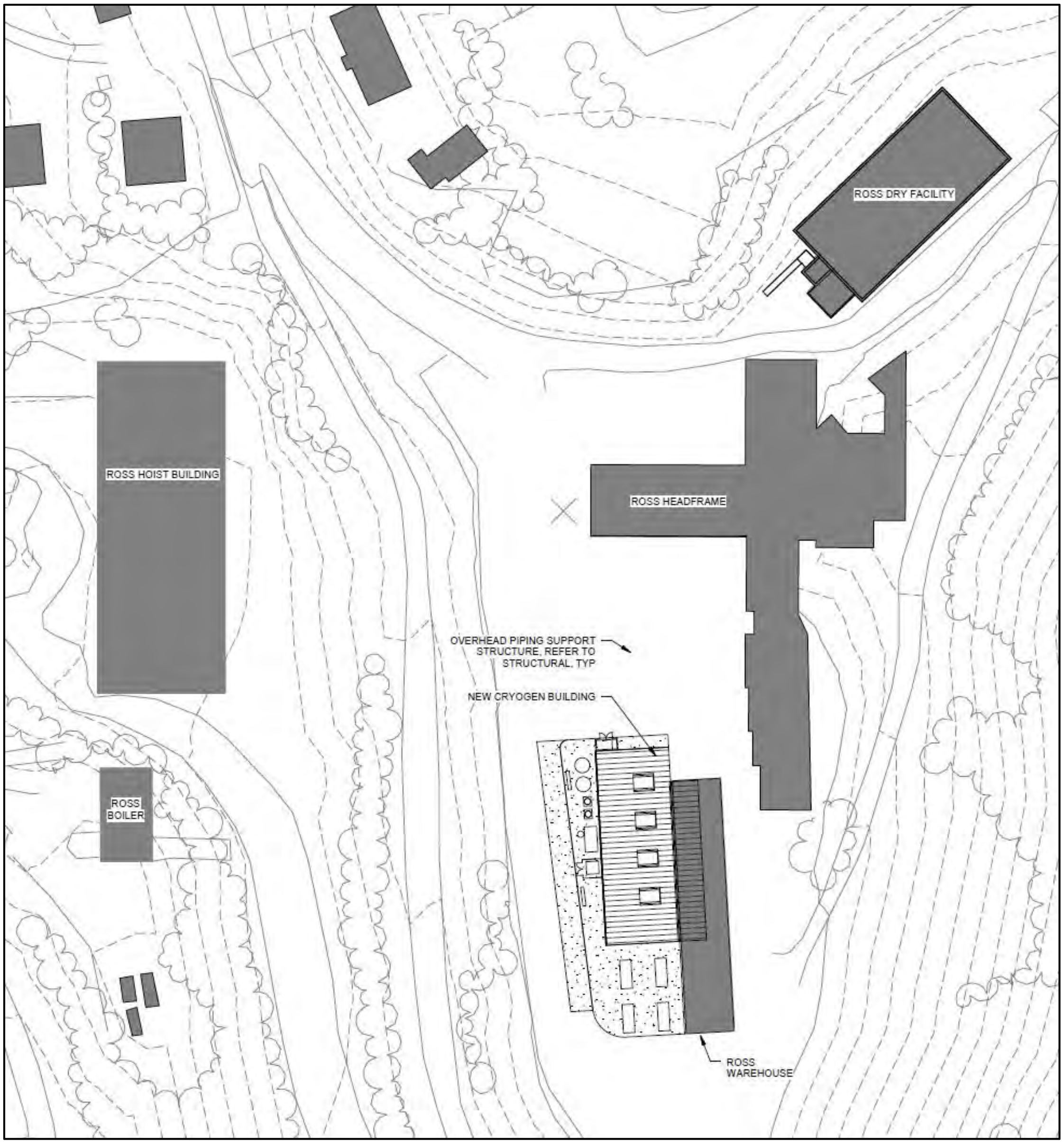

Figure 3-7: Ross Complex architectural site plan (Arup)

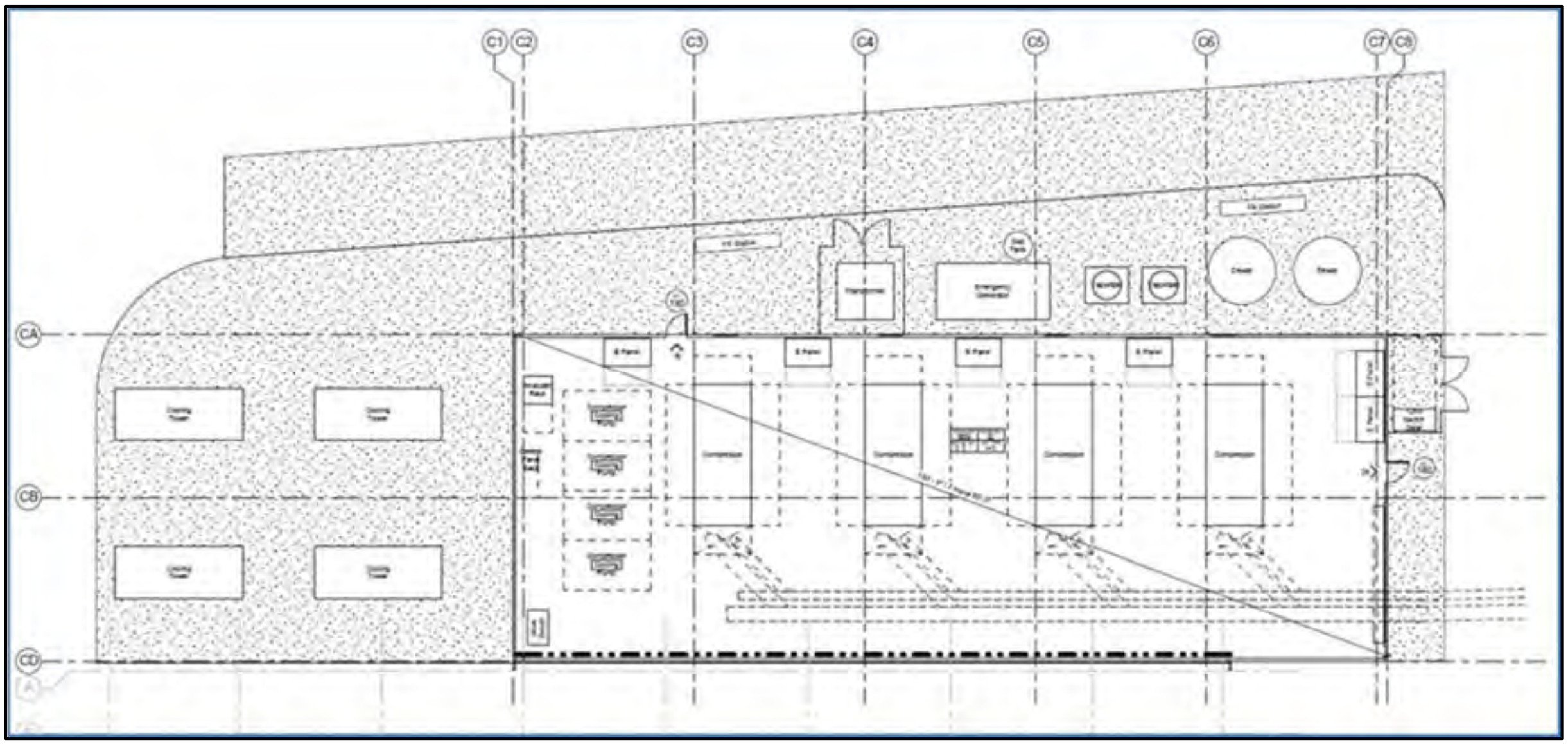

Figure 3-8: Architectural layout of LBNF Cryogenics Compressor Building

### 3.3.2.1 Ross Dry

The Ross Dry building is in use by the Sanford Laboratory to provide office and meeting space in addition to men’s and women’s dry facilities. A portion of an existing meeting space within this building will be modified to allow the installation of a control room for both facility and experiment control.

The exterior of the Ross Dry is shown in Figure 3-9. The location of the new command and control center is shown in Figure 3-10.

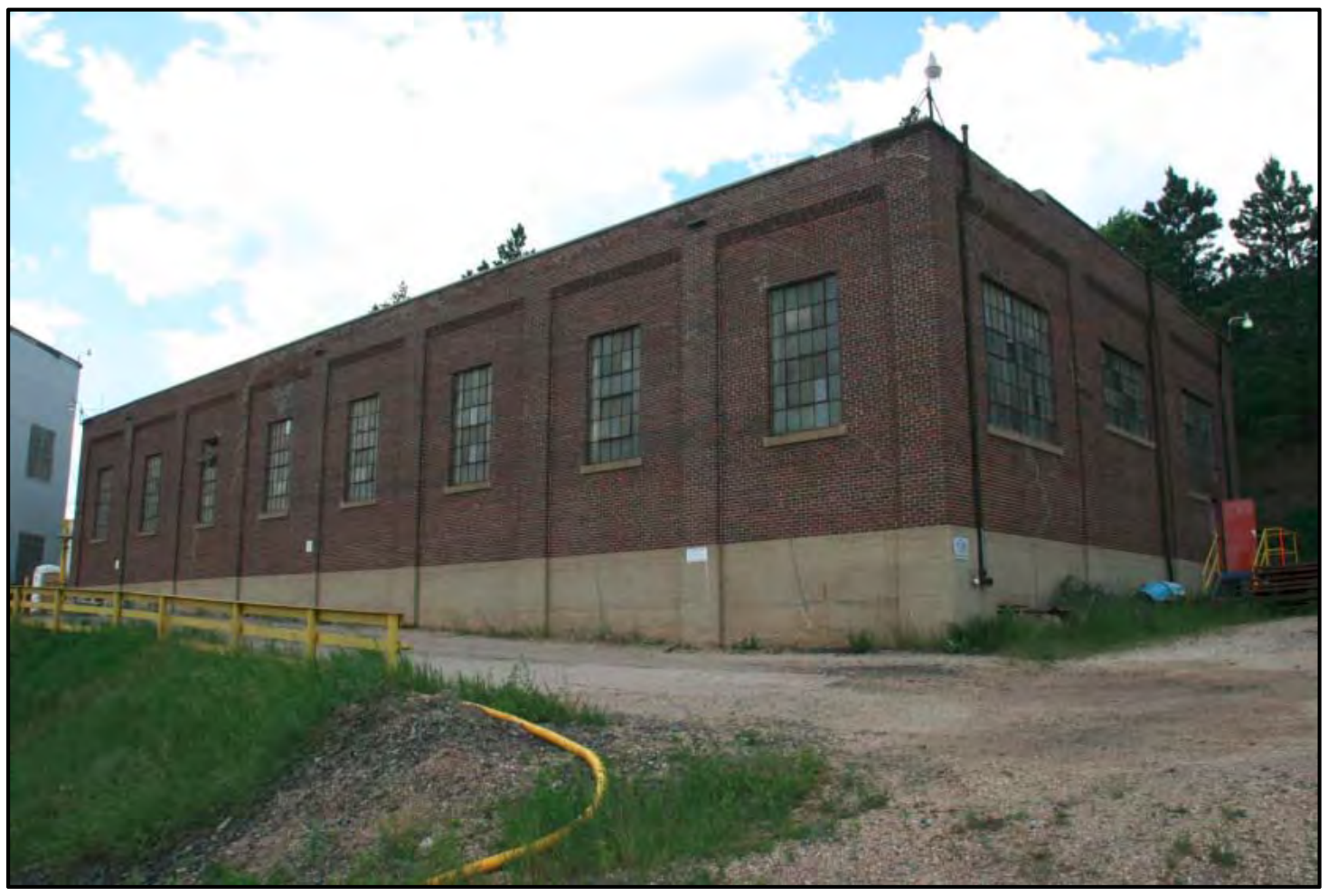

Figure 3-9: Photo of Ross Dry exterior (HDR)

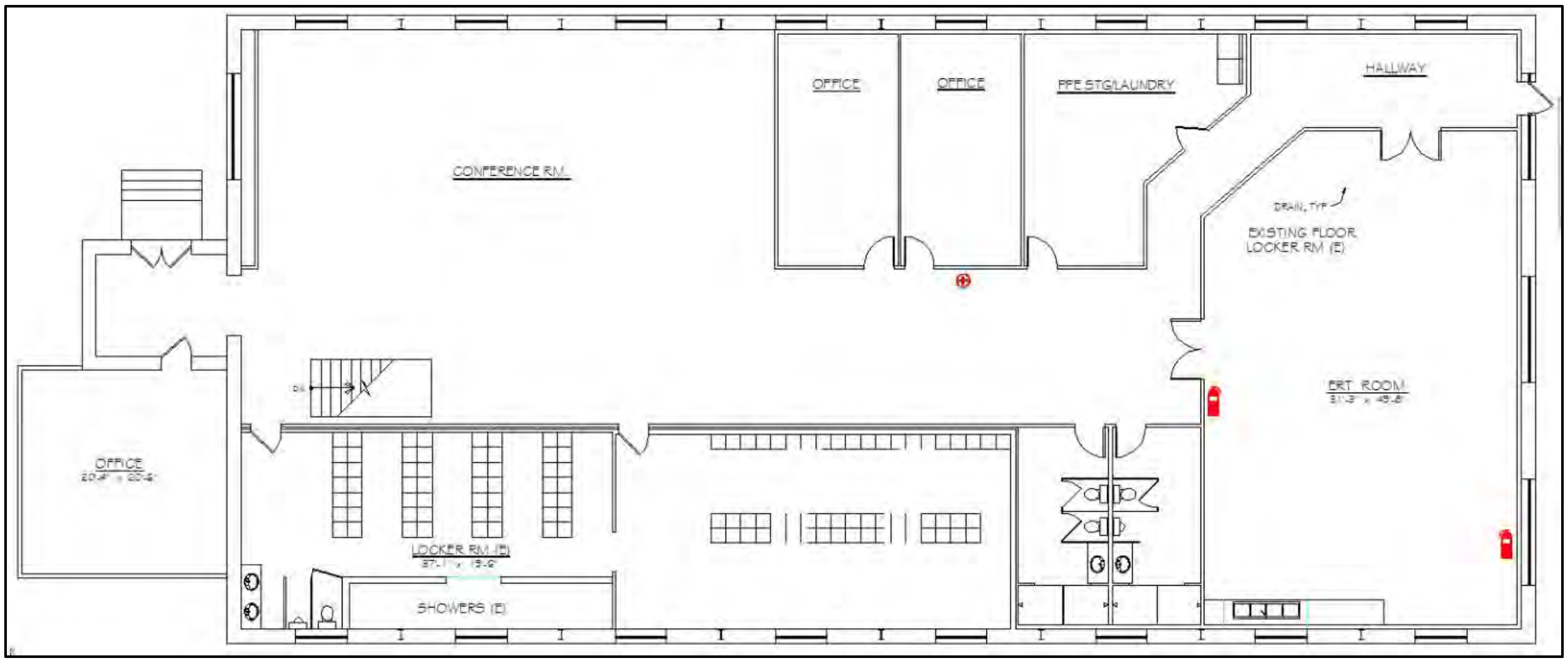

Figure 3-10: Location of new Command and Control Center (Sanford Lab)

### 3.3.2.2 Ross Headframe and Hoist Buildings

The headframe and hoist buildings at the Ross Campus provide services for LBNF use. The Ross Headframe Building will be the main entry point for construction activities as well as the ongoing operations and maintenance functions. Gas pipe from the LBNF Cryogenics Compressor Building will pass through this building to get to the shaft.

### 3.3.2.3 Ross Crusher Building

The existing Ross Crusher Building is a high bay space that contains rock crushing equipment that will be used for construction operations. The exterior of the building will be repaired to create a warm, usable shell. The upgrade of the existing crusher equipment is part of the waste rock handling work scope and not part of the building rehabilitation.

## 3.3.3 New Surface Infrastructure

Surface infrastructure includes surface structures such as retaining walls and parking lots, as well as utilities to service both buildings and underground areas. Existing infrastructure requires both rehabilitation as well as upgrading to meet code requirements and LBNF needs. The experiment needs were documented in the requirements found in LBNF Requirements Document [13] and combined with facility needs for the design detailed in the Arup 30% Preliminary Design Report.

No new roads or parking lots are required for LBNF at the Sanford Laboratory. The Ross Complex site will require minor demolition of power lines and a fire hydrant that are no longer used to provide adequate accessibility for truck traffic to the new Cryogenics Compressor Building. An existing space will be designated for handicap parking adjacent to the Ross Dry Building. Additional road work is required for truck transportation of waste rock, as described in the waste rock handling section.

# 3.4 Underground Excavation

The main excavated spaces necessary to support the LBNF experiment are a combination of excavations required for the experiment and those required for constructability. Experimental spaces on the 4850L include the detector cavities, several drifts for access and utility routing, and the Central Utility Cavern. Spaces identified as likely necessary for the excavation subcontractor include mucking drifts connected to the Ross Shaft to enable waste rock handling and equipment assembly shops to provide space to assemble and maintain excavation equipment underground. In addition, a spray chamber is provided for heat rejection from the chilled water system. All spaces are identified on the 30% Preliminary Design excavation drawings produced by Arup [12]. The spaces are shown below in Figure 3-11.

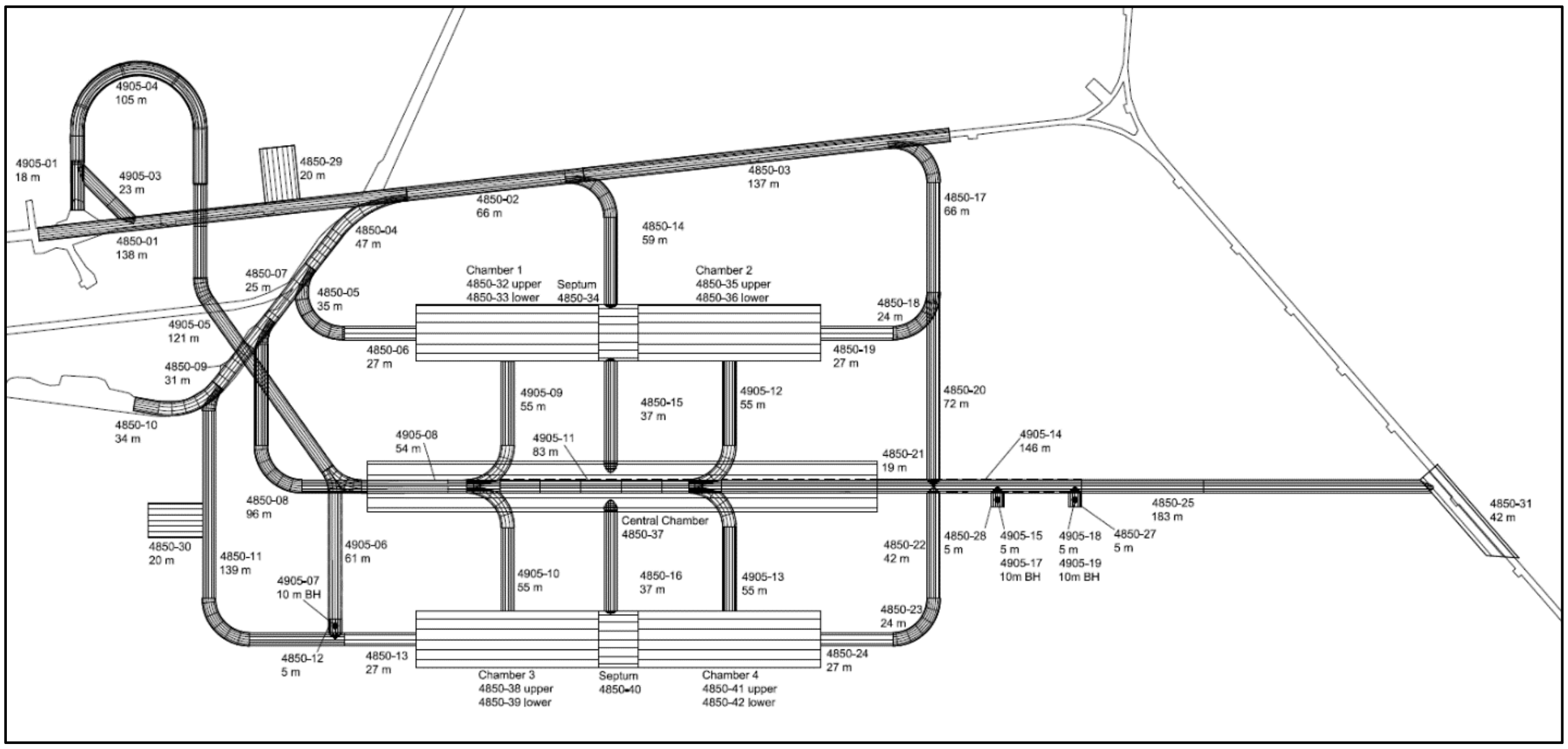

Figure 3-11: Spaces required for LBNF at 4850L (Sanford Lab)

## 3.4.1 LBNF Cavities

### 3.4.1.1 Detector Cavities

The required experimental spaces were defined through interaction with the DUNE design team and are documented in *LBNF Requirements Document* [14]. The size and depth of the LBNF cavities were prescribed to suit the scientific needs of the experiment. The overall main cavern sizes are shown graphically in Figure 3-12. The DUNE experiment will be housed in four detector pits within two main caverns at the 4850L. Siting deep underground is required to shield from cosmic rays, as detailed in *Report on the Depth Requirements for a Massive Detector at Homestake* [15]. The 4850L is deeper than what is absolutely required, but is used because of existing access at this level.

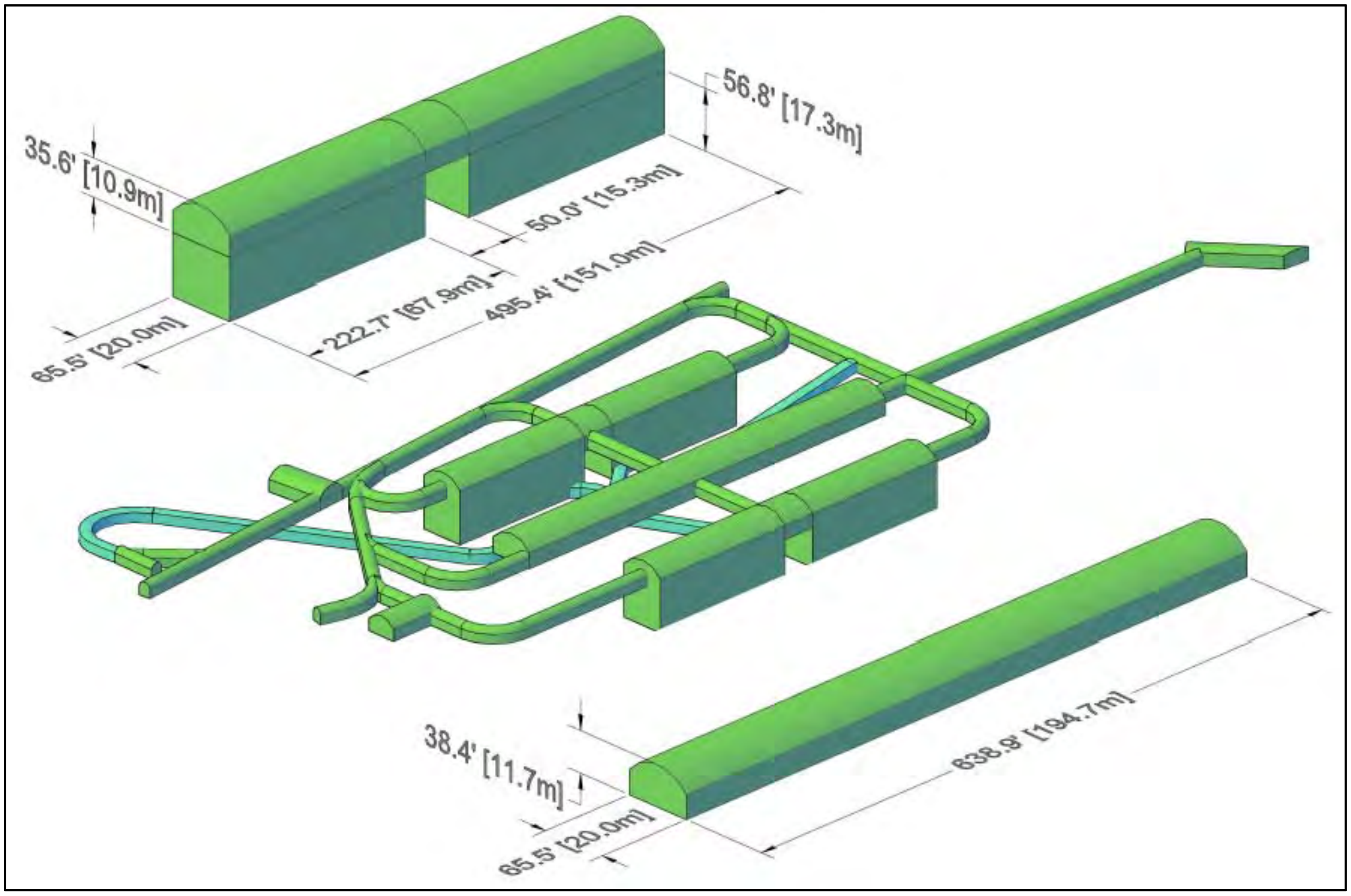

Figure 3-12: Dimensions of the main LBNF cavern excavations (final dimensions will be slightly smaller). (Sanford Lab)

The limits on size for the detector are determined by rock strength and the limits on the ability to produce large dimension anode and cathode plane arrays. Space occupied by the vessel liner, and an intentional exclusion zone reduce the fiducial volume of the detector below the volume of the excavation. Current assessment of rock quality indicates that a cavity of this size is reasonable with the rock quality assumed for this formation.

Preliminary modeling of the proposed excavations included 2D and 3D numerical modeling. The intact rock strength and joint strength had the greatest impact according to the 2D modeling, and 3D modeling confirmed that the complex geometry is possible.

The far detector cavity and drifts will be supported using galvanized rock bolts/cables, wire mesh, and shotcrete for a life of 30 years. The floor of the cavity is not anticipated to require support.

A groundwater drainage system will be placed behind the shotcrete in the arch and walls of the far detector cavity rock excavation. This drain system will collect groundwater (native) seepage and eliminate the potential for hydrostatic pressure build-up behind the shotcrete. Channels will be placed in the concrete invert to drain groundwater to the sump system.

#### 3.4.1.2 Structure and Cranes

The LBNF caverns require monorail cranes to facilitate the construction of the detector components. Rock bolts will be coordinated with the excavation contractor to provide anchorage to support these monorails.

### 3.4.2 LBNF Central Utility Cavern

LBNF requires spaces for cryogenics equipment outside of the detector caverns. These requirements have been combined with that for the conventional facilities utilities in an independent Central Utility Cavern. This area will house the experiment's cryogen system, electrical equipment to supply power for facility and experiment needs, sump pump access and controls, fire sprinkler room, air handling units (AHUs), chilled water system, and exhaust ducting. The centralized location minimizes overall utility distribution costs. Isolating the utilities from the experiment simplifies electrical ground isolation to avoid interference with sensitive detector electronics, and also provides the opportunity to optimize ventilation to control heat emanating from the equipment in the Central Utility Caverns.

### 3.4.3 Access/Egress Drifts

In order to accomosdate deliveries, the drift connections from the Ross Shaft to new excavations required for LBNF will be optimized to accommodate the maximum load size possible through the shaft plus the utilities required to service the facility. At the writing of this document, an assumed size of 5m wide by 6m tall is used for all access and egress drifts. All new excavations, or drifts enlarged for LBNF will be provided with a shotcrete wall (rib) and ceiling (back) and a concrete floor (sill).

### 3.4.4 Excavation Sequencing

A key goal of both LBNF and DUNE is to complete construction of one 10 kt detector as soon as possible. To facilitate this, the excavation will be sequenced to allow DUNE to begin installation of a cryostat in the first detector pit while excavation continues. A temporary wall will be built in the detector installation laydown space between detector pits to isolate one area from another. This wall must be of sturdy construction to withstand air shock waves associated with drill and blast type construction. Further evaluation of vibration limits and controls must be considered as the design advances to avoid damaging the cryostat during assembly.

In addition to controlling the impacts from blasting, logistical coordination is a key concern with a sequenced excavation allowing cryostat construction concurrent with excavation. Most experiment components will be delivered through the Yates Shaft, leaving the Ross shaft dedicated for excavation and other construction. The area between Governor's Corner and the experiment, however, will require interfacing between delivery of experiment equipment, delivery of ground support and explosives, and transportation of excavated material. The contractor will be tasked to determine whether "traffic control" at this intersection is performed with electronic signaling or personnel.

Most excavated material will travel through a mucking ramp starting at the base of each detector pit and ending at the waste dump near the Ross Shaft. This route is completely independent of all other traffic and includes a separate ventilation stream to keep diesel exhaust from other occupied spaces. During times when excavation is establishing the upper sections of the caverns and developing a means of dumping excavated material to this lower elevation, material will need to be transported at the 4850L. This not only introduces concerns with interference between experiment deliveries and rock haulage, but also means that both diesel exhaust and dust can impact the “clean” experiment pit where construction is occurring for the first cryostat. Dust will need to be controlled using water throughout all excavation, but during these times it will be more critical. Diesel fume controls will be further evaluated as the design progresses to determine the best means of controlling them.

Delivery of cryostat components to the individual pits can be accomplished in one of two ways. All materials are delivered through the shafts to the 4850L, which is ~15m above the base of the pits. During construction of the first cryostat, while excavation continues in the other areas, all materials will be delivered to the detector installation laydown area between the first and second detector pits. An overhead crane will be used to lower this material into the pits. This crane is required for installation of detector component within the cryostat, so is not additional equipment. Further schedule analysis will determine whether the construction of the second cryostat will also be performed in this way. All excavation will be completed before any construction is required in the third and fourth detector pits, providing the opportunity to use the excavation mucking ramp for delivery of cryostat components. This ramp has been designed at a steep 15% grade as of the 30% preliminary design to get it deep enough to pass below existing excavations in the shortest distance possible. Further coordination will be required as the design advances to determine if the slope should be reduced to make the ramp more useful for operations other than excavation.

## 3.4.5 Interfaces between DUNE, Cryogenics and Excavation

There are several points at which the experiment and the facility interface closely. These are managed through discussions between DUNE design team, the Cryogenics Infrastructure design team, and the Conventional Facilities design team and design consultants.

- The LBNF cryostat is a freestanding structure requiring infrequent access for inspection around the vessel. Low tolerance control in excavation will impact the cost of providing access to inspect this vessel.

- The utility spaces to house the cryogen system are directly influenced by the size of the cryogen system equipment.

- The size and construction sequencing of the detector pits are critical to the experimental strategy.

# 3.5 Underground Infrastructure

The requirements for underground infrastructure for the LBNF Project will be satisfied by a combination of existing infrastructure, improvements to those systems, and development of new infrastructure to suit specific needs. The Project must consider the other tenants underground at Sanford Laboratory for which infrastructure is required, including both the existing Davis Campus experiments and the Ross Campus Experiments. The Ross campus experiments in particular are in relatively close proximity (~150m) to LBNF.

The systems must support the LBNF Conventional Facilities (CF) construction activities, Cryogenics Infrastructure, DUNE experiment installation, and operations of both CF Equipment and the experiment. These three scenarios were analyzed and the most demanding requirements chosen from each situation were used to define the requirements for design.

Some of the Sanford Laboratory infrastructure that requires upgrading for LBNF will be rehabilitated prior to the beginning of LBNF construction funding. This includes Ross Shaft rehabilitation, Yates Shaft focused maintenance and repair, and ground support activities at the 4850L between the Yates and Ross Shafts. Additional discussion of this work is included in section 3.5

The conceptual underground infrastructure design for LBNF has been performed by several entities. The primary designer referenced in this document is Arup, USA. Arup's scope includes utility provisions and fire protection- life safety (FLS) strategy, covering infrastructure from the surface through the shafts and drifts, to the cavity excavations for the experiment. Utility infrastructure includes fire/life safety systems, permanent ventilation guidance, HVAC, power, plumbing systems, communications infrastructure, lighting and controls, per the experimental utility requirements provided by DUNE and through coordination with LBNF, Sanford Laboratory and the excavation and surface design teams. The design is described in Arup's *LBNF 30% Preliminary Design Report* and in the conceptual design drawings [12]. This chapter summarizes the work done by Arup and utilizes information from that report.

Shaft rehabilitation and waste rock handling design were previously provided by Arup for the DUSEL PDR. This chapter uses excerpts from the DUSEL Preliminary Design Report, Chapter 5.4 [10]. The research supporting this work took place in whole or in part at the Sanford Underground Laboratory at Homestake in Lead, South Dakota. Funding for this work was provided by the National Science Foundation through Cooperative Agreements PHY-0717003 and PHY-0940801. The assistance of the Sanford Underground Laboratory at Homestake and its personnel in providing physical access and general logistical and technical support is acknowledged.

## 3.5.1 Fire/Life Safety Systems

Life safety is a significant design criterion for underground facilities, focusing on events that could impact the ability to safely escape, or if escape is not immediately possible, isolate people from events underground. Design for fire events includes both preventing spread of fire and removing smoke and/or cryogenic gasses through the ventilation system. The evaluation and establishment of requirements for cryogenic gas removal is performed by the cryogenics group and provided to CF.

Life safety requirements were identified and the design developed by Arup, utilizing applicable codes and standards, including NFPA 520: Standard on Subterranean Spaces, which requires adequate egress in the event of an emergency. Facility fire detection and suppression systems, as well as personnel occupancy requirements are defined in accordance with NFPA 101: Life Safety Code. The design was reviewed by Aon Risk Solutions and the recommendations documented in *Fire Protection/Life Safety Assessment for the Conceptual Design of the Far Site of the Long Baseline Neutrino Experiment* [16]. Due to the unique nature of the experiment and its' location, a number of potential variances will require approval from the authority having jurisdiction (AHJ). Significant examples include use of elevators for egress and use of drifts as air "ducts". The AHJ for Lead, SD is familiar with the facility and the project, and is expected to provide reasonable and timely feedback for proposed variances.

Based on data provided by Sanford Laboratory the maximum occupant load of the 4850L will be controlled 144 occupants following completion of the Ross Shaft Rehabilitation. This can support the anticipated 42 Underground Operations staff, 50 science staff for LBNF (during installation), and 20 science staff associated with the existing experiments.

Compartmentation will be needed for egress routes to separate them from adjacent spaces to limit the horizontal and vertical spread of fire and smoke. Use of compartmentation will help to reduce the likelihood of fire and smoke spreading from the area of fire origin to other areas or compartments. Compartmentation will also help limit the spread of other materials such as cryogenic gases, leaks and spills. This results in design criteria of minimum 4-hour fire separation between the LBNF cavities and adjacent drifts, while all rooms that connect directly to the egress drift at 4850L, as well as the shafts, will have 2-hour minimum fire separation.

## 3.5.2 Shafts and Hoists

The Ross and Yates Shafts provide the only access from the surface to the underground, and are therefore critical to the function of the Facility. Both shafts provide service from the surface to the 4850L, though not every intermediate level is serviced from both shafts. The shafts also provide a path for all utilities from the surface to the underground.
The Ross and Yates Shafts were both installed in the 1930s and have operated since installation. These shafts, along with their furnishings, hoists, and cages, were well maintained during mining operations, but have experienced some deterioration as described in this section. A complete assessment of the Ross and Yates shafts was conducted for the DUSEL Project, and is documented in the Arup Preliminary Infrastructure Assessment Report (DUSEL PDR Appendix 5.M [10]).

### 3.5.2.1 Ross Shaft

The Ross Shaft will be used for facility construction, including waste rock removal, routine facility maintenance, and secondary egress path for the finished underground campuses. It will also be used for LBNF experiment primary access. Excavation for LBNF cannot begin until the Ross Shaft is rehabilitated by the Sanford Lab.

The Ross Shaft is rectangular in shape—14 ft 0 in (4.27 m) by 19 ft 3 in (5.87 m), measured to the outside of the set steel. The shaft collar is at elevation 5,354.88 ft (1,632.17 m) and the 5000L is the bottom level at elevation 277.70 ft (84.64 m) above sea level. Service is provided to 29 levels and five skip loading pockets. The shaft is divided into seven compartments: cage, counterweight, north skip, south skip, pipe, utility, and ladder way. See Figure 3-13 below showing shaft layout.

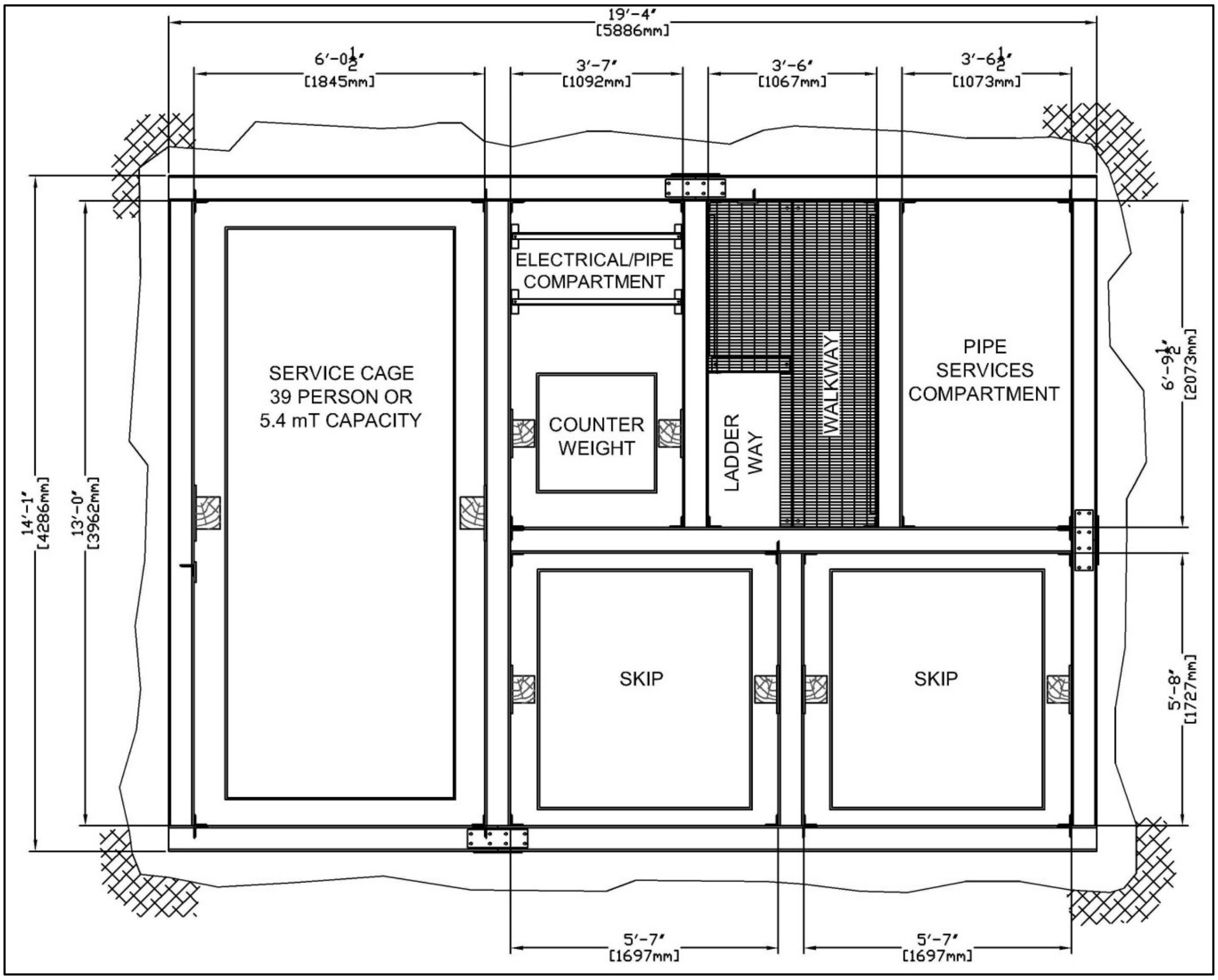

Figure 3-13: Ross Shaft, typical shaft set (SRK, Courtesy Sanford Laboratory)

The Ross Shaft was in operation until the Homestake Gold Mine closed in 2003. Deterioration through corrosion and wear on the shaft steel, including studdles (vertical steel members placed between steel sets), sets, and bearing beams, prompted a full “strip and re-equip” project being performed by the Sanford Lab. The Ross Shaft layout will not be significantly modified from the existing configuration. The set spacing is being increased from 6 ft to 18 ft, but the general configuration is remaining the same to allow for emergency egress during rehabilitation. The shaft was installed with limited ground support, electing to utilize lacing to prevent spalled rock from reaching the personnel conveyances. The new design replaces this system with a pattern bolting system to control rock movement. The requirements for this shaft are safety, performance, and code driven and defined by the existing configuration. Most of the shaft rehabilitation and headframe work is planned to be executed by Sanford Laboratory with non-LBNF Project funds prior to LBNF construction beginning. The rehabilitation is just over 50% complete as of this report and is planned to be completed in 2017. Some items specific to LBNF, such as the development of the skip loading pocket for waste rock handling, are part of the scope of the LBNF project.

The production and service hoists at the Ross Shaft are located on the surface in a dedicated hoistroom west of the shaft. The service hoist operates the service cage and the production hoist operates the production skips. The DUSEL PDR describes the condition assessment of the electrical and mechanical hoisting systems which are described in detail in the Arup Preliminary Infrastructure Assessment Report (DUSEL PDR Appendix 5.M [10]). These electrical and mechanical systems will have standard maintenance performed on them to make them in like new condition, but will not be modified from the existing design. The Ross Headframe steel requires some strengthening and modifications to meet code requirements. Responsibility and timing for this is still being discussed as of this report.

#### 3.5.2.2 Yates Shaft

The Yates Shaft is rectangular in shape—15 ft-0 in (4.572 m) by 27 ft-8 in (8.433 m) measured to the outside of the set timbers. There are two cage compartments and two skip compartments as shown in Figure 3-14. In addition to the cage and skip compartments, there are two other compartments in which shaft services are located. The shaft collar is at 5,310.00 ft (1,618.49 m) elevation and the 4850L is the bottom level at elevation 376.46 ft (114.75 m) above sea level. Service is provided to 18 levels plus four skip-loading pockets. Sets are made up of various length and size timbers located to maintain compartment spaces. The Yates Shaft is timbered except for a fully concrete-lined portion from the collar to the 300L. Recent repairs include full set replacement from the concrete portion to the 800L and additional set repair below this level where deemed critical.

The Yates Service Hoist and Production Hoist are planned to be used as existing, with maintenance performed to bring them into like new condition. Further details regarding the condition of the Yates Hoists' electrical and mechanical condition can be found in Section 2.2 of the Arup *Preliminary Site Assessment Report* (DUSEL PDR Appendix 5.M) [10].

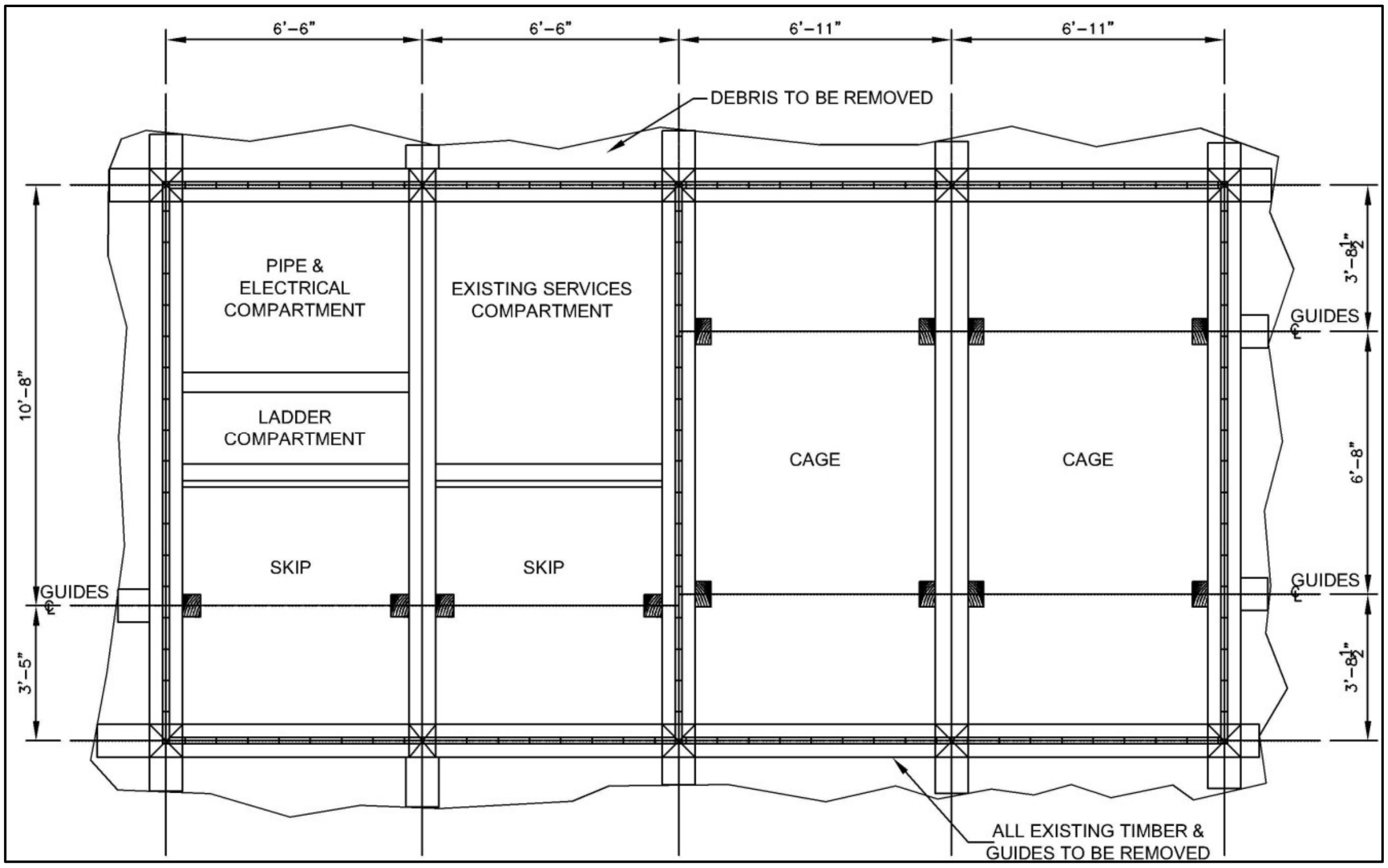

Figure 3-14 Existing Yates Shaft layout (Adapted from SRK, Courtesy Sanford Laboratory)

## 3.5.3 Ventilation

The ventilation system will utilize the existing mine ventilation system as much as possible with minimal modifications. Fresh air for the LBNF cavities and the utility drifts will be provided by pulling air directly from the existing drifts, which is supplied from the Yates and Ross Shafts. Air will be exhausted from the LBNF cavities and utility drifts through a spray chamber rejecting heat from the LBNF chilled water system into an existing exhaust route. 143,000-290,000 cfm design is required for heat extraction depending on intake temperature. 45,000-cfm passes through the each main experimental area (one air change per hour), with the balance of the air required for heat rejection coming directly from the shafts through existing drifts. A full ventilation modelling exercise will be completed for the 60% preliminary design to understand the implications of these volumes. It may become necessary to split the exhaust and/or add booster fans at the 4850L to manage the volumes required. The environmental design criteria for LBNF underground spaces is shown in Table 3-1.

.

Table 3-1: Environmental design criteria (Arup)

| Room | Internal Temperature | Humidity range | Min Ventilation rate/Fresh Air Changes | Occupancy (during assembly) | Additional Information |
|---|---|---|---|---|---|
| LBNF Cavities | 40-82 ˚F (10-28 ˚C) | 15-85% | 1 | 20 (50) | See note 1 below |
| Access Drifts | Min 50˚F (10˚C) | Uncontrolled | | Transient space | |
| Utility spaces / Electrical rooms | 50-95 ˚F (10-35 ˚C) | Uncontrolled | 1 | | |
| Storage Rooms | 59-104 ˚F (15-40 ˚C) | Uncontrolled | Min 15 cfm/person | Room dependent | |

*Occupancy of 10 during operations.
Note 1: Temperature, humidity and filtration requirements in localized areas of this space may differ, dependent on requirements. This will be provided by the experiment installation design team. The internal conditions stated above will be used to inform the design of plant and services for each space unless specific requirements that differ from this are provided by LBNF/Sanford Laboratory or the lab experiment design teams.

Per historical data, outdoor temperatures can drop below -20°F; therefore, the intake air requires heating to prevent ice build-up in the shafts which could potentially disrupt hoisting operations and damage shaft support members, cables and piping. The existing shaft heaters are expected to be adequate for normal operation, but temporary supplemental heating may be necessary during excavation due to higher demands.

## 3.5.4 Electrical

The underground facilities at the 4850L will have electrical power for normal operations as well as standby power for emergency occupant evacuation. LAr experiment power requires standby power for circulation of cryogens to avoid rapid boil-off and loss of argon.

### 3.5.4.1 Normal Power

The estimated electrical loads for both the far detector experiment and the underground infrastructure serving the experimental spaces are included in the facility load determination and design.

Power to serve the far detector experiment will originate from the Ross substation and be routed down the Ross Shaft to the 4850L. One 15kV mining cable shall be installed down the Ross Shaft to the 4850L and shall be cable rated for mine use, highly flame retardant, low smoke toxicity with high tensile strength and self-supporting. At the 4850L, the 15-kV mining cable will terminate in 15-kV switchgear located in a new Ross underground substation. Facility power will be provided in a similar manner, with a dedicated power supply through the Ross shaft terminating in the new electrical substation near the Ross Shaft. This will be provided early in the construction process to allow it to be used for construction.

### 3.5.4.2 Standby and Emergency Power

A 300kW emergency/standby diesel generator will be provided in the Central Utility Cavern to serve standby and emergency loads. 48 hours of diesel fuel will be provided to operate the generator when surface power is inoperable. The following 4850L electrical loads are anticipated to be installed to the emergency/standby power system:

- Security
- IT System for communications
- Smoke control fans
- Mono rail
- Cryostat system controls

### 3.5.4.3 Fire Alarm and Detection

The 4850L will have notification devices installed to alarm the occupants of a fire. Notification devices will consist of speakers and strobe lights. Manual pull stations will be provided within 200 ft of egress. Phones will be installed in the liquid argon chambers and every 400 ft along the access drifts to communicate with the surface level command center.

An air sampling and gas detection system will be installed in the drifts and liquid argon detector chamber as an early detection of a fire condition. The air sampling system will be connected into the fire alarm system.

The fire alarm system will also interface with the oxygen deficiency hazard (ODH) system to activate the fire alarm system and initiate an alarm at the respective level fire alarm panel and at the surface level command center. Specific sounds and strobe colors will be identified based on the type of alarm (fire, ODH, etc.).

### 3.5.4.4 Lighting

Suspended lights mounted at a height just below the lowest obstruction will be provided for all drifts and ramps. Mounting is to be coordinated with conduit and supports of other systems running overhead. Maintained average illumination of approximately 24 lux (2.4 foot candles) at floor level will be provided throughout the drifts. Lighting control in drifts will be via low voltage occupancy sensors and power packs suitable for high humidity environments.

### 3.5.4.5 Grounding

The grounding system will be designed to provide effective grounding to enable protective devices to operate within a specified time during fault conditions, and to limit touch voltage under such conditions.

The grounding system will be designed for a maximum resistance of 5 ohms where possible based on Mine Safety and Health Administration (MSHA) recommendations for ground resistance in mines. Ground beds, consisting of an array of ground rods, will be installed at each substation to provide low impedance to ground.

Electrical separation between the cryostat detectors and cavern utilities will be achieved by separating the metal components (rebar, structure support, etc.) from each other. Inductors will be installed between grounding systems to control noise between systems while also controlling touch potential for safety.

## 3.5.5 Plumbing

Plumbing provided by CF, but specific to DUNE, includes plumbing for the cooling systems and gas piping for nitrogen and argon delivery from the Cryogenics Compressor Building on surface to the Central Utility Cavern. Beyond this the facility requires supplies of both potable and industrial water, as well as a means to remove water inflows and sewage.

### 3.5.5.1 Industrial Water

An existing 4-inch industrial water riser will be used for construction and as a secondary fire service. It is not feasible to run an uninterrupted main water supply line from grade level down to serve the lower levels due to the extremely high hydrostatic pressure that would occur in the system. A series of pressure reducing stations are located at regular intervals in intermediate levels and at the 4850L in order to maintain the pressure within the capability of readily available piping.

### 3.5.5.2 Potable Water

Potable water is not required in large quantities for LBNF. The Sanford Lab experience has been that plumbing potable water through the shafts for low volumes is not effective, as the pressure reducing systems have the potential to introduce biological contaminants that result in the water no longer meeting drinking water standards, especially in low flow situations. To address this, local filters and ultraviolet treatment is done at the 4850L to make industrial water meet drinking water standards. This system has been used successfully for several years at the Sanford Lab.

### 3.5.5.3 Chilled Water

The DUNE equipment will produce a significant amount of heat which will be removed by LBNF-provided chillers. Three chillers at 50% each have been selected to provide N+1 redundancy to allow for maintenance. Heat from the chillers and various process loads will be rejected using a spray chamber located at the east end of the 4850L LBNF caverns immediately before exhausting into the existing exhaust route. This location maximizes the available air flow by capturing air from both LBNF, the Davis Campus, and directly from either the Yates or Ross Shaft (or both). The ventilation air is a mixture of air (170,000-3500,000 CFM) from the Yates and Ross Shafts at approximately 68°F. This volume of air is such that the total heat rejected (2.5 MW or 740 Ton) will raise the air temperature to no more than 95°F. The dry coolers exhaust ductwork is arranged in a header and is ducted to the ventilation borehole.

### 3.5.5.4 Fire Suppression

The source of fire water main will be the existing 4-inch industrial water main at Ross Shaft. The connection to this line will be at the 4100L, where a new sump with at least 90,000 gallons capacity will be built using sump walls in an existing drift to provide 90 minutes of capacity even if the supply were cut off. The fire protection system at the 4850L Campus will be a gravity fed system. There will be a connection to an existing 6” industrial water main in the west drift fed from Yates Shaft, where a similar, but slightly smaller at 50,000 gallons, sump has been built by the Sanford Lab. This provides redundant supply from surface.

### 3.5.5.5 Drainage

Drainage [17] from the drifts, mechanical electrical rooms (MERs), and any areas where spillage is likely to occur will be collected locally in open sumps. Sumps will be located every 500 in any areas where drainage to the drifts is not practical. Sumps will be equipped with sump pumps in a staged configuration where each pump discharging to the adjacent sump until water is discharged to the #6 Winze, where it flows to the primary facility pool approximately 1,000 feet below the 4850L. From there, the existing Sanford Lab dewatering system pumps the water in stages to the surface where it is treated before discharge into a nearby stream.

### 3.5.5.6 Sanitary Drainage

No sanitary drainage is included in the requirements for LBNF. Existing Sanford Laboratory facilities are planned to be used.

### 3.5.5.7 Chilled Water

The DUNE equipment will produce a significant amount of heat which will be removed by LBNF-provided chillers. Three chillers at 50% each have been selected to provide N+1 redundancy to allow for maintenance. Heat from the chillers and various process loads will be rejected using a spray chamber located at the east end of the 4850L LBNF caverns immediately before exhausting into the existing exhaust route. This location maximizes the available air flow by capturing air from both LBNF, the Davis Campus, and directly from either the Yates or Ross Shaft (or both). The ventilation air is a mixture of air (140,000-290,000 CFM) from the Yates and Ross Shafts at approximately 68°F. This volume of air is such that the total heat rejected (2.1 MW) will raise the air temperature to no more than 95°F. The dry coolers exhaust ductwork is arranged in a header and is ducted to the ventilation borehole.

### 3.5.5.8 Nitrogen and Argon Gas Piping

Three 12” and three 8” mild steel pipes are provided by CF from the surface Cryogenics Compressor Building to the shaft, through the shaft, and across the 4850L to the Central Utility Cavern west entrance. The design and specifications of this piping are the responsibility of the Cryogenics Infrastructure Project team. The supply and installation within the Cryogenics Compressor Building and the central Utility Cavern is also the responsibility of the Cryogenics Infrastructure Project.

## 3.5.6 Cyberinfrastructure

The Structured Cable System design will be based on uniform cable distribution with a star topology. New fiber connections will be extended to the 4850 level from the Ross Dry Building, and will be dedicated to the use of LBNF experiments at the 4850 level. There is currently no requirement for redundancy in the connections to the experiments and the connection to the Ross Hoist presents a single point of failure. The design provides one (1) 96 –strand single mode armored fiber optic cable from the Control room dedicated to the experiments. Figure 3-15 shows the fiber distribution network for LBNF.

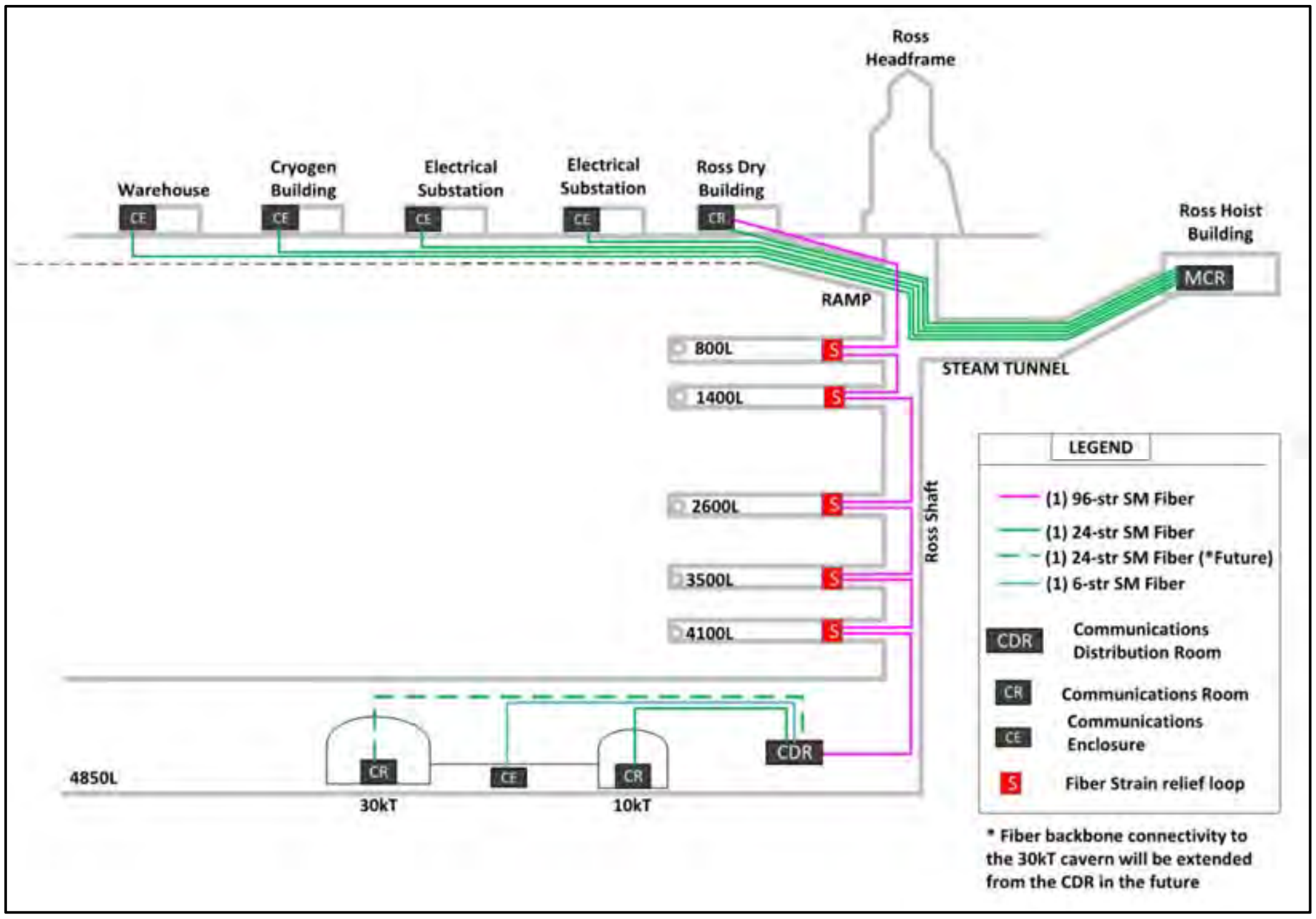

Figure 3-15: Fiber distribution system for LBNF (Arup)

Voice communications are provided via two-way radios and phones distributed throughout the underground spaces (in every room as well as every 500 ft in drifts). Two-way radios utilize a leaky feeder system to ensure communications over long distance without line of site. These leaky feeders are cables that act as antennas installed the length of all drifts and shafts. Phones utilize Voice over Internet Protocol (VoIP) to provide communication though the fiber optic data backbone.

The data system is designed to provide 10-Gigabit Ethernet in the backbone and 1-Gigabit Ethernet to connected systems (computers). This system is intentionally left at a lesser level of design due to the continuous progression and advancement of technology that will almost certainly result in more advanced technologies than are currently available being utilized at the time of construction.

## 3.5.7 Waste Rock Handling

Prior to the commencement of any excavation activities, it will be necessary to establish a waste rock handling system. The capacity of this system will be equivalent to what was in place during mining.

There are a number of components to the waste rock handling system, including refurbishing the Ross Shaft hoisting system, the Ross Shaft crushers, and a new conveying system to transport rock downhill to the Kirk Road, as seen in Figure 3-16.

The systems utilize experience and equipment from the former Homestake Mining Company legacy, where rock was removed to the surface using skips in both the Yates and Ross Shafts. At the headframe of each shaft, the material was crushed to a nominal ¾ in, passed through ore bins, and was transported via underground rail to the mill system. All systems from the underground to the crushers will be rehabilitated from the original systems.

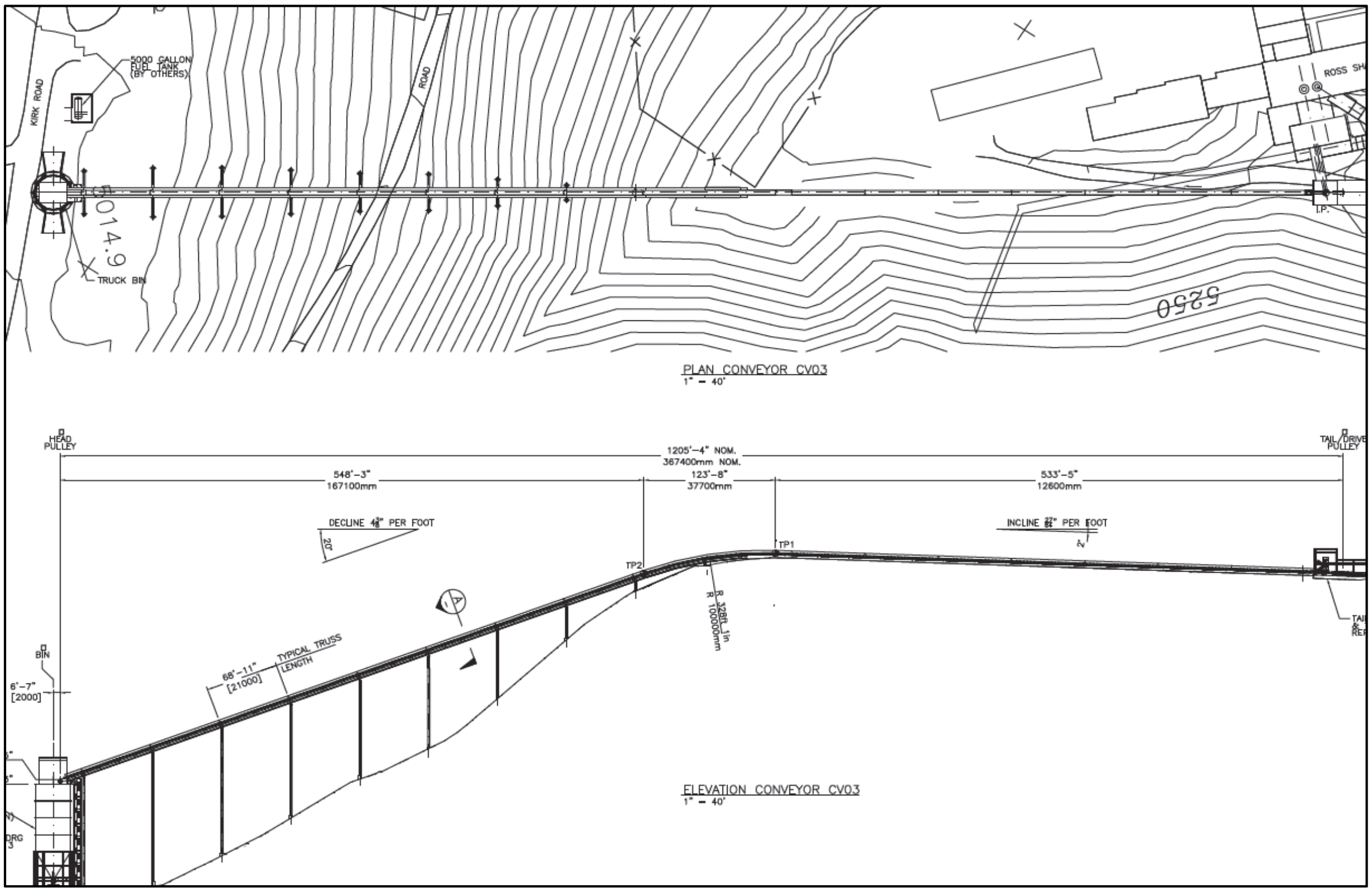

Figure 3-16: Waste rock handling system route (SRK, Courtesy Sanford Laboratory)

# 4 Far Site Facilities: Cryogenics Infrastructure

## 4.1 Overview, Development Program, and ES&H

The scope of the LBNF Cryogenics Infrastructure includes the design, procurement, fabrication, testing, delivery, and installation oversight of the following components:

1. Four identical cryostats to contain the liquid argon (LAr) and the time projection chambers.
2. A comprehensive Cryogenics System that meets the performance requirement for
   - purging, cooling down and filling the cryostats
   - acquiring and maintaining the LAr temperature within ±1 K around nominal temperature (88.3 K)
   - purifying the LAr

The designs of the cryogenic infrastructure are summarized in the following sections. A more complete description is given in *Annex 3D: Detailed Report on the LBNF Cryostat and Cryogenics System* [18].

The reference-design for the LBNF cryogenics Infrastructure encompasses the following components:

- Four identical 10-kt (fiducial mass) membrane cryostats
- Receiving facilities for LAr and LN2 tanker trucks
- Transfer system to deliver gas argon (for liquefaction) and nitrogen (for a closed-loop refrigeration system) from the surface to the underground cavern area
- Closed loop LN2 refrigeration system for condensing GAr
- Boil-off gas re-liquefaction equipment
- LAr-purification facilities
- Cryostat-purge facilities

The development of the LBNF cryogenics infrastructure from conceptual to preliminary design includes a prototyping program. The most significant issue to resolve is whether a membrane cryostat in the size of LBNF can achieve the required electron drift lifetime. The Liquid Argon Purity Demonstrator (LAPD) was an off-project prototype, built to study the concept of achieving LAr purity requirements in a non-evacuated vessel [19]. The purge process accomplished in the LAPD was repeated on the 35 ton membrane-cryostat prototype developed as an LBNE effort, which confirmed that initial evacuation of the cryostat is unnecessary and that a LAr purity level sufficient to enable the electron lifetime required in a membrane cryostat can be achieved [20]. Further prototyping program aiming to test and demonstrate this technology at the 1 kt scale is foreseen over the next two years as part of the CERN Neutrino Platform program.

During the course of the LBNF project and the proposed prototypes, Fermilab ES&H standards and Sanford Underground Research Facility (SURF) ES&H codes and standards will guide the design, procurement and installation phases of the project. Particular attention will be paid to the critical sections of FESHM [21] Chapter 4240 relating to ODH and Chapter 5000 standards for piping construction and vessel design. The planned work process will provide for reviews through all phases of the project to guarantee stringent adherence to the safety requirements. Requirements on the membrane-cryostat materials and their fabrication will be strictly outlined in the specification documents. Close communication between the vendors, Fermilab and CERN's cryogenics and process engineers, and Fermilab and SURF ES&H personnel will be maintained at all times.

## 4.2 Steel Cryostat

Each of the four identical cryostats consists of two major components: a steel outer frame (warm vessel) and a membrane cold vessel. The membrane cold vessel is based on the technology used for liquefied natural gas (LNG) storage and transport ships. It consists of an inner stainless steel, corrugated thin membrane in contact with the liquid and thermal insulation surrounding it. Details of this technology are presented in section 4.3. The main idea behind this concept is that the cold membrane vessel represents a fully contained vessel with two independent barriers.

The function of a steel warm vessel is to contain the membrane vessel and provide mechanical support to it, while providing also a gas barrier towards the outside. Figure 4-1 below presents the layout of such an assembly, which consists of a modular self-supporting steel structure.

The warm vessel consists of outer supporting profiles, interconnected through a steel grid and a 10 mm thick stainless steel (type 304L), continuous plate inside in contact with the membrane insulation.

The material used is S460ML structural carbon steel with yield strength of 430 MPa and tensile strength of 510 MPa. The main profile used is HL 1100 x 548 or its ASTM alternative W 44 x 16 x 368. Four profiles are bolted together by four bolting connections, forming a structural "portal". Each bolting connection consists of 16 bolts (M42). The additional grid is made of the IPE300 profile. The total self-weight of the structure is approximately 2000 T.

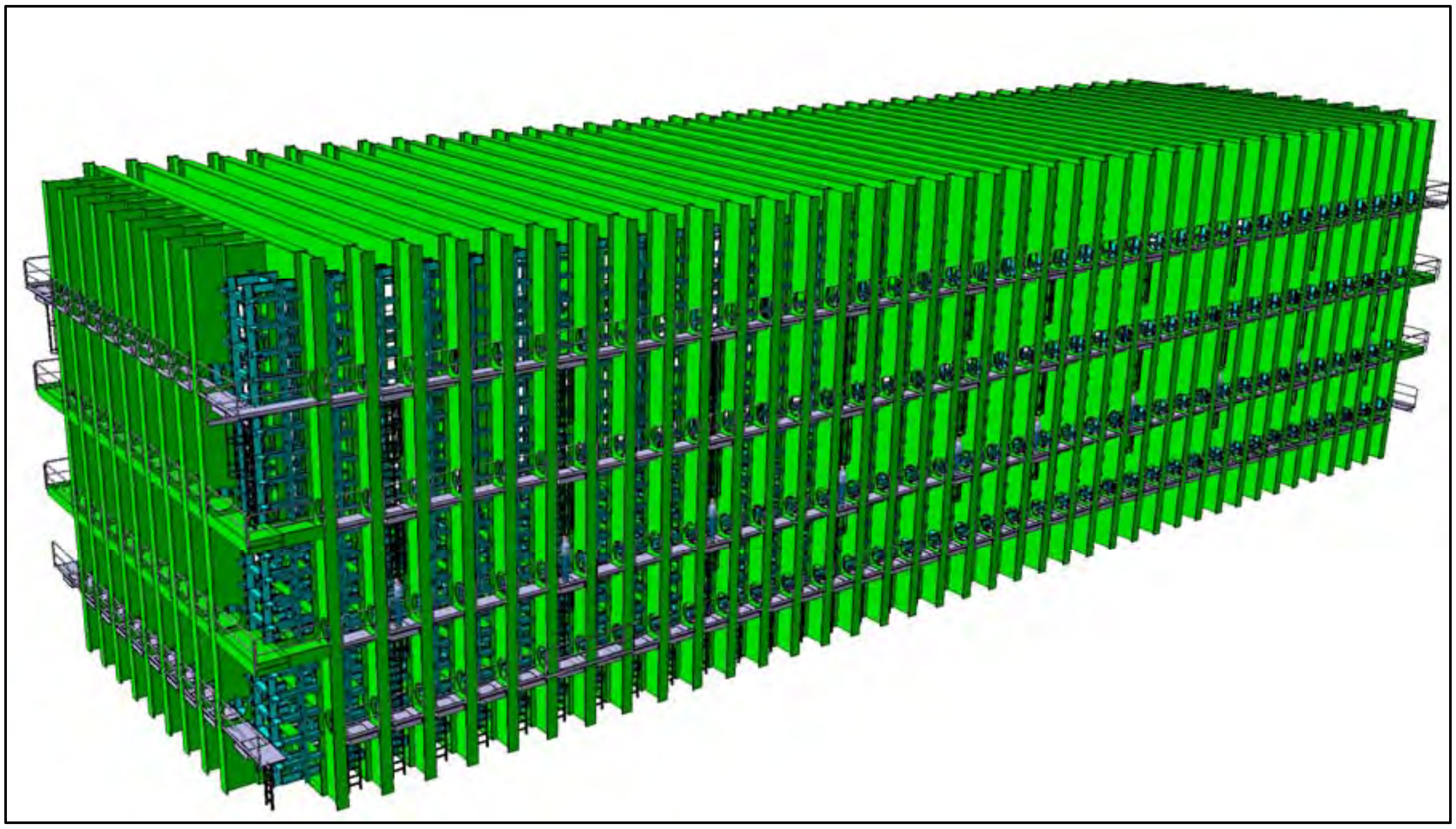

Figure 4-1: Outer layout of steel warm vessel

The structure is positioned on a firm surface with no additional structural connections necessary for either the cavern floor or the cavern walls. The internal (external) dimensions of the structure are approximately 16.9 (19.0) m in width, 15.8 (18.0) m in height and 63.8 (66.0) m in length.

The main advantage of this design is the fact that such a structure can be fully decoupled from the civil engineering work related to the excavation and finishing of the four caverns. All components can be procured and prepared on the surface, ready to be lowered through the Ross shaft. Underground installation will take 4 months for each of the cryostats and can be done sequentially. The warm vessel will be fully accessible from outside and can be inspected at any moment. A net of stairs and gangways is included in the design at the allocated space. No requirements are put on the distance of the warm structure to the cavern walls. Typically this value might vary between 200 mm and 500 mm. The warm cryostat is positioned inside the cavern pit with enough air space reserved for convection or forced air circulation, which maintains the rock temperatures above freezing.

Finite Elements Analysis Methods using the commercial ANSYS code have been employed as the main design technique. The safety codes used are the Eurocode III and American Society of Mechanical Engineers (ASME) Boiler and Pressure Vessel code Section VIII, Rules for Construction of Pressure Vessel, Division II. The most conservative requirements among the two codes have been adopted. The structure has been treated as a low pressure vessel (<500 mbarg).

Approximately 18000 T of LAr is acting as load on the floor, i.e., around 20 T/m$^2$. Approximately 8000 T of hydrostatic force is acting on each of the long walls, with triangular distribution over the height, and around 2000 T of hydrostatic force is acting on each of the short walls. Additionally, a normal ullage operational pressure of 75 mbarg (0.75 T/m$^2$) is considered acting on every wall. The structure has also been verified to accidental overpressure of 350 mbarg (3.5 T/m$^2$), which is the maximum allowable working pressure of the cryostat. The weight of the detector itself, as well as seismic action, has been taken into account in the calculations.

The following models and analyses methods have been utilized:

- For evaluation of the global behavior of the entire structure, a beam model has been developed.
- Analytical models of a single portal, i.e., 4 main beams connected together (roof, floor and the side walls) have been used.
- To study in more details the main elements of the structure, an additional shell model of single cell, i.e., one portal and 8 additional grid beams, has been also developed.
- In order to evaluate the stability of the structure specific analyses, i.e., linear (eigenvalue) and nonlinear buckling, have been performed on the following parts of the structure:
    - A single portal by utilizing beams elements
    - A single cell (one portal and two additional grid beams, one on the left and the other on the right) on two different FE models:
        - One consisting of beam & shell (using ANSYS Workbench)
        - On another one containing only shell elements (using ANSYS APDL)
- Very detailed models on the connections (bolting and/or welding) on a single portal utilizing solid and contact elements have been further developed and used.

The maximum stress levels at the main profiles at the location of the maximum moment are in the range of 125 MPa, which allows a safety factor of 4 with respect to the tensile strength of the chosen material. Additional bracing of the main profiles increases the stability of the structure by factor 2.5, as verified with stability analyses. An additional optimization work aiming at reducing the global external dimensions, as well as the self-weight of the structure is also ongoing.

## 4.3 Membrane Cryostat

The conceptual reference design for the DUNE far detector specifies four rectangular cold vessels, each measuring 15.1 m internal width, 14.0 m internal height and 62.0 m internal length. Each vessel contains a total mass of 17.1 kt of LAr and between 3 and 5% of Ar gas operating at a pressure of 75 mbarg, depending on the final time projection chamber (TPC) design. The membrane design is commonly used for liquefied natural gas (LNG) storage and transport tanker ships and has been proven to be a viable option for LArTPC experiments. A membrane tank is made of a stainless-steel liner to contain the liquid cryogen. The pressure loading of the liquid cryogen is transmitted through rigid foam insulation to the surrounding structural support (see Section 4.2), which provides external support for the liner. The membrane liner is corrugated to provide strain relief resulting from temperature-related expansion and contraction (see Figure 4-2: Corrugated stainless steel primary barrier).

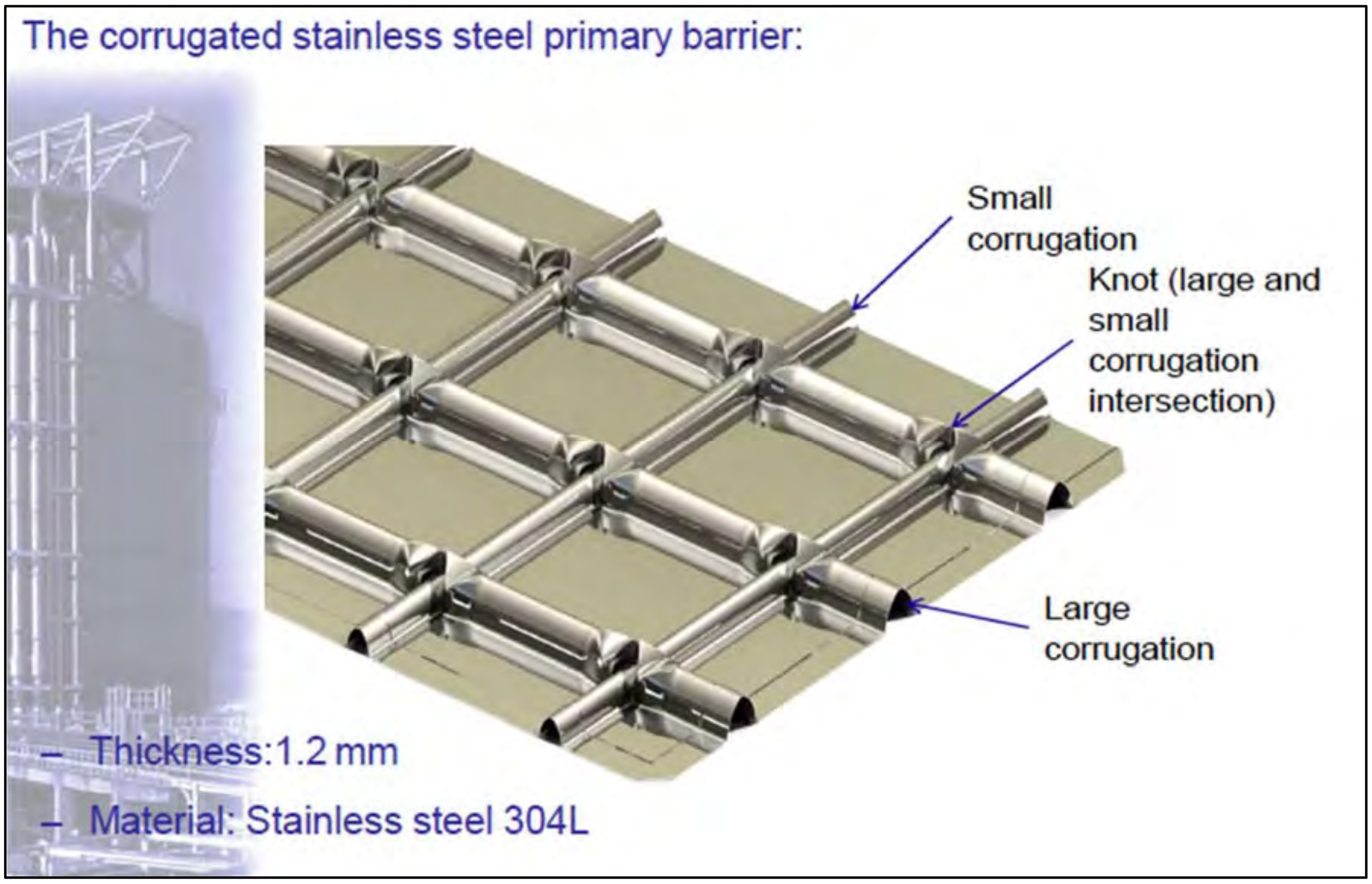

Figure 4-2: Corrugated stainless steel primary barrier

## 4.3.1 Sides and Bottom of Cryostat

The membrane cryostat is a sealed container that relies on external support from the surrounding steel frame (outer warm vessel) to resist the hydrostatic load of the contents. From the innermost to outermost layers, the side walls of the membrane cryostat consist of

- the stainless-steel primary membrane
- an insulation (inner layer)
- a secondary barrier (a thin aluminum membrane that contains the LAr, in case of a leak in the primary membrane)
- more insulation (outer layer)
- a barrier to prevent water-vapor ingress to the cryostat
- the steel frame (outer warm vessel)

The basic components of the membrane cryostat are illustrated in Figure 4-3. The cryostat is positioned inside the rock pit with enough air space reserved for convection or forced air circulation, which maintains rock temperatures above freezing.

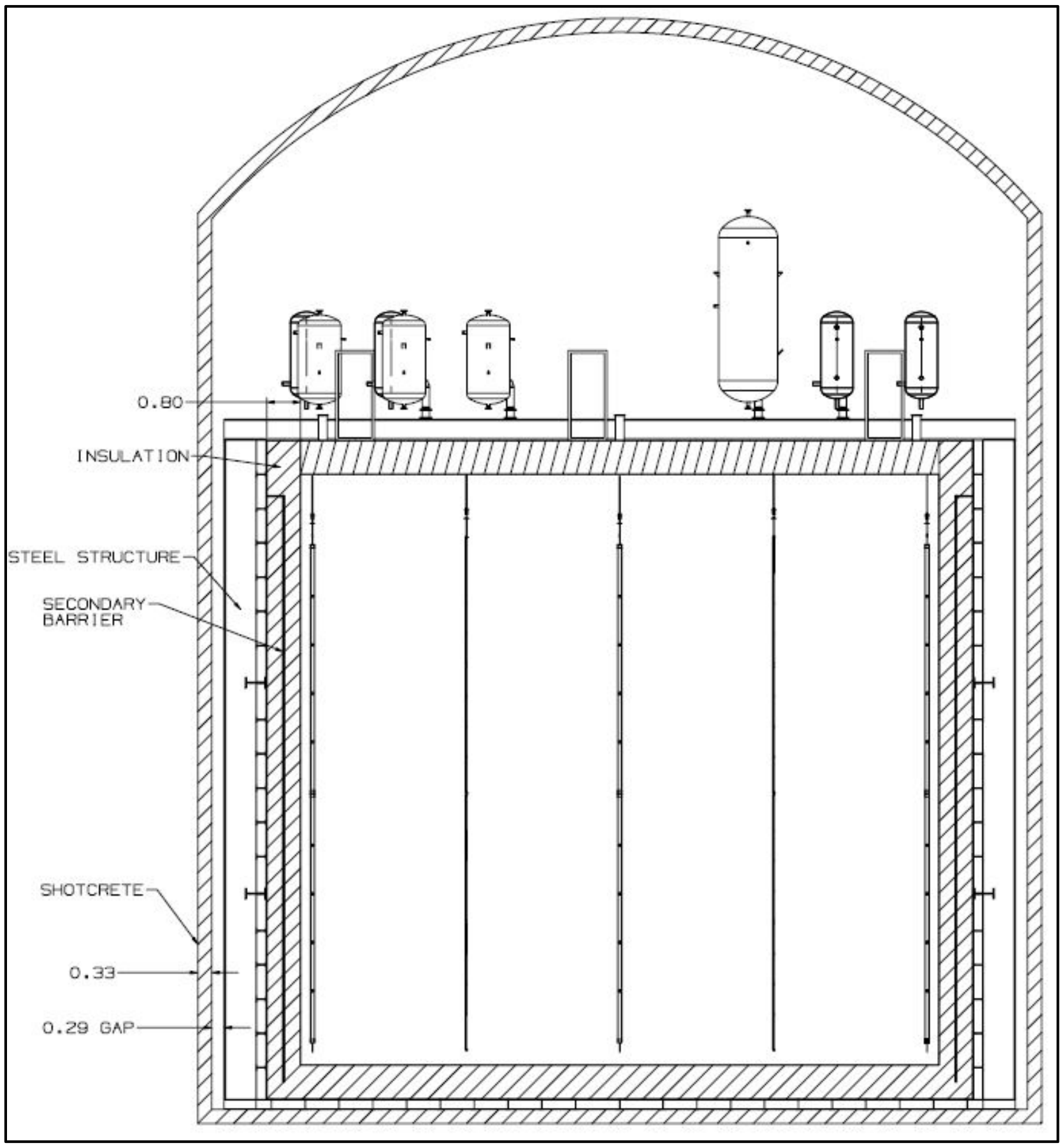

Figure 4-3: Composite system as installed for the LBNF reference design

## 4.3.2 Steel Frame and Vapor Barrier

A vapor barrier is required on all internal surfaces of the steel frame (base, side walls, and end walls) and the roof to prevent the ingress of any water vapor into the insulation space. If water vapor were permitted to migrate into the insulation space, it could freeze and degrade the thermal performance of the insulation. The barrier must also reliably absorb the stresses and strains from all normal loading conditions. The selected vapor barrier material is a stainless steel plate of 10 mm applied to the sides, top and bottom surfaces.

## 4.3.3 Insulation System and Secondary Membrane

The membrane cryostat requires insulation applied between the primary stainless steel membrane and the moisture barrier and to the roof to minimize the heat ingress and the required refrigeration load.

Choosing a reasonable insulation thickness of 80 cm, given an average conductivity coefficient for the insulation material of 0.0283 W/m-K, the heat input is expected to be 32.1 kW per cryostat. The insulation material, a solid fiberglass foam, is manufactured in 1 m × 3 m composite panels. The panels are laid out in a grid with 3 cm gaps between them (these will be filled with loose fiberglass) and fixed onto anchor bolts embedded into the steel outer structure at about 3 m intervals. The composite panels contain an outer insulation layer, the secondary membrane and an inner insulation layer. After positioning adjacent composite panels and filling the 3 cm gap, the secondary membrane is spliced together by epoxying. All seams are covered so that the secondary membrane is a continuous liner. A corner detail is shown in Figure 4-4**.** The secondary membrane comprises a thin aluminum sheet and fiberglass cloth. The "fiberglass aluminum fiberglass" composite is very durable and flexible with an overall thickness of about 1 mm. The secondary membrane is placed within the insulation space. It surrounds the internal tank on the bottom and sides, and it separates the insulation space into two distinct, leak-tight, inner and outer volumes. The outer insulation separates this membrane from the steel frame. This secondary membrane is connected to embedded metal plates in the vertical steel wall at the upper edge of the tank. In the unlikely event of an internal leak from the cryostat's primary membrane into the inner insulation space, the liquid cryogen will be contained in the secondary membrane volume.

## 4.3.4 Cryostat Layers as Packaged Units

Membrane tank vendors have a "cryostat in a kit" design that incorporates insulation and secondary barriers into packaged units. See Figure 4-5.

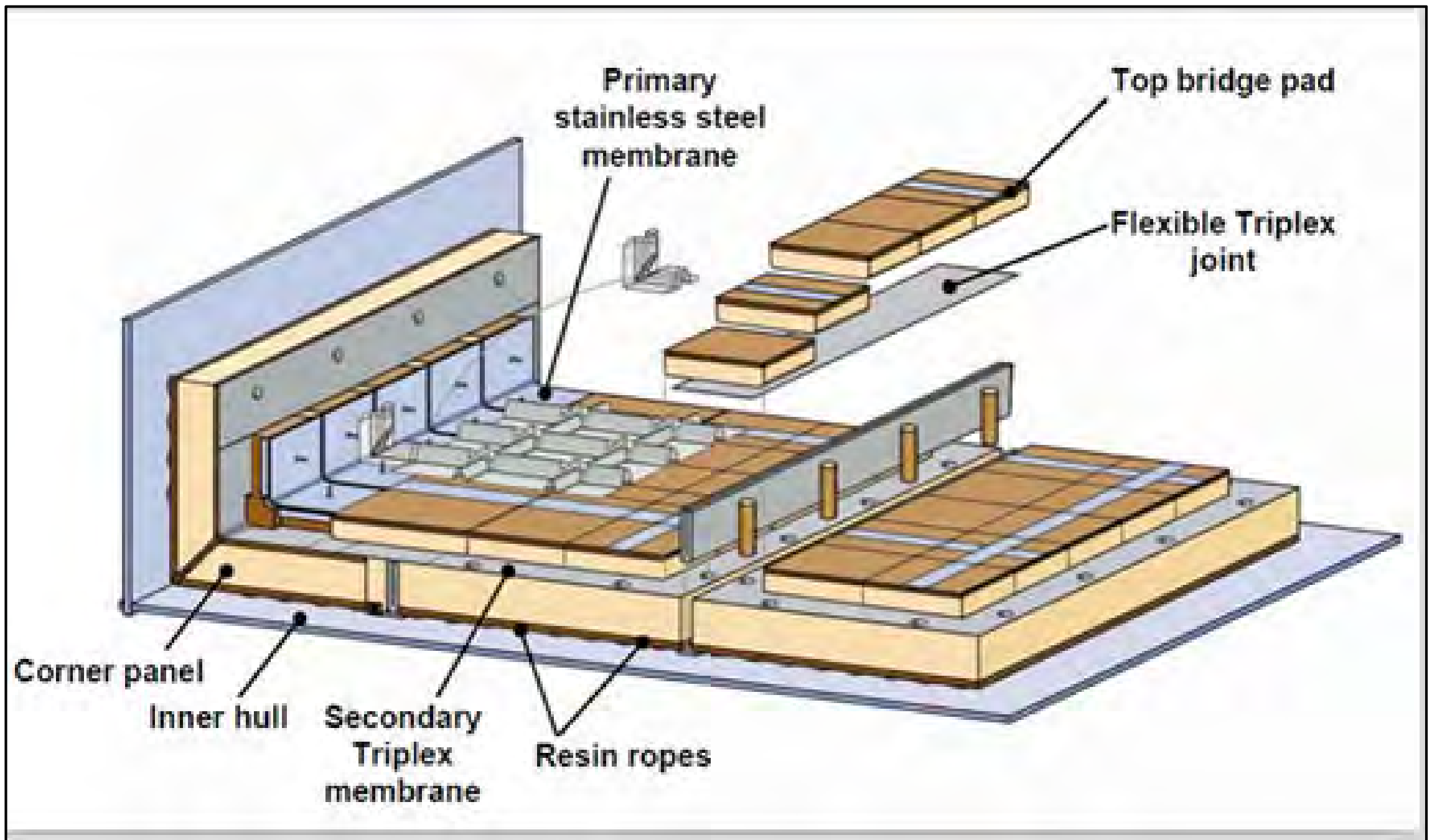

Figure 4-4: Membrane corner detail

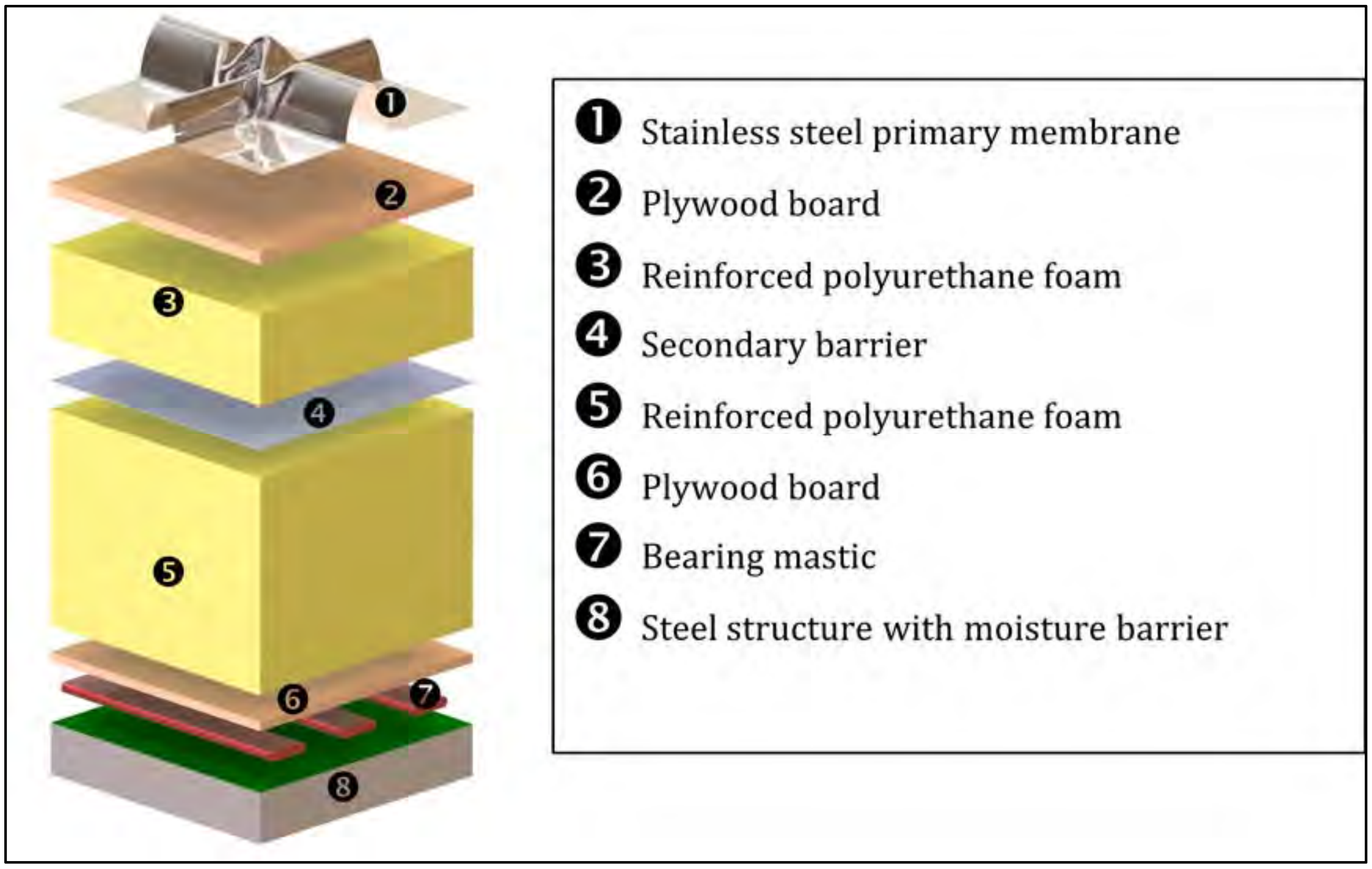

Figure 4-5: GST (Composite system from GTT)

## 4.3.5 Top of Cryostat

The stainless-steel primary membrane and the intermediate layers of insulation and water-vapor barrier continue across the top of the cryostat, providing a vapor-tight seal. **Note:** No secondary membrane is required for the cryostat top.

The hydrostatic load of the LAr in the cryostat is carried by the steel frame on the sides and bottom. Everything else within the cryostat (TPC planes, electronics, sensors, cryogenics- and gas-plumbing connections) is supported by the top of the cryostat. All piping and electrical penetrations into the interior of the cryostat (except for sidewall penetrations from the external liquid argon recirculation pumps) are made through this top plate to minimize the potential for leaks.

Studs are welded to the underside of the steel plates to bolt the insulation panels to the steel plates. Insulation plugs are inserted into the bolt-access holes. The primary membrane panels (also manufactured in smaller sheets) are first tack-welded then fully welded to complete the inner cryostat volume. Feed-through ports located at regular intervals within the corrugation pattern of the primary membrane to accommodate TPC hangers, electrical and fiber-optic cables, and piping are shown in Figure 4-6.

Some equipment, such as monitoring instrumentation, will be installed within wells extending through the roof structure. All connections into the cryostat (except for sidewall penetrations from the external liquid argon recirculation pumps) will be made via nozzles or penetrations above the maximum liquid level and mostly located on the roof of the cryostat. See Figure 4-6 for a typical roof-port penetration.

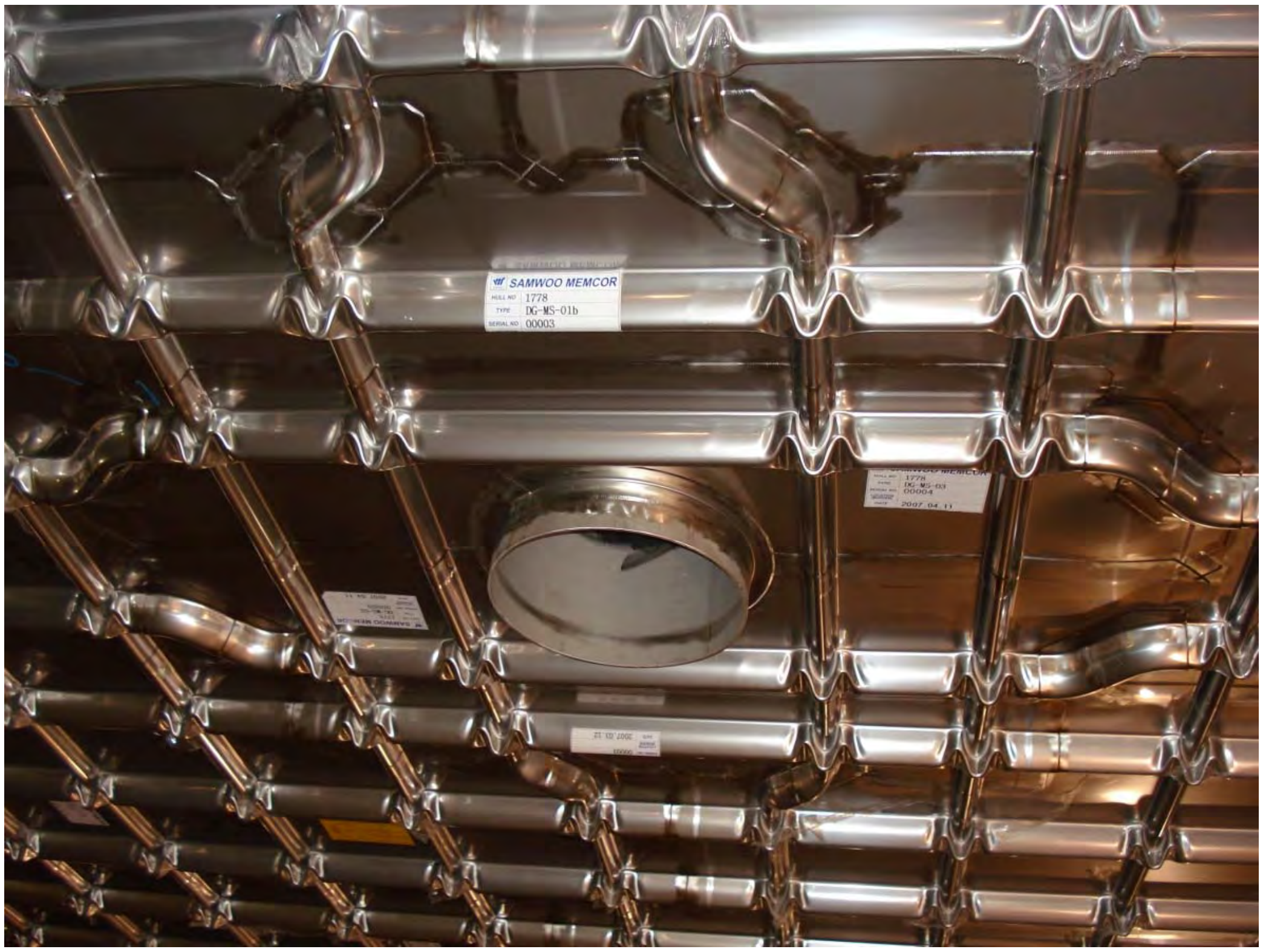

Figure 4-6: Nozzle in roof membrane cryostat (Figure courtesy GTT)

## 4.4 Cryogenics Systems Layout

Cryogenics system components are located in and around the surface building, in the Ross shaft, and within the underground caverns. On the surface near the Ross shaft, there will be a cryogen receiving station. A 50 $m^3$ (69 tons of LAr capacity) vertical dewar will have two LAr truck connections to allow for the receipt of LAr deliveries for the initial filling period. This liquid argon dewar serves as a buffer volume to accept liquid argon at a pace of about 5 LAr trailers (18 tons per trailer) per day during the fill period. An analyzer rack with instruments to check water, nitrogen, and oxygen content of the trailers will also be located in the vicinity. A large 280 kW vaporizer at the surface is used to vaporize the liquid argon from the storage dewar and warm up the resulting gas to room temperature prior to the argon gas being transferred by uninsulated piping down the Ross shaft.

Another 50 $m^3$ vertical dewar and fill connection will be available near the liquid argon dewar. This dewar is used to accept nitrogen deliveries for the initial charging and startup of the nitrogen refrigerator. It is also used for pressure control of the liquid argon storage dewar. A large vaporizer for the nitrogen circuit is located nearby to vaporize liquid nitrogen to nitrogen gas and to warm the gas to room temperature. This gas is used as the feed for the compressors of the nitrogen refrigerator. Four compressors are located in a compressor building on the surface near the Ross shaft and cryogen receiving area. The compressors require a set of nitrogen gas buffers to regulate the nitrogen compressor low and high pressure loop, a set of oil separators and coalescers to clean the nitrogen gas after the pressurization loop, and a closed loop water cooling circuit. The closed loop water-cooling circuit has recirculation pumps in the compressor building and an evaporative cooling tower located outside near the vaporizers. The compressors are the only refrigerator components located on the surface. The compressors discharge high pressure (1.14 MPa) nitrogen gas into pipes that run down the Ross shaft. The compressors were chosen to be located on the surface because the electrical power requirement and cooling requirement is much cheaper to provide at the surface rather than at deep depth in the mine. Each compressor is a 1500 horsepower machine running at 4160 volts. Four running compressors will require a total of 2.6 MW of electrical power at the surface.

The Ross shaft contains the vertical pipelines connecting the surface equipment with the equipment in the cavern area. The piping run consists of a gas argon transfer line and the compressor suction and discharge lines. At the bottom of the Ross shaft at the 4850 level, the piping exits the shaft and runs along a drift to the detector caverns.

The Central Utility Cavern at the 4850 level contains the rest of the nitrogen refrigerator (cold boxes), liquid nitrogen storage vessels, and filtration equipment. The nitrogen refrigerator equipment is located at the far end of the Central Utility Cavern, away from Ross shaft. Fresh ventilation air is supplied down the Ross and Yates shafts, enters the detector cavern and flows over the cryostats or the refrigerator equipment in the Central Utility Cavern before being exhausted out via the Oro Hondo exhaust shaft.

There are four argon recondensers per cryostat. They are placed above each cryostat. Four 23 $m^3$/hr recirculation pumps per cryostat are used to circulate liquid from the bottom of the cryostat through the LAr filters and are set close to each cryostat as well. Each pump will have sidewall penetration to the membrane cryostat.

The piping between the surface and cavern is located in a utility chase down the Ross shaft. See Figure 4-7. The piping material is carbon steel coated with a corrosion barrier, a single layer of fusion epoxy.

Table 4-1 lists the piping and its duty and size. The frictional pressure drop for the supply pipes match the pressure gained due to the static head from elevation change. All the piping connections will be made with Victaulic fittings. The nitrogen and argon being transferred in the Ross shaft piping will be at ambient temperature, in the gas phase. Having gas phase only in the 1.5 km vertical piping is an advantage over liquid transfer because the hydrostatic head for gas only piping is on the order of 0.05 MPa, whereas for the liquid transfer it is 20 MPa. If liquid phase fluid was transferred it would require on the order of seven pressure reducing stations evenly spaced along the vertical drop. In fact, using liquid cryogen delivery was considered in the March 2012 LBNE CDR. However, the cost of providing the excavated spaces and pressure reducing stations was costly as compared to using gas only transfer. For a liquid transfer option with pressure reducing stations, the piping would need to be routed down the Oro Hondo ventilation shaft that would need rehabilitation. With gas only transfer to the cavern, straight piping can be run down the Ross shaft. The drawback of gas only transfer is that one must provide the liquid nitrogen refrigeration in the cavern to fill the cryostats by liquid argon condensation. Filling each cryostat with liquid argon in a reasonable period of time is a driving factor to determine the size of the refrigerator and condenser.

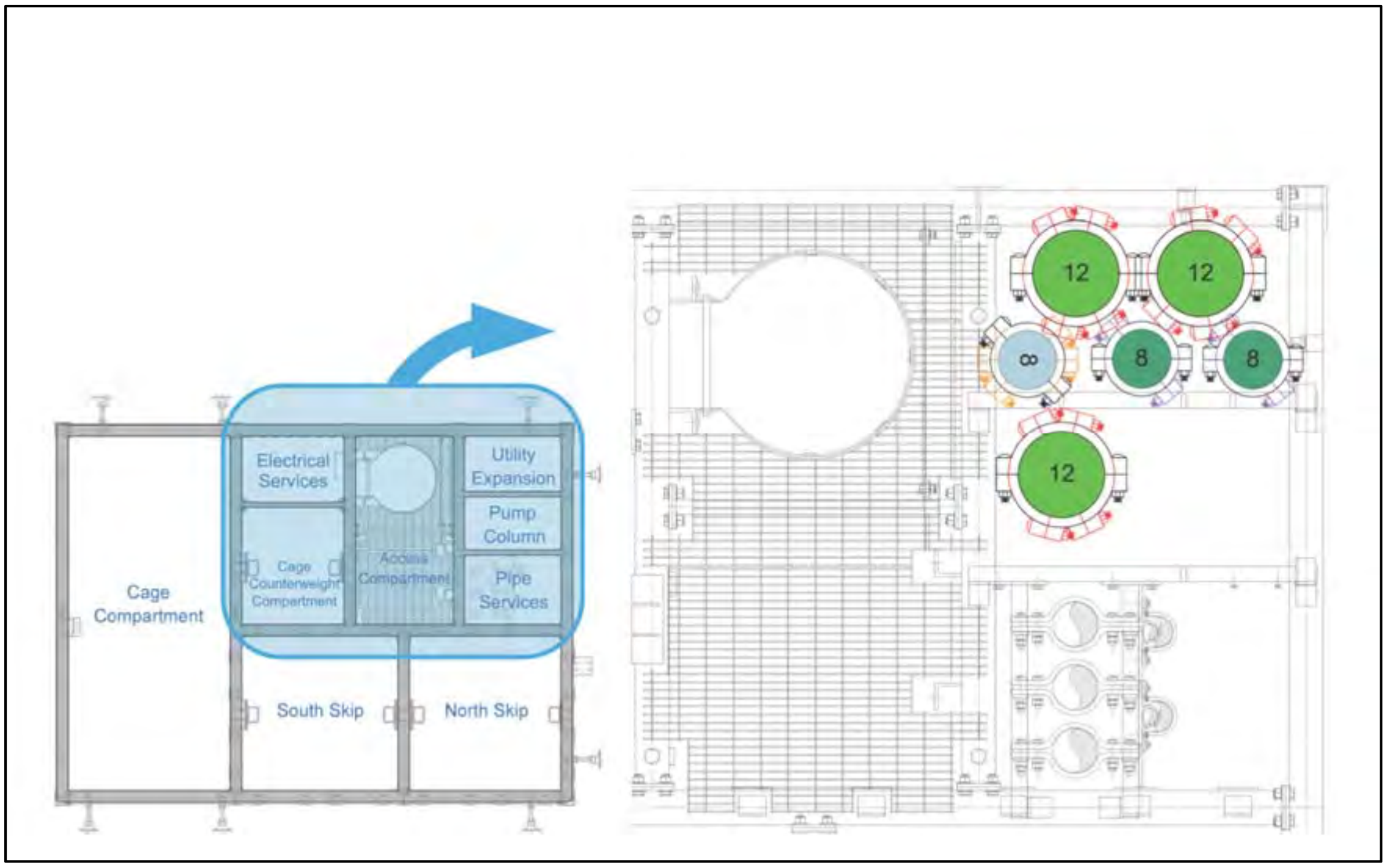

Figure 4-7: The framing of the Ross Shaft is shown on the Left. The utility area in the upper right corner contains the piping associated with the cryogenics system

Table 4-1: Piping between surface and cavern; description, duty, size and pressure required

| Description | Duty | [Number of Pipes] Size | Gas Pressure |
|---|---|---|---|
| GAr Transfer | During filling and emptying | [1] 8" SCH. 40 | 0.24 MPa |
| N2 Compressor Discharge | Continuous | [2] 8" SCH. 40 | 1.14 MPa |
| N2 Compressor Suction | Continuous | [3] 12" SCH. 40 | 0.19 MPa (down);<br>0.11 MPa (up) |

The facility is designed that the fresh air is drawn in through Ross and Yates shafts and the exhaust air is drawn out of mine through Oro Hondo shaft. The loss of mine ventilation for more than a few hours risks mine safety even without oxygen deficiency hazard (ODH) conditions. The response for unplanned loss of ventilation is to evacuate.

A preliminary ODH assessment for the piping in the Ross shaft has been done. If any of the pipes for the cryogenics system ruptured in the shaft, they would only not be able to reduce the oxygen content to a level that would create an oxygen deficiency concern.

There are four independent 85 kW (maximum capacity: 20% up above nominal) nitrogen refrigerators in the Central Utility Cavern area. The nitrogen refrigerator heat exchangers and expander sets are located at the east end of the Central Utility Cavern, see Figure 4-8. The heat exchangers (1.2 m diameter × 9.1 m long) will be in a horizontal orientation in order to fit them within the cavern. The liquid nitrogen produced by the refrigerators is stored in six horizontal 8.3 $m^3$ (1.2 m diameter × 11 m long) liquid nitrogen vessels per cryostat that are mounted in the Central Utility Cavern. These liquid nitrogen vessels feed the argon condensers that are connected to the cryostat. The returning nitrogen gas from the condensers is routed through the refrigerator heat exchangers and warmed to ambient temperature. The nitrogen gas is then boosted by four independent 120 kW compressors located in the Central Utility Cavern to 0.19 MPa and returned in the nitrogen suction piping in the Ross shaft.

Four argon condensers (0.8 m diameter × 2.0 m long) are located above each of the cryostats. The full power of the argon condensers is used during the initial cool down and filling phase of the cryostats in order to condense the gas argon transferred down the Ross shaft. The fill process is expected to take between 6 – 16 months. The fill time durations given here assume three refrigeration units available for the first and second cryostat fill and all four units available for the third and fourth cryostats (where steady state operations are maintaining liquid in the full cryostats).

Purification filters are located in the Central Utility Cavern. The filters (1.0 m diameter × 4.3 m high) contain dual media, a molecular sieve for removal of water and a copper coated catalyst media for oxygen removal. There are four gas filters used during the argon filling phase and four liquid filters for each cryostat. Associated with the filters, there will be regeneration equipment such as heaters, gas blowers, and a hydrogen generator also located in the Central Utility Cavern.

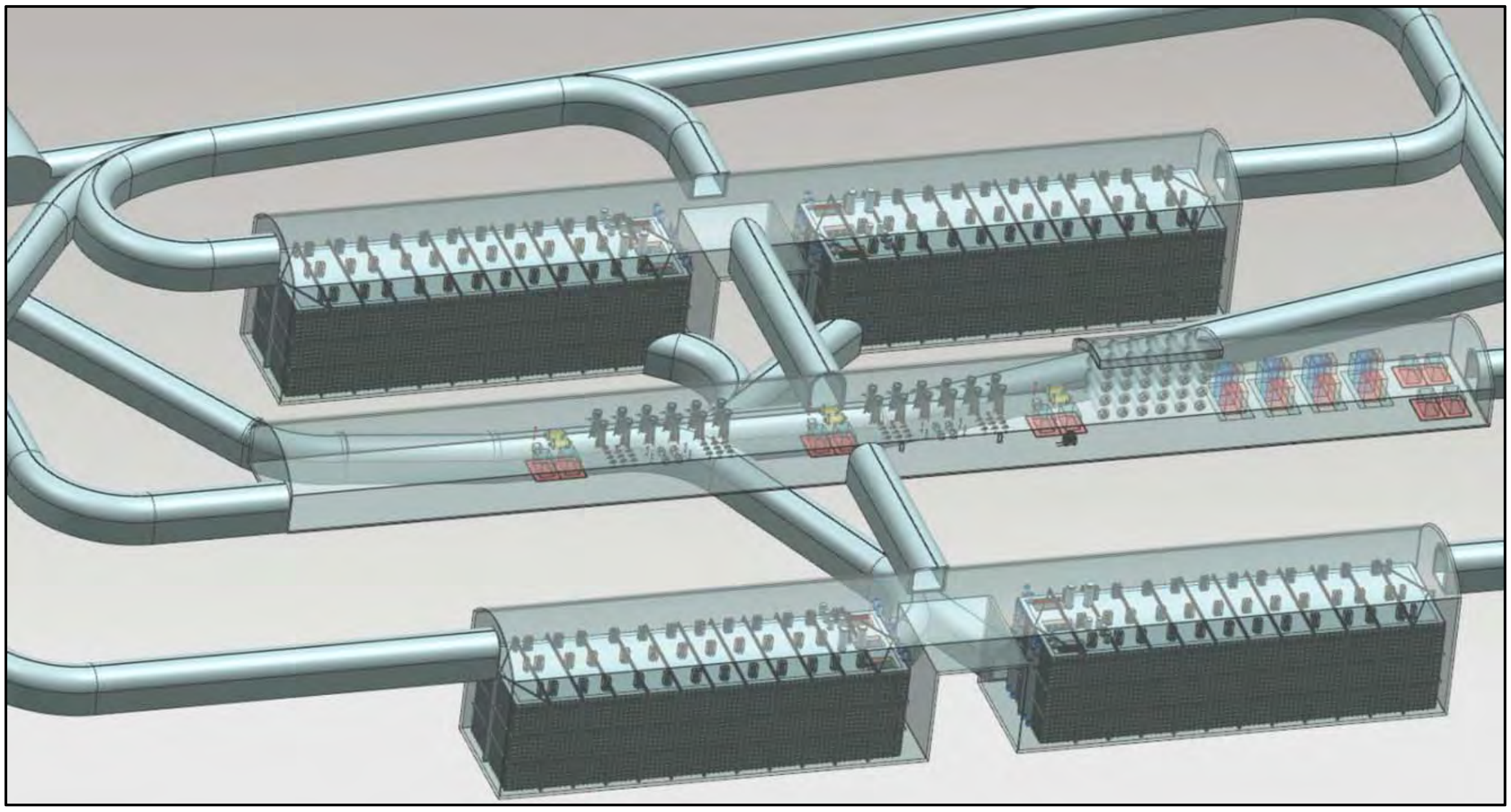

Figure 4-8: Isometric view of the underground cavern layout

# 4.5 Cryogenics System Process

The entire cryogenics system process is summarized with functions of every cryogenics system component in Figure 4-9. The major functions serving the cryostat are cryogen supply for cool down and fill, gas filtration, argon condensing, liquid circulation and filtration, and argon-purity analysis. The methods presented in this section are motivated by experience from the cryogenics systems of other LArTPC experiments, such as ICARUS, LAPD, and the prototype 35 ton membrane cryostat. The piping connections between major cryogenics system components are shown in Figure 4-10.

## 4.5.1 Cryostat Initial Purge and Cool Down

After cryostat construction and following installation of all scientific equipment, the cryostat will be cleaned, purged and cooled. Construction procedures leading up to this point will ensure that the completed cryostat does not contain debris and is free of all loose material that may contaminate the LAr.

Internal piping is positioned within the cryostat to support the purge and cool down procedure. Heavy argon vapor, which is a result of cooling down the membrane bottom with liquid, will promote purging after it rises from the base of the cryostat and is vented from the roof level. The LAr-supply pipework will have nozzles spaced along its length to distribute equal liquid-delivery flow rates across the bottom of the cryostat. The flow nozzles will be directed downward or to the side so that the injection velocity will not cause local vertical gas plumes or turbulent mixing but rather will spread across the bottom of the cryostat and produce a stable, upwardly advancing argon wave front. The vertical velocity of 1.2 m/hr for the gas purge includes a contingency for some level of turbulent mixing.

Main gas returns, used for pressure control, will be distributed along the cryostat roof. All nozzles and dead-end (stagnant) volumes located at the top of the cryostat will have gas-exhaust lines for the initial purge and for continuous sweep-purge of those volumes during normal operations. The sweep-purge during the initial stage of purging will be vented outside of the cavern. After all the air is expelled, except trace amounts, the gas returns will be routed to the recondensers and then to the filtration system before being returned to the cryostat. It is in this step that the cool down of the cryostat begins. The vent gas will be cooled down and returned to the cryostat in a temperature controlled scheme, allowing for the cool down requirements of the cryostat manufacturer and TPC to define these criteria. During this cool down the gaseous argon is continuously drawn into the condensers, liquefied, filtered and returned to the cryostat as purified argon. This recirculation further purifies the gas. As the temperature of the cryostat/TPC mass approaches the liquid temperature of LAr, the fill process of the cryostat will begin.

### 4.5.1.1 Initial Purge

Argon piping will be isolated, evacuated to less than 0.1 mbar absolute pressure and backfilled with high-purity argon gas. This cycle will be repeated several times to reduce contamination levels in the piping to the ppm level. The reference-design choice for removing air from the membrane cryostat is argon flow/piston-purge, introducing the heavier argon gas at the bottom of the cryostat and removing the exhaust at the top. The bottom field cage (part of the TPC) serves an additional role as a flow diffuser during the initial purge. A matrix of small holes in the field cage, approximately 10 mm diameter at a 50 mm pitch, will provide a uniform flow.

The flow velocity of the advancing argon-gas volume will be set to 1.2 m/hour. This velocity is high enough to efficiently overcome the molecular diffusion of the air so that the advancing pure argon-gas wave front will displace the air rather than just dilute it. A 2D Computational Fluid Dynamics (CFD) simulation of the purge process on a 5 kt fiducial-mass cryostat for LBNE shows that after 20 hours of purge time and 1.5 volume changes, the air concentration will be reduced to less than 1%. At 40 hours of elapsed time and three volume changes, the purge process is complete with residual air reduced to a few ppm. This simulation includes a representation of the perforated field cage at the top and bottom of the detector. The cathode planes are modeled as non-porous plates although they will actually be constructed of stainless-steel mesh.

The CFD model of the purge process has been verified in multiple arrangements: (1) in an instrumented cylinder of 1 m diameter by 2 m height, (2) Liquid Argon Purity Demonstrator (LAPD), a vertical cylindrical tank of 3 m diameter by 3m height, taking gas-sampling measurements at varying heights and times during the purge process, (3) within the 35 ton membrane cryostat, a prototype vessel built for LBNE in 2013, of which the results are found at [20], and (4) within MicroBooNE cryostat, a horizontal cylindrical tank of 3.8 m diameter by 12.2 m length.

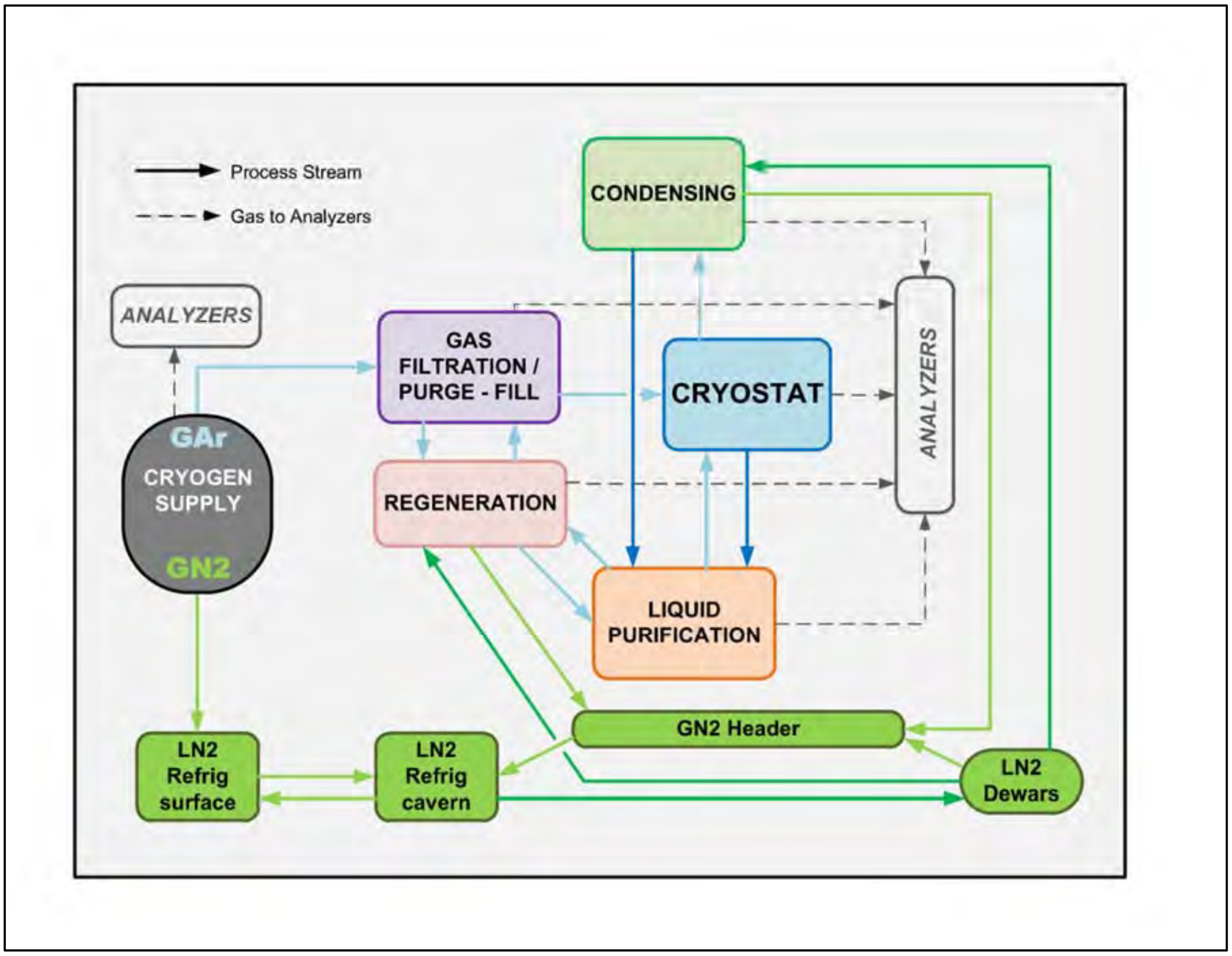

Figure 4-9: Cryogenics system functions

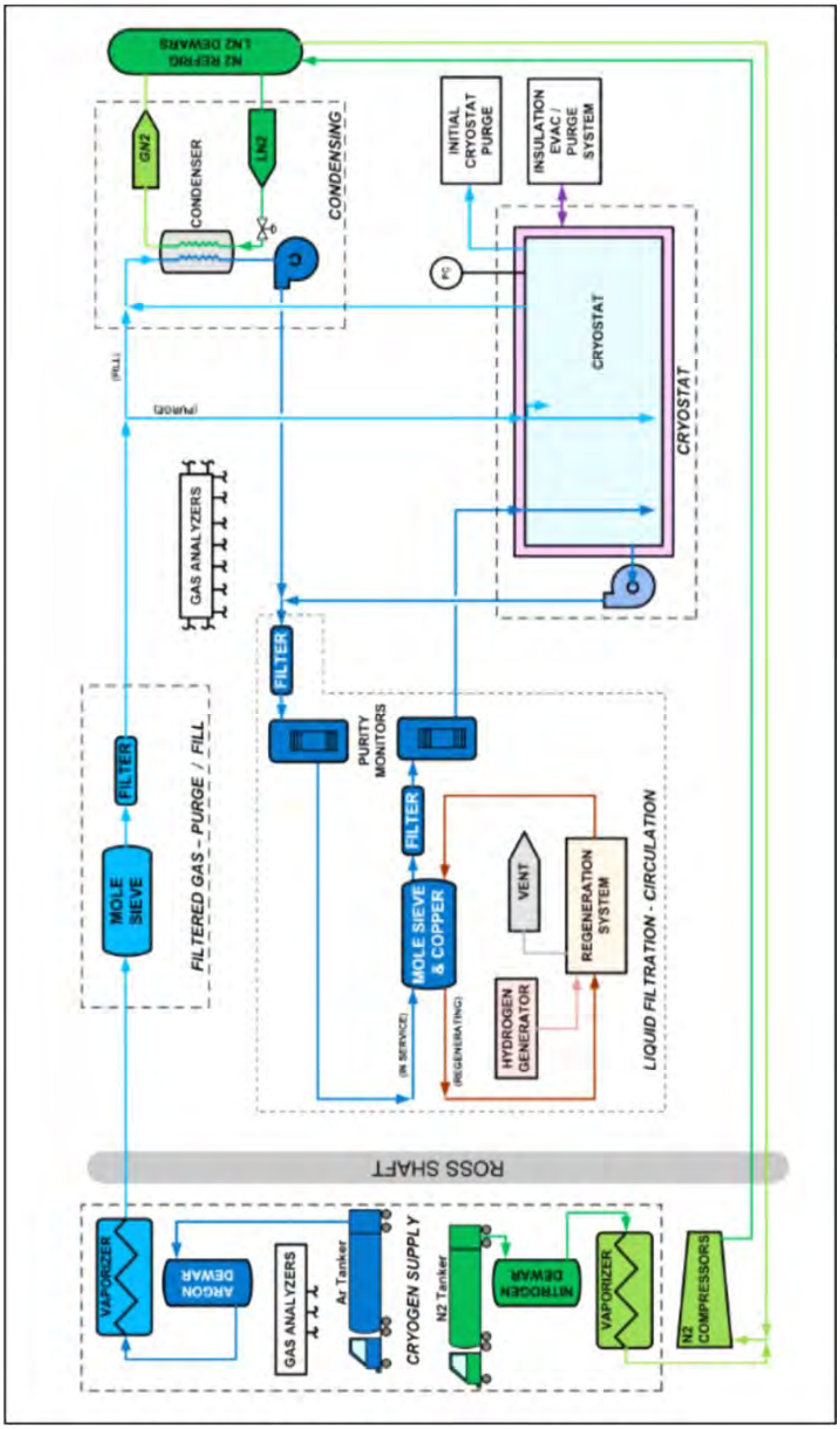

Figure 4-10: Cryogenics system block flow diagram

### 4.5.1.2 Water Removal via Gas Flow

Water and oxygen will continue to be removed from the system for several days following the initial purge. At this stage, the gas will be routed to the recondensers, filtered and recirculated to the cryostat. Each cryostat contains five tons of FR4 circuit-board material and a smaller inventory of plastic-jacketed power and signal cables. These somewhat porous materials may contain as much as 0.5% water by weight. Water-vapor outgassing from these materials will be entrained in the gas flow exiting the top of the cryostat and will be removed from the gas stream by filters. Adsorbed water will also be removed from the metallic inner surfaces of the cryostat and piping system. Water deep within porous materials will remain; this is not a problem since the water diffusion rate in FR4 at room temperature is already quite low (0.3 $\mu m^2/s$) and the FR4 assemblies are relatively thick (1 cm).

### 4.5.1.3 Initial Cool Down

The liquefaction rate in the recondensers is increased so that liquid argon can be returned to the cryostat. This purified LAr will be distributed near the bottom of the cryostat to cool down the cryostat in a controlled spray. The boil-off gas will flow through the volume of the cryostat, recondensers and liquid-filtration system. Simulation has shown that the liquid cool down method can be controlled to stay within the available recondenser capacity. The required cooling rate is determined by the maximum stress that detector components can tolerate. For example, the 150 μm APA wires will cool much more rapidly than the APA frames. A mass flow control system with temperature-monitoring system will be used to control the temperature difference across the cryostat. The exact temperature difference required is yet to be determined; it will be based on input from the cryostat designer and the requirements of the TPC components and structure.

## 4.5.2 Liquid Argon Receipt

Each 10 kt fiducial mass LAr detector cryostat will hold an inventory of 17.1 kt of liquid argon. Considering that some product will also be lost in transit, approximately ($17.1 + \alpha$) kt of LAr will need to be procured to fill the first cryostat. Planning the logistics and supply of LAr to the facility requires the consideration of the following issues:

- Total capacity of commercial air-separation plants within freight distance of the facility (the peak delivery potential)
- Extent of boil-off that will occur in transit (that is the biggest contribution to determine $\alpha$)
- Number of vehicle movements required and their impact on the local community
- Costs and benefits associated with stockpiling LAr at the facility ahead of commencing the purge, cool down and fill procedure
- Provision of a temporary air-separation plant at the facility to generate liquid argon
- Availability and cost associated with the delivery of high-purity LAr as opposed to lower-quality, commercial-grade argon combined with on-site, coarse purification

The current total argon capacity in the United States is approximately 5.2 kt per day, whereas the demand is about 4.7 kt per day, which means 90% capacity utilization in 2015. Argon demand slowed down during recession (2008 – 2009), but has been recovering strongly since 2010, especially in electronics and welding industries, in a pace faster than capacity growth. Some capacity was taken offline in recession and has not come back. The trend of growing demand at a rate of 3.4% per year, faster than capacity, is expected to continue for at least the next five years and will cause argon supply to be tight and prices to rise. Therefore, creating clusters of existing argon capacity that can provide argon to LBNF, rather than using one supplier, or identifying new argon capacity (preferably closer to SURF site) can be options to consider for economic and reliable supply. At the time of market research [22], air separation plants (ASPs) in Chicago and the Gulf Coast were identified as the most important supply source, but this option will require significant delivery cost.

The standard grade specification for argon is a minimum purity of 99.995%, allowing a maximum concentration of 5.0 ppm for $O_2$ and 10.5 ppm for $H_2O$. This is designated as Grade 4.5 in the gas-supply industry. Requiring higher-purity product would significantly reduce the volume of product available to the experiment, increasing cost and pushing out the schedule. Therefore, the standard product will be procured from multiple vendors.

The most efficient mode of argon delivery is over-the-road tank truck with a maximum capacity of 18.7 metric ton (MT). The expected number of such deliveries per cryostat is about 1000 over six to sixteen months. Rail delivery is not cost-effective as there are no rail spurs leading to the site. This mode would require transfer of product from rail tanker to a tank truck, introducing cost that exceeds the benefit.

Surface facilities for offloading $LN_2$ and LAr road tankers are required. It will be necessary to procure approximately four trailer loads of liquid nitrogen (about 40 tons) for the initial filling of the $LN_2$ refrigeration dewar and charging of a single refrigeration plant. Vehicle access and hard-surfaced driving areas are required adjacent to the $LN_2$ and LAr dewars. An interim LAr storage dewar will hold the contents of several road tankers in order to minimize off-loading time. Road tankers will connect to a manifold and will use their on-board pumps to transfer the LAr to the storage dewar. Each tanker will be tested to ensure that the LAr meets the purity specification. The LAr will be stored in the surface dewar and vaporized before transporting by pipe feed to the underground cavern for liquefaction.

#### 4.5.2.1 Cryostat Filling

Liquid argon will be delivered to the cryostat through the cryostat-filling pipework. Argon will be piped to the cavern in gas form from the surface and condensed/liquefied via the $LN_2$ exchange in the condenser units. The filling process will take place over many weeks due to the delivery schedule of liquid argon described in the previous section and the need to condense gaseous argon. Liquid-argon purification can begin once the liquid depth reaches about 0.5 m in the cryostat. At this depth, the recirculation pumps can be safely turned on and it will direct up to 93 $m^3$/hr (411 gpm) of liquid argon through the purification system.

## 4.5.3 Argon Reliquefaction and Pressure Control

The high-purity liquid argon stored in the cryostat will continuously be evaporating due to the unavoidable heat ingress. The argon vapor (boil-off gas) will be recovered, chilled, recondensed and returned to the cryostat. A closed system is required to prevent the loss of the high-purity argon.

During normal operation, the expected heat ingress of approximately 69.9 kW to the argon system will result in an evaporation rate of 1571 kg/hr and expanding in volume by a factor of 200 when it changes from the liquid to vapor phase. This increase in volume within a closed system will, in the absence of a pressure-control system, raise the internal pressure.

In LBNF, argon vapor will be removed from the top of the cryostat through the cryogenic feedthroughs. As the vapor rises, it cools the cables and feedthrough, thereby minimizing the outgassing. The exiting gaseous argon will be directed to a heat exchanger (a recondenser) in which it is chilled against a stream of liquid nitrogen and condensed back to a liquid. As the argon vapor cools, its volume reduces and in the absence of pressure control further gas would be drawn into the heat exchanger, developing a thermal siphon. Therefore, a pressure-control valve on the boil-off gas lines will control the flow to the recondenser to maintain the pressure within the cryostat at 0.113 MPa ± 0.008 MPa. The liquid nitrogen stream (serving as the coolant for the recondenser) will be supplied from a closed-loop $LN_2$ refrigeration plant. The commercial refrigeration plant uses compression/expansion and heat rejection to continuously liquefy and reuse the returning nitrogen vapor. The estimated heat loads within the cryostat are listed in Table 4-2.

Each cryostat has a dedicated nitrogen-refrigeration plant and all four will be used for the initial cool down and filling of the cryostats if possible because of the large volume of gas that must be cooled from 300 K to liquid argon temperature (88.3 ± 1 K). Further, each cryostat will have four 85 kW condensers to provide the cooling power needed during initial cool down and filling operations where warm GAr is cooled and reliquefied to fill the cryostat. After filling, only one condenser is needed with the others providing redundancy. This will ensure high availability of the recondensing system, minimize the need to vent high-purity argon and allow down-time for maintenance of the recondensers and the refrigeration plants.

Table 4-2: Estimated heat loads within the cryostat

| Item | Heat Load (kW) |
|---|---|
| Insulation Heat Loss | 32.1 |
| Electronics Power | 23.7 |
| Recirculation-pump Power | 10.4 |
| Misc. Heat Leaks (Pipes, Filters, etc.) | 3.7 |
| Total | 69.9 |

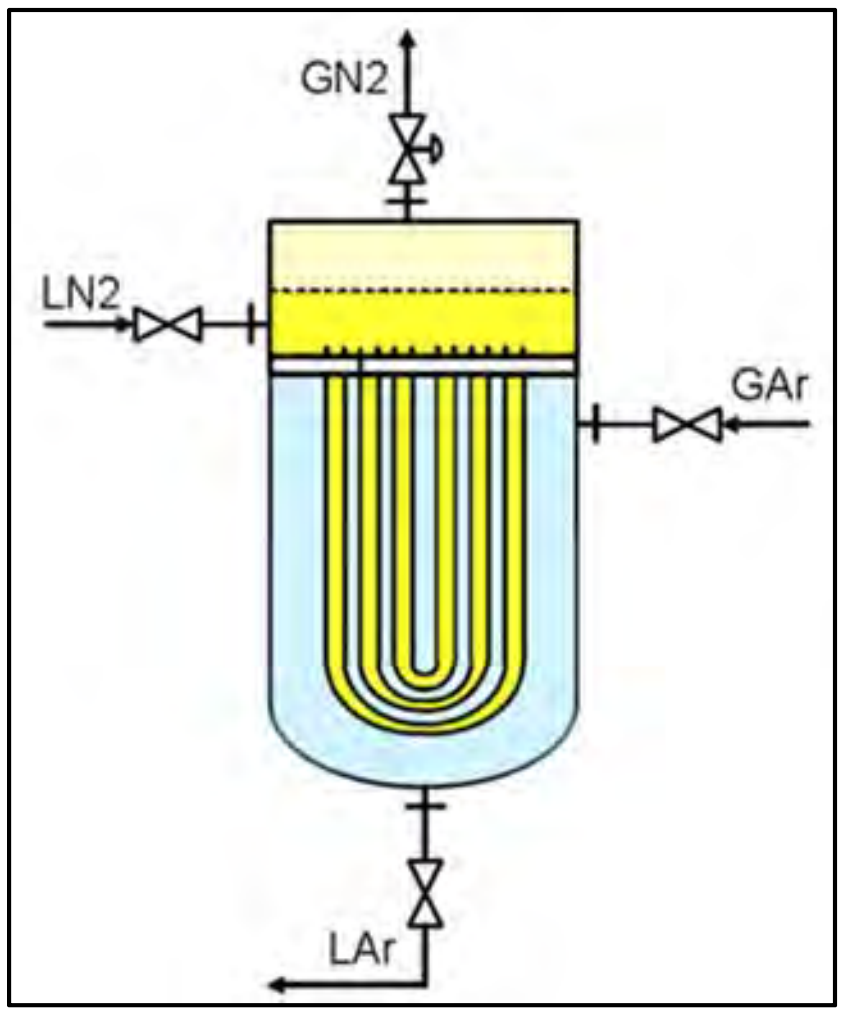

Figure 4-11: Liquid argon recondenser

## 4.5.4 Argon Purification

The cryostat is to be designed with side penetrations below the liquid level for external recirculation pumps used to continuously filter the cryostat's LAr. Figure 4-12 illustrates this mechanism: a vertical pump inserted into vacuum insulated pump wells, and external to the cryostat. The pump suction will be located at a distance below the lowest liquid level to prevent cavitation and vapor-entrapment. The base of the pump tower will be placed close to the cavern floor and below the inner floor of the cryostat. The pump suctions could be staggered at different elevations to allow flexibility in drawing liquid from different elevations. Vertical cryogenic pumps are supplied by manufacturers such as Ebara and Carter Cryogenic Products.

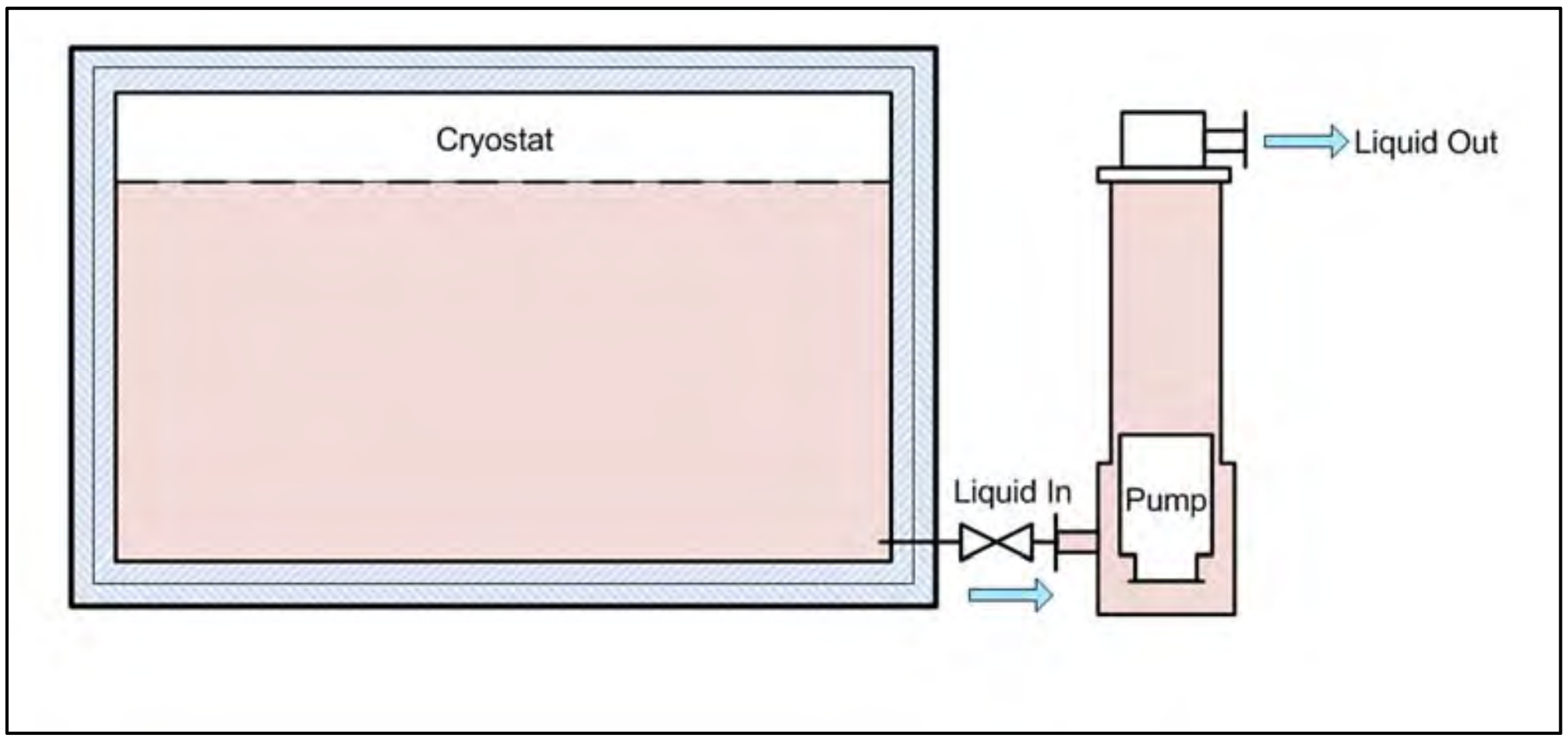

Figure 4-12: Liquid argon recirculation mechanism using external pumps

The required flow rate of liquid argon to be sent for purification is expected to decrease over time. The initial maximum flow rate will be 93 $m^3$/hr (411 gpm). The liquid-argon volume in one cryostat will turn over every 5.5 days at this rate. Longer term, the rate will decrease to 46 $m^3$/hr with a turn-over rate of 11 days. As a point of comparison, ICARUS T600 has a maximum turn-over rate of eight to ten days. The purification skids are located in the Central Utility Cavern. The multiple-pump arrangement will provide a very high level of redundancy, which will extend the maintenance-free operating period of the cryostat.

The purification system consists of two types of filter vessels containing molecular-sieve and copper media filters. The filter is 1.0 m in diameter by 4.3 m tall. The filters are sized to provide effective media usage at low pressure drop (2 kPa or 0.3 psi) over the expected range of flow rates. One filter is for gas filtration during filling; the other type is for liquid filtration. After filling is complete, the gas filter can be repurposed for cryostat liquid filtration.

The cryostat liquid argon inventory is circulated through a purification filter to achieve and maintain the required purity. The purification filter, containing molecular sieve media to remove water and copper media to remove oxygen, will become saturated. The nearly saturated purification filter is regenerated to vent the contaminants. The liquid argon flow is switched to another purification filter for uninterrupted filtration.

A purity monitor after the purification filter will monitor the filter effectiveness. Purity monitors measuring electron lifetime will also be in the LAr bath and resident in the cryostat. It is a requirement that purity levels reach < 100 ppt oxygen equivalent to match the required electron lifetime of >3 ms for the detector.

The regeneration of a filter is done in several steps. A saturated purification filter is first warmed with heated argon gas to an elevated temperature driving the captured water into the gas. Hydrogen gas is generated and mixed with the circulating argon gas up to 1.5% hydrogen by volume. The hydrogen reacts with the oxygen and makes water that is also released into the circulating argon gas. Argon gas is vented to purge water from the hot circulating gas.

The hot filter full of regenerated media is cooled by circulating chilled argon gas. The circulating argon gas is chilled down to cryogenic temperatures by circulating argon gas chilled by a heat exchanger with liquid nitrogen coolant. This completes the regeneration steps for a purification filter. The filter is now ready to be switched into service or held cold until needed. Two spare purification filters are used with separate heating and cooling loops to reduce the usage rate of electricity and liquid nitrogen. This also reduces the stresses on heat exchangers by decreasing their temperature swings.

## 4.5.5 Pressure Control

### 4.5.5.1 Normal Operations

The pressure-control valves are sized and set to control the internal cryostat pressure under normal operating conditions to the nominal design pressure of 0.113 MPa. Fluctuations between 0.105 MPa (50 mbarg) and 0.121 MPa (200 mbarg) are intended to be the normal operating range. Ten percent excursions above or below these levels will set off alarms to alert the operator to intervene. Further excursion may result in automatic (executive) actions. These actions may include stopping the LAr circulation pumps (to reduce the heat ingress to the cryostat), increasing the argon flow rate through the recondenser, increasing the $LN_2$ flow through the recondenser vessel, powering down heat sources within the cryostat (e.g., detector electronics). Eventually, if the pressure continues to rise, it will trigger the pressure-relief valves to operate. Table 4-3 gives important pressure values.

Table 4-3: Important pressure values

| Pressure | Value |
|---|---|
| Vessel ullage maximum operating pressure | 0.121 MPa, 200 mbarg, 2.9 psig |
| Cryostat Design Pressure; Relief valve set pressure | 0.135 MPa, 350 mbarg, 5.1 psig |

The ability of the control system to maintain a set pressure is dependent on the size of pressure deviations (due to changes in flow, heat load, temperature, atmospheric pressure, etc.) and the volume of gas in the system. The reference design has 0.66 m of gas at the top of the cryostat. This is 5% of the total argon volume and is the typical vapor fraction used for cryogenic storage vessels. Reaction time to changes in the heat load is slow, on the order of an hour. At the expected heat-load rate of 69.9 kW, and for an isolated or un-cooled cryostat, the rate of pressure rise would be 393 mbarg (5.7 psi) per hour. Two redundant pressure control valves will maintain the required pressure range, each sized to handle at least 1300 kg/hr of argon flow to the recondenser to handle the cooling and reliquefaction of warm GAr during cryostat filling.

### 4.5.5.2 Overpressure Control

In addition to the normal-operation pressure-control system, it is planned to provide a cryostat overpressure-protection system. This must be a high-integrity, automatic, failsafe system capable of preventing catastrophic structural failure of the cryostat in the case of excessive internal pressure.

The key active components of the planned system are pressure-relief valves (PRVs) located on the roof of the cryostat that will monitor the differential pressure between the inside and the outside of the cryostat and open rapidly when the differential pressure exceeds a preset value. A pressure-sensing line is used to trigger a pilot valve, which in turn opens the PRV. A pressurized reservoir of power fluid is provided to each valve to ensure that the valves will operate under all deviations and/or shutdown scenarios. The PRVs are self-contained devices provided specially for tank protection; they are not normally part of the control system.

The installation of the PRVs will ensure that each valve can periodically be isolated and tested for correct operation. The valves must be removable from service for maintenance or replacement without affecting the overall containment envelope of the cryostat or the integrity of the over-pressure protection system. This normally requires the inclusion of isolation valves upstream and downstream of the pressure-relief valves and at least one spare installed relief valve (n + 1 provision).

When the valves open, argon is released, the pressure within the cryostat falls and argon gas discharges into the argon vent riser. The valves are designed to close when the pressure returns below the preset level.

#### 4.5.5.3 Vacuum-Relief System

The cryostat vacuum-relief system is a high-integrity, automatic, failsafe system designed to prevent catastrophic structural failure of the cryostat due to low internal pressure. The vacuum-relief system protects the primary membrane tank. Activation of this system is a non-routine operation and is not anticipated to occur during the life of the cryostat.

Potential causes of reduced pressure in the cryostat include operation of discharge pumps while the liquid-return inlet valves are shut, gaseous argon condensing in the recondenser (a thermo-siphon effect) or a failure of the vent system when draining the cryostat. Vacuum-relief valves are provided on LNG/LPG storage tanks to protect the structure from these types of events.

The key active components of this additional protection system are vacuum-relief valves located on the roof of the cryostat that will monitor the differential pressure between the inside and the outside of the cryostat and open when the differential pressure exceeds a preset value, allowing cavern air to enter the cryostat to restore a safe pressure.

### 4.5.6 $LN_2$ Refrigeration System

Four commercial $LN_2$-refrigeration plants will be procured for LBNF. After achieving the required purity and completing the initial fill, each cryostat will have a dedicated $LN_2$ plant for steady-state operations. The plants will be located in the Central Utility Cavern at the 4850L. Each will be a closed-loop system supplying $LN_2$ to the argon recondenser. The nominal rating of the quoted refrigerators is in the range of 71 kW.

Two-phase nitrogen is delivered from the cold end of the refrigerator into a farm of $LN_2$ storage vessels with a total capacity of 50 $m^3$. Pure liquid is withdrawn from the $LN_2$ storage vessels and is supplied via a transfer line to a pressure-reducing valve and phase-separator tank also located within the Central Utility Cavern. $LN_2$ is then withdrawn from the bottom of the phase-separator tank, at a pressure of 2.0 bar and temperature of 84 K, and directed to the recondenser. This results in a 5 K temperature difference relative to the 89 K argon recondenser temperature. The six 8.3 $m^3$ $LN_2$ vessels will allow for greater than forty hours of refrigeration time. This time window is adequate to cover most power outages, refrigerator performance problems and refrigerator switch-overs.

The refrigeration system operation, illustrated in Figure 4-13, is based on a screw compressor package and three turbo expanders. This system is expected to be capable of running continuously for at least a year, and then require only minor servicing. The system will be equipped with automatic controls and a remote monitoring so that no operator will be required during normal operation. Estimated maximum power requirement is 1500 hp (1119 kVA), not taking into account the power generated by the expanders. The LBNF reference design places the nitrogen compressor in a surface-level equipment building. A closed-loop water system with evaporative cooling tower removes heat from the compressor.

Compression is carried out at close-to-ambient temperature. A compressor after cooler is provided to reject heat.

The fluid is next routed to a 'cold box' consisting of four heat exchangers. This series of exchangers provides staged heat transfer from a cooling nitrogen stream to a warming one. The expanders are connected between the heat exchangers to progressively reduce the pressure of the cooling nitrogen stream to isentropically reduce the pressure and temperature of the nitrogen stream, eventually leading to a large liquid-nitrogen fraction at the coldest end of the cold box.

The main cold box shell is 1.22 m (4 ft) in diameter and 8.2 m (27 ft) tall. The expanders are adjacent to the cold box at three elevations and extend about 1 m to the side of the cold box shell. The cold box will weigh 5670 kg. The compressors are located at the surface inside an equipment building. The compressor skid (frame) is 4.3 m long, 1.8 m wide and 2.7 m tall and will weigh approximately 3630 kg.

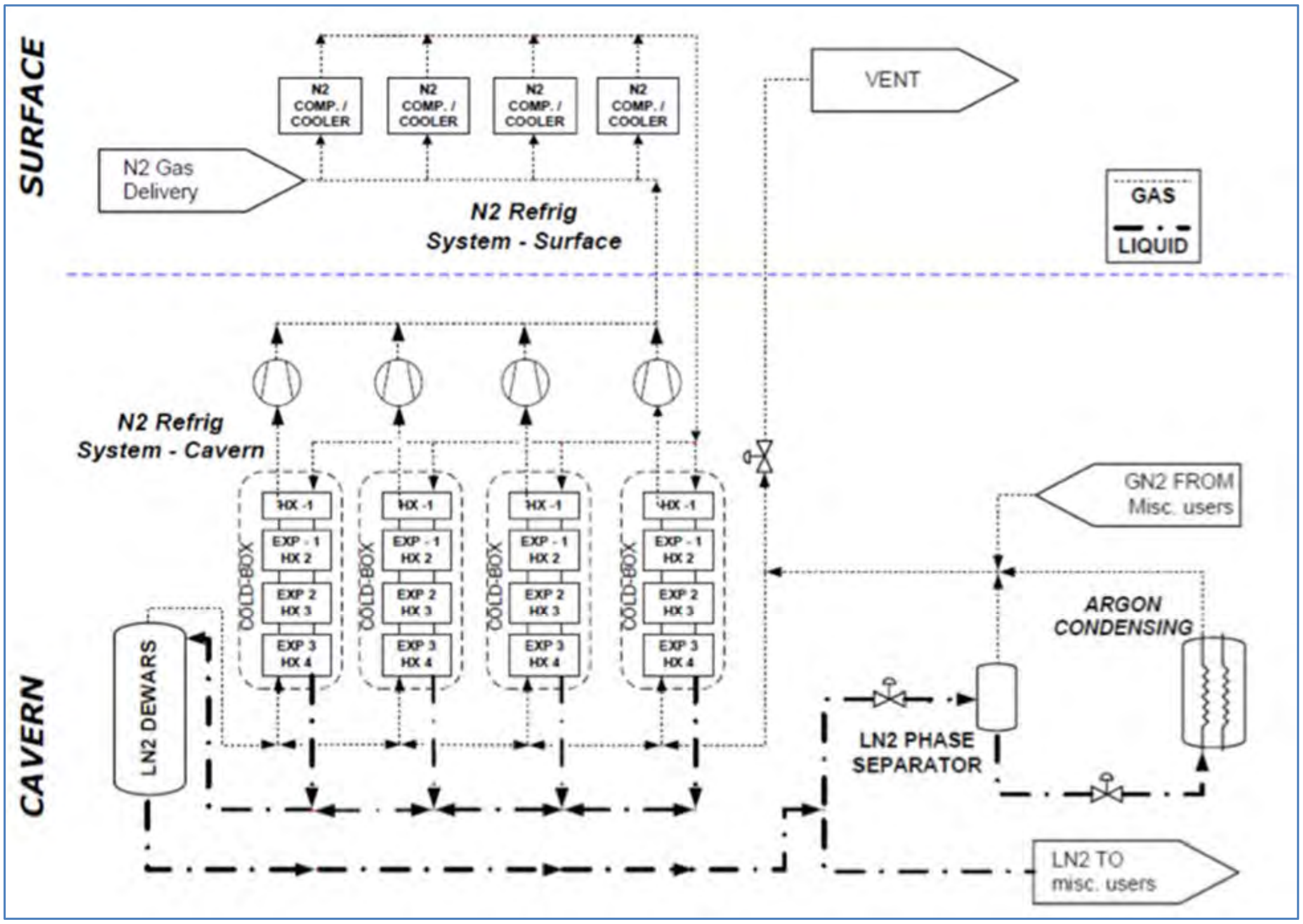

Figure 4-13: Nitrogen refrigeration plant flow diagram

### 4.5.7 Liquid Argon Removal

Although removing the LAr from cryostats is not in the scope of project, it is part of final dispostion of the facility components. A method to remove LAr from the cryostats at the end of life has been conceptualized here. The LAr is assumed to be resold to suppliers at half the supply cost.

It is foreseen that storage dewars, sized for the task, can be carried up and down the skip compartments of the shaft initially used to haul up waste rock from the mine. Because there are two skip compartments, an empty vessel can simultaneously be lowered to the 4850L in one skip while a full vessel is raised to the surface in the other. The physical dimensions of skip compartment will accommodate the dewar size upto about 3000 L at maximum inside. If the vessel is pressurized to 50 psig, it will contain roughly 4.2 tons of LAr. The pumps already present at the cryostats can be used to transfer the LAr from cryostat to the storage dewar.

With assumption of having crews working concurrently at the surface and at the 4850L, one optimized conveyance cycle can be fit in approximately 34 minutes, including 4 minutes of skip transit time up and down. This will allow for at least 31 cycles in an 18 hour day, corresponding to 130.2 tons per day of LAr delivered to the surface. To empty each cryostat will require about 126 days.

It is expected that 15.6 kts of LAr per cryostat can be recovered in this process, or better than 91% of the total, considering the amount of liquid below the lowest liquid height that can not be taken out using pumps and 5% loss in the transfer of remaining liquid to the dewars.

# 5 Near Site Facilities: Near Site Conventional Facilities

## 5.1 Overview

This chapter presents the Near Site Conventional Facilities (NSCF) required to house the Beamline and near detector technical facilities at the Fermilab site (also referred to as the Near Site). Figure 5-1 shows the overall NSCF layout on the Fermilab site. The NSCF provides the infrastructure to house the Beamline technical systems from the extraction point, through the target and absorber to the near detector, as well as support buildings for the underground facilities.

The planning and development of the Near Site facilities are summarized in the following sections. For more information on NSCF, refer to *Annex 3B: CF at the Near Site* [23].

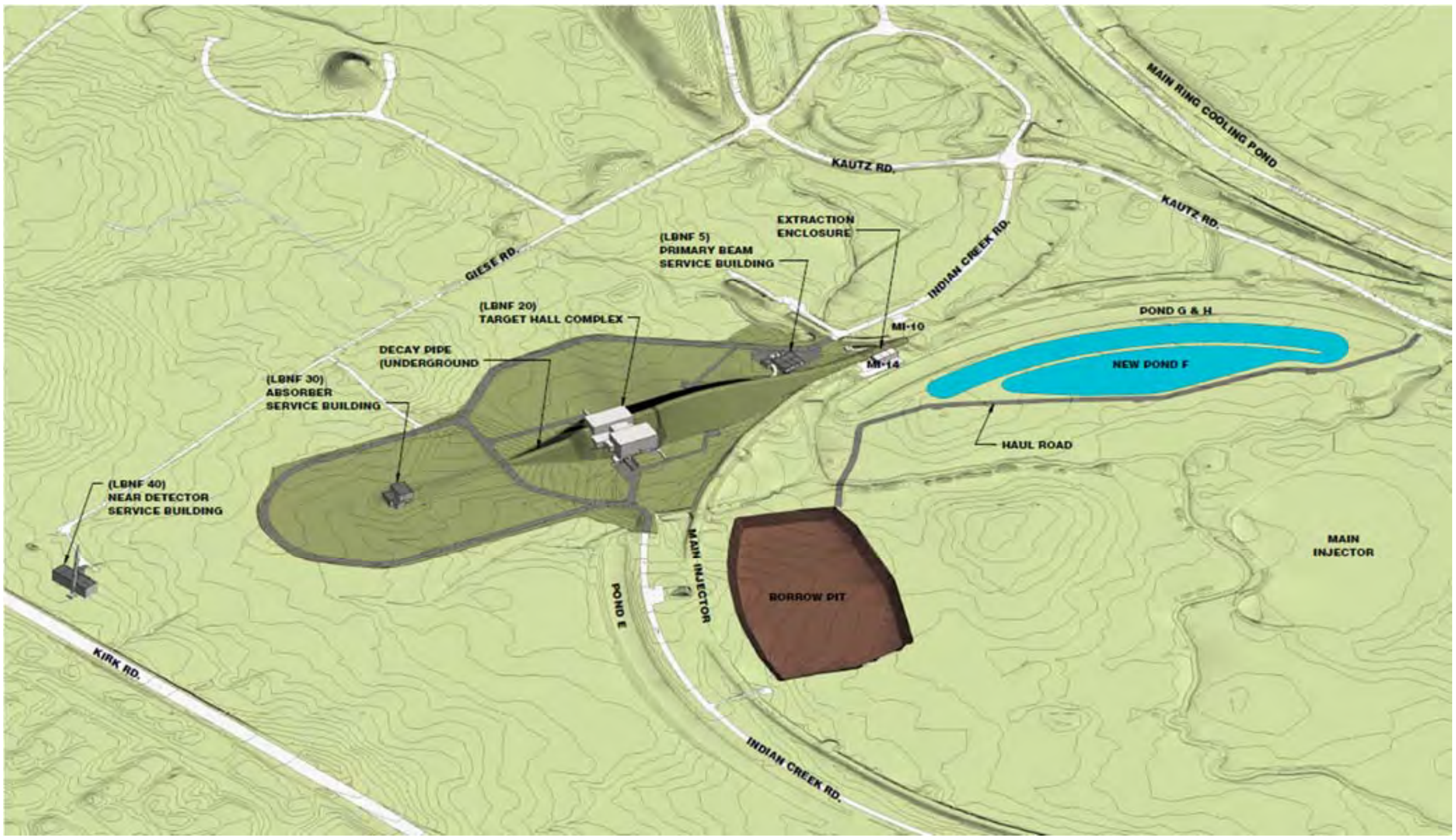

Figure 5-1: High level view of the Near Site facilities

# 5.2 Existing Site Conditions

The LBNF Project area is located in the western portion of the Fermilab site in Batavia, Illinois. The sections below describe the known and anticipated surface conditions at the site and also include site geology, groundwater conditions, and natural gasses.

## 5.2.1 Surface Development, Topographic and Environmental Conditions

The site is partially developed with existing surface and underground structures for the support of ongoing research at the laboratory. Existing underground structures include building foundations, buried utilities, shallow tunnel enclosures constructed by cut-and-cover methods, the associated remnants from previously constructed braced excavation structures, and the existing Neutrinos at the Main Injector (NuMI) tunnel, which was excavated in the same rock units that LBNF underground enclosures will encounter.

Existing facilities on or adjacent to the Fermilab property that will interface with or constrain the development of the Project are the Main Injector, Kautz Road, Indian Creek Road (also known as the Main Injector Road), the Main Injector Cooling Pond F, and Kirk Road.

The site surface topography is predominantly flat with areas of prairie grass, heavy brush, woodlands, wetlands and developed sites. Surface elevations within the Project area range from about 740 ft to 760 ft above mean sea level (MSL). The topography in the Project area will not be an impediment to the development of the construction sites or the use of standard heavy equipment for construction.

## 5.2.2 Overview of Site Geology

Subsurface conditions at the Near Site comprises glacial, glaciofluvial and glaciolacustrine deposits, along with flat lying bedrock strata of the middle to lower Silurian period. In descending order, the Silurian rock formations include the Markgraf and Brandon Bridge Members of the Joliet Formation, the Kankakee Formation and the Elwood Formation.

Glacial processes during the Wisconsin glaciation resulted in the deposition of a thick blanket of glacial tills, lacustrine silts and clays, and outwash sands and gravels across the Project area. The total thickness of these overburdened sediments in the Project area ranges from about 50 ft to 85 ft. The majority of the sediments are over consolidated glacial till deposits consisting of silt, sand, gravel, cobbles and boulders in a predominantly clay matrix.

Additional descriptions of subsurface materials, geologic profiles and boring logs along the Near Site Project alignment are provided in the *LBNE Geotechnical Data Report* [24] prepared by AECOM, dated October 2013, the LBNE Site Investigation Geotechnical Engineering Services Report [5] prepared by Groff Testing Corporation, dated February 26, 2010. An additional independent evaluation is documented in the *Geotechnical Investigation*, *Heuer Review* [25].

### 5.2.3 Overview of Site Groundwater Conditions

The groundwater regime within the Project area is controlled by the glacial drift aquifer, bedrock aquifers and aquitards. The glacial drift aquifer can be categorized as buried and basal drift aquifers. The buried aquifers occur as isolated lenses or layers of permeable silt, sand and gravel outwash, separated by relatively impermeable clayey and silty tills. The basal aquifers are associated with localized lenses or layers of permeable silt, sand and gravel.

The upper bedrock aquifer consists of the upper weathered and jointed bedrock regardless of stratigraphy or lithology but dominantly the Silurian dolomite formations. The aquifer has a low primary permeability and a much higher secondary permeability consisting of local flow systems mostly associated with discontinuities in the rock mass. The upper bedrock aquifer is a groundwater source for many private and public wells in the Batavia area, including the main water supply for Fermilab. The potentiometric surface of the upper bedrock aquifer is approximately 10 ft below the bedrock surface; however, groundwater elevations may vary.

## 5.3 The Facility Layout

The LBNF Conventional Facilities on the Near Site consist of six functional areas – three surface buildings, a near-surface, shallow-buried Target Hall Complex located in an embankment constructed of engineered fill, and three underground facility enclosures. Construction will be executed and packaged in a logical sequence based on programmatic and funding driven limitations.

Figure 5-2 shows the Project site aerial view with LBNF facilities highlighted. The Project limits are bounded by Giese Road to the north, Kautz Road to the east, Main Injector Road to the south, and Kirk Road to the west. The Beamline is shown in red in Figure 5-2. The four surface and near-surface buildings consist of:

- Primary Beam Service Building (LBNF-5)
- Target Hall Complex (LBNF-20)
- Absorber Hall Service Building (LBNF-30)
- Near Detector Service Building (LBNF-40)

The four underground facilities consist of:

- Beamline Extraction Enclosure and Primary Beam Enclosure
- decay pipe
- Absorber Hall, Muon Alcove and support rooms

- Near Detector Hall and support rooms

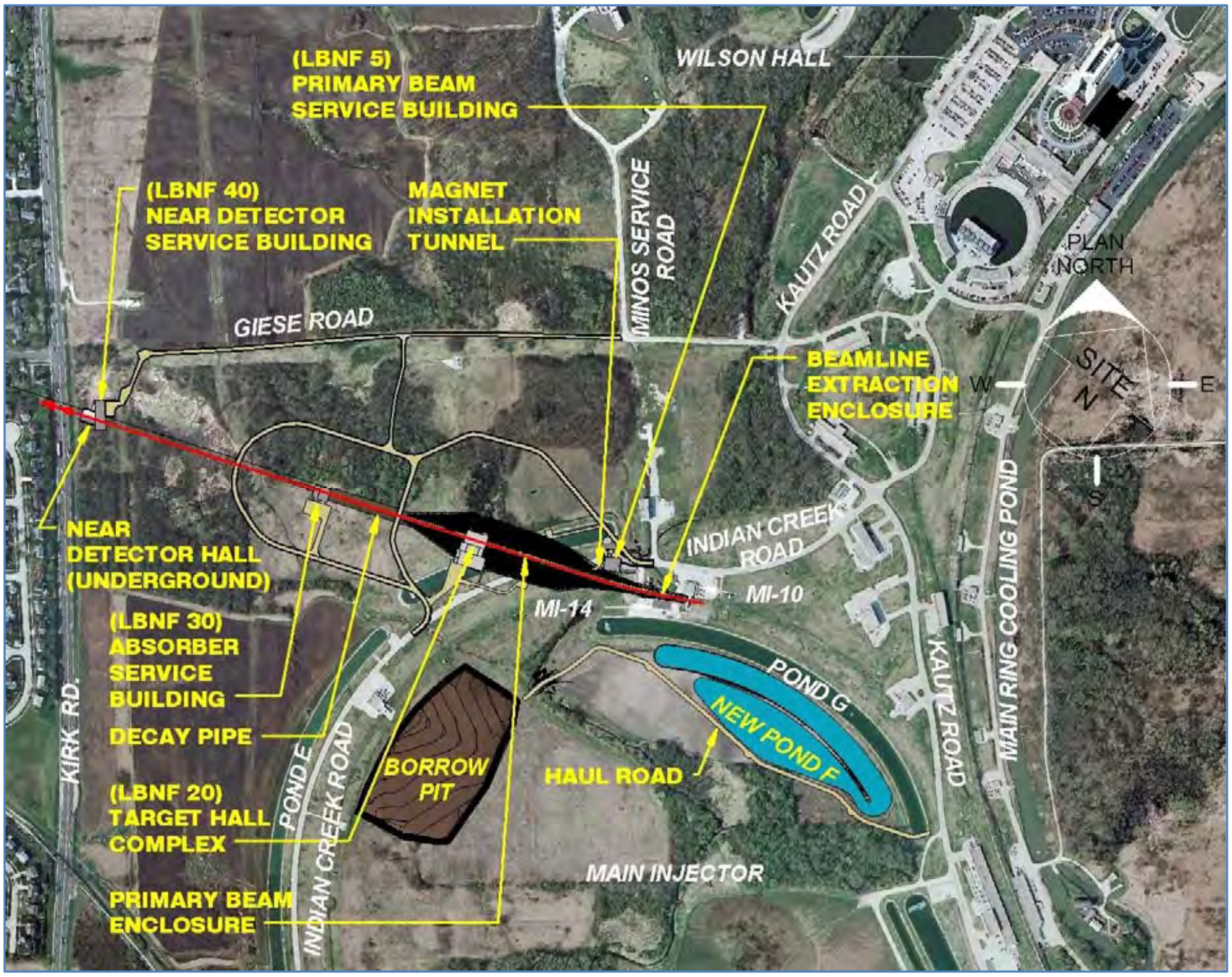

Figure 5-2: LBNF overall project layout at Fermilab

Each underground facility has a surface/above-ground service building that functions as a conveyance conduit for conventional and programmatic (for the technical systems) utilities as well as a location for equipment conveyance and personnel access and egress from the underground enclosures. Note that the shallow above and below grade Target Hall are included in the surface based Target Hall Complex.

Figure 5-3 shows the beamline facilities longitudinal section view and how the surface facilities relate to their corresponding underground facilities.

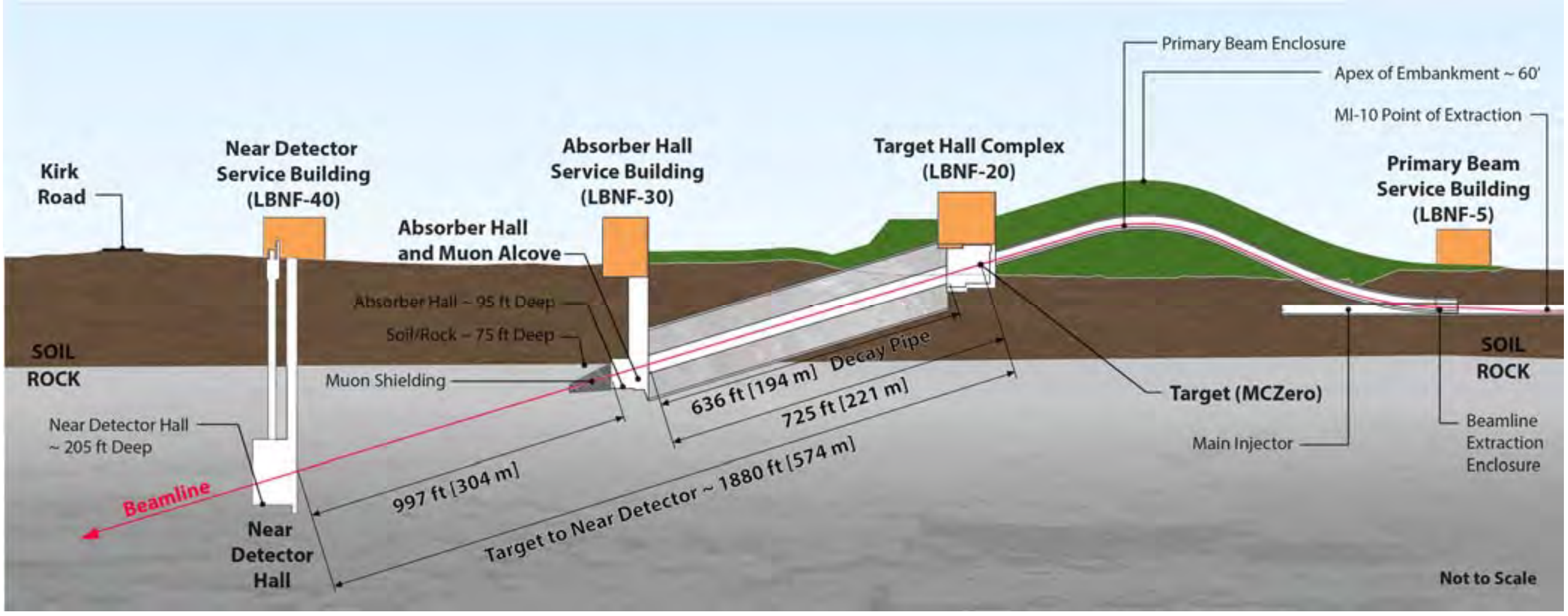

Figure 5-3: LBNF overall project schematic longitudinal section view

The conceptual design drawings, LBNF Conventional Facilities at Fermilab [26], depict the general Project layout, transverse and longitudinal cross sections of surface and underground facilities, overall Project and individual surface and underground facilities plan views, single line diagrams of mechanical, electrical, plumbing, and fire protection routing and layout, as well as details and general conceptual specifications of the Conventional Facilities as required by the technical system groups.

The LBNF Conventional Facilities at Fermilab drawings are augmented by a set of drawing sheets called LBNE Advanced Site Prep (ASP) [27]. The ASP was prepared in 2014 and represents the preliminary design of the first anticipated phase of construction of the Conventional Facilities at Fermilab. Refer to Section 5.3.1.8 for a discussion of the planned construction phasing.

## 5.3.1 Project-Wide Considerations

There are several design considerations that apply to many of the facilities that will be constructed for LBNF and are not necessarily specific to any single structure or system. These considerations include the structural and architectural treatment of surface structures, structural and excavation approaches to underground or shallow buried structures, environmental protection, fire protection and life safety systems, safeguards and security, emergency shelter provisions, energy conservation, and DOE space allocation. These Project-wide considerations are addressed in this section.

### 5.3.1.1 Structure and Architecture for Surface Structures

The structural building and construction systems for the Near Site Conventional Facilities will be constructed utilizing conventional methods similar to systems established at Fermilab. The architectural features of the Near Site Conventional Facilities will include four surface or near-surface buildings:

- Primary Beam Service Building (LBNF-5)

- Target Hall Complex (LBNF-20)
- Absorber Service Building (LBNF-30)
- Near Detector Service Building (LBNF-40)

The Primary Beam Service Building and the Absorber Service Building will be constructed as a braced-frame, steel and concrete construction with prefinished metal siding. The construction type and style will be consistent with similar adjacent facilities on the Fermilab campus. The Target Complex support service rooms will be constructed of pre-cast and cast-in-place concrete and braced-frame, steel construction with prefinished metal siding as well as natural concrete finish. A Project-specific style of architecture will be developed to unify and mitigate the presence of new buildings upon the surrounding environment.

The Near Detector Service Building at LBNF-40 is architecturally significant because of its proximity to Kirk Road and visible to Fermilab's residential neighbors to the west. Therefore the architectural style of the near detector building needs to be complimentary to the surroundings. A landscape/screening embankment is planned between the construction/building site and Kirk Road to shield the neighbors from construction noise and to minimize the visual impact of the building.

### 5.3.1.2 Structure and Excavation for Underground Structures

The construction systems for the underground portion of the LBNF Conventional Facilities will be constructed utilizing conventional underground excavation methods.

Most of the below-grade facilities to be built will be constructed using standard open cut methods. This includes much of the Extraction Enclosure, the Primary Beam Enclosure, the Target Complex, the decay pipe, and the Absorber Hall.  Some of the Primary Beam Enclosure and all of the Target Complex will be constructed in an embankment constructed of engineered fill that reaches a maximum height of about 60 ft above existing grade. The toe of the embankment is shown as a red dashed line in Figure 5-4**.** The extent and height of the embankment will cause consolidation (settlement) of native in situ soils resulting in potential adverse impacts to existing facilities including the Main Injector. Figure 5-4 also shows the locations, limits, and types of braced excavation, retaining wall systems, and slurry trenches that are planned to provide protection of the Main Injector.

Open cuts for the decay pipe and the Absorber Hall will require excavation 70 feet in soil and another 25 feet or so into the bedrock underlying the project site. Rock will be excavated using quarry-type drill-and-blast techniques. The lower half of the downstream one-third of the decay pipe and a portion of the Absorber Hall will be excavated in rock.

The Extraction Enclosure and Primary Beam Enclosure will be covered with the required minimum 25 ft of earth-equivalent shielding.

The Near Detector Hall will be constructed using conventional underground rock excavation and tunneling methods. Shaft excavations will employ an earth retention system in the soils. The rock portions of shaft and underground cavities and halls will be excavated using drill-and-blast methods.

Rock support of excavations in rock will be provided by rock bolts or rock dowels. Shotcrete will be applied to exposed rock faces.

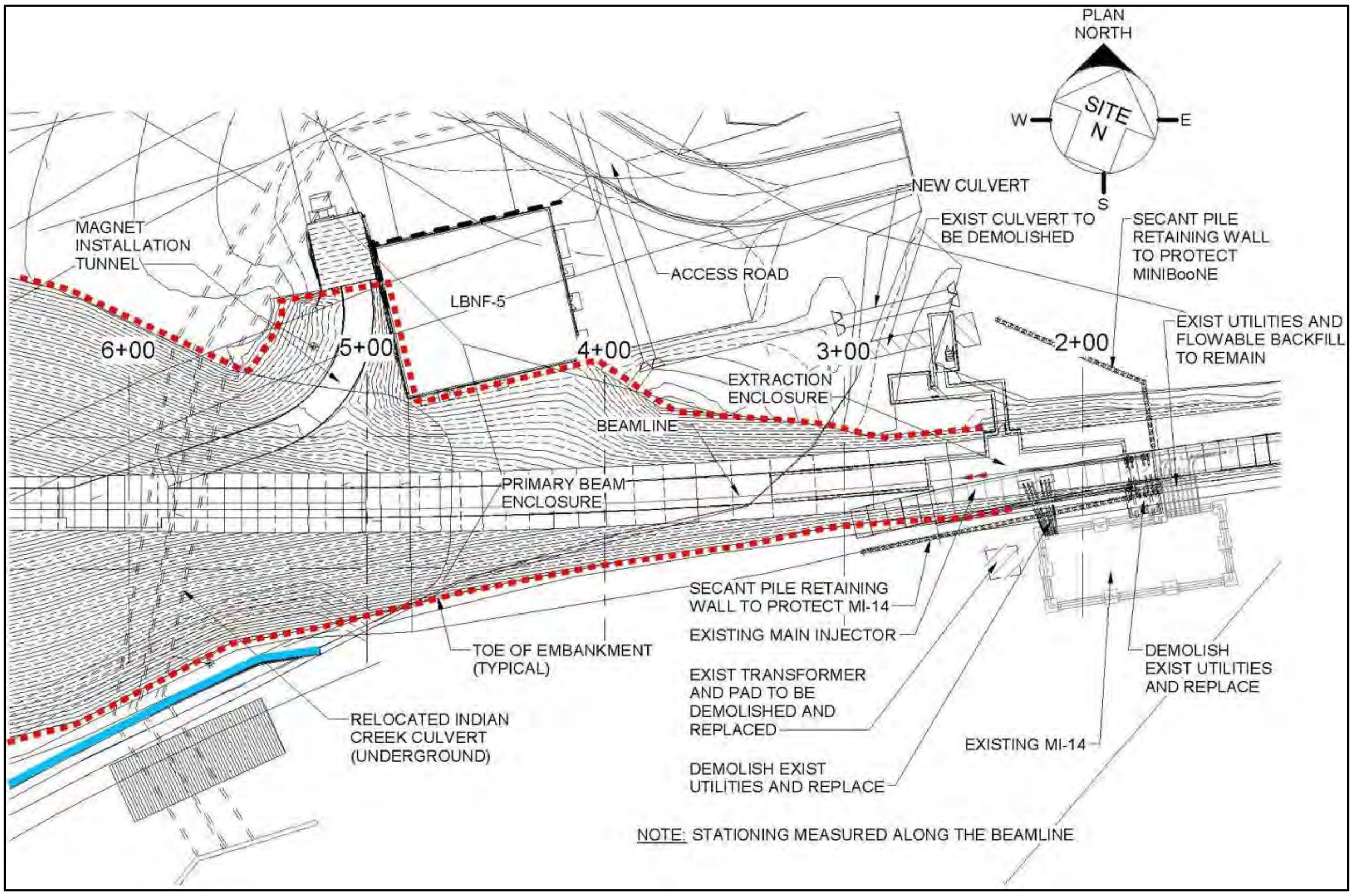

Figure 5-4: Braced excavation and retaining wall systems. The red dashed line shows the toe of the embankment and blue line is the slurry trench.

### 5.3.1.3 Environmental Protection

The overall environmental impact of this Project will be evaluated and reviewed as required to conform to applicable portions of the National Environmental Policy Act (NEPA). All required permits will be obtained prior to the start of construction. During the ASP Preliminary Design phase of the Project, environmental consultants located and defined the limits of all areas impacted by the project including wetland areas, floodplain and storm water management areas, sites of archaeological concern, and any other ecological resource areas. This provided input to the Fermilab NEPA Program Manager and DOE in the preparation of an Environmental Assessment.

### 5.3.1.4 Fire Protection/Life Safety Systems

A fire protection-life safety (FLS) Assessment/Report was completed by Aon/Shirmer Fire Protection Engineering Corp. for the LBNF Project [28].

Consistent with the FLS report, facility access and egress will be designed and provided in accordance with all applicable National Fire Protection Association (NFPA) Life Safety Codes and Standards including NFPA 520: Standard on Subterranean Spaces, which requires adequate egress in the event of an emergency. Egress paths for surface (service buildings) and underground facilities (tunnels and halls) has been designed to limit the travel paths to egress shafts, stairways, and safe/fire rated corridors to the exterior and surface to a safe gathering location. The specific egress routes are described in several sections that follow.

### 5.3.1.5 Safeguards and Securities

Direction for security issues related to the design of this Project is taken from the current operating procedures for the Fermilab site.

Service buildings and facilities will be accessible to required Fermilab personnel and contractors during normal work hours. Access to the controlled areas during normal working hours will be controlled internally by the appropriate technical division occupants of each respective building or underground enclosure.

### 5.3.1.6 Emergency Shelter Provisions

Required provision for occupant protection in the event of tornadoes or other extreme weather conditions will be incorporated into the design of the service buildings. Guidelines established by the Federal Emergency Management Agency (FEMA) in publications TR-83A and TR-83B will be used to assess the design of the buildings to insure safe areas within the buildings for the protection of the occupants. These protected areas will also serve as dual-purpose spaces with regard to protection during a national emergency in accordance with the direction given in Section 0110-10, DOE 6430.1A.

### 5.3.1.7 Energy Conservation

Early 2012, in a position paper titled "Fermilab Strategy for Sustainability," the likelihood of attaining the LEED Gold certification for LBNF facilities was discussed. The context for that analysis was the requirement by DOE that all new buildings and major building modification over $5M must obtain LEED Gold certification from the U.S. Green Building Council (USGBC). That paper was written in part to support a Fermilab request to DOE for an exemption from the LEED Gold requirement for LBNF. The LEED Gold requirement was first articulated in a memo by then-Secretary Samuel Bodman on February 29, 2008, and was subsequently incorporated into DOE's Strategic Sustainability Performance Plans (SSPPs) for 2011 and 2012. The Bodman memo has subsequently been rescinded, and DOE removed the LEED Gold requirement for new projects.

The LBNF Project has already begun evaluating LBNF facilities using the federal Guiding Principles for High Performance and Sustainable Buildings [29], and credit has been provisionally taken for 11 of the 34 total required items, based on the near site. The requirements in the Guiding Principles are more policy/process oriented, whereas the LEED credits are more building/site oriented. For the near site, it is believed that LBNF can eventually meet almost 50% of the Guiding Principles requirements simply by citing Fermilab-level policies, or overall Lab performance. Of the remaining individual requirements of the Guiding Principles, almost all of them are easily incorporated into the design of the project. Examples are the use of water conserving fixtures, energy efficient lighting, metering, and using materials with recycled and/or bio-based content.

### 5.3.1.8 Construction Phasing

Conventional Facilities construction is planned in two phases. The primary objective of the first phase, referred to as the “Advanced Site Preparation” (ASP) phase, is to construct a preload embankment necessary to induce settlement in the elevated Primary Beam and Target Hall Complex regions before the remaining facilities are constructed in order to minimize beamline settlement after they are constructed. To accomplish this objective, existing facilities in the region must be relocated or replaced, including the relocation of Indian Creek Road, the relocation of the utilities serving the Main Injector, the decommissioning and replacement of Cooling Pond F, new culverts for Indian Creek where it crosses the Main Injector and Indian Creek Road. During this phase, electrical and communications ductbanks, Industrial Cooling Water (ICW), and sewer intended to serve the future LBNF operational spaces will be constructed along the alignment of the relocated Indian Creek Road to avoid disturbing the new alignment later. To provide the needed fill material for the preload embankment, a borrow pit will be opened within the Main Injector infield. Excess topsoil will be stockpiled near Butterfield Road on the Fermilab site.

The second phase of construction, referred to as the “Beamline and Near Neutrino Detector Conventional Facilities”, or “Beamline Conventional Facilities” (BCF) and “Near Detector Conventional Facilities” (NDCF) for short, includes the balance of the work including the Project’s underground enclosures and surface-level service buildings, utilities, roads, hardstands, and other conventional facilities features needed for the Project.

## 5.3.2 Project Site Infrastructure

The locations of the six Near Site Conventional Facility components define the LBNF Conventional Facilities at the Near Site. Facility locations were selected based on the programmatic requirement for extracting beam from the existing Main Injector near MI-10 and the planned location of the far detector. A significant portion of the Project site infrastructure is provided for the benefit of the entire Project and is not provided in response to a requirement of a specific site address, such as LBNF-20. This section describes Project site infrastructure systems that apply Project-wide.

### 5.3.2.1 Roads and Infrastructure

The existing Giese Road will be improved and extended to provide access to the Main Injector Road. This reroute will also allow access to LBNF-20 and LBNF-30 from the north. The new road will be suitable for all-weather and emergency access during construction and beamline operations.

Parking and staging areas will be incorporated into the design of each surface building. Designs will provide for the operation phase, with parking and hardstand requirements for construction and installation designed to be temporary.

### 5.3.2.2 Electrical

Fermilab is supplied electrical power through the northern Illinois bulk power transmission system that is operated by a local investor-owned utility. The site interconnects with the bulk transmission system at two locations. Service connections, at 345 kV voltage, are made to one of two transmission lines at each location. At the interconnection sites, Fermilab takes power and delivers it along Fermilab owned and operated transmission lines to two separate electrical substations where it is transformed to 13.8 kV for site-wide distribution.

Fermilab maintains two separate types of power systems, pulsed power and conventional power. The technical systems pulsed power loads are large and can cause power quality issues for the conventional facilities if interconnected. Therefore two separate systems are maintained. The electrical systems located throughout the LBNF Project will conform to the National Electric Code (NEC) and applicable sections of the Fermilab Engineering Standards Manual.

The electrical power requirements for the LBNF Project are significant and will require the extension and expansion of the existing 13.8-kV electric distribution facilities. The improvements include electrical substation modifications, the extension of existing 13.8-kV distribution feeders from a nearby feeder for pulsed power and the expansion of Kautz Road substation for the conventional power. The LBNF Project will also require the relocation of the existing electrical power ductbank system around the proposed facilities. Existing ductbanks will be rerouted along the proposed roadways to the LBNF facilities and continue to reconnect to the existing ductbanks to maintain the existing infrastructure.

##### *5.3.2.2.1 Pulsed Power System*

The pulsed power system for LBNF will be served from the existing Kautz Road substation, feeder 96/97 at 13.8 kV. The pulsed power system requires a harmonic filter to maintain power quality. The existing feeder 96/97 is connected to a harmonic filter at the Kautz Road substation.

The feeder system from the substation to the LBNF Project will be provided by connecting to an existing pulsed power feeder that is currently serving the Main Injector. The feeder is configured as a loop for operational flexibility. The LBNF pulsed power services to LBNF-5 and LBNF-20 will be inserted into the loop, maintaining the configuration.

Sectionalizing switches, 13.8 kV, 600 A, will be installed at various locations along the feeder route. A switch will be installed at each location where pulsed power is required. The switches will serve to provide operational flexibility in load and fault isolation.

##### *5.3.2.2.2 Conventional Power System*

The conventional power system for LBNF will be served from Kautz Road substation at 13.8 kV where two new feeders will be constructed. The feeder equipment will include the cable and a new 13.8-kV Kautz Road substation circuit breaker. The feeder system from the Kautz Road substation to the LBNF Project will be provided by installing the new underground feeder cables in existing spare ductbank. The feeder will be constructed in a looped configuration to enable cable segment isolation, fault clearing and service restoration without cable repair or replacement and extended service outage.

Sectionalizing switches, 13.8 kV, 600 A, will be installed at various locations along the feeder route. A switch will be installed at each location that conventional power is required. The switches will serve to provide operational flexibility in load and fault isolation.

### 5.3.2.3 Mechanical and HVAC

The heating, ventilation and air conditioning (HVAC) systems located throughout the LBNF Project will conform to ASHRAE 90.1, ASHRAE 62, applicable NFPA requirements and applicable sections of the Fermilab Engineering Standards Manual. The design parameters for the general HVAC are summarized below:

- Air conditioned facilities will be maintained between 68° to 78°F.
- The relative humidity in air conditioned spaces will be maintained below 50%. There is no minimum requirement.
- Ventilated spaces will be maintained at maximum approximately 10ºF above ambient.
- HVAC for the target chase, Target Hall, decay pipe, Absorber Hall, and Near Neutrino Detector Hall are specially designed systems that are detailed in other sections of this document.

All HVAC systems will be design and installed with Metasys automated building controls capable of local and remote monitoring, control and operation optimization. Direct Digital Controls will be further investigated during subsequent phases in accordance with the applicable codes and Federal life cycle costing analysis.

### 5.3.2.4 Plumbing and Cooling Systems

The industrial cooling water (ICW), domestic water service (DWS), sanitary sewer, and other related utility services for the Project will be extended from existing services found along the Main Injector Road utility corridor to LBNF-5, LBNF-20, and LBNF-30. The residential subdivision directly west of Kirk Road has City of Batavia domestic water and sanitary sewer capacity to accommodate the Near Detector facility (LBNF-40) needs. All domestic plumbing work will be installed in accordance with the Illinois Plumbing Code and Standard Specifications for Water and Sewer Main Construction in Illinois, and applicable sections of the Fermilab Engineering Standards Manual.

### 5.3.2.5 Data and Communications

The existing Fermilab data, telephone communications and controls network will be extended from existing sources at the MI-8 service building to the LBNF-5, Target Complex (LBNF-20), Absorber Hall (LBNF-30) and Near Detector facility (LBNF-40), to provide normal telecommunication and controls communication support to the new LBNF facilities. Connections will be included in the form of new stub-ups for future expansion of the existing fiber network. The required communications ductwork and manholes will be designed, estimated and constructed as part of the conventional facility portion of the Project. The Conventional Facilities work will also include supplying and installing (pulling and connecting) the required fiber-optic lines.

# 5.4 New Surface Buildings

The LBNF Conventional Facilities on the Near Site will include surface buildings at LBNF-5, LBNF-30, and LBNF-40. A near-surface, shallow-buried structure will be located in an embankment constructed of engineered fill at LBNF-20. This section provides additional details regarding these surface-based structures.

## 5.4.1 Primary Beam Service Building (LBNF-5)

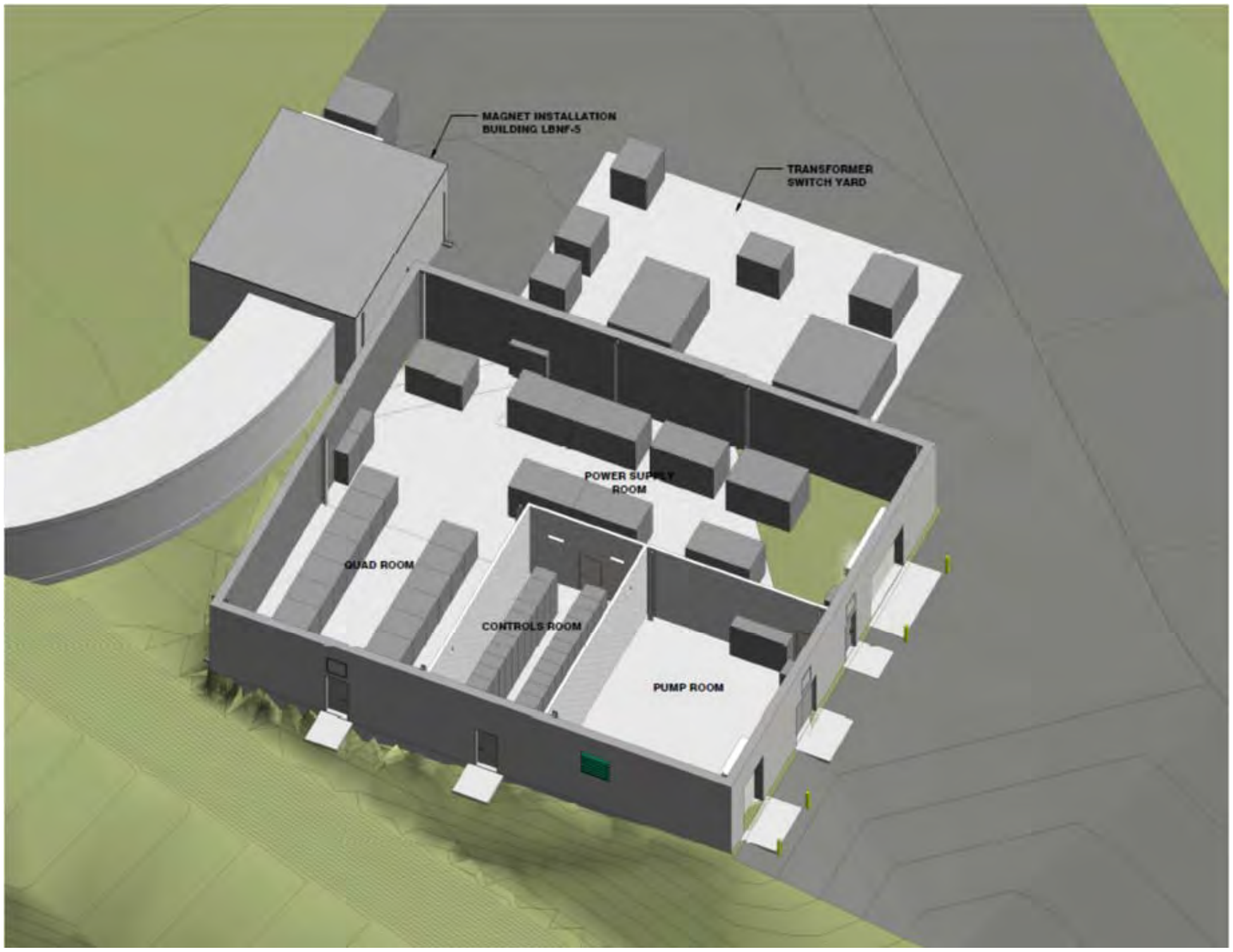

Figure 5-5: Primary Beam Service Building (LBNF-5)

One of four-at-grade service buildings is the Primary Beam Service Building (LBNF-5). A facility layout of LBNF-5 is shown in Figure 5-5 and Figure 5-6. The function of LBNF-5 is to provide housing for primary beam support equipment and utilities, and access for personnel and light equipment to the Primary Beam Enclosure below. This single-story, steel-framed, metal-sided service building will have approximately 4,800 sf of floor space (60 ft by 70 ft main building plus an approximately 500 sf Magnet Installation Tunnel structure), with a minimum 12-ft interior clear height. The building will be positioned off center and north of the beamline and interface with the Magnet Installation Tunnel to provide a path for technical and conventional utilities routing to the Primary Beam Enclosure. The Magnet Access Tunnel structure is a small building constructed similar to the main LBNF-5 building and includes the facilities needed to offload magnets as well as serve as a convenient location to shed foul-weather clothing and to prepare tools and equipment outside the interlocked entry to the Primary Beam Enclosure.

Utilities conveyed to the Primary Beam Enclosure will consist of low conductivity water (LCW), technical (pulsed) and conventional power, and communication/control lines. Access to and egress from the Primary Beam Enclosure will be provided through the Magnet Installation Tunnel as described later in this section.

Space is provided in the Power Supply Room for installation of dipole and quadrupole power supplies within the building as well as water-cooling lines and related equipment. A Pump Room is provided for LCW to cooling pond water (CPW) heat exchangers, and LCW pumps; and a Controls Room is provided for technical system controls. Electrical switchgear and power transformers will be installed on the transformer pad adjacent to the building, as shown in Figure 5-6.

### 5.4.1.1 Mechanical

Heat rejection for the LCW system will utilize CPW from the modified Main Injector Cooling Pond network. The building will be heated with electric unit heaters to maintain a minimum winter temperature of 68ºF. The power supply room and the pump room will be ventilated using outside air fans to maintain summer maximum temperatures of about 10ºF above ambient outdoor temperature.

The electronics/control room will be cooled by wall mounted cooling units to maintain summer maximum temperatures of 78ºF. There will be no natural gas or sanitary sewer service. Restroom facilities are available in the existing MI-12 Service Building located about 350 ft to the north. One of the building’s roof support beams will be sized and configured to serve as a maintenance monorail for equipment removal and replacement.

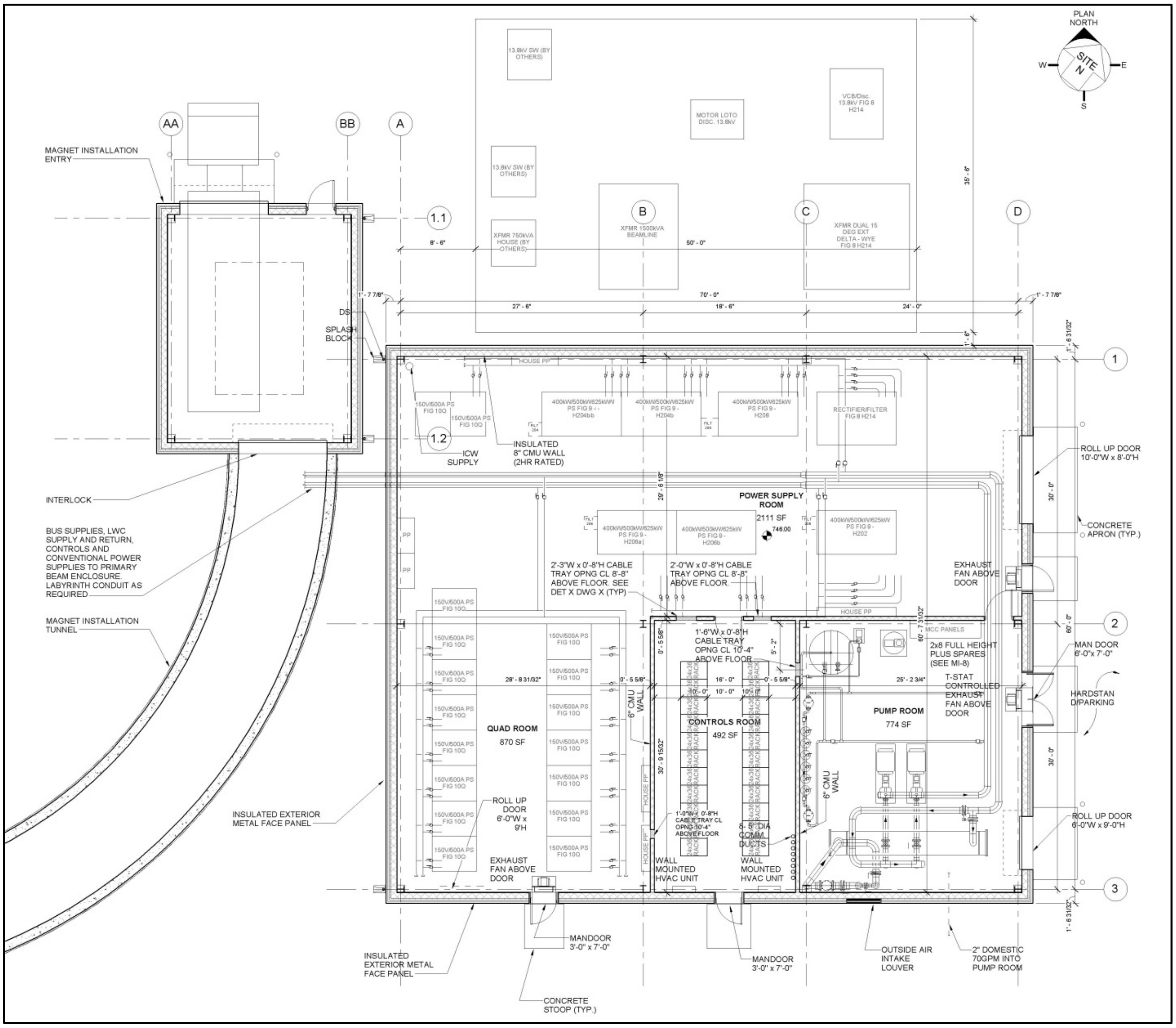

Figure 5-6: Primary Beam Service Building (LBNF-5) and exterior transformer pad

## 5.4.1.2 Electrical

The electrical facilities provided at LBNF-5 will support the requirements of the Beamline in accordance with the Fermilab standards, NEC, and other applicable codes. The building site will have two separate electrical services. A conventional power system will be provided for electrical service to the building systems, lighting, HVAC, crane, miscellaneous loads, and low power technical system components such as electronics racks and computers. The electrical power loads are summarized in Table 5-1. A technical system power system space will be provided for electrical service to large technical systems such as beam power supplies. Conventional Facilities will provide 13.8-kV primary electrical power to outdoor switchgear and prepare a space at LBNF-5 for the technical systems transformers and components to be installed by the technical systems. An emergency/standby power system with generator will be installed to serve critical loads for life safety and technical system components.

Table 5-1: Primary Beam Service Building (LBNF-5) electrical power loads

| Equipment Description | Normal Power (kw) | Standby Power Generators (kw) |
|---|---|---|
| Packaged AHU | 73 | |
| Electric Heater (2 heaters/room; 3 rooms = 6 total) | 210 | 210 |
| Lighting & Receptacle | 42 | 11 |
| Pump Room Ventilating Fan (3 hp; 5,000 cfm) | 3 | 3 |
| Power Supply Room Ventilating Fan (5 hp; 10,000 cfm) | 5 | 5 |
| Total Connected Load | 333 | 229 |

### 5.4.1.3 Plumbing

A 2” domestic water service will be extended from Indian Creek Road for LCW make-up water. There is no sanitary sewer or natural gas services provided to this building.

### 5.4.1.4 Fire Protection/Life Safety Systems

The Primary Beam Service Building (LBNF-5) will be equipped with a wet pipe sprinkler system served from the site-wide industrial cooling water (ICW) network that is extended to the building from the existing nearby system. Egress paths for surface (service buildings) and underground facilities (tunnels and halls) have been conceptually designed to limit the travel path distances to egress shafts, stairways, and safe/fire rated corridors to the exterior and surface to a safe gathering location. See Section 3.1.4 of this volume for a general overview of fire protection and fire life/safety requirements.

## 5.4.2 Target Hall Complex (LBNF-20)

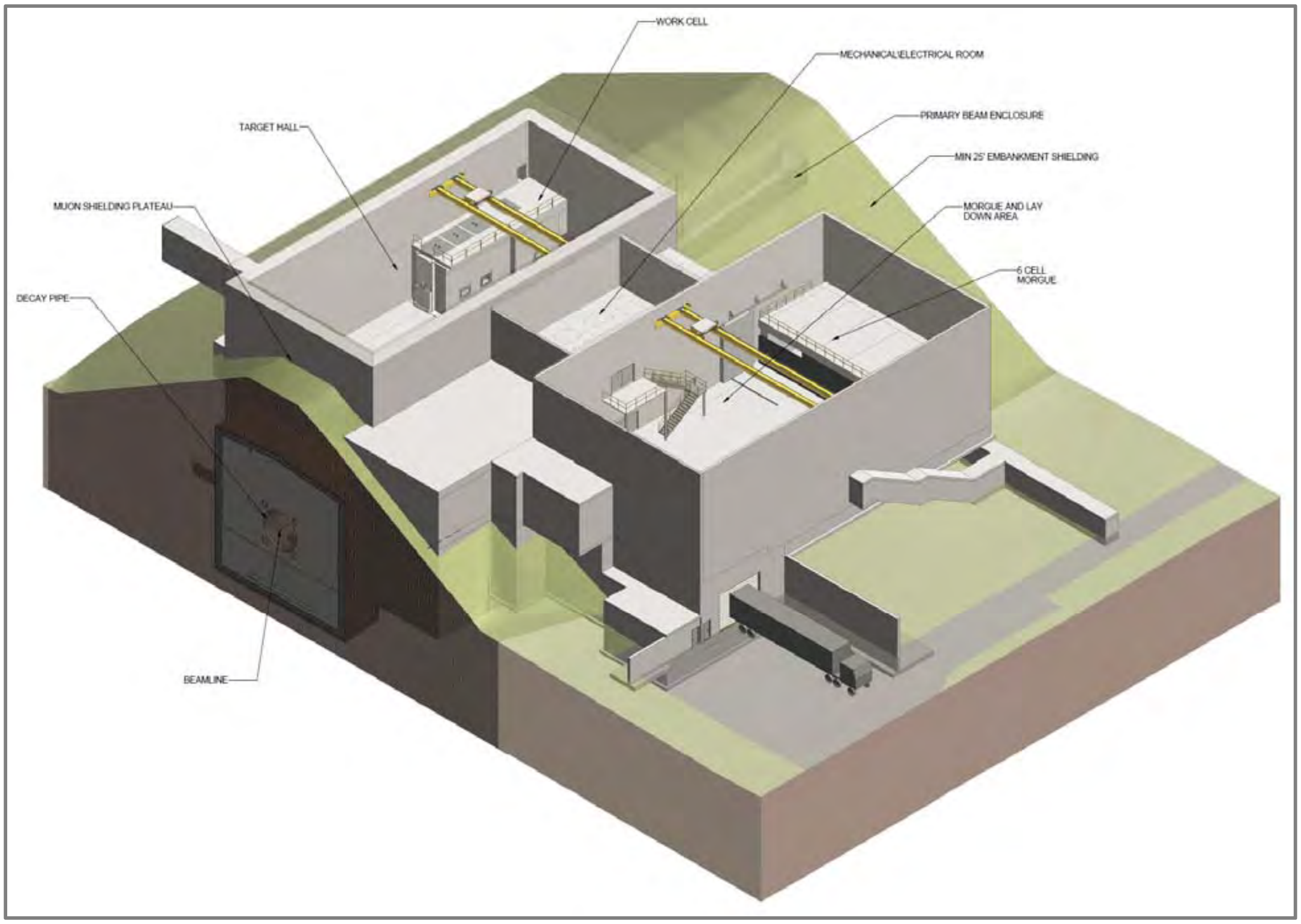

Figure 5-7: Target Hall Complex

The Target Complex (LBNF-20) will be a 31,000 net square feet near-surface facility constructed in the engineered fill embankment, that combines the Target Hall, corresponding beam-on and beam-off support rooms (power supply, piping system carrying radioactive water (RAW), and air handling rooms), and the Target Hall service rooms (a truck bay with a lay down/staging area, morgue area, mechanical/utilities rooms, and rest room). Beam-on spaces are accessible by personnel while the beamline components are energized, and access to beam-off spaces is only permitted when the beamline components are not energized. These support and service rooms accommodate the support equipment and utilities and provide access needed to assemble and operate the equipment and conventional and programmatic/technical components for the Target Hall. An overview is shown in Figure 5-7.

The Target Hall, shown in Figure 5-8 is located at the north end of the Target Hall Complex, with beam-on and beam-off service and support rooms adjacent on the beam left side (as one faces downstream), which is also the south side of the Target Hall.

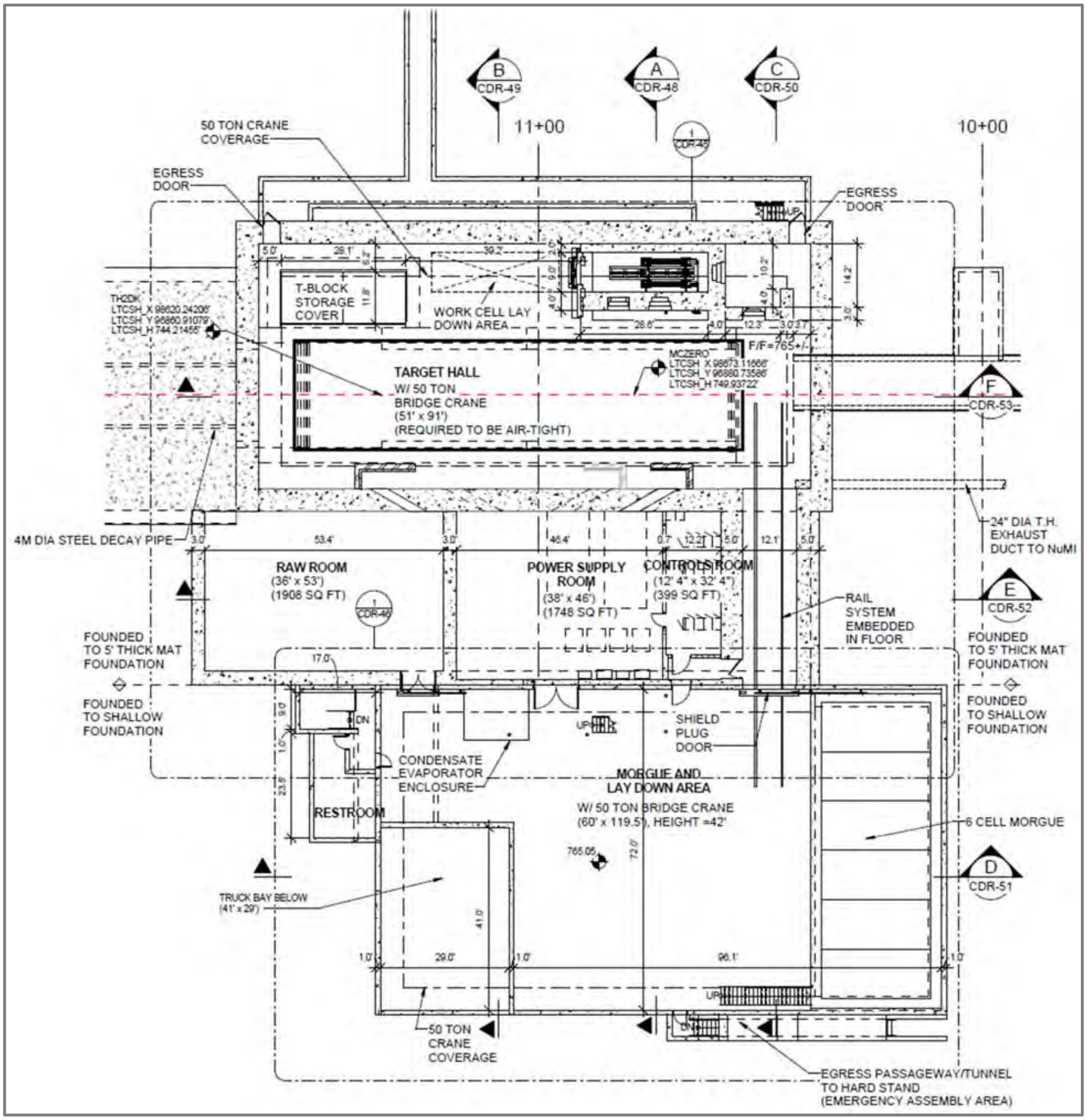

Figure 5-8: Target Complex: main level floor plan

The Target Hall will house the target and focusing horns in a shielded target chase enclosure below the floor. The required concrete and steel shielding will be provided around these beamline target components. The Target Hall and target chase (section views shown in Figure 5-9) consist of steel shielding blocks, and the target and horns, along with the associated power feeds, cooling water channels, and gaps/spaces for air cooling. Target pile bulk shielding steel will be provided by the Beamline Level 2 Project and installed by Conventional Facilities. A portion of the steel shielding blocks at the upstream end of the chase will be permanently cast into concrete. Other shielding blocks will be stacked within the chase.

The target chase is longer and wider than the minima necessary to accommodate the reference design target-horn system described in Sections 6.4.2 and 6.4.3. This will provide flexibility to accommodate more advanced designs which have the promise of enhancing the neutrino flux and substantially increasing the capability of the experiment. This is discussed in more detail in Section 6.7.

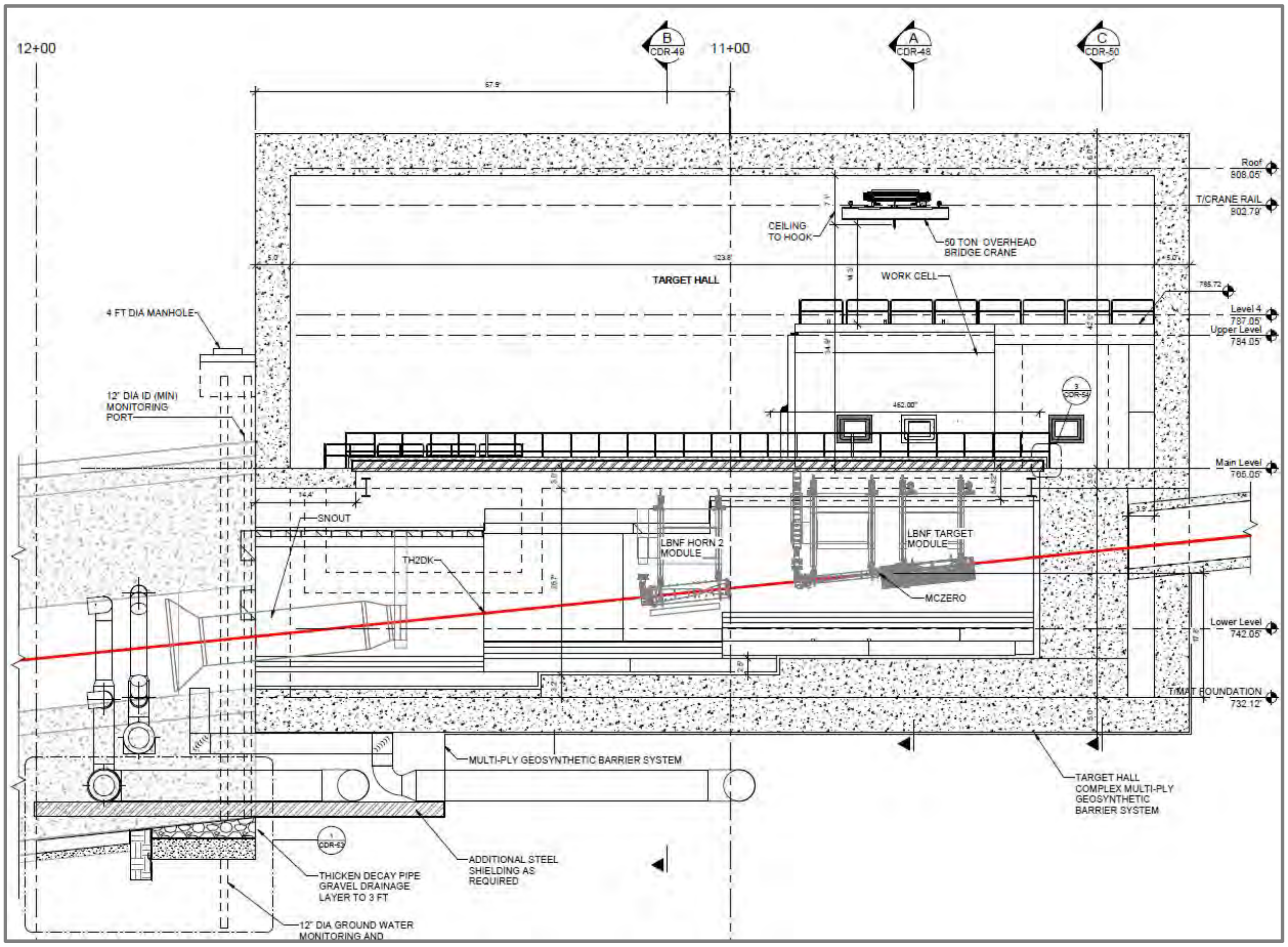

Figure 5-9: Target Hall and target chase longitudinal section

## 5.4.2.1 Mechanical

The entire Target Hall Complex (LBNF-20) shown in Figure 5-7 will be conditioned to 68°F (winter) or 78°F (summer) using chilled water/hot water (CHW/HW) air handling units (AHUs). Additional HW unit heaters will be strategically located as necessary to insure winter time minimum temperatures. CHW will be provided by packaged air-cooled chillers located on the LBNF-20 hardstand. The chilled water system will contain an appropriate level of propylene glycol to prevent freezing damage to all associated components. HW will be provided by natural gas hot water heaters.

The Target Hall chase and decay pipe will each be served by separate 35,000-CFM (cubic feet per minute) custom built air conditioning units capable of removing heat and moisture. The supply air for both regions shall be at a temperature of 59°F+/-2°F and 12 grains +/- 2 grains of moisture per pound of dry air. The return air condition will be in the range of 90°-100°F and 30 grains of moisture per pound of dry air. The units will utilize CHW, HW desiccant wheels for dehumidification, and bag-in/bag-out

High Efficiency Particle Arrestor (HEPA) air filter systems. All materials of the unit that come in contact with the airstream or condensate will be resistant to corrosion from the radio-chemically induced nitric acid that is present in the air. The AHUs will be constructed minimizing single points of failure. Ductwork to and from the target chase and to and from the decay pipe will be routed through passageways/ducts between the air handling room and the Target Hall and the upper end of the decay pipe. Duct materials will be welded steel pipe and welded steel plate constructed to Sheet Metal and Air Conditioning Contractors Association (SMACNA) 10 in water gage (wg) pressure class.

### 5.4.2.2 Electrical

The electrical facilities provided at LBNF-20 Target Hall Complex, will support the requirements of the technical systems in accordance with the Fermilab standards, NEC, and other applicable codes. A conventional 13.8-kV power system will be provided for electricity service to the building systems, lighting, HVAC, crane, misc. loads, and low power technical system components such as racks and computers. The conventional systems group will provide the 13.8-kV primary electrical power to an outdoor switchgear, one for each power system. Conventional Facilities will provide a prepared space at LBNF-20 for the technical system transformers and components to be installed by the Beamline Level 2 Project. Two transformers for Conventional Facilities will be installed at LBNF-20, one to serve the building loads and the other to serve the 4.16-kV chillers. An emergency/standby power system with generator will be installed to serve critical loads for life safety and technical system components as well as the crane systems.

Table 5-2 shows the Target Hall Complex (LBNF-20) electrical power loads, for both normal power and the standby power generators.

### 5.4.2.3 Plumbing

Fire protection systems and plumbing systems, including a restroom, are included. The wet pipe sprinkler system for this complex is served from the sitewide ICW network that is extended to the building from the existing system. Target Pile and decay pipe AHU cooling water is supplied from the air cooled chiller in the LBNF-20 mechanical room which is a closed-loop glycol system. Natural gas is routed to this building from the site-wide network to be used for hot water heating for domestic water and building heat.

### 5.4.2.4 Fire Protection/Life Safety Systems

Egress paths for surface (service buildings) and underground facilities (tunnels and halls) have been conceptually designed to limit the travel path distances to egress shafts, stairways, and safe/fire rated corridors to the exterior and surface to a safe gathering location.

Table 5-2: Target Hall Complex (LBNF-20) Electrical Power Loads

| **Equipment Description** | **Normal Power (kw)** | **Standby Power Generators (kw)** |
|---|---|---|
| Chiller (400 ton) (3 ea.) | 1123 | |
| Chilled water pumps [total]<br>(2200 gpm, 100 hp) (2 ea., 1 full time) | 70 | |
| Condensate pumps (3 total) | 3 | |
| Sump pump (3 total) | 3 | 3 |
| Hot water pump | 19 | |
| MAU (makeup air):<br>Support, morgue, truck dock/control rooms (3 total) | 15 | |
| MAU Desiccant | 12 | |
| RCU Desiccant (Target Hall) | 203 | |
| RCU Desiccant (other) | 24 | |
| RCU: Utility Room, Power Supply Room (2 total) | 16 | |
| AHU: service, MEP, Target Hall (3 total) | 43 | |
| Dehumidifier | 5 | |
| Exhaust fans:<br>morgue, truck dock, Target Hall, target chase (4 total) | 22 | |
| Lighting & Receptacle & Experimental | 233 | 58 |
| 50 ton bridge crane -Target Hall, morgue/laydown (2 total) | 98 | 98 |
| **Total Connected Load** | **1889** | **159** |

## 5.4.3 Absorber Service Building (LBNF-30)

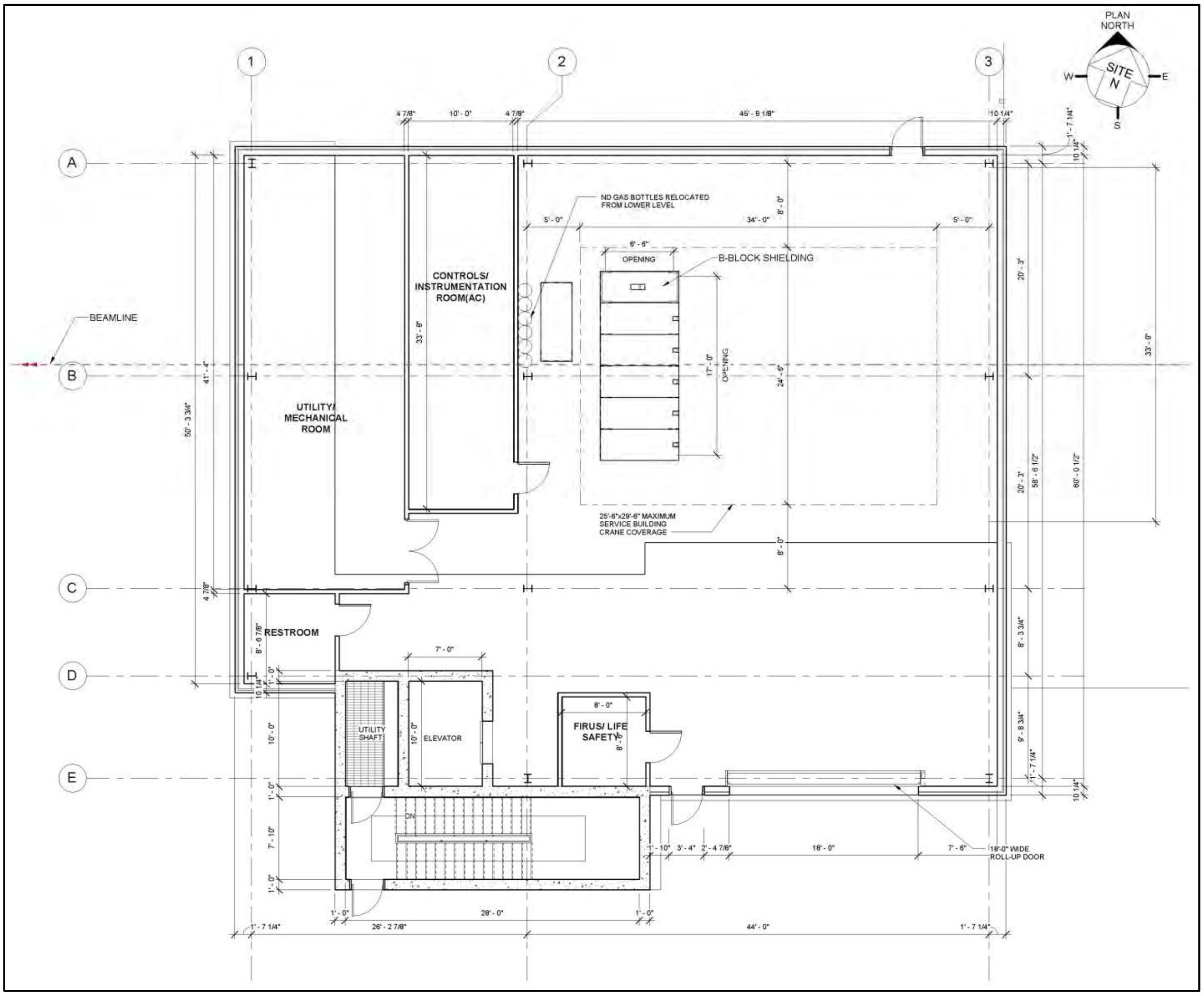

Figure 5-10: Absorber Service Building (LBNF-30) floor plan

The 4,300 sf Absorber Service Building (LBNF-30), shown in Figure 5-10, consists of a 70-ft wide by 62-ft long above-grade concrete and steel-framed building with metal siding which accommodates the support equipment and provides access needed for the assembly and operation of the equipment and technical components of the Absorber Hall, Muon Alcove, and support rooms located beneath the service building floor. Personnel will access the lower levels using an elevator. Secondary egress is provided by a stairwell.

This building has a 35-ft clear crane height for the truck bay area and will house the open truck bay and laydown area. A portion of the LBNF-30 floor slab is a 9-ft thick cast-in-place concrete slab providing shielding from the Absorber Hall which lies directly underneath. A 6.5-ft wide by 17-ft long opening provides equipment direct access to the Absorber Hall and Muon Alcove 90-ft below using the 30-ton bridge crane. This opening will be filled with shielding blocks and sealed while beam is in operation.

Much of the rest of the LBNF-30 floor slab overlies support rooms which are adjacent to the Absorber Hall including the Absorber Air Handling Room, Instrumentation Room, Controls Room, RAW Room.

### 5.4.3.1 Mechanical

The air in the Absorber Service Building (LBNF-30) will be conditioned to 68ºF (winter) and 78ºF (summer) using CHW/gas-fired AHUs. Additional gas unit heaters will be located, as necessary to ensure winter time minimum temperatures. CHW will be provided by packaged air cooled chillers located on the LBNF-30 hardstand. The chilled water system shall contain propylene glycol to prevent freeze damage to all associated components. Also located in the mechanical area are CHW/gas-fired dedicated outdoor air AHUs with desiccant dehumidification that provide dry neutral temperature ventilation air to the below grade general areas and exit passageways.

### 5.4.3.2 Electrical

The electrical facilities provided at LBNF-30 Absorber Service Building will support the requirements of the Beamline technical systems in accordance with the Fermilab standards, NEC, and other applicable codes. The building site will have one electrical service. A conventional 13.8-kV electrical power service will be provided to the building systems, lighting, HVAC, crane, miscellaneous loads, and low power technical system components such as racks and computers. A beamline technical system (pulsed power), 13.8-kV power system is not required. The Conventional Facilities scope of work will provide the 13.8-kV primary electrical power to an outdoor 600-A switchgear. Two transformers for Conventional Facilities will be installed at LBNF-30, one to serve the building loads and the other to serve the 4.16-kV chillers. An emergency/standby power system with generator will be installed to serve critical loads for life safety and technical system components. A dedicated separate emergency/standby power system will be provided for the three Absorber Hall sump pump systems.

Table 5-3 shows the Absorber Service Building (LBNF-30) electrical power loads, both normal power and standby power generators.

### 5.4.3.3 Plumbing

Fire protection systems and plumbing systems, including restrooms, are included. The wet pipe sprinkler system for this complex is served from the site wide ICW network that is extended to the building from the existing system. Natural gas is routed to this building from the site-wide network to be used for hot water heating for domestic water and building heat.

### 5.4.3.4 Fire Protection/Life Safety Systems

Egress paths for surface (service buildings) and underground facilities (tunnels and halls) have been designed to limit the travel path distances to egress shafts, stairways, and safe/fire rated corridors to the exterior and surface to a safe gathering location. See Section 3.1.4 of this volume for a general overview of fire protection and fire life/safety requirements.

Table 5-3: Absorber Hall and Absorber Service Building (LBNF-30) electrical power loads

| Equipment Description | Normal Power (kw) | Standby Power Generators (kw) |
|---|---|---|
| Chiller (300 tons) (2 ea.) | 562 | |
| Chilled water pump (1,400 gpm, 50 hp) (2 ea., 1 full time) | 37 | |
| Condensate pump (3 total) | 3 | |
| Hot water pump | 5 | |
| Refrigerated Dehumidifier | 12 | |
| Sump pump (6 total, 4 running full time) | 163 | 25 |
| Manual Pumps – Decay pipe water (2 total) | 31 | 31 |
| Holding tank pump (2 total, 1 running full) | 38 | |
| Fan coil | 1 | |
| Elevator | 37 | 37 |
| AHU Surface Building | 27 | |
| AHU desiccant | 6 | |
| Dehumidifier | 5 | |
| Exhaust fan | 2 | 2 |
| Lighting & receptacle & Experimental | 85 | 21 |
| 30-ton bridge crane | 49 | 49 |
| **Total Connected Load** | **1060** | **302** |

## 5.4.4 Near Detector Service Building (LBNF-40)

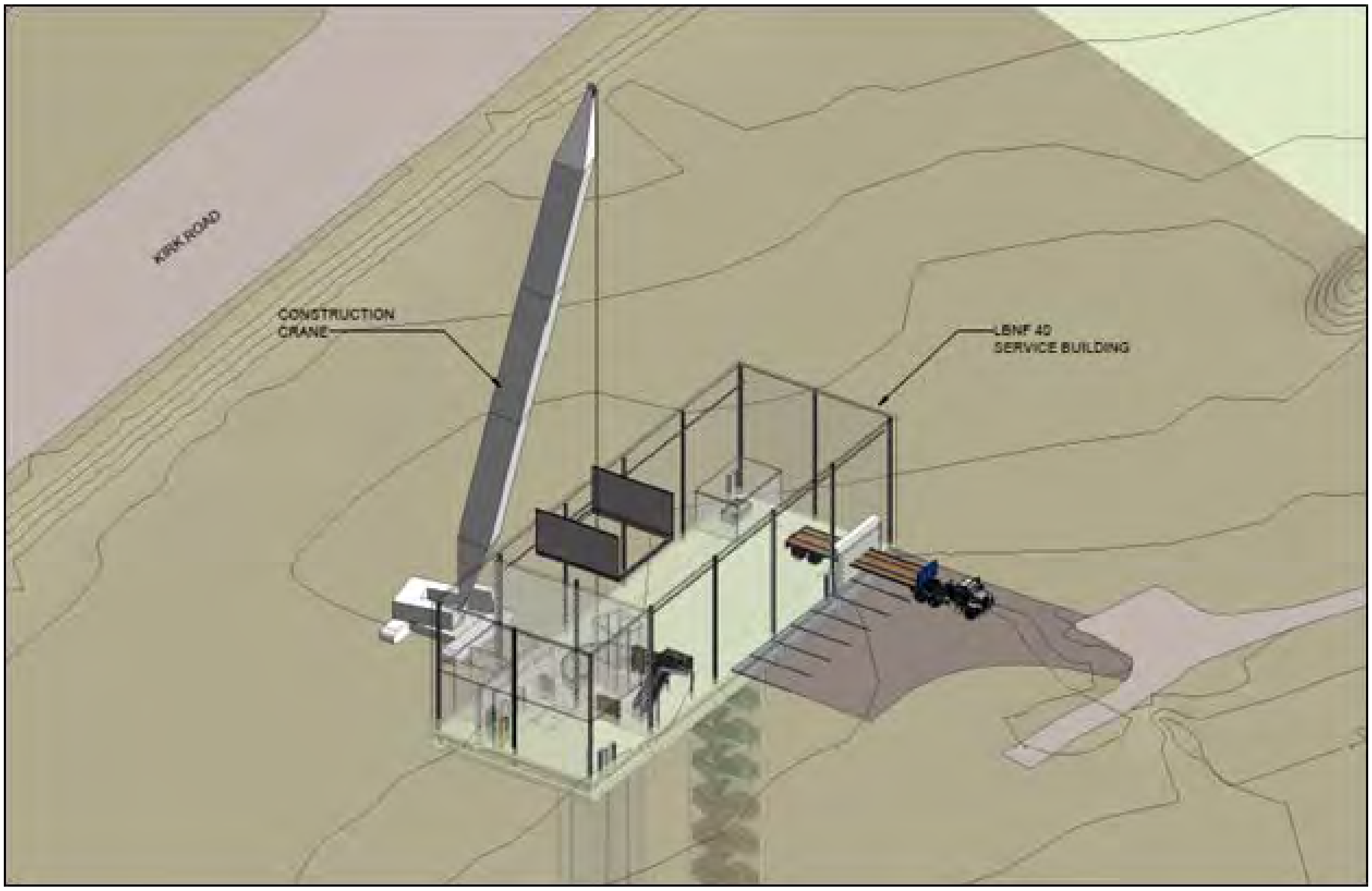

Figure 5-11: Near Detector Service Building (LBNF-40)

The Near Detector Service Building (LBNF-40), as shown in Figure 5-11 and Figure 5-12 is a 45-ft wide by 136-ft long by 42-ft high grade-level building. It will be used to house the support equipment and truck bay/lay down area needed for the assembly and operation of the equipment and technical components for the Near Detector Hall and support rooms. The truck bay/staging area portion of the building has a 35-ft interior clear ceiling height.

The truck bay will be provided for equipment to be unloaded using the 15-ton overhead bridge crane. Equipment and detector components will be lowered down the 22-ft diameter shaft to the Near Detector Hall approximately 210 ft below grade. This equipment can then be moved, using carts or portable hoists, where required. Because the initial detector installation and possible future removal are isolated events, the building crane capacity is not designed for these infrequent loads. Instead, a removable service building roof hatch cover over the 22-ft diameter shaft will be required for initial installation and future removal or replacement of Near Detector Hall components by use of a rented 200-ton crawler crane.

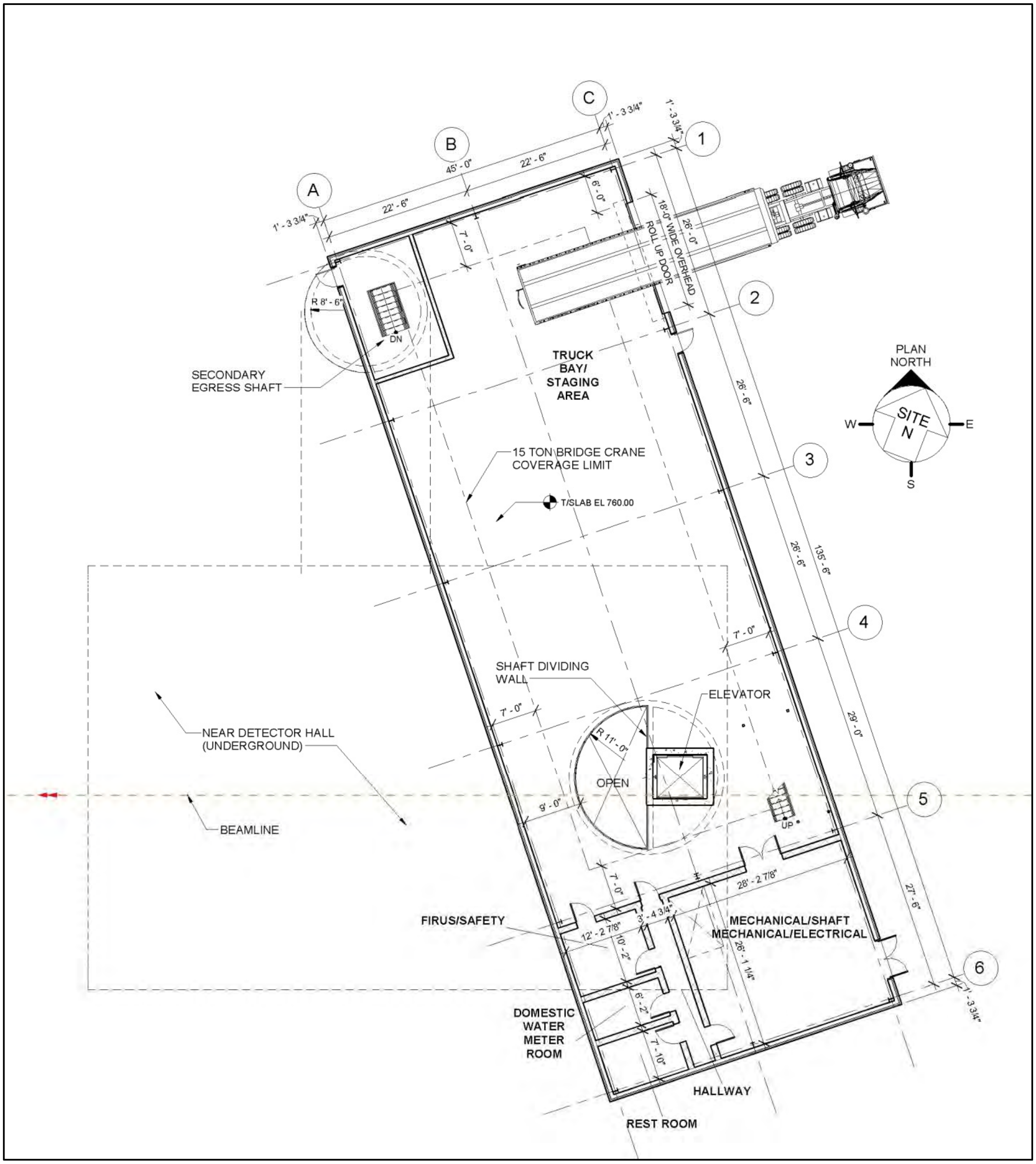

Figure 5-12: Near Detector Service Building (LBNF-40) floor plan

## 5.4.4.1 Mechanical

The entire Near Detector Service Building (LBNF-40) will be conditioned to 68ºF (winter) and 78ºF (summer) using CHW/gas-fired AHUs. Additional gas-fired unit heaters will be located, as necessary to insure winter time minimum temperatures. CHW will be provided by packaged air cooled chillers located on the LBNF-40 hardstand. The chilled water system shall contain propylene glycol to prevent freeze damage to all associated components. Also located in the mechanical area are CHW/gas-fired dedicated outdoor air AHUs with desiccant dehumidification that provide dry neutral temperature ventilation air to the below grade general areas and exit passageways.

### 5.4.4.2 Electrical

The electrical facilities provided at LBNF-40 Near Detector Service Building will support the requirements of the technical systems in accordance with the Fermilab standards, NEC, and other applicable codes. The building site will have one electrical service. A conventional 13.8-kV power system will be provided for electricity service to the building systems, lighting, HVAC, crane, miscellaneous loads, and low power technical system components such as racks and computers. A technical system (pulsed power), 13.8-kV power system is not required and will not be provided. The Conventional Facilities scope of work will provide the 13.8-kV primary electrical power to an outdoor 600-A switchgear. Two transformers for Conventional Facilities will be installed at LBNF-40, one to serve the building loads and the other to serve the 4.16-kV chillers. An emergency/standby power system with generator will be installed to serve critical loads for life safety and detector technical system components.

Table 5-4 shows the Near Detector Service Building (LBNF-40) and Near Detector Hall electrical power loads for both Normal Power and the Standby Power Generators.

### 5.4.4.3 Plumbing

Fire protection systems, plumbing, and a restroom for occupants are included. The wet pipe sprinkler system for this complex is served from the LBNF-40 domestic water system. The domestic water and sanitary sewer services will be supplied from the city of Batavia water and SS mains located on the west side of Kirk Road. Natural gas is routed to this building from the site-wide network to be used for hot water heating for domestic water and building heat.  A connection to the ICW system will be provided to allow the capture of water discharged from the Detector Hall sumps.

### 5.4.4.4 Fire protection/Life Safety Systems

Egress paths for surface (service buildings) and underground facilities (tunnels and halls) have been conceptually designed to limit the travel path distances to egress shafts, stairways, and safe/fire rated corridors to the exterior and surface to a safe gathering location.

Table 5-4: Near Detector Hall and Near Detector Service Building (LBNF-40) electrical power loads

| Equipment Description | Normal Power (kw) | Standby Power Generators (kw) |
|---|---|---|
| Chiller (300 tons) (2 ea.) | 243 | |
| Chilled water pump (700 gpm, 25 hp) (2 ea., 1 full time) | 19 | |
| Hot water pump | 5 | |
| Fan coil (2 total) | 4 | |
| AHU desiccants (2 total) | 11 | |
| AHU (2 total) | 27 | |
| Elevator | 37 | 37 |
| Exhaust fan (2 total) | 17 | 17 |
| Sump pump (2 total) | 38 | 38 |
| Lighting & Receptacle & Experimental | 85 | 21 |
| 15-ton bridge crane – Near Detector Hall, LBNF-40 (2 total) | 310 | 310 |
| **Total Connected Load** | **795** | **423** |

# 5.5 New Underground Structures

The LBNF Conventional Facilities on the Near Site will include new underground structures including the Beamline Extraction Enclosure and Primary Beam Enclosure, the decay pipe, and the Absorber Hall and Support Rooms, and the Near Detector Hall and Support Rooms. This section provides additional details regarding these facilities.

## 5.5.1 Beamline Extraction Enclosure and Primary Beam Enclosure

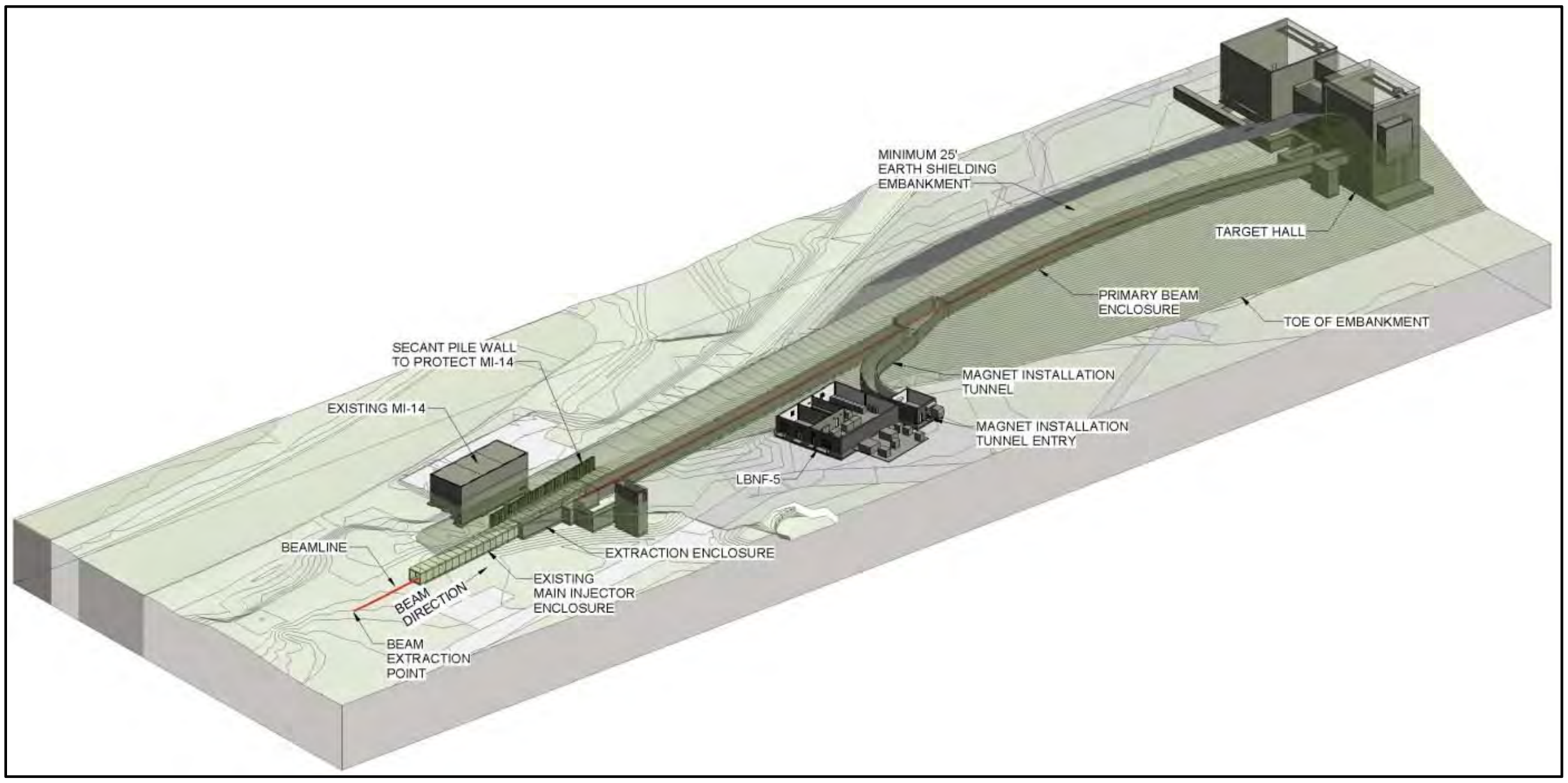

Figure 5-13 Overview of Beamline Extraction Enclosure and Primary Beam Enclosure

The first underground functional area is the Beamline Extraction Enclosure and Primary Beam Enclosure. This area, consisting of the upstream Beamline Extraction Enclosure and the downstream Primary Beam Enclosure are required to allow the beam to be extracted from the Main Injector and transported 328 m (1078 ft) to the LBNF Target Complex (to the target). An overview of these regions is shown in Figure 5-13. The construction of the upstream portion of this area will consist of a cut-and-cover below-grade section. The downstream portion of this area includes a transition from the below-grade section to an above-grade section located within an embankment. Some of these areas will be constructed using cast-in-place concrete sections and other areas will be pre-cast concrete enclosure sections. The Beamline Extraction Enclosure and Primary Beam Enclosure are shown in aerial view in Figure 5-14.

The Beamline Extraction Enclosure portion, shown in plan view in Figure 5-15, will start from the existing Main Injector near MI-10, and extend approximately 43 ft to the start of the Primary Beam Enclosure section.

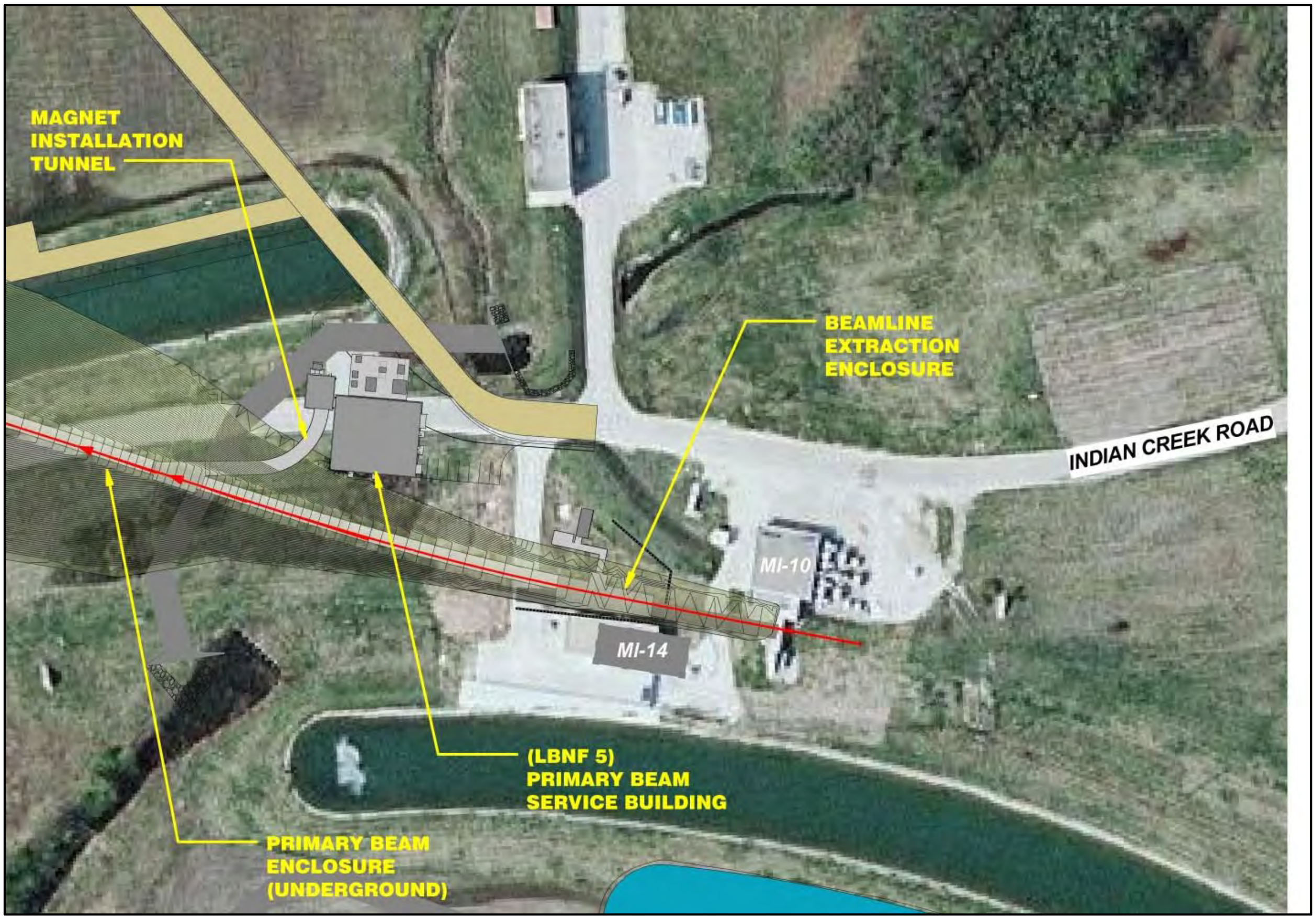

Figure 5-14: Beamline Extraction Enclosure and Primary Beam Enclosure – aerial view

The 778-ft long Primary Beam Enclosure will start at the end of the Extraction Enclosure and continue along a 15% incline into and through the above-grade embankment, and then at a 10% decline to and through the Target Hall. The depth of the enclosure will be 33 ft from the top of the soil shielding fill to the invert/floor. This will provide a minimum of 25 ft of soil and concrete shielding (measured radially outward from the center of the beamline) for both the 1.2-MW and 2.4-MW beam power levels. The apex of the embankment over the Primary Beam Enclosure will be approximately 60 ft above existing grade. With the required minimum 25 ft of soil shielding, the apex of the beamline will be about 30 ft above existing grade as shown in Figure 5-3.

The Primary Beam Enclosure has interior dimensions measuring 10 ft wide and 8 ft high. These dimensions match that of the existing Main Injector enclosure. Figure 5-16 shows a typical cross section of the Primary Beam Enclosure that shows the locations of the technical components and technical and conventional utilities.

### 5.5.1.1 Mechanical

Ventilation air is drawn from the Target Hall support area and the Extraction Enclosure region then exhausted through a vent shaft at the high mid-point of the enclosure.

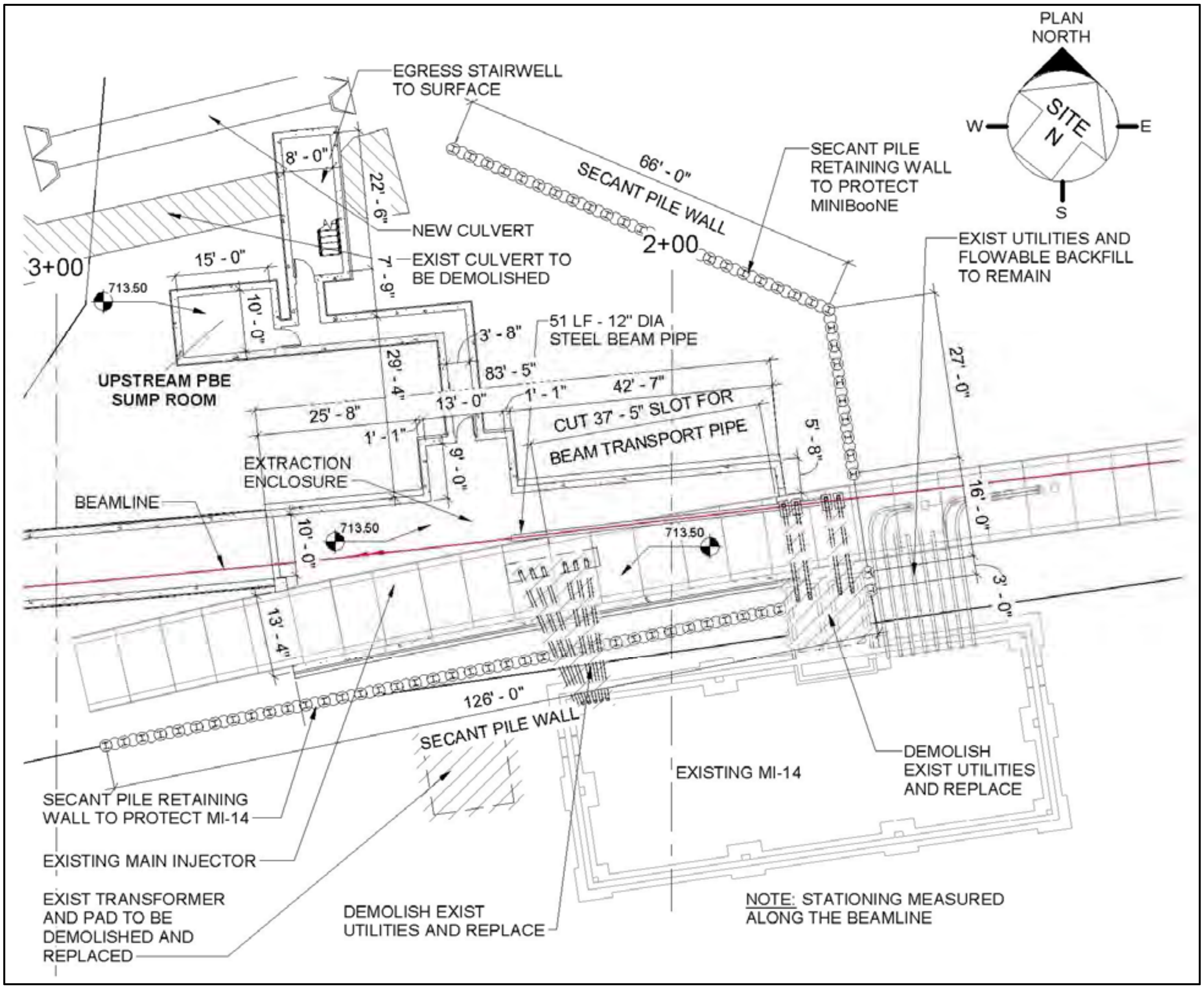

Figure 5-15: Beamline Extraction Enclosure and Primary Beam Enclosure with section A-A

### 5.5.1.2 Electrical

The Primary Beam Enclosure will be outfitted with electrical facilities to support the small programmatic equipment and periodic maintenance tasks. Conventional Facilities will provide lighting and electrical facilities to support all mechanical systems, small programmatic loads, and power receptacles needed for maintenance. The power will be delivered from the nearest surface building to 480-V panels in the enclosure. Dry type transformers with 208/120-V panelboards will be provided in the enclosure for power devices and receptacles.

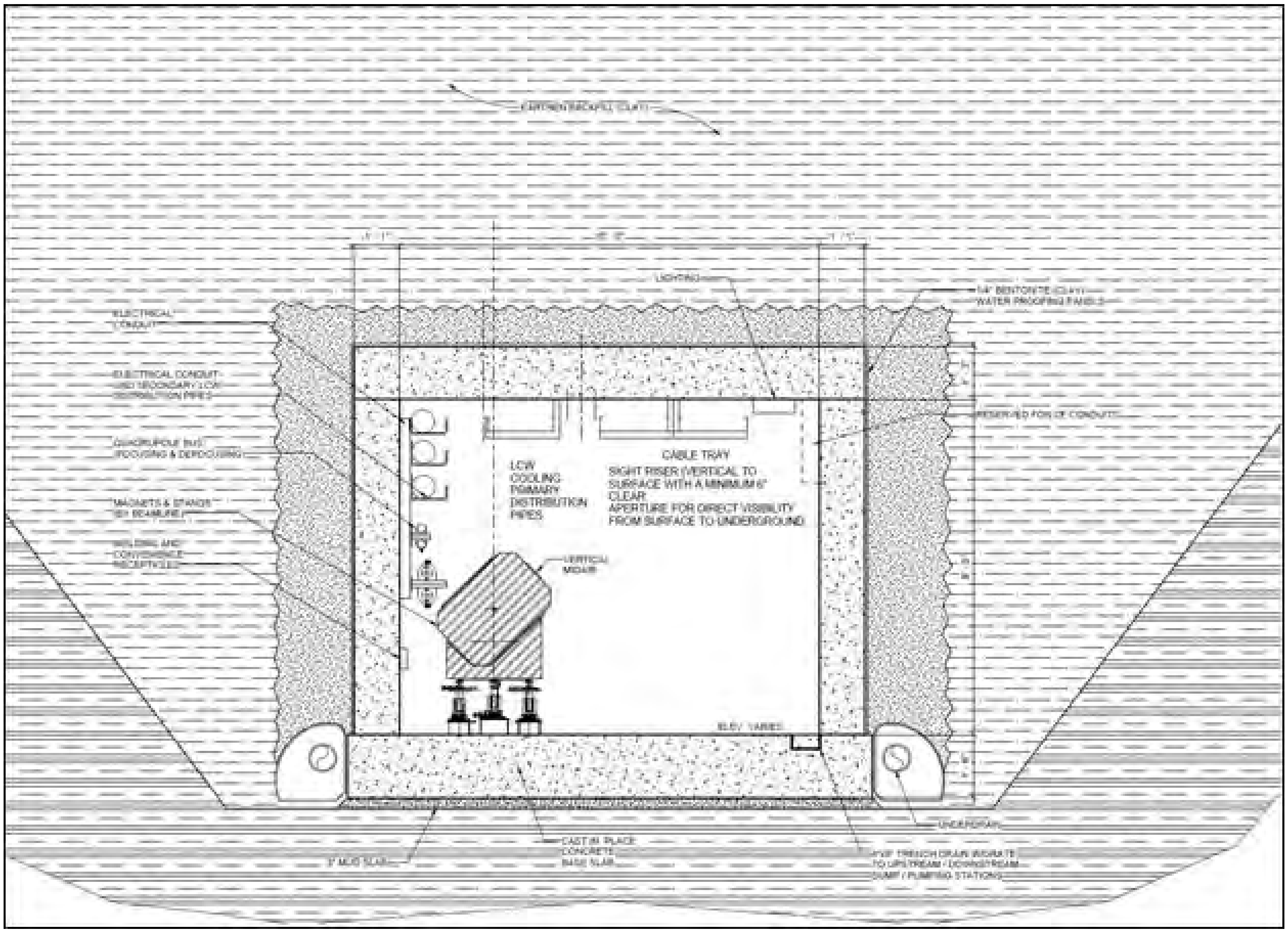

Figure 5-16: Primary Beam Enclosure showing technical components – typical enclosure section

### 5.5.1.3 Plumbing

The Primary Beam Enclosure is equipped with a floor trench drain and exterior underdrains, a fire department stand pipe and hose connection fire suppression system, as well as LCW cooling lines, and power and control lines for the equipment. Site risers to the surface, will be constructed at the required spacing along the Primary Beam Enclosure, and will provide a minimum 6-in clear aperture for magnet/beam alignment. Enclosure underdrain water collection is provided at both low ends of the enclosure. These duplex sump pumps discharge to grade where they will flow into existing ditches or cooling ponds.

### 5.5.1.4 Fire Protection/Life Safety Systems

Conventional Facilities is responsible for the design, cost/scheduling, and construction of the fire protection and life safety systems including the mechanical (emergency ventilation), electrical (emergency generator for lighting, ventilation, and sump pumping, fire alarms, and communication), and plumbing (fire suppression/sprinkler piping and fixtures, and emergency sump pumping). Any space where the application of water could constitute a radiation-related risk as determined by LBNF and the AHJ will not have sprinkler systems.

## 5.5.2 Decay Pipe

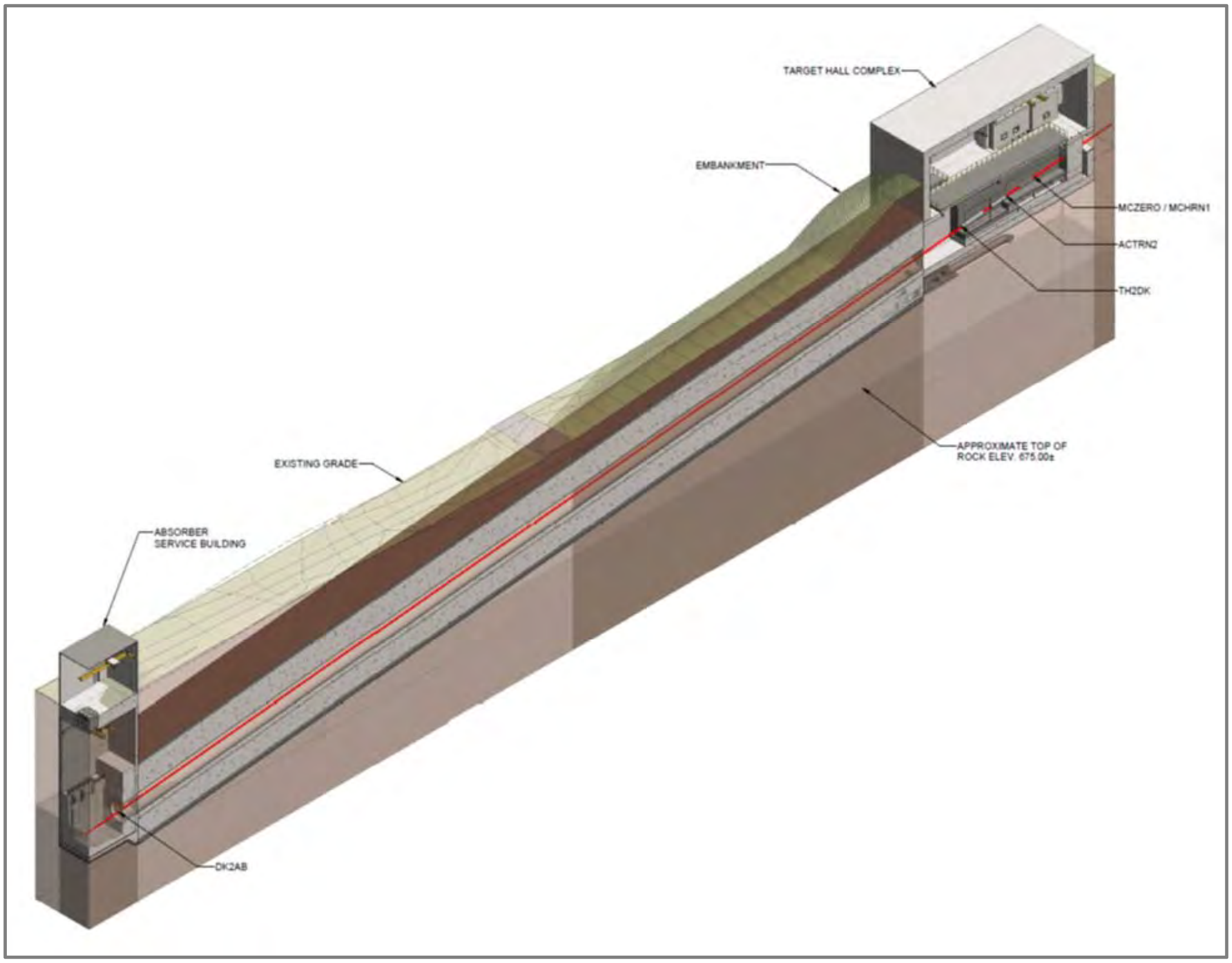

Figure 5-17: Longitudinal section through the decay pipe

The LBNF decay pipe (longitudinal section shown in Figure 5-17) begins at the downstream end of the target chase/Target Hall and continues 636 ft (194 m) at a decline of approximately 10% to the Absorber Hall, which is approximately 95 ft below grade.

The decay pipe consists of two concentric steel pipes; the inner pipe is a 13 ft-2 in (4-m) diameter pipe and the outer pipe is a 14 ft-5 in (4.4-m) diameter pipe. See Figure 5-18, which shows a tranverse section of the decay pipe. The walls of both pipes are ½-inch thick. The outer pipe is surrounded by 18 ft-4.5 in (5.6-m) of cast-in-place concrete shielding.

The ends of both the inner and outer pipes are sealed. The inner pipe will be filled with heliumby the Beamline subproject. The annulus between the inner and outer pipes is used to convey air for air-cooling the decay pipe steel and surrounding shielding concrete. Supply air from the decay pipe's air handling unit located in the Target Hall Complex air handling room is directed through the 20-cm annulus between the two concentric decay pipes towards the Absorber Complex and returns back to the Target Hall

Complex through four 28-in diameter pipes that run parallel to the decay pipe and then back to the air handling room. (See Figure 5-19.) Also shown in Figure 5-19 is a 12-in diameter, gasketed concrete or HDPE pipe (cast into the decay pipe concrete shielding backfill) air duct connection from the interior of the Absorber to the Target Hall. This will allow a negative air pressure to be maintained in the Absorber Hall.

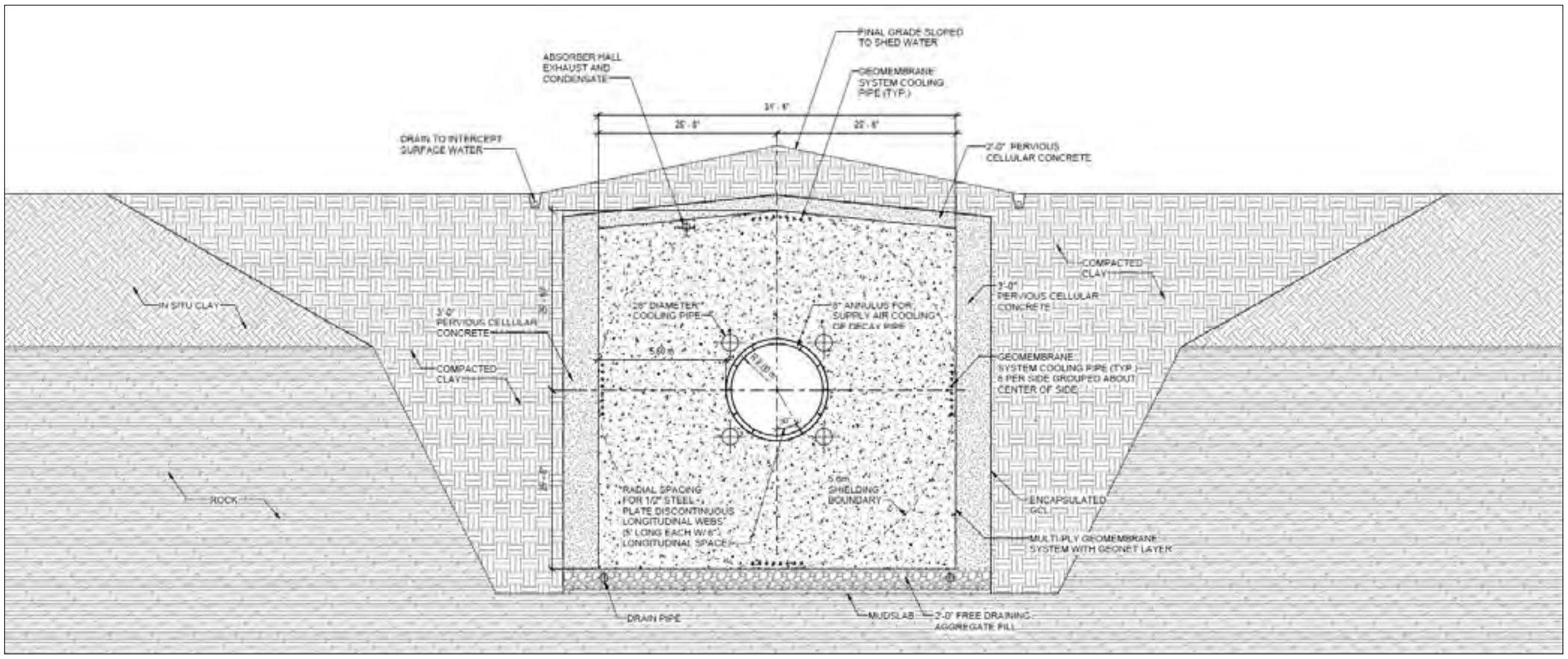

Figure 5-18: Decay pipe cross section

### 5.5.2.1 Decay Region Geosynthetic Barrier System

A multiply geosynthetic barrier system, as shown in Figure 5-20 will surround the decay pipe structure to protect the decay region from potential groundwater infiltration and also to protect the surrounding groundwater from any possible tritiated water being created and escaping from the decay region. The use of the geosynthetic system in the decay region is a unique application of standard and common practices and materials used for decades in the landfill industry. This system will create a three-dimensional barrier system between the decay region and the environment.

The proposed geosynthetic barrier system concept includes an encapsulated geosynthetic clay liner surrounding a several-feet-thick drainage region constructed of free-draining porous cellular concrete that surrounds an inner multi-ply geosynthetic barrier in contact with the decay pipe shielding concrete. The inner multi-ply system will incorporate a minimum of one membrane barrier and a geonet leak detection layer placed on the inside of the membrane barrier.

A secondary cooling pipe system consisting of approximately 32 – 3-in diameter pipes located along the perimeter of the decay pipe shielding concrete can be employed using a one-pass air system to provide additional cooling if the geosynthetic barrier system temperature rises above acceptable levels.

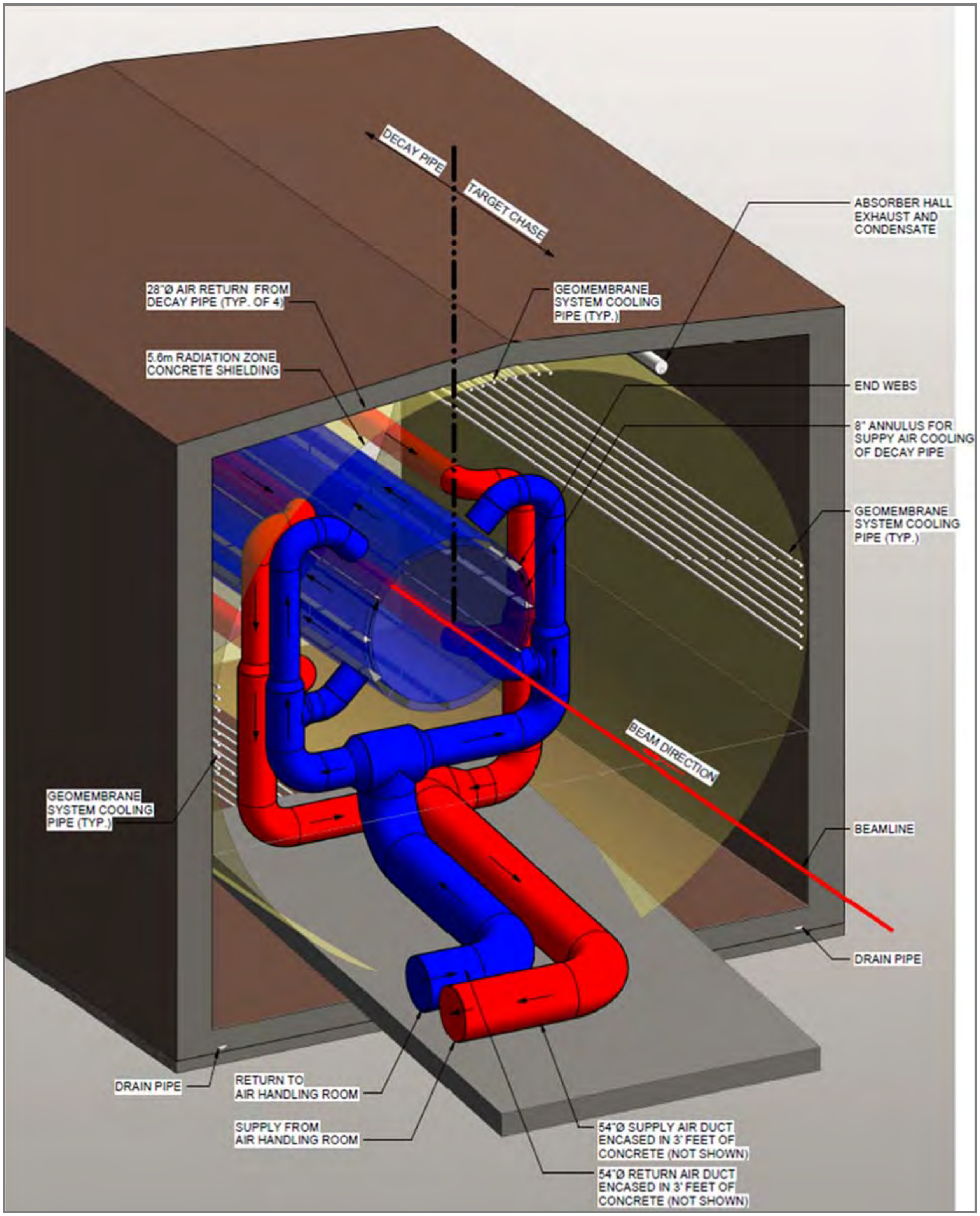

Figure 5-19: Decay pipe cooling ducts

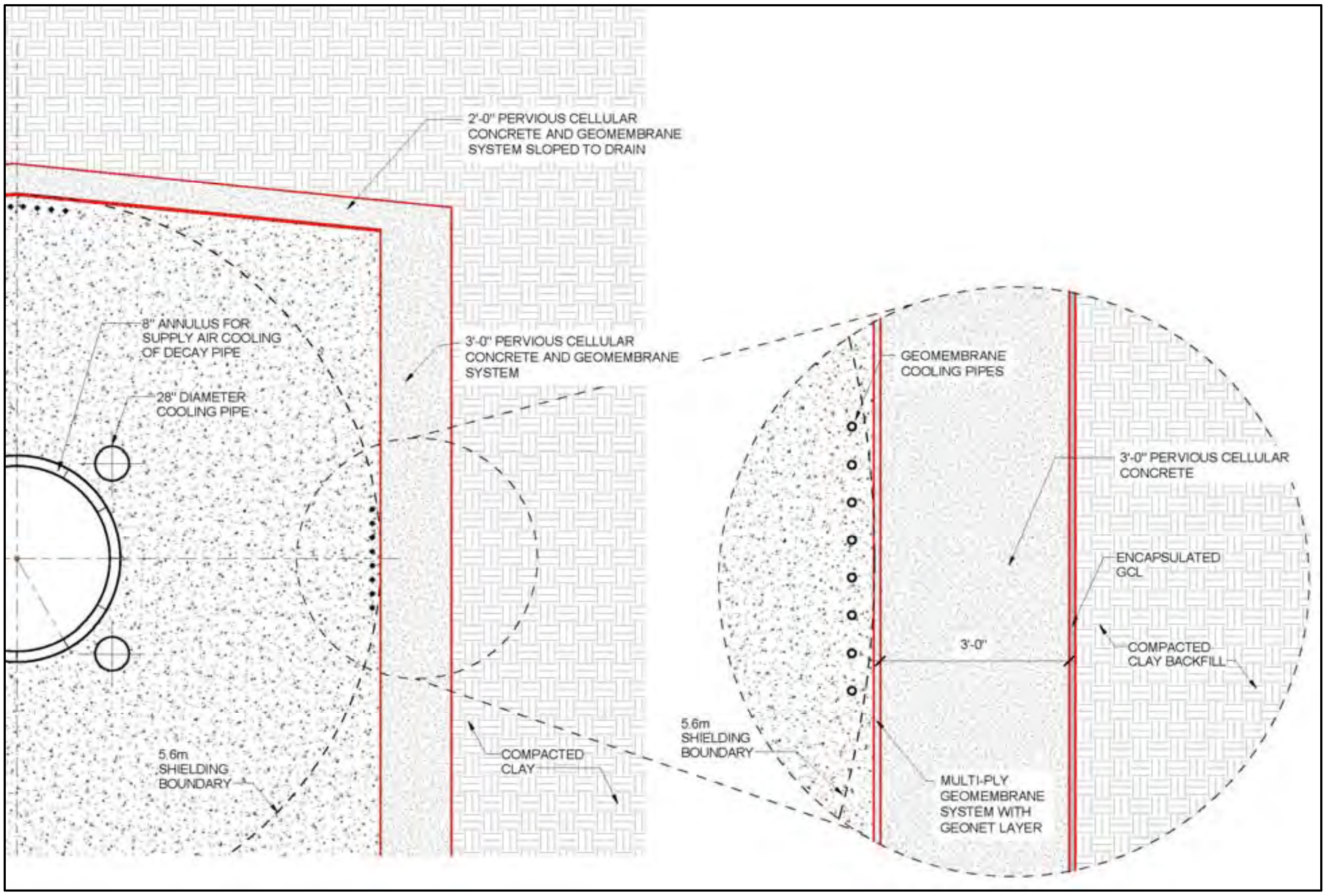

Figure 5-20: Geomembrane system section view

# 5.6 Absorber Hall and Support Rooms

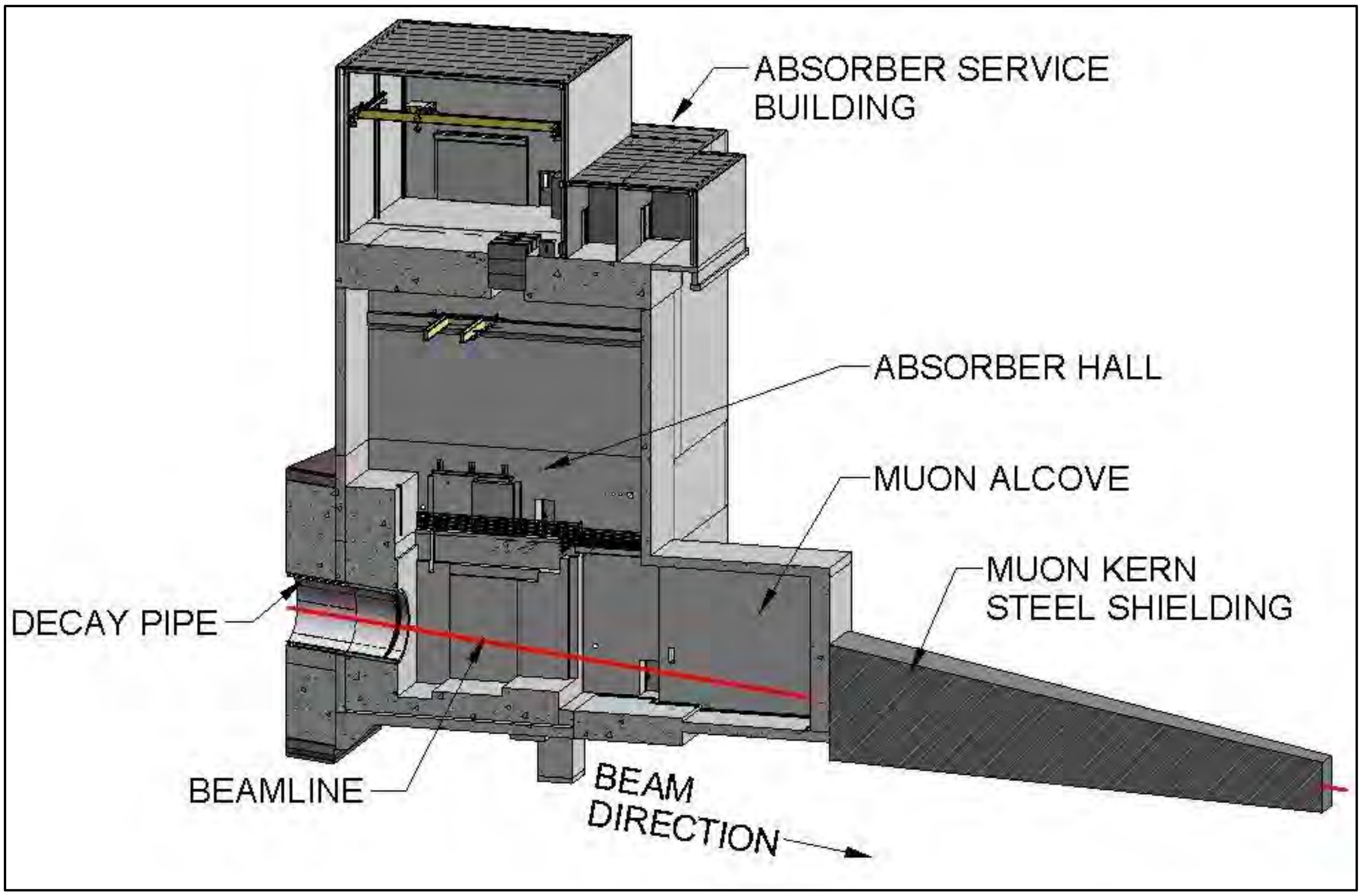

Figure 5-21 Absorber Complex including the Absorber Service Building (LBNF-30)

An overview of the Absorber Hall and Support Rooms is shown in Figure 5-21. The Absorber Hall will be approximately 90 ft below grade. The Absorber Hall will house the concrete shielded hadron absorber and monitor, the Muon Alcove, and the absorber support rooms, all constructed in an open cut soil excavation to bedrock and then a drill-and-blast rock excavation to the base of the underground facility. Also housed in the Absorber Hall and Muon Alcove is the DUNE Project's beamline measurement system (muon detectors) and a Global Data Acquisition system. A geosynthetic barrier system will be included between the rock/open cut earth excavation and the internal concrete structure to seal the facility from groundwater infiltration. The underground Absorber Hall will be a three-level cast in place concrete structure. The lower level of the Absorber Hall will house the absorber pile/enclosure/hadron monitor, Muon Alcove, and sump and pump systems, and will provide access to the base of the egress shaft. The middle level of this underground facility will house the top of the absorber pile and the RAW Room. The upper level of this underground facility will house the Instrumentation and Air Handling rooms.

As shown in Figure 5-22**,** the 9-ft thick roof/ceiling of the underground portion of the facility will serve to provide concrete shielding between the Absorber Hall and the above ground Absorber Service Building (LBNF-30). A portion of the concrete foundation walls for the above ground service building serves as the shaft walls to the underground facilities' equipment and utilities access corridor, which is accessed via a 6.5-ft wide by 17-ft long opening in the truck bay floor of the above ground service building, which has 30-ton bridge crane coverage. This opening will have a 9-ft thick concrete shield block air sealed hatch cover that will be provided by the Beamline Project.

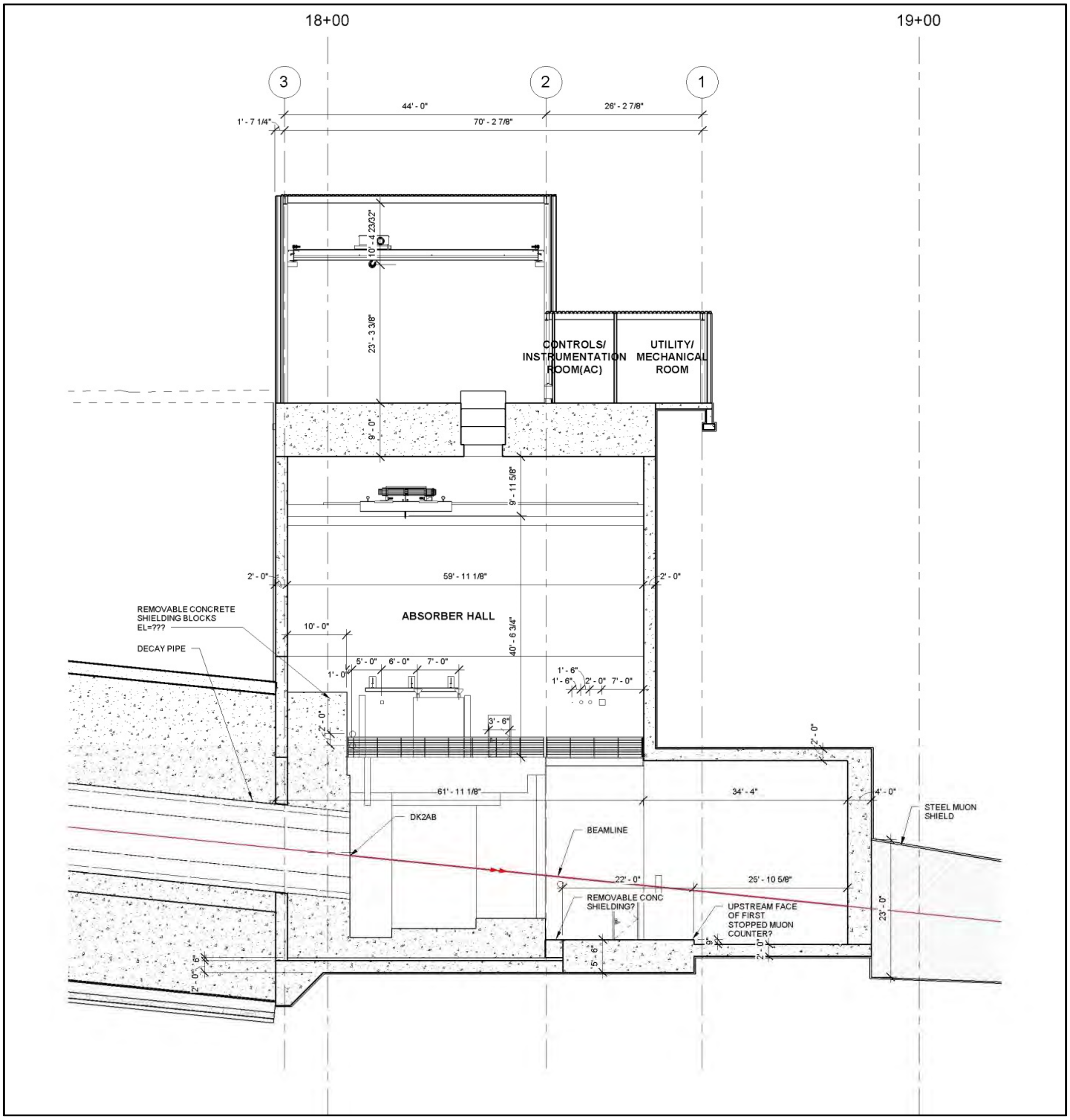

Figure 5-22: Absorber Hall Longitudinal cross section cut along the decay pipe centerline

### 5.6.1.1 Grouting of the Rock Mass in the Decay/Absorber Region

The downstream end of the Decay Region and the base of the Absorber Hall will penetrate the top of rock to a depth of up to about 25 ft. The soil/rock interface and the upper portion of the rock mass is regionally known as a water bearing zone or aquifer. Due to the importance of providing as dry a Decay

Region and Absorber Hall as possible, a systematic program to grout the rock mass to seal off fractures and bedding planes is included in the conceptual design. This grouting program will be executed prior to any excavation and will augment the groundwater barrier system installed between the rock face and the internal concrete structure.

### 5.6.1.2 Mechanical

Ventilation for this area is to be provided by a dedicated outside air system (DOAS) located in the LBNF-30 service building mechanical area. The DOAS shall provide adequate personnel ventilation and dehumidified neutral air to the underground space for humidity control and positive pressurization with respect to the Absorber Hall. Maximum final space humidity shall be 50% RH.

A 2,400-CFM combined refrigeration/desiccant air conditioning unit will be provided that is capable of removing of heat and moisture from the Absorber Hall. The space condition of the Absorber Hall will be kept at 80ºF +/- 5ºF. All materials of the unit that come in contact with the airstream or condensate will be resistant to corrosion from the level of nitric acid that is anticipated to be present in the air. The AHU shall be constructed minimizing single points of failure.

### 5.6.1.3 Electrical

The Absorber Hall will be outfitted with electrical facilities to support the small programmatic equipment and periodic maintenance tasks. Conventional Facilities will provide lighting and electrical facilities to support all mechanical systems, small programmatic loads and power receptacles needed for maintenance. The power will be delivered from the Absorber Hall Service Building main panelboard to 480 V panels in the lower Absorber Hall. Dry type transformers with 208/120 V panelboards will be provided in the hall for small power devices and receptacles.

### 5.6.1.4 Plumbing

The underground absorber area sump pump system will consist of three duplex pump systems. The first will receive any drainage from the decay pipe enclosure system. This system will be provided a dedicated monitoring sump with switchable automatic/manual controls and a holding tank so that contaminating drainage can be held and monitored. The second system receives drainage from the decay enclosure under drainage and is discharged to the third and main sump pump system. This system will be sized to serve the entire upstream underground facilities with redundant back up pumps and emergency back-up power. The system shall be designed to a 0.9999 reliability level. This system shall discharge to a surface holding tank near LBNF-30. Duplex pumps within the holding tank shall discharge to the site-wide ICW system.

### 5.6.1.5 Fire Protection/Life Safety Systems

Conventional Facilities is responsible for the design and construction of these systems including the mechanical (emergency ventilation), electrical (emergency generator for lighting, ventilation, sump pumping, fire alarms, and communication), and plumbing (fire suppression/sprinkler piping and fixtures, and emergency sump pumping). Any space where the application of water could constitute a radiation-related risk as determined by LBNF and the AHJ will not have sprinkler systems.

## 5.6.2 Near Detector Hall and Support Rooms

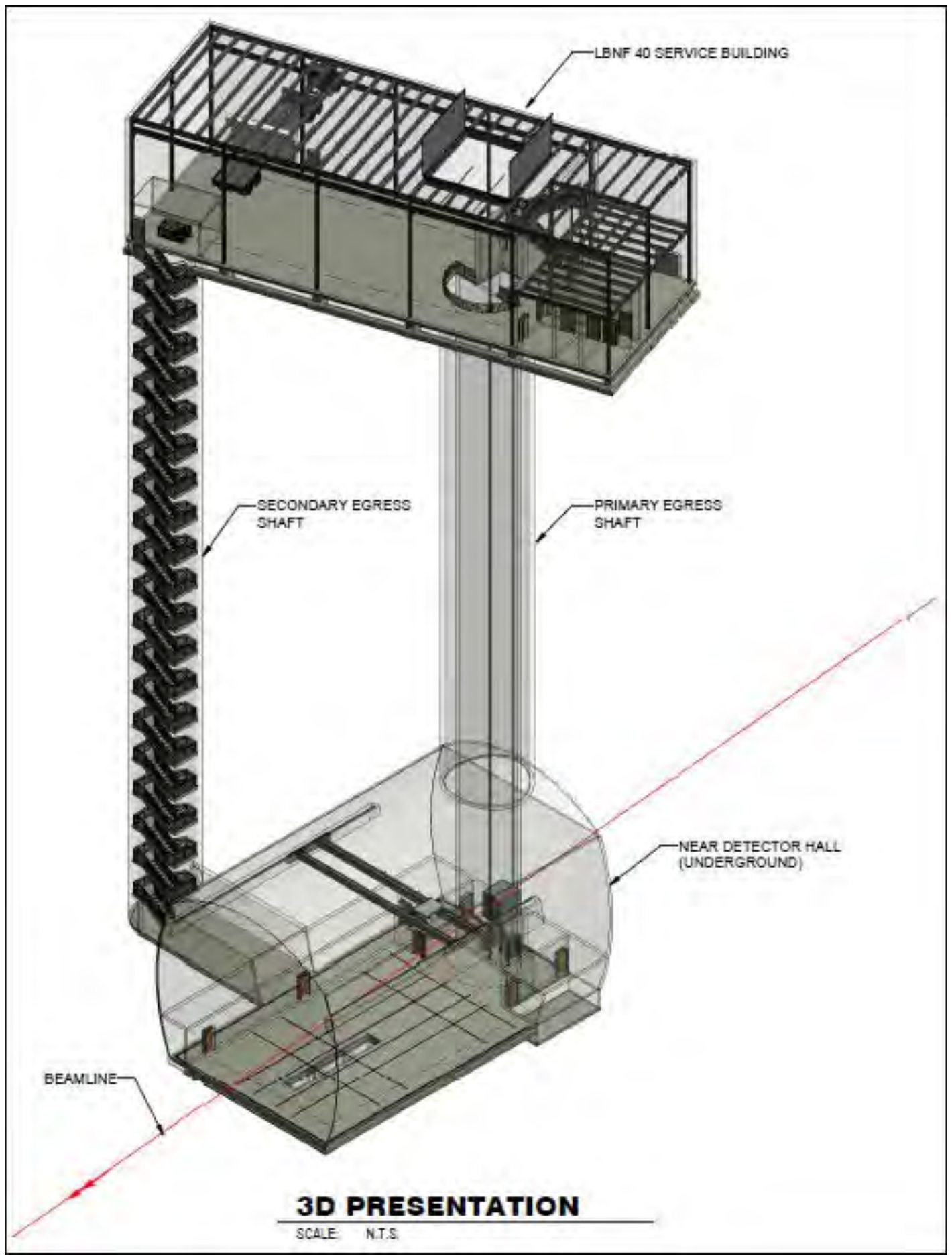

Figure 5-23: Near Detector Complex

The Near Detector Hall and support rooms (see Figure 5-23 and Figure 5-24) will house the LBNF near detector and related components, and is located a minimum of 689-ft (210-m) downstream of the Absorber Hall approximately 185 ft below grade. The Near Detector Hall has been sized to accommodate either of two detector technologies that were considered during Conceptual Design; a Straw Tube Tracker design, and, alternatively a design similar to the MicroBooNE liquid argon (LAr) detector. At Conceptual Design, the Conventional Facilities design as presented is generic enough to accommodate either detector technology. The LAr detector will require special Oxygen Deficiency Hazard (ODH) emergency ventilation, and egress design criteria, which has been included in the reference design.

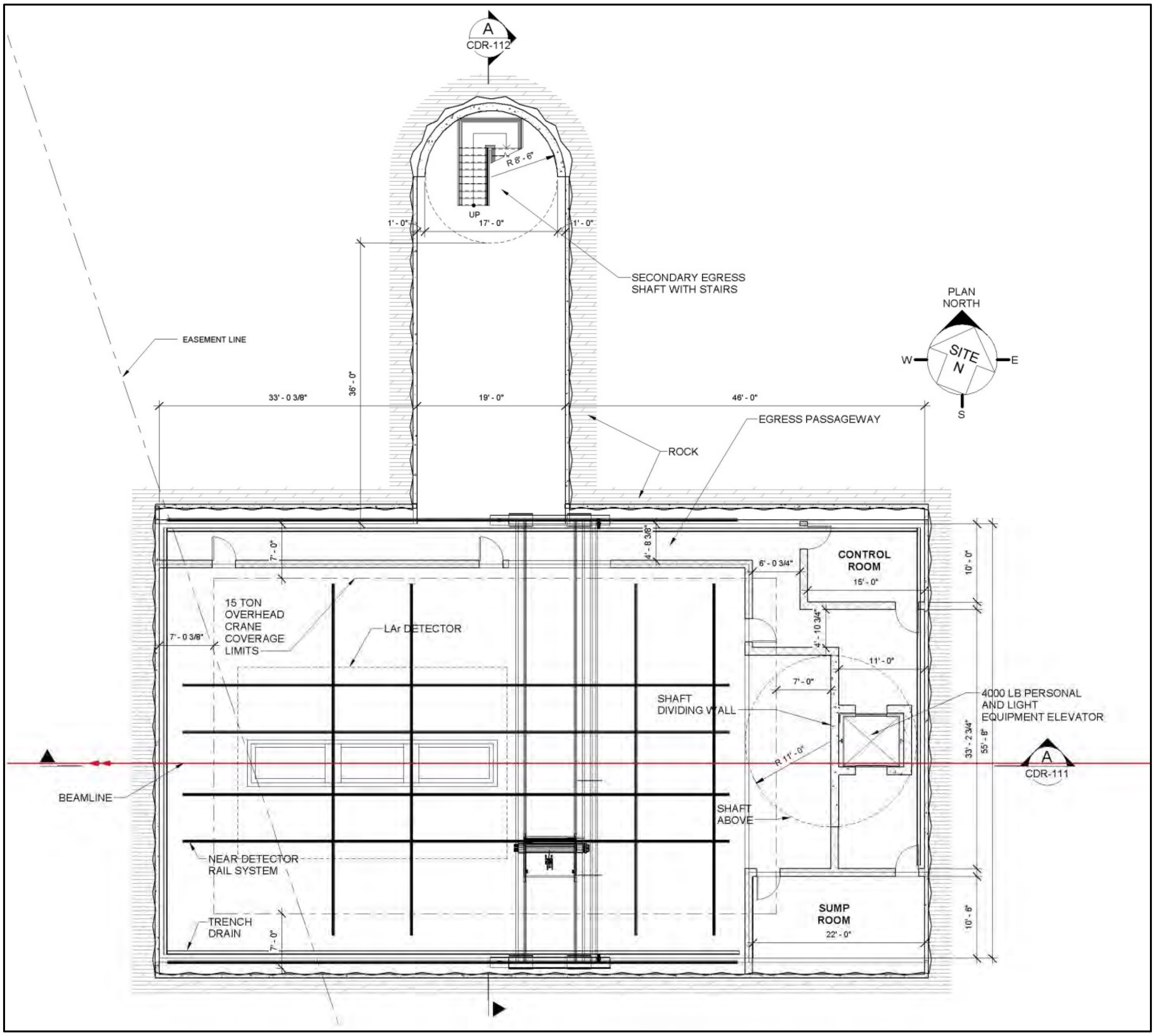

Figure 5-24: Near Detector Hall plan view

The 5,510-ft2 Near Detector Hall and support rooms will include access to a 22-ft diameter shaft which will be used for equipment handling and primary egress that includes utilities and has a 4,000-lb capacity elevator. This shaft serves as the major equipment access corridor from LBNF-40 above to the Near Detector Hall and support rooms below grade. The Near Detector Hall will have a 15-ton bridge crane running the length of the hall that will be used for installing detector components and related equipment lowered down from the surface building to the Near Detector Hall. A secondary egress shaft which is 17 feet in diameter will allow for personnel to egress the hall via a stairway to the surface.

A scope increase option is being considered to double the length of the Detector Hall to provide space for additional detectors in the future. There is adequate distance between the Absorber Complex and the present upstream end of the Detector Hall to accommodate additional length without compromising the 210-m muon range-out distance.

### 5.6.2.1 Mechanical

Ventilation for the underground enclosure area is to be provided by a CHW/HW/Desiccant dedicated outside air system (DOAS) located in the Near Detector Service Building (LBNF-40) mechanical area. The DOAS shall provide adequate personnel ventilation and dehumidified neutral air to the underground space for humidity control and positive pressurization with respect to the Near Detector Hall. Maximum final space humidity shall be 50% RH. The two ventilation systems serving the Near Detector Hall/Support Rooms and the emergency egress corridor (including stairway and elevator areas also) are provided by a 2000-CFM and a 1000-CFM AHU respectively that are located in the surface service building. Additional local cooling in the Near Detector Hall/Support Rooms will be provided by a 6,000-CFM CHW AHU that will be located below grade in the Near Detector Hall. Local cooling for the Near Detector Hall control room will be provided by small fan coil units. A separate ODH emergency ventilation exhaust system will be required for the LAr detector. This system will draw air from the LAr spill containment and exhaust directly to the outside through a separate dedicated 36-inch exhaust duct.

### 5.6.2.2 Electrical

The Near Detector Hall will be outfitted with electrical facilities to support the small programmatic equipment and periodic maintenance tasks. Conventional Facilities will provide lighting and electrical facilities to support all mechanical systems, small programmatic loads and power receptacles needed for maintenance. The power will be delivered from the main panelboard in the Near Detector Service Building to 480 V panels in the below grade Detector Hall. Dry type transformers with 208/120 V panelboards will be provided for small power devices and receptacles. Lighting and emergency signage will be provided with remote or isolated ballast and alternate power sources.

### 5.6.2.3 Plumbing

The near detector area sump pump system shall have redundant back up pumps and emergency back-up power. The system shall be designed to a 0.9999 reliability level. This system shall discharge to a surface holding tank near LBNF40. Pumps within the holding tank discharge to the site wide ICW system. This underground complex is provided with a wet pipe sprinkler system served from the LBNF-40 domestic water system.

### 5.6.2.4 Fire Protection/Life Safety Systems

Conventional Facilities is responsible for the design and construction of these systems including the mechanical (emergency ventilation), electrical (emergency generator for lighting, ventilation, sump pumping, fire alarms, and communication), and plumbing (fire suppression/sprinkler piping and fixtures, and emergency sump pumping.

# 6 Near Site Facilities: Beamline

## 6.1 Introduction

The LBNF Beamline at Fermilab is designed to provide a neutrino beam of sufficient intensity and appropriate energy range to meet the goals of the DUNE experiment with respect to long-baseline neutrino-oscillation physics. It aims a wide band neutrino beam toward detectors 1,480m underground, at the Sanford Underground Research Facility in South Dakota about 1,300 km away.

The Beamline chapter of this CDR volume provides a comprehensive summary of the beamline design. A more detailed description of the design can be found in *Annex 3A: Beamline at the Near Site* [30].

The design is a conventional, horn-focused, sign selected neutrino beam. A proton beam is extracted from the Fermilab Main Injector (MI) and transported to a target area. The secondary particles produced when the beam strikes the target are focused and aimed toward the far detector, and are allowed to decay to generate the neutrino beam. At the end of the decay pipe, an absorber pile removes the residual hadrons. (See Figure 6-1.)

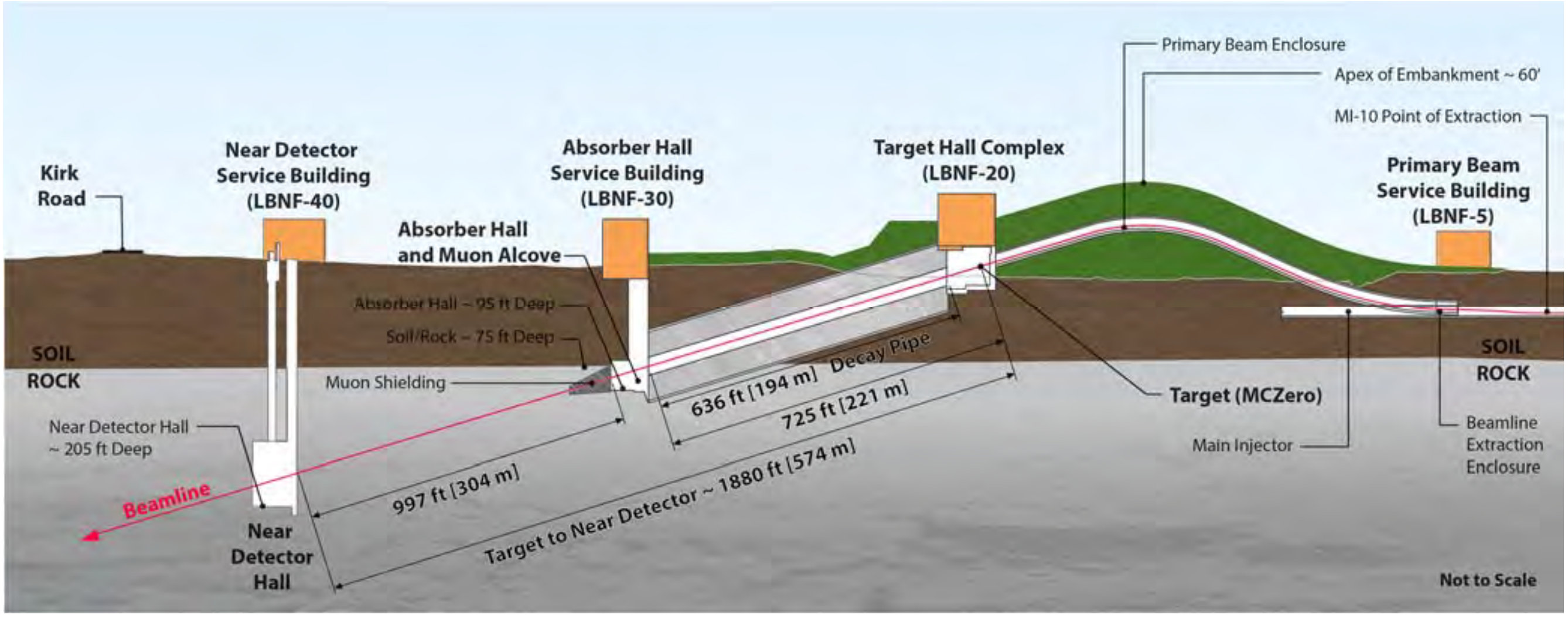

Figure 6-1: Longitudinal section of the LBNF beamline facility at Fermilab.

The proton beam (60 – 120 GeV) is extracted at MI-10, a new installation. The extraction and transport components send the proton beam through a man-made embankment/hill whose apex is at 18.3 m from the ground and with a footprint of ~21,370 $m^2$. The beam is bent downward towards a target located at grade level. The overall bend of the proton beam is 7.2$^o$ westward and 5.8$^o$ downward to establish the final trajectory towards the far detector.

The general primary-beam specifications and beam characteristics are listed in Table 6-1 and Table 6-2, corresponding to the parameters expected following the PIP-II [31] upgrade, which will provide 1.2 MW at 120 GeV and 1.0 MW at 60 GeV.

Table 6-1: Summary of principal primary proton beam design parameters

| Parameter | Value | |
|---|---|---|
| Energy | 60 GeV | 120 GeV |
| Protons per cycle | $7.5\times10^{13}$ | $7.5\times10^{13}$ |
| Spill duration | $1.0\times10^{-5}$ sec | $1.0\times10^{-5}$ sec |
| Protons on target per year | $1.9 \times 10^{21}$ | $1.1\times10^{21}$ |
| Beam/batch (84 bunches) | $12.5\times10^{12}$ nominal; ($8\times10^{11}$ commissioning) | |
| Cycle time | 0.7 sec | 1.2 sec |
| Beam Power | 1.03 MW | 1.2 MW |

Table 6-2: Primary proton beam characteristics

| Parameter | Value |
|---|---|
| Beam size at target | 1.5 to 1.7 mm |
| $\Delta$ p/p | $11\times10^{-4}$ 99% ($28\times10^{-4}$ 100%) |
| Transverse emittance | 30π μm 99% (360π μm 100%) |
| Beam divergence (x,y) | 17 to 15 μrad |

Neutrinos are produced after the protons hit a solid target and produce mesons that are subsequently focused by magnetic horns into a decay pipe where they decay into muons and neutrinos. The distance from the nominal target postion (MCZero) to the end of the decay pipe is 221 m, of which 27 m is in the target chase and 194 m is in the decay pipe itself. See Figure 6-1. The focusing structure is optimized to provide a wide band neutrino beam with an energy range between 0.5 and 5 GeV so as to cover the first and second neutrino oscillation maxima, which for a 1300 km baseline are approximately 2.4 and 0.8 GeV.

The facility is designed for initial operation at proton beam power of 1.2 MW with the capability to support an upgrade to 2.4 MW. The Beamline systems that are designed from the beginning for 2.4 MW operation are those which cannot be replaced later and include:

- The enclosures (primary proton beamline, target chase, Target Hall, decay pipe, Absorber Hall)
- The radiological shielding of the enclosures, the only exception being the roof of the Target Hall that can be easily upgraded later for 2.4 MW
- The primary proton beamline components
- The water cooled target chase shielding panels
- The decay-pipe and its cooling and the decay pipe downstream window
- The beam absorber
- The remote handling equipment
- The Radioactive Water (RAW) system piping

None of these can be upgraded after exposure to a high-intensity beam.

The LBNF Beamline is designed for twenty years of operation, while the Beamline Facility, including the shielding are planned for a lifetime of 30 years to allow for cooldown of irradiated components before decommissioning. A conservative plan is that for the first five years, the Beamline will operate at 1.2 MW of beam power and for the remaining fifteen years at 2.4 MW.

## 6.1.1 Scope

For organizational purposes, the LBNF Beamline Project has four principal systems:

- **Beamline Management:** Management and oversight, modeling effort, radiation physics and radiation protection activities
- **Primary Beam:** Components required for the initial, high-intensity proton beam
- **Neutrino Beam:** Components used to create a high-intensity neutrino beam from the initial proton beam.
- **System Integration:** Integration across the Beamline sub-project, controls, interlocks monitoring, alignment and installation coordination.

## 6.1.2 Neutrino Flux Spectra with the Reference Design

The goal for accumulating 120-GeV protons at the neutrino target with beam power of 1.2 MW is $1.1\times10^{21}$ protons-on-target (POT) per year. This assumes $7.5\times10^{13}$ protons per MI cycle of 1.2 sec [31] and a total LBNF efficiency of 0.56. The total LBNF efficiency used in the POT calculation and discussed below includes the total expected efficiency and up-time of the accelerator complex as well as the expected up-time of the LBNF Beamline.

The neutrino flux at the far detector site in the absence of oscillations is shown in Figure 6-2 and Figure 6-3 with the horns focusing positive and negative particles respectively. The calculations are for a 120 GeV proton beam, NuMI horns at 230 kA and 6.6 m apart, distance between Horn 1 and the decay pipe of 27 m, and decay pipe of 194 m length and 4 m diameter.

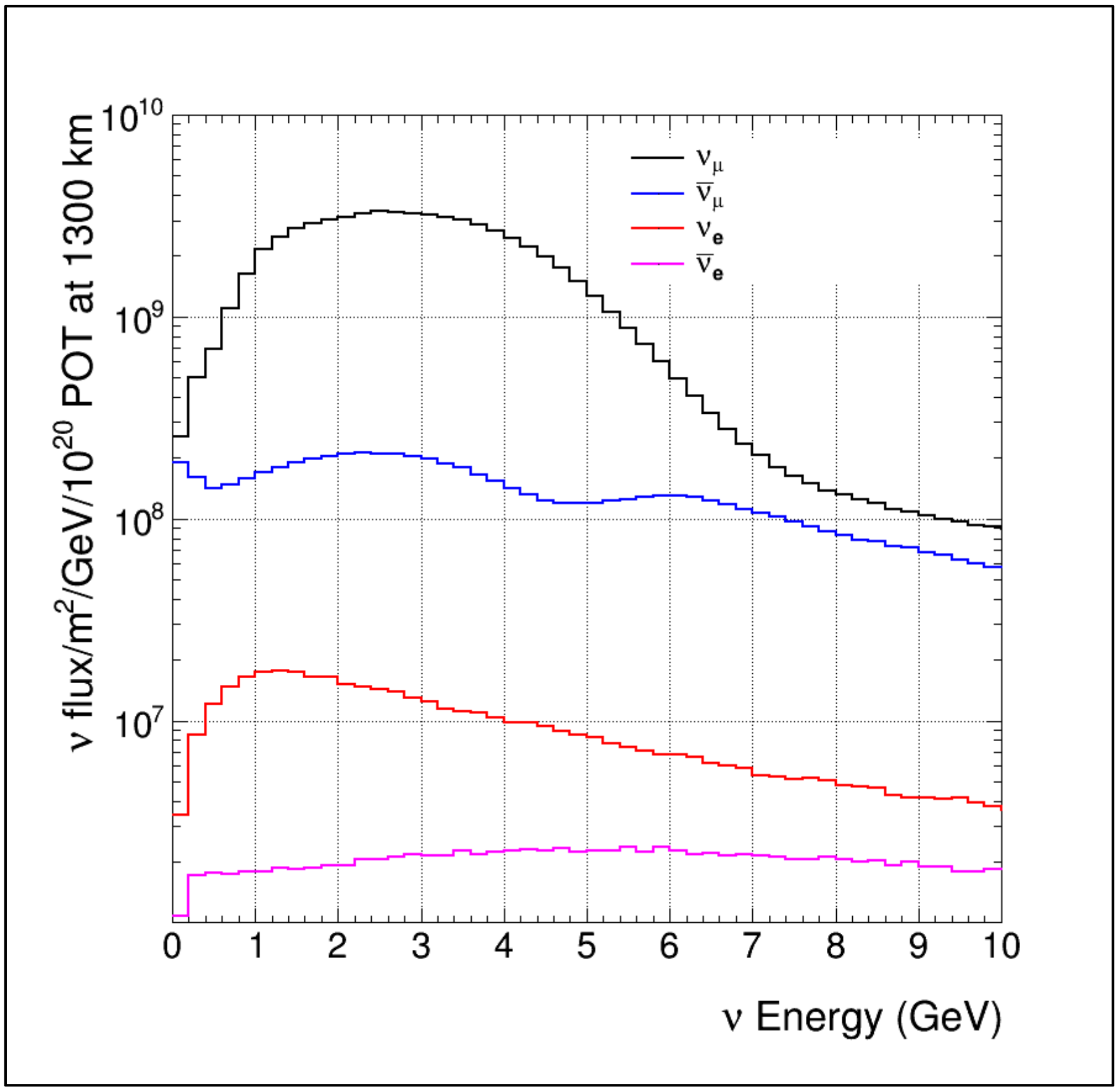

Figure 6-2: Neutrino Fluxes at the Far Detector as a function of energy in the absence of oscillations with the horns focusing positive particles. In addition to the dominant $\nu_\mu$ flux, the minor components are also shown.

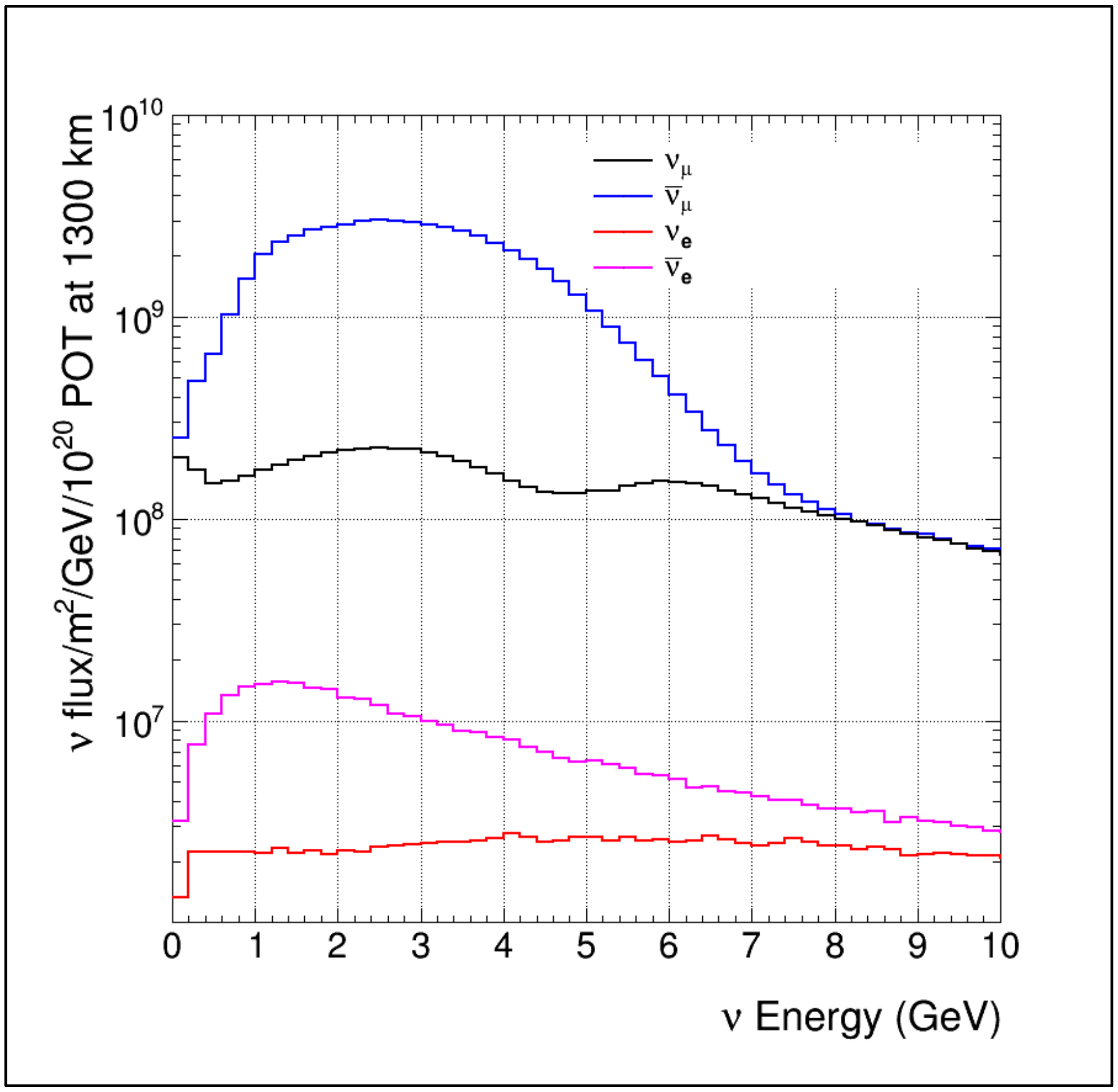

Figure 6-3: Antineutrino Fluxes at the Far Detector as a function of energy in the absence of oscillations with the horns focusing negative particles. In addition to the dominant $\bar{\nu}_\mu$ flux, the minor components are also shown.

# 6.2 Primary Beam

## 6.2.1 Introduction

This chapter describes the reference design for the LBNF primary (proton) beamline. This system will extract protons from Fermilab's Main Injector (MI) synchrotron, using a single-turn extraction method, and transport them to the target in the LBNF Target Hall. The nominal range of operation is from 60 to 120 GeV.

The principal components of the primary beamline include specialized magnets at the MI-10 extraction point to capture all of the protons in the synchrotron and redirect them to the LBNF beamline, a series of dipole and quadrupole magnets to transport the proton beam to the target, power supplies for all the magnets, a cooling system, beamline instrumentation and a beam-vacuum system for the beam tube.

All of the LBNF primary-beam technical systems are designed to support sustained, robust and precise beam operation. Careful lattice optics design and detailed beam-loss calculations are essential for the proper operation of the primary-beam system. A comprehensive beam-permit and control systems is being developed.

## 6.2.1.1 Design Consideration

### *6.2.1.1.1 Length and Elevation*

Primary-beam extraction using the “MI-10 Shallow” design was chosen after a thorough value-engineering process evaluating shallow and deep configurations at both MI-10 and other locations. See the *LBNF/DUNE Alternatives Analysis* [32]

The shallow beam design offers a significant cost savings for the neutrino beam facility, plus significant advantages with tritium mitigation for the near-grade Target Hall relative to the fully underground design used for NuMI [33].

The target elevation of a few feet above natural grade elevation is chosen to optimize overall facility-construction technical and resource requirements. It was selected as the best balance between minimizing Absorber Hall depth in the rock and the extent of the primary beam transport line above natural grade level. Additional constraints include limiting the maximum primary-beam enclosure angles to 150 milliradians, and achieving the required trajectory for transport of the neutrino beam to the far detector site. A profile-view schematic of the primary-beam transport with Target Hall and decay region is shown in Figure 6-1.

### *6.2.1.1.2 Existing Infrastructure and Shielding*

The choice of a shallow beam extracted at MI-10 avoids beamline crossings, and allows for a simpler extraction enclosure, enabling a cost-effective facility design of the entire extraction region. It interferes minimally with existing beam systems in this region, reduces magnet count, and also provides some shielding separation from accelerator-tunnel beam losses at the beginning of the LBNF primary-beam transport enclosure.

### *6.2.1.1.3 Beam Control*

Techniques, hardware and control applications for accomplishing primary-beam control at the required level were developed for the NuMI proton beamline. These features have been demonstrated to perform very well during the decade long operation of the NuMI beamline and are therefore being used in the design for the LBNF primary beam.

## 6.2.1.2 Reference Design Overview

The LBNF primary beamline is extracted using single-turn extraction, in which all the protons accelerated in the Main Injector is diverted to the LBNF beamline within one evolution. After extraction, the beam is controlled by a series of dipole (bending) and quadrupole (focusing) magnets collectively called the “lattice optics.” The LBNF lattice optics directs the proton beam toward the target and resulting neutrinos toward the far detector.

In the last section of the primary beamline, the beam size and divergence are tailored to the desired distribution for hitting the production target. This is accomplished by a set of eight quadrupoles in the final-focus section, which can be tuned to produce a wide range of beam spot sizes while maintaining a narrow angular spread. This section also cancels any dispersion introduced in the beamline.

Some magnets will be grouped into a single "magnet loop" and powered by a single power supply, whereas others will be powered individually, according to the lattice optics design. In order to maintain the lowest possible power consumption, all of the larger magnet loops will be ramped. A primary water system will feed cooling water to the magnets and power supplies. Beam instrumentation will monitor important beam parameters, such as beam positions, stability, losses, and intensity. A vacuum system will maintain a vacuum of better than $10^{-7}$ torr residual gas pressure in the beam tube in order to reduce the beam loss due to proton-gas interaction.

## 6.2.2 Lattice Optics

### 6.2.2.1 Overview

LBNF will implement a modular optics design comprises three distinct lattice configurations in series: the specialized MI-to-LBNF matching section, the transport section and the final focus of the beam on the production target. The beam will be transported with very low losses.

### 6.2.2.2 Optics

The initial section extracts beam from the Main Injector and matches the phase space to the transport section. A series of six fast-pulsed kicker magnets in the MI ring extract the beam. The kicker magnets have a fast ramp-up to the required current followed by a ramp-down; the ramps occur during gaps in the circulating MI beam. The kickers are followed by a set of three Lambertson magnets that bend the extracted beam away (upward) from the MI trajectory. A string of individually controlled quadrupole and dipole magnets match the MI optics to the optics of the transport section. The magnets are summarized in Table 6-3; the beamline, and its interface with the accelerator is shown in Figure 6-4; a detailed write-up can be found in [34].

The transport section steers the beam from the extraction/matching section up to and over the hill toward the Target Hall. This section consists of an optical lattice composed of a series of six periodic focusing units, "FODO" cells, which terminates 119 m upstream of the target. The lattice rises 64 feet above the Main Injector elevation, then returns to final focus elevation. This large change in elevation introduces a vertical dispersion, which must be removed in the final focus section.

The final focus section produces the required spot size at the target and corrects the dispersion. It is composed of eight independently tunable quadrupoles interspersed by five dipoles. The final focus is tunable to produce a spot-size ($\sigma$) from 1.00 to 4.00 mm over the entire momentum range 60 to 120 GeV/c; additionally, the final focus is achromatic.

### 6.2.2.2.1 Corrections

Experience with the MI-style 3Q120 magnets has shown that these magnets are very high quality, with a spread in gradient errors on the order of $\sigma(\Delta G/G) \sim 0.08\%$ or less. For nominal beam parameters at 120 GeV/c, a simple thin-lens calculation predicts that the largest error-wave expected in the 99% beam envelope (±3.89 mm nominal at $\beta = 64.5$ m) would be less than 75 microns.

Alignment tolerance is assumed to be 0.25 mm in each plane and 0.50 mrad for roll. The effect of misalignment has been modeled by randomly misaligning all elements according to a Gaussian distribution and recording the beam excursions; this process is repeated several times.

New corrector elements have been designed to compensate for alignment and gradient errors. Calculations have shown that the errors can be kept to less than 1mm while running the correctors at no more than 60% design current.

### 6.2.2.2.2 Low Loss

A 360π mm·mrad (100%, normalized) emittance with $28.0\times10^{-4}$ (100%) momentum spread can be transmitted cleanly through the beamline. This corresponds to the Main Injector admittance at transition. This admittance, in turn, is determined by the aperture of the Lambertson magnets as seen by the circulating MI beam. Thus, the Main Injector acts as a collimator for the LBNF primary beam.

Although a rigorous argument can be made that the beamline will accept anything the Main Injector can send, an additional study was performed. The front end of the extraction was modeled in a Monte Carlo code (MARS) to allow a ray-tracing study. In the study, the boundary of a 360π mm·mrad phase space was populated in both horizontal and vertical planes. 10,000 rays were tracked through the front end. No tracks were lost on any aperture. Details of this study can be found in [35].

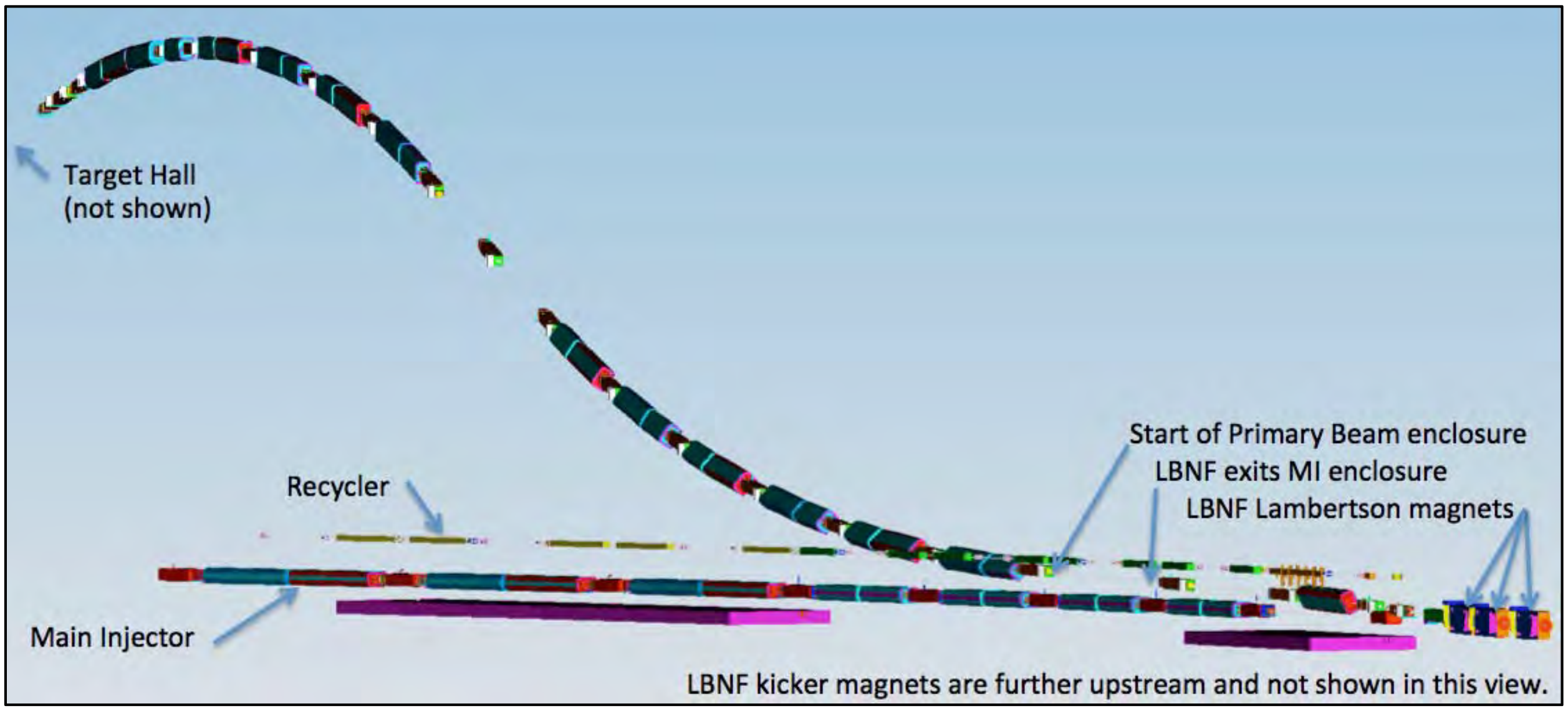

Figure 6-4: Overview of the primary beamline

## 6.2.3 Magnets

### 6.2.3.1 Introduction

This section discusses the magnets that will be used in the primary beamline to steer and focus the beam. The set of magnets includes six extraction kickers, three Lambertson magnets, one current septum C-magnet (the first magnet that is external to the MI ring), 25 main dipole magnets, 21 quadrupole magnets and 23 dipole corrector magnets for fine-tuning. From the extraction point, the lattice optics have to transport the primary beam to the target with the highest possible intensity. The magnet counts are summarized in Table 6-3.

Table 6-3: Summary of primary-beam magnet specifications

| Magnet | Description | Steel Length | Nom. Strength at 120 GeV | Count |
|---|---|---|---|---|
| RKB type Kicker | LBNF extraction | ~1.3 m | 0.058 T | 6 |
| ILA | MI Lambertson | 2.800 m | 0.532 / 1.000 T | 3 |
| ICA | MI C Magnet | 3.353 m | 1.003 T | 1 |
| IDA | MI Dipole 6 m | 6.100 m | 1.003 - 1.604 T | 13 |
| IDD | MI Dipole 4 m | 4.067 m | 1.003 - 1.604 T | 12 |
| 3Q120 | 120 inch quadrupole | 3.048 m | 9.189 - 16.546 T/m | 17 |
| 3Q60 | 60 inch quadrupole | 1.524 m | 11.135 - 17.082 T/m | 4 |
| IDS | LBNF trim dipoles | 0.305 m | Up to 0.365 T | 23 |

### 6.2.3.2 Design Considerations

There are two technical considerations for the beamline magnets beyond providing the integrated dipole field and quadrupole gradient to establish the design lattice. First, the magnet apertures must be large enough to allow for an upgrade of beam power to 2.4 MW and alignment of the magnets should be sufficiently precise so as to not require any further enlargement due to the relative placement of the apertures. Secondly, the magnets must support rapid ramping of excitation. Beam only passes through the magnets for 10 μs of spill time out of each approximately 1 s beam acceleration cycle, so the current between spills can be turned down to save power. This reduces the cost of the magnets (by reducing the amount of conductor needed), the cost of the power supplies, and the cost of the cooling systems, though the ramping does impose additional requirements. The rates at which the magnets can be ramped affect the average power consumption, which in turn, affects the heat load and operating cost of the beamline.

The intention is to make use of existing magnets and designs as much as possible for both cost containment and a general commitment to recycling. For each magnet function, the existing uncommitted magnets available at Fermilab and elsewhere have been reviewed. Suitable candidates have been identified for the Lambertsons, C-magnet, 3Q120 and 3Q60 quadrupoles; they will be refurbished or rebuilt as needed for use in the primary beamline. Existing designs to which additional magnets can be built will be used without any design changes except to the mechanical-support system. The main dipoles and quadrupoles fall into this category. The trim dipole magnets will be constructed according to a new design based heavily on existing corrector magnets. The kickers will be a minor modification to existing NOvA kicker design. Existing tooling will be used to the extent possible for all magnets.

## 6.2.4 Magnet Power Supplies

### 6.2.4.1 Introduction

This section describes the power supply system for the magnets that comprise the lattice optics of the primary beamline. Fermilab has a long history of developing and procuring power supplies for large magnet systems and this experience will guide the design of the LBNF magnet power systems. Some magnets will be grouped and powered by a single "magnet loop" (many magnets powered with one set of supplies), the rest will be powered individually, according to the lattice optics design. The power supply system design seeks to minimize power consumption, and reuse existing supplies from the Tevatron and the NuMI beamline, whenever possible, to better manage the cost.

### 6.2.4.2 Design Considerations

Power consumption is a cost driver during operation, and thus a design driver. In order to maintain the low power consumption, all of the magnet currents will be ramped. Each power-supply design will be selected to provide the best balance between the voltage stresses on the magnet and average power consumption. Also, each power supply will be constructed to use the maximum voltage necessary to reach the peak current and settle into regulation before the beam is extracted from the MI.

### 6.2.4.3 Reference Design

#### *6.2.4.3.1 Power-supply Loops*

The primary beamline will contain a kicker supply, three extraction power-supply loops, five major bending-magnet loops, one large quadrupole loop, eleven minor quadrupole loops and a series of corrector-magnet power supplies.

A kicker system, two extraction Lambertson magnet loops and a C-magnet will be placed at the beginning of the beamline. The power supplies for these magnets will need to be powered from the MI-10 service building and will be part of the MI electrical safety system. This ensures that during access to the MI that these supplies are de-energized using the normal MI procedures. The magnets and the power supplies will be removed from the NuMI beam line extraction and installed at MI-10. The large magnet supplies will be located in the MI-10 service building and will be powered using a relocated, existing transformer from the MI pulse power feeders.

Magnets located in the LBNF Primary Beam Enclosure will be powered from the LBNF-5 Service Building.

#### *6.2.4.3.2 Power Supply Topology*

The ramped power supplies will be constructed using 12-pulse rectifiers with a passive filter connected to the output. Only supplies using 12-pulse rectifiers will be connected to the pulse power feeder because a tuned harmonic filter is installed on the feeder to reduce the voltage stress on the 13.8-kVAC components. The feeder will be extended to connect to the MI beamline feeder system (under the Conventional Facilities scope), which has a harmonic filter with the capacity to power the LBNF beamline. The feeder will need to be extended from MI-10 to the new LBNF-5 service building. The details of the feeder and filter construction are given in Volume 5 of this CDR. The on/off switch for the 13.8-kVAC feeder system will be controlled locally at the LBNF-5 Service Building using a motor-driven disconnect to make access to the MI and LBNF easier for the operation crews. This will allow for access into the LBNF enclosure without turning off the MI.

## 6.2.5 Primary Water System

The primary water system will feed cooling water to the magnets, power supplies and other equipment of the primary beamline. The system will include a heat exchanger, filtration systems, pumps, expansion tank, instrumentation, busswork, and piping, valves, fittings and other hardware.

This system will supply low-conductivity water (LCW) of a resistivity in the range of 16 to 18 MΩ*cm, at a nominal supply temperature of 95°F. (Operationally, when conditions allow most of the year, the system will be operated at 90F, due to the savings in lower electrical resistive losses caused by heat.) The majority of the system's components will be located in the pump room at ground-level in LBNF 5. From there, LCW will be fed to components in LBNF 5, as well as into and throughout the beamline enclosure, and finally to LBNF 20 Service Building and Target Hall horn power supplies. This system may be used to supply the make-up water to the Target Hall radioactive water (RAW) systems. Beamline components at the extraction point in the Q-100 area of the MI will be fed from the MI Global LCW System.

## 6.2.6 Beam Instrumentation

The LBNF primary beamline includes instrumentation and diagnostics to characterize important beam parameters, for example, beam positions, stability, losses, and intensity. It also continuously monitors the operation of all the beamline elements under operating conditions, i.e., with a high-power beam. During the first commissioning and machine studies, the diagnostics systems also have to operate with a low-intensity beam (approximately $3 \times 10^{11}$ protons per batch).

The four core instrumentation systems for the primary beamline are as follows:

1. **Beam-Position Monitors (BPMs):** 26 dual-plane BPMs for beam-trajectory measurement, based on button-style pickups and digital-receiver read-out electronics

2. **Beam-Loss Monitors (BLM):** 30 ion-chamber BLMs for local beam-loss monitoring, and four long (approx. 250-ft) total-loss monitors (TLM)

3. **Beam-Intensity Monitors:** two toroidal transformer-based beam-intensity monitors

4. **Transverse-Beam Profile Monitors:** six dual-plane secondary emission monitors (SEM) to measure the transverse beam profile (effectively a 2D intensity plot of the beam at a given location), from which the beam emittance can be calculated.

BPMs and BLMs are part of an integrated machine-protection system (MPS), where a beam-based technical interlock is used to prevent damage from a mis-steered or out-of-control, high-power beam.

## 6.2.7 Primary Vacuum

### 6.2.7.1 Introduction

The Primary Beamline vacuum system will maintain better than 1 x $10^{-7}$ Torr residual gas pressure in the beam tube in order to reduce beam losses from proton-gas interaction. The entire system is approximately 1,200 feet long from the extraction at MI-10 to the Target Hall, and it will be divided into several independently evacuated sections to accommodate the physics requirements, civil structures, and the overall pumping scheme. No section shall exceed 400-ft, and each will have about 20 ion pumps to achieve and maintain the required pressure level. The downstream end of this system concludes with a beam window inside the Target Hall, which is covered by the Neutrino Beam WBS.

### 6.2.7.2 Design Considerations

LBNF's vacuum system design is typical for a single-pass beamline. The specified residual gas pressure for the beam tube is not technically challenging to achieve, and the design will be similar to existing systems installed for the Main Injector transfer lines and NuMI. Experience from operating and maintaining these vacuum systems validates that a highly reliable, low-maintenance vacuum system is critical for minimizing outgassing and the potential for leaks, and thus for improving the overall operational efficiency of the beamline.

## 6.2.8 Magnet Installation

The lattice design specifies installation of 56 major magnets (listed in Table 6-4), six kicker magnets, and 23 corrector magnets. The total length of the Primary Beamline is approximately 1,200 feet from the extraction at MI-10 to the Target Hall. An image from the preliminary 3-D model is shown in Figure 6-4 and depicts an overview of the Primary Beamline. Based on experience with NuMI and other MI projects, LBNF will use a combination of magnet installation methods. Although the methods for transporting and positioning magnets will vary by location, the scheme for supporting and adjusting them will be the same. Each magnet will have a stand that provides three-point support and six degrees of freedom for precise adjustment. The magnet installation parameters are also shown in Table 6-4

This section focuses on the technical aspects of installation. Section 6.6 System Integration, explains how to best sequence the installation steps relative to each other and all related tasks. Because magnet installation is a significant part of the beamline installation, it must be integrated into the overall plan to ensure it is done safely and efficiently, but the detailed process of magnet installation is developed within its own WBS.

Table 6-4: Magnet installation parameters

| | **Main Injector Enclosure** | **Primary Beam Enclosure** |
|---|---|---|
| Section Length | ~300 ft | ~900 ft |
| Enclosure Notes | Co-existing with other beamlines | New enclosure with sloped floor up to 150 mrad |
| Major Magnets | 15 | 41 |
| Support/Adjustment | Mixture of new and modified designs | Modified MI designs |
| Transportation Method | Tugger and dolly via MI | Winch and dolly via LBNF access tunnel |
| Positioning Method | Existing MI equipment | MI hydraulic carriage with modification |

## 6.3 Primary Beam Loss Calculations

The high-intensity proton beam needs to modeled and understood to a very high precision to ensure that beamline components are kept to low activation levels and to be confident that accidental losses are rare and not damaging. The MARS and STRUCT codes are used exclusively for the calculations. The model is as complete as possible, from extraction through the Target Hall, where beam particles (protons) and their interactions are tracked individually and in full detail. Simulated magnets are controlled in groups appropriate to the designed power-supply bus configuration. Errors in magnetic fields for individual magnets can be inserted as random manufacturing errors and as simulated current fluctuations from power-supply errors. Beam loss studies provide one key input in the requirements for the magnet power-supply stabilities.

In addition to providing validation of operational beam-loss control for environmental and component protection, the simulations can provide a level of confirmation for the design of the beam-interlock systems 6.6.3.

The main design criteria for the primary beamline are (1) the transmission of high-intensity beam with minimum losses, (2) precision of targeting and (3) minimization of component activation. Mitigation of groundwater activation is relatively straightforward for the above-grade beam, however protection from prompt radiation, such as muon plumes, requires mitigation. An example of component activation from constant, low-level losses is shown in Figure 6-5.

Serious consideration must also be given to accidental beam losses that, within just a few beam pulses, can cause beamline-component damage. Significant sources of beam-position instability on the target as well as increased beam loss along the beam line include the power supplies for the extraction kicker, and the quadrupoles and dipole magnets. Variations in the element strength that occur over a period of minutes or hours can be corrected. However, for variations on shorter time scales, such as pulse-to-pulse jitter, the specification on beam instability would have to be met directly at the power supply. Detailed studies have been done for (1) losses under normal conditions, (2) accidental total beam loss on components and (3) losses due to magnet power supply errors. A complete description is found in the Annex 3A [30].

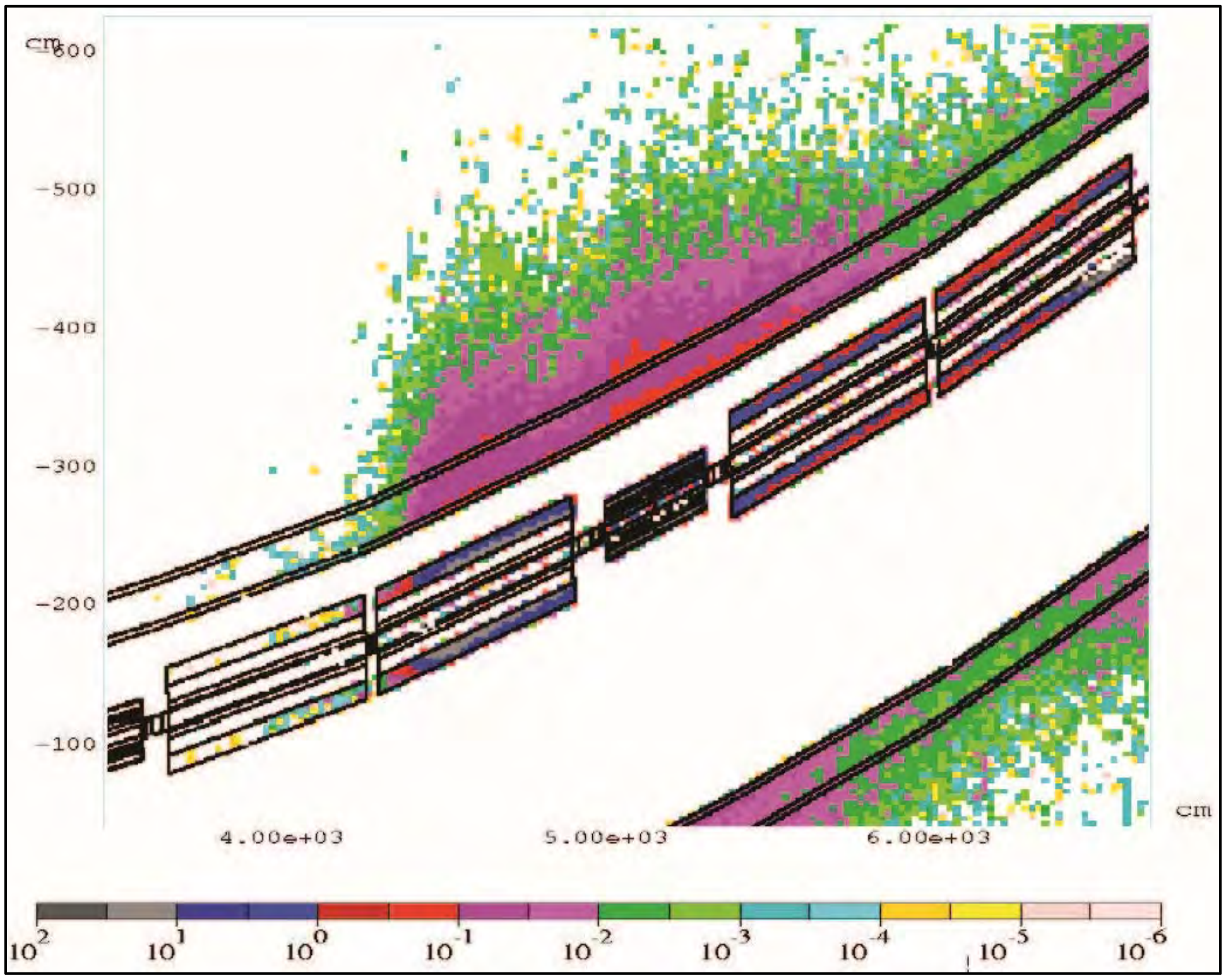

Figure 6-5 : Residual dose rates along the beamline and magnets following a loss of 0.3% of the beam for 30 days followed by one day cool-down. The color-coded logarithmic scale has units of mSv/h (1 mSv/h is equivalent to 100 mrem/h).

# 6.4 Neutrino Beam

## 6.4.1 Introduction

This section discusses the conceptual design of the second main system within the LBNF Beamline, the Neutrino Beamline, which refers to the set of components and enclosures designed to efficiently convert the initial proton beam into a high-intensity neutrino beam aimed at the far detector, 1,300 km away. Primary design considerations include the need to provide a wide-band beam to cover the first and second neutrino-oscillation maxima. The proton beam power is expected to be 1.2 MW at start-up, and increasing after some years of operation to 2.4 MW.

A 3-D cutaway model of the target hall complex showing the key elements necssary to produce the neutrino beam is shown in Figure 6-6. A proton-beam pulse from the primary-beam system enters the neutrino beamline system (from the right in through a beryllium "window." This window seals off the evacuated beam pipe of the primary beamline, and the protons enter the air-filled target chase (the volume surrounding the target and focusing system). Initially they pass through a small aperture in a 1.5-m-long graphite cylinder, called a baffle, which protects equipment downstream from mis-steered beam. Sixty-eight cm past the end of the baffle, they reach the target, a long, thin set of graphite segments in which about 85% of the protons interact and produce secondary particles. The target is partially surrounded by the first horn, a magnetized structure which provides initial focusing for the secondary particles, predominantly pions and kaons. A second horn, a few meters downstream, provides additional focusing for the secondary particles before they enter a He-filled 194 m long decay pipe, where a large fraction of the pions will decay to neutrinos, forming the neutrino beam. Horns are supported and positioned by support modules. Space is provided between the end of the second horn and the beginning of the dceay pipe to allow flexibility for more advanced target-horn systems or ones optimized for different physics goals that may become relevant during the multi-decade lifetime of this facility. (See Section 6.7.) The final portion of the neutrino beamline is the absorber, downstream of the decay pipe. The absorber stops the protons that failed to interact in the target and the secondary particles that failed to decay to neutrinos; it is designed to sustain the beam energy deposition under expected normal operational conditions as well as under accident situations.

The designs for the neutrino beam components detailed in this Chapter are appropriate for a primary beam energy range from 120 GeV down to 60 GeV/c, except for the horns and the associated stripline, where the analysis has been performed only in the range from 80 GeV to 120 GeV.

All neutrino beam subsystems have been designed for 1.2 MW beam power. Subsystems which are difficult or impossible to upgrade to a higher beam power have already been designed for the potential beam power upgrade to 2.4 MW. These include the target pile, the decay pipe and the absorber as well as the main elements of the associated cooling systems.

Radiological concerns have been extensively modeled, and have been addressed in the system design. All simulations of activation, dose rates and beam energy deposition use the MARS package. The MARS model of the reference design describes the target, horns, decay pipe and absorber, as well as all of the shielding.

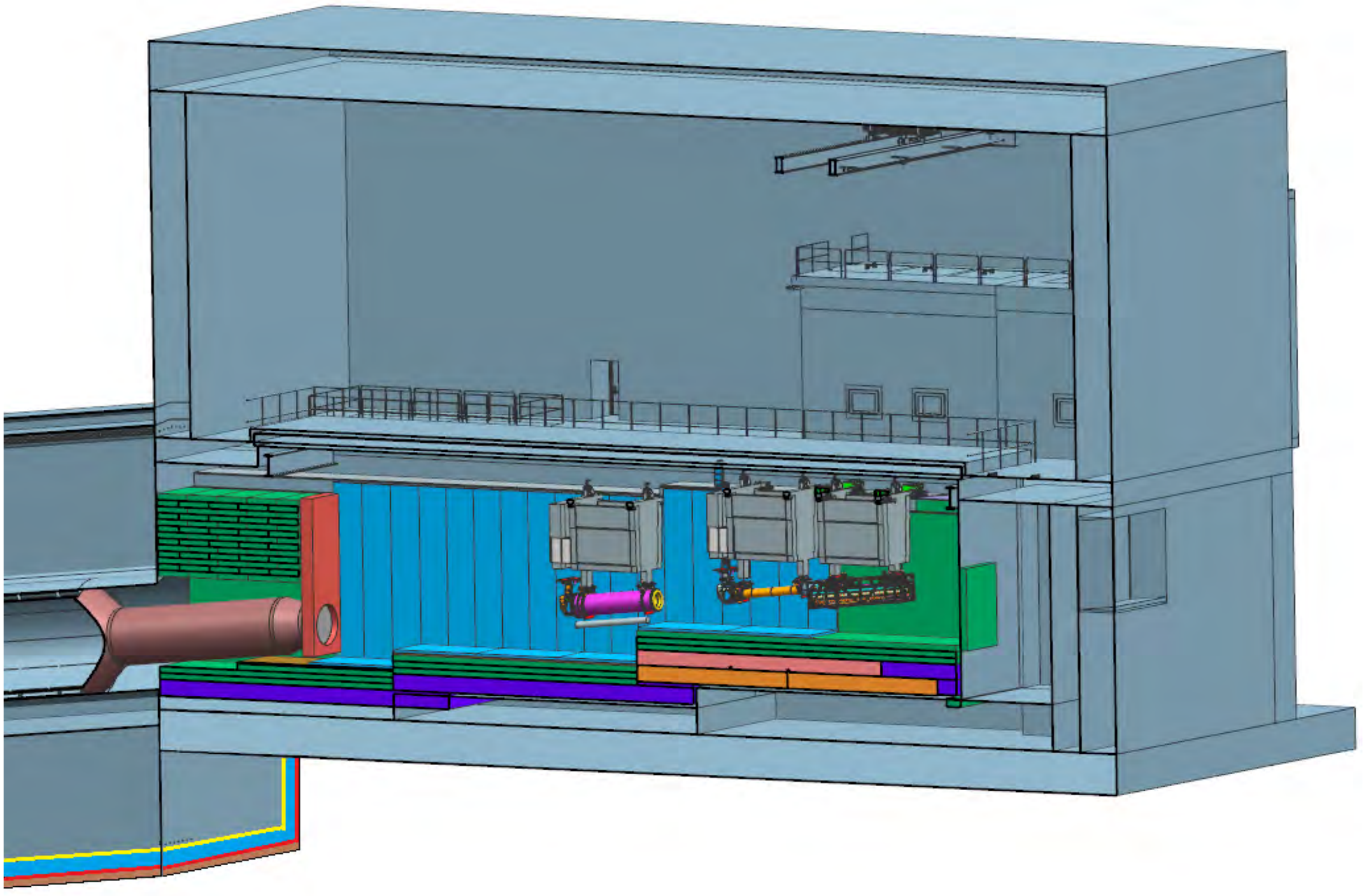

Figure 6-6: Schematic of the upstream portion of the LBNF neutrino beamline showing the major components of the neutrino beam. The target chase bulk steel shielding is shown mainly in green. Inside the target chase from right to left (the direction of the beam) pointing downwards: the beam window, horn-protection baffle and target mounted on a carrier, the two toroidal focusing horns and the decay pipe. At the upstream end of the decay pipe is a “snout” which holds the upstream decay pipe window. Above the chase and to the right is the work cell for horn and target system repairs. The grey areas around the decay pipe indicate concrete shielding. The yellow and red lines indicate multi-ply geosynthetic barriers, separated by a drainage layer (blue).

## 6.4.2 Targetry

The scope of targetry includes the target, the horn protection baffle, their support structures, and the accompanying instrumentation for commissioning, alignment and monitoring of the target and focusing system in the beam.

### 6.4.2.1 Baffle and Target

The baffle, just downstream from the primary proton window, is a passive device that works similar to a collimator to prevent any mis-steered beam pulse from causing damage. The baffle baseline design consists of ten 57 mm O.D., 17 mm I.D., 150 mm long graphite R7650 grade cores which are enclosed by a 61 mm, 3 mm thick , 150 cm long aluminum tube after annealing.

The target design is determined by balancing the ideal production of mesons for neutrino production and the survivability of the device for tens of millions of beam pulses. The target must have the following features:

- Adequate material to convert the protons into mesons, while not absorbing too many of the produced particles
- The ability to withstand the instantaneous thermal and mechanical shocks due to the beam
- The ability to withstand the sustained thermo-mechanical stresses and temperatures
- A cooling system to remove the heat deposited by the beam interaction (approximately 40 kW, or 3% of the beam energy)
- Resistance to the effects of radiation damage so as not to encounter substantial change in mechanical properties during the run

The target width must be sufficient to cover the beam spot, but is otherwise minimized, except for the practical concerns of heat removal and mechanical integrity. The primary target material must have high mechanical strength, high specific heat, high thermal conductivity, a low coefficient of thermal expansion, and good radiation properties. Although a number of single-element materials generally fit the above requirements, the two materials best fitting those parameters for neutrino beams are beryllium and graphite.

The LBNF target is based on the NuMI target design that has operated since 2005 with some modifications to accommodate higher beam power. The target core is graphite segmented into short rectangular segments oriented vertically, with the short dimension horizontal to the beam. The heat from the core is removed by dual titanium water lines brazed to the top and bottom of the graphite. The entire assembly is encased within a titanium containment tube. The segments are 10 mm in width and 20 mm in length. A total of 47 segments, each 2 cm long and spaced 0.2 mm apart, result in a total target core length of 95 cm, corresponding to two interaction lengths. A cross section of the target is shown in Figure 6-7. The target will be cantilevered into the horn and inserted via a carrier similar to that used in NuMI which also carries the baffle.

The NUMI LE target was designed for $4\times10^{13}$ protons per pulse; this will increase to $7.5\times10^{13}$ for LBNF 1.2 MW beam power. Design studies calculated the stress in the graphite segments at these intensities. The results of this study indicate that increasing the beam spot size from the 1.0 mm RMS used in the original NUMI calculations to 1.7 mm RMS is adequate to lower the stress in the graphite for the more intense beam.

The expected lifetime based upon MINOS experiences is $4.5\text{x}10^{20}$ protons on target. For LBNF, taking into account both graphite degradation and mechanical failures, this implies approximately 2.5 targets per year will need to be replaced.

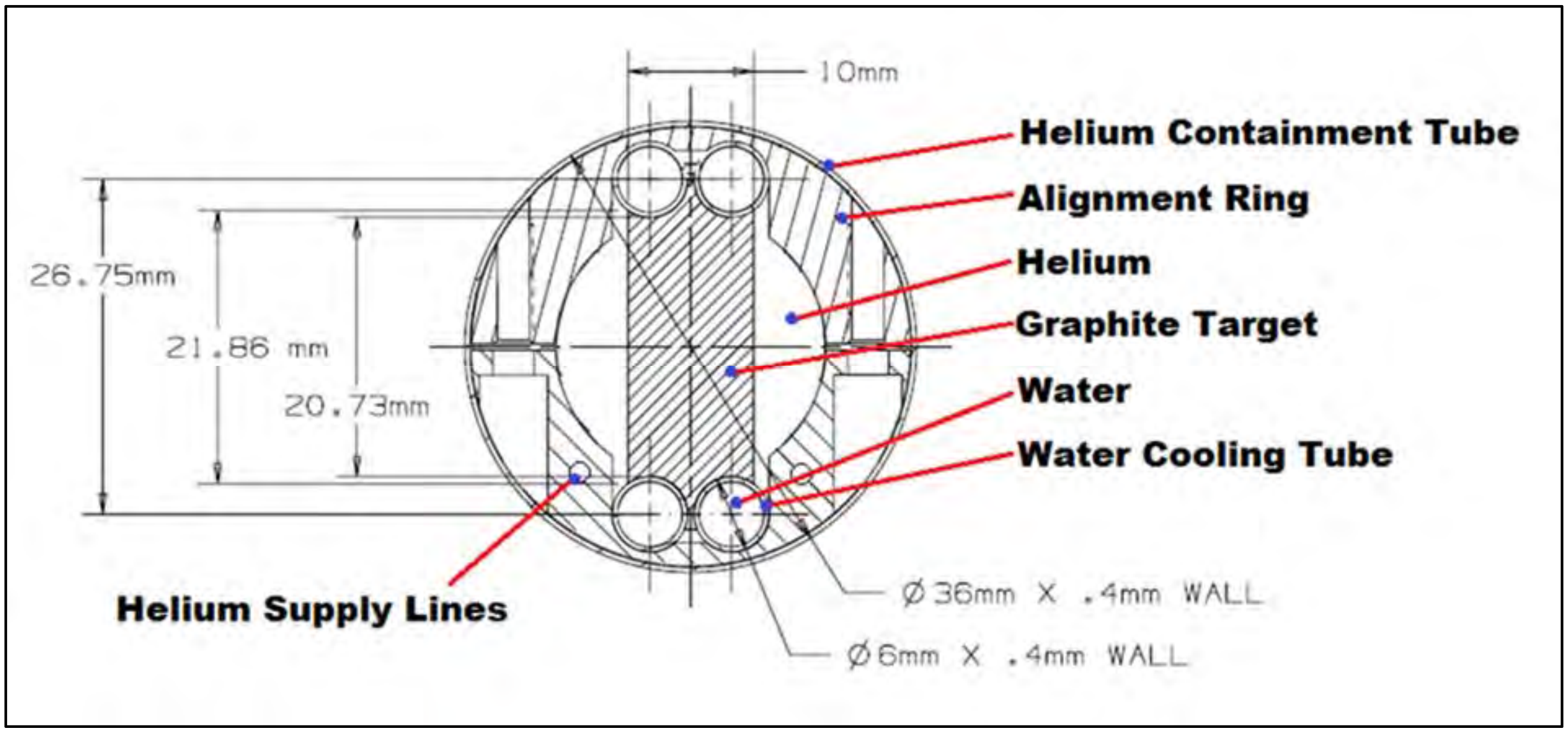

Figure 6-7: Cross section of LT target for LBNF. The alignment rings do not run the full length of the target.

Target longevity is a major issue for the performance of the LBNF facility. A target R&D program will explore options of target material, geometry, cooling, and other design issues.

The Radiation Damage In Accelerator Target Environments (RaDIATE) collaboration is a program to investigate various materials of interest in the high energy proton irradiation regime primarily using the Materials for Fission and Fusion Power group at Oxford University. Testing materials in the high intensity pulsed beam available at HiRadMat at CERN will validate simulations of response of solid materials to short pulses of proton beam and failure criteria, potentially proving beryllium as a valid alternative to graphite.

### 6.4.2.2 Module and Carrier

The baffle is mounted rigidly to the target, and moves with it, relative to a carrier that hangs off of shafts extending through the support module (see Figure 6-8). The carrier supplies motion of the target along the beam direction that allows (i) insertion of the target into the horn for standard running and (ii) the flexibility to move the target up to 2.5 m upstream of its standard location for special runs.

## 6.4.3 Horns

The horns are focusing and sign selecting devices for secondary particles produced by the interaction of the primary proton beam on the target. This focusing of particles is achieved through a pulsed toroidal magnetic field, which is present in the inert gas volume between the co-axial inner and outer conductors that form the horn structure. The horns are identical to those were developed for the NuMI neutrino beam, but they will be operated at 230 kA current and subjected to a beam power of 1.2 MW. To compensate for the larger current and higher beam power, a new horn power supply (see Section 6.4.4) is necessary to provide a half-sine wave current pulse width of 0.8 ms, instead of the 2 ms width of the NuMI power supply.

### 6.4.3.1 Horn Focusing system

The LBNF focusing system consists of two horns, with the upstream end of the second horn (Horn 2) located 6.6 m from the reference point MCZERO (near the upstream end of Horn 1). Both horns consist of an inner conductor, an outer conductor, a current-supply stripline, a cooling system and a support structure.

The inner conductor of Horn 1 has a parabolic upstream section that surrounds the target tube up to the neck of the horn (see Figure 6-9). This neck is followed by a parabolic downstream profile that ends at the downstream face where the stripline is mounted. Horn 2 follows this layout, although varies in parabolic lengths, placement, and wall thicknesses. Both Horn 1 and Horn 2 inner and outer conductors are of aluminum 6061-T6 construction. The horn parameters are summarized in Table 6-5.

The outer surfaces of the inner conductor will be cooled by water spray nozzles distributed along the beam axis and 120° azimuthally. Nozzles at the top of the outer-conductor cylinder will spray water to form a film running down from both sides. To resist water erosion, the surface of inner conductors will be coated with electroless nickel.

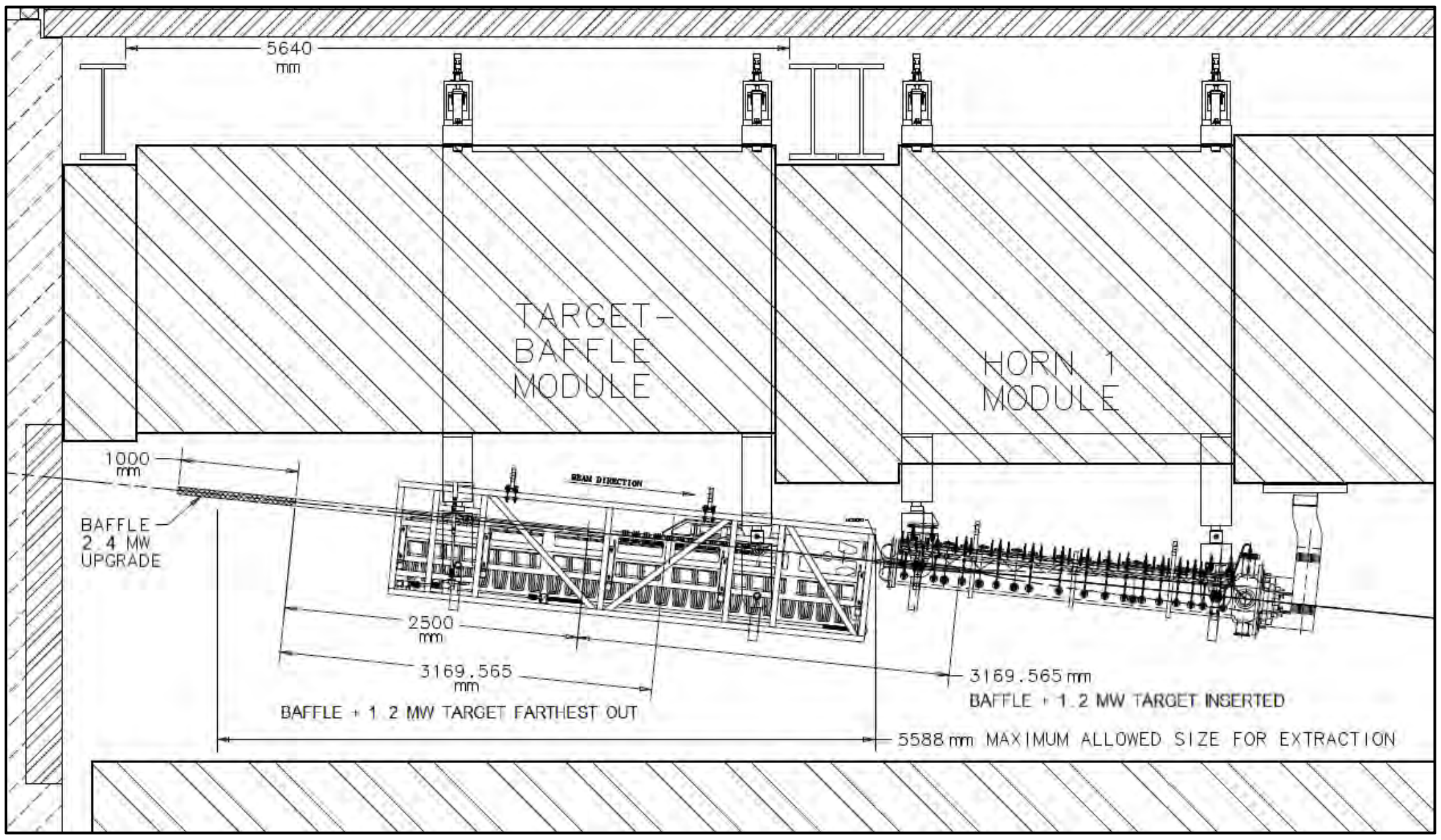

Figure 6-8: Target carrier in target pile shielding. The length of the baffle plus target assembly is shown in the fully inserted downstream position, and also in the furthest out position 2.5 m upstream of that. The extra 1000 mm length of a baffle for 2.4 MW operation is also sketched in, showing that the usable range of target motion may be modestly reduced by that upgrade.

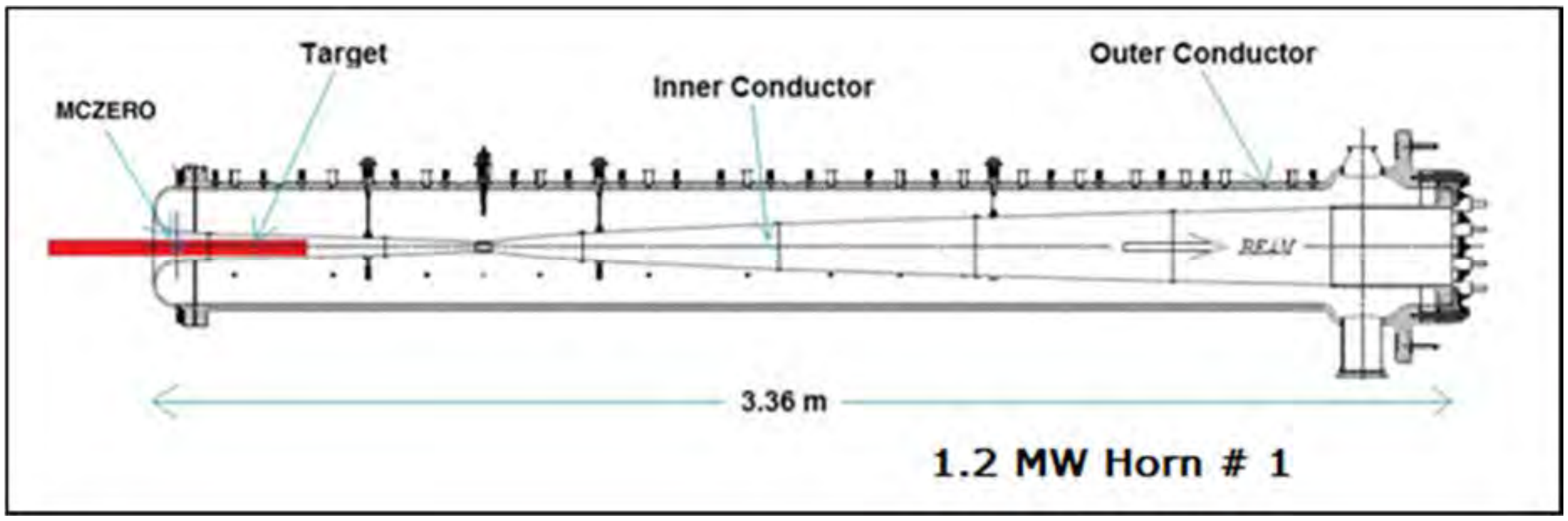

Figure 6-9: Horn 1 section. The reference "MCZERO" is the point along the beam that sets the coordinate system origin for Monte Carlo simulations. The red segment represents the target, which begins 45 cm upstream of MCZERO.

The electrical connection between the power supply and the horn is provided by a planar-design stripline, which has minimal inductance and resistance, and allows thermal expansion/contraction of the horns and transmission lines. Attached to each horn and supported by the module mainframe, the stripline block (see Figure 6-10 left) provides radiation shielding, as well as a containment structure for the striplines that supply current to the horn conductors. The blocks must have an integral labyrinth for these aluminum layers due to radiation shielding concerns, and also must remotely attach and un-attach from the horn stripline through use of a remote clamp which is mounted to the lower end of the block. The stripline connection to the horns consists of eight layers of aluminum 6101-T61 bus bars that are spaced by zirconia ceramic insulators, as shown in Figure 6-10 right.

Table 6-5: Horn parameters.

| | Horn 1 | Horn 2 |
|---|---|---|
| Material | Al 6061-T6 | |
| Peak Current | 230 kA | |
| Min. aperture "neck" radius | 9 mm | 39 mm |
| Inner Conductor Thickness | 2 mm | 3 mm |
| Length | 3.36 m | 3.63 m |
| Outer Conductor radius (outer) | 165 mm | 395 mm |
| Outer Conductor Thickness | 16 mm | 25 mm |

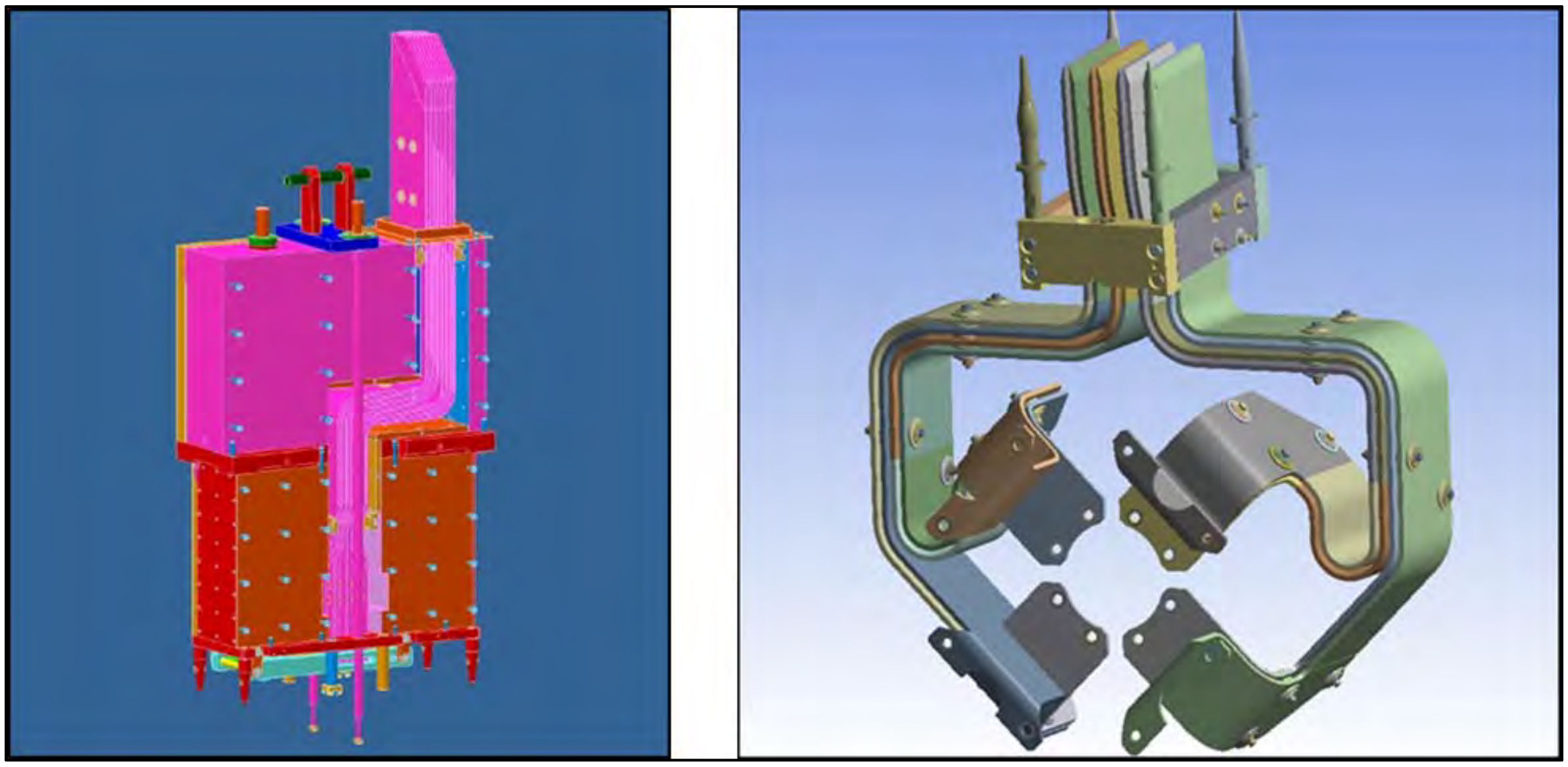

Figure 6-10: Left: Conceptual horn stripline block. Right: Horn stripline connection

The heating sources on the horn conductors and on the stripline include electrical-resistive heating by current and instantaneous beam heating due to secondary particle interactions in the material. Heating of the horn conductors produces thermal stresses, and electromagnetic forces generate magnetic stresses on the inner conductor during current pulsing. Thermal and structural finite element analysis (FEA) have been completed to verify the design and study the fatigue strength of the inner conductors, the alignment stability of the horns, and the temperature profile of the striplines.

Final safety factors for conductor construction were found to be sufficient to meet the expected horn lifetime of 3 years or $1\times10^8$ pulses, whichever is greater.

Current analysis of the stripline shows high temperatures present at the interior layers where air cooling is currently not sufficient. Efforts are underway to increase air cooling in the inner layers, by means of roughly doubling the gap between segments, and adding forced air cooling through the top of the stripline block.

### 6.4.3.2 Horn Support Module

Horns will be supported and positioned by support modules. The intensely radioactive environment of the target chase requires that the horn-support module be adjustable and serviceable by remote control. The horn-support modules provide radiation shielding, and allow the mounting and dismounting of feed-through connections for the stripline, cooling water and instrumentation cabling from the top of the module mainframe, away from the most highly activated areas.

Horn-support modules are rectangular boxes open at the top for shielding block insertion, and are constructed from plate steel. The modules fix the horn with respect to the module in the horizontal degrees of freedom, but not in the vertical. The module is adjusted with respect to the beam for transverse horizontal position and yaw. The horn is adjusted with respect to the module for vertical and pitch alignment.

# 6.4.4 Horn Power Supplies

The scope of the Horn Power Supplies WBS includes the design and construction of a horn power supply and the electrical connection to the horns by means of a stripline.

## 6.4.4.1 Horn Power Supply

The horn power supply will be designed to supply two horns with a half-sine wave current pulse width of 0.8 ms, with a maximum of 300 kA peak, at least every 0.7 s.

A damped LC discharge circuit, as shown in **Figure 6-11**, will achieve the peak current when the silicon controlled rectifier (SCR) switch releases stored energy from the capacitor bank to the horns via a planar transmission line ("stripline"). The power supply will be able to drive a maximum inductance of 2.6 µH and a resistance of 0.87 mΩ.

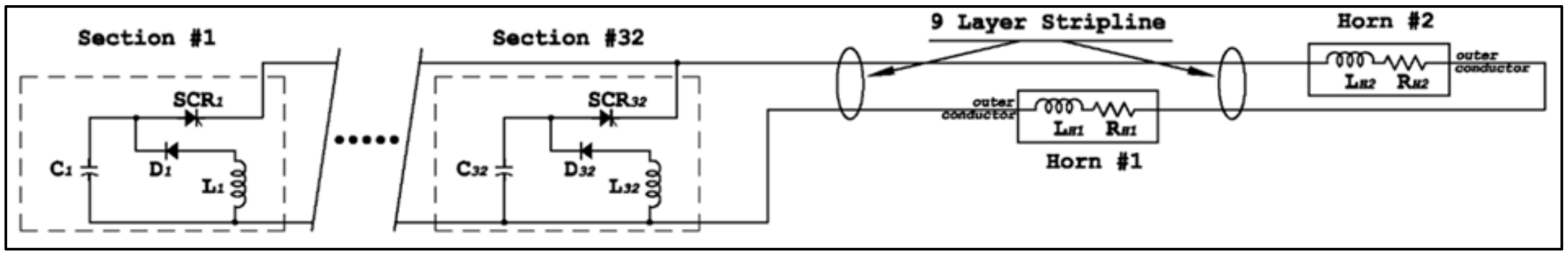

Figure 6-11: LBNF horn Power Supply simplified circuit diagram

The capacitor bank will be charged by a custom charging supply. Operating voltage for the capacitor bank will be nominally 3.5 kV. Calculated DC-power consumption during operation of the focusing horns is 60 kW.

Based on the inductance and resistance value estimates, the capacitance required for the bank is 23 mF. This will be made up of an array of individual capacitors connected in parallel, but electrically separated into 32 cells. The number of capacitors in each cell will be chosen to limit the amount of energy per cell to a value that can be safely contained within an individual capacitor case in the event of an internal fault.

Passive current transformers installed within each capacitor bank cell monitor the cell performance to 0.4% accuracy. These 12 signals are also summed to provide individual stripline currents plus total load current for over-current monitoring and readout display.

## 6.4.4.2 Stripline

A stripline consisting of a nine-layer assembly of parallel aluminum electrical bus conductors will connect the power supply to the two series connected horns. The aluminum alloy of choice is 6101-T61, having nearly the conductivity of pure Al but with enhanced mechanical properties. Of the successful designs presently in service for the NuMI and MiniBooNE horn systems, the MiniBooNE design is best scaled to the higher LBNF peak current. Its balanced configuration offers much reduced electromagnetic, vibration and mechanical stress.

## 6.4.5 Target Hall Shielding

Target Hall shielding (also called the target pile) is designed to (1) keep the accumulated radionuclide concentration levels in the surrounding soil below standard detectable limits; (2) keep prompt radiation levels low enough for electronics in the Target Hall to have adequate lifetimes; and (3) keep residual radiation rates on top of the shield pile low enough to allow personnel to access the top of the steel shielding pile for maintenance with beam off. Target pile size cannot be modified or upgraded after completion. Therefore, this part of the neutrino beam has been designed for 2.4 -MW beam-power operation, corresponding to the maximum anticipated power.

### 6.4.5.1 Target Hall Shield Pile

The Target Hall shield pile refers to the steel shielding surrounding the beamline components (baffle, target, Horn 1, Horn 2, and the decay pipe upstream window) installed in the target chase. The target chase is the central rectangular open volume that runs the 34 m entire length of the steel shield pile. The chase is 2.0 m wide at the water-cooling panels in the region of the horns and 2.2 m wide elsewhere.

The shield consists of two main layers. An inner, steel layer will absorb all of the stray particles from interactions of the primary beam, except neutrons below a few MeV energy. The outer layers are used to moderate and absorb most of the neutrons that escape from the steel layer. These outer layers consist of concrete, marble, or borated polyethylene plates, depending on location. The shielding is illustrated in Figure 6-12, which is the lateral cross section of the target pile at MC-ZERO.

### 6.4.5.2 Target Chase Cooling

Energy deposited in the shield pile and the beamline components by the beam is removed by water-cooled shielding, an air-cooling system and cooling systems on the beamline components. The water-cooled shielding, i.e., carbon steel chase panels, T-blocks and module bottoms, intercept approximately half of the beam energy leaving the chase. The air-cooling system and cooling systems on the beamline components remove the balance of the deposited beam energy. The air cooling flow rate is 35,000 scfm. The flow schematic for the target pile air cooling system is illustrated in Figure 6-14.

## 6.4.6 Helium-filled Concentric Decay Pipe

The scope of work of this WBS includes specifying (1) the length, material, diameters and wall thicknesses for the concentric decay pipe, (2) specifying the cooling parameters, and (3) designing and providing the downstream window.

The helium-filled decay pipe is the volume in which the pions and kaons decay into neutrinos. The length is determined by the distance at which most of the pions decay, producing neutrinos near the maximum energy required by the physics goals of DUNE. The pipe must be of sufficient diameter to allow for decay of the lowest-energy pions required by the experiment. The decay-pipe length is 194 m and the diameter of the inner pipe is 4 m. The total decay length, from the nominal target postion (MCZero) to the end of the decay pipe is 221 m, of which 27 m is in the target chase.

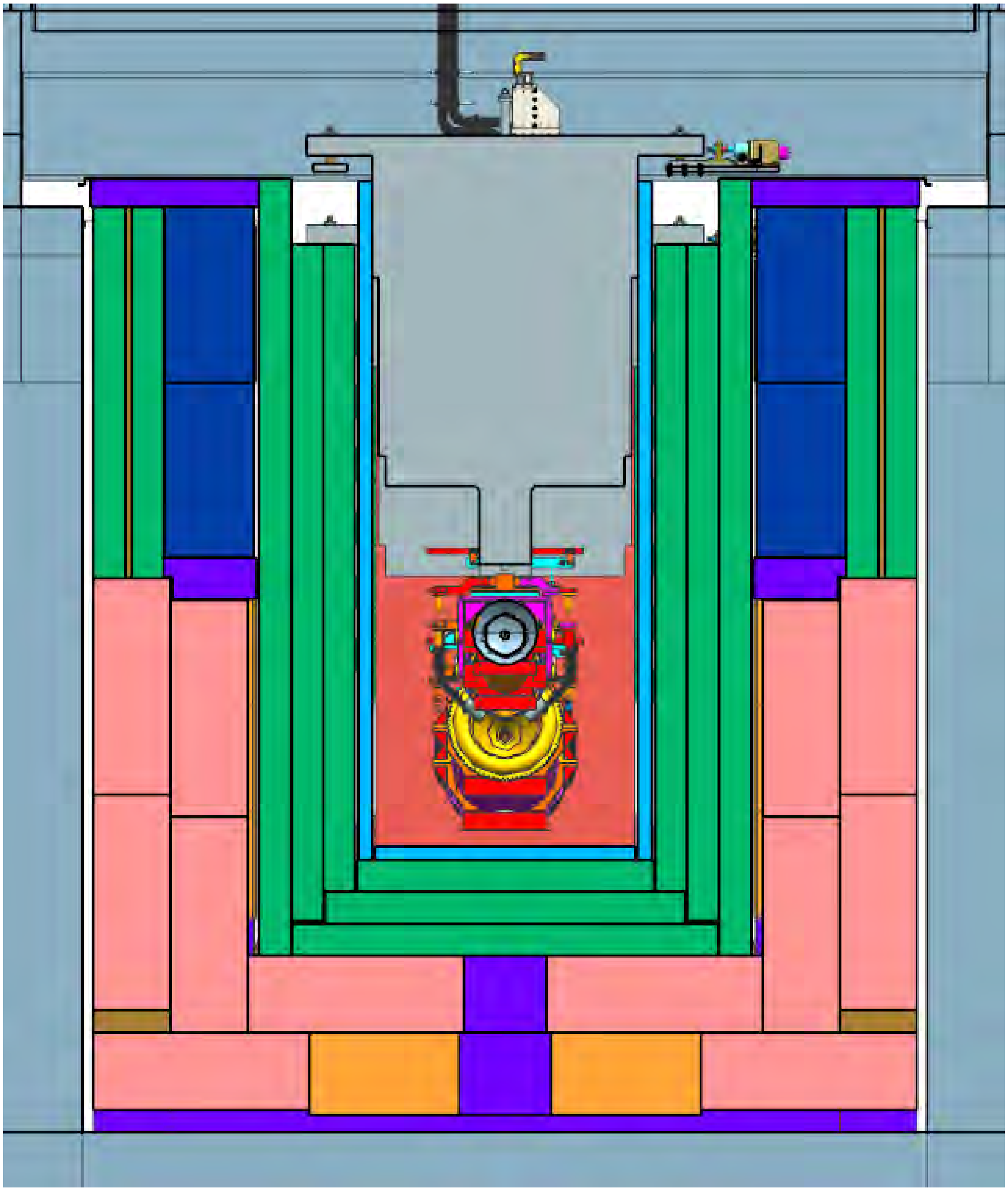

Figure 6-12: Cross section of target chase steel shielding at the upstream end of horn 1.

Concrete radiation shielding surrounds the decay pipe to mitigate activation of surrounding ground water. Heat generated in materials due to beam reactions will be removed by airflow through an annular duct surrounding the decay pipe. A multi-ply geosynthetic system surrounds the decay-pipe concrete to act as a barrier for minimizing ground-water inflow. A second set of air-cooling pipes just inboard of the geosynthetic system keeps the geosynthetic at low temperature to extend its lifetime.

### 6.4.6.1 Decay Pipe Structure and Shielding

The concentric decay pipe and shielding concrete are illustrated in Figure 6-13, which shows a typical cross-section of the system designed as part of the Near Site Conventional Facilities to satisfy the Beamline requirements.

The decay pipe and shielding have the following specifications:

- 194 m length
- 4 m inside-diameter steel pipe installed concentrically in a 4.43-m inside diameter steel pipe; the radial annular gap between the pipes is 0.2 m
- commercial-grade pipe with thickness of 12.5 mm
- spacers welded between the two pipes to maintain concentricity and to not interfere with the airflow
- a multi-ply geosynthetic membrane to keep water away from the decay pipe as part of the overall tritium-mitigation strategy
- alignment accuracy maintained at 20 mm
- external and internal corrosion protection
- concrete radiation-shielding thickness of 5.6 m

The decay-pipe region begins 27 m downstream of the beam-sheet coordinate MC-ZERO which defines Horn 1 position. A standard pressure vessel head is welded to the upstream end of the inner decay pipe. The head has a 2.1 meter diameter opening at its center. A 2.1 meter diameter pipe is welded at the center of the head to cover the opening. The pipe extends 5 m into the target pile chase and is referred to as the decay pipe snout. The decay pipe upstream window is installed on the upstream end of the snout.

Heat generated by beam interaction is distributed non-uniformly along the length of the decay pipe. Approximately half of this heat is generated in the inner steel pipe, with the remainder generated in the outer steel pipe and concrete. An air cooling system is employed to cool the decay pipe. The cooling airflow schematic is shown in Figure 6-14. For the cooling airflow of 35,000 scfm and an air supply temperature of 15 °C, maximum temperatures at the point of peak energy deposition are estimated to be 90 °C for the steel pipes, 95 °C for the shielding concrete, and 42 °C for the multi-ply geosynthetic waterproof barrier.

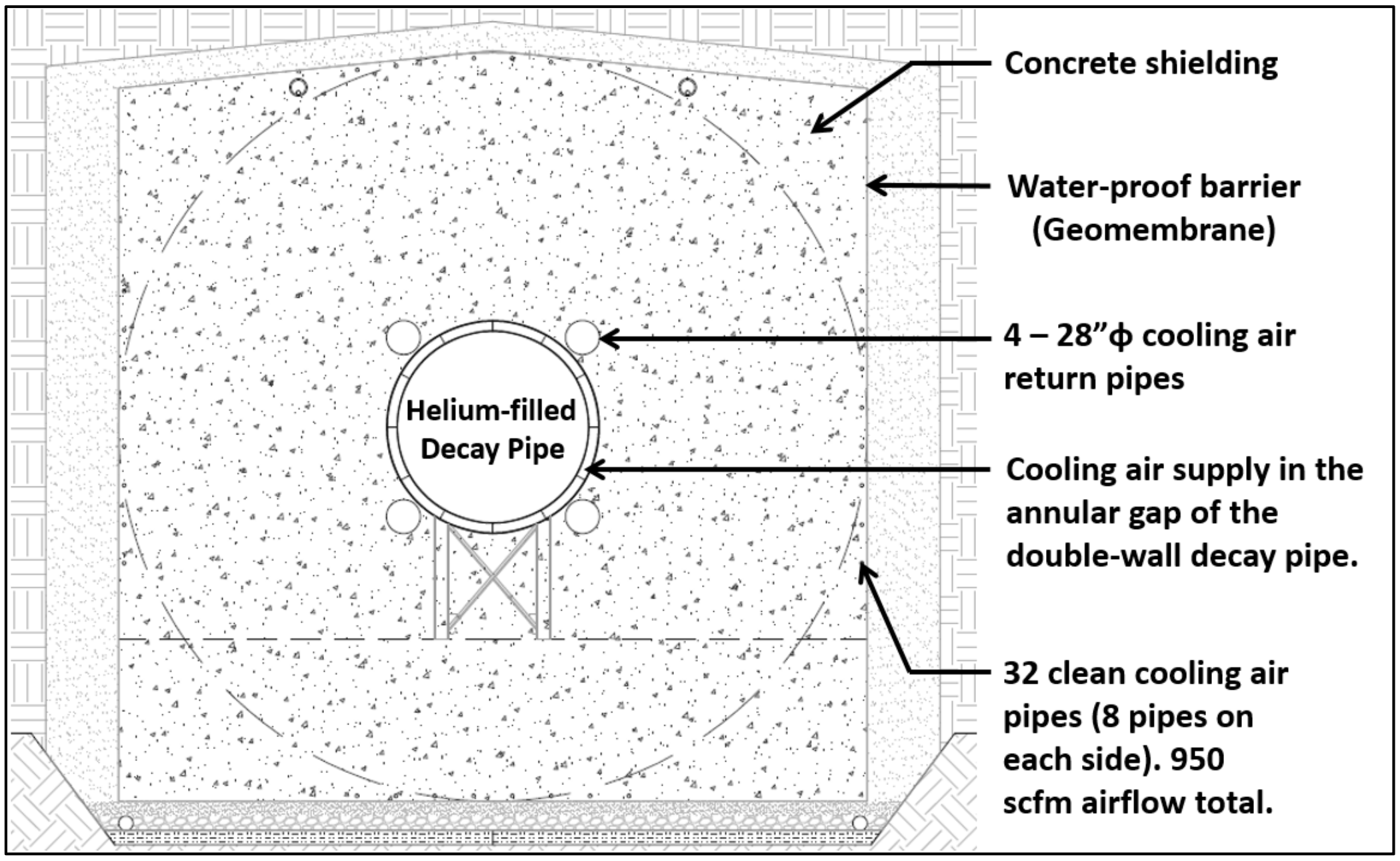

Figure 6-13: Typical cross section of concentric decay pipe and shielding concrete

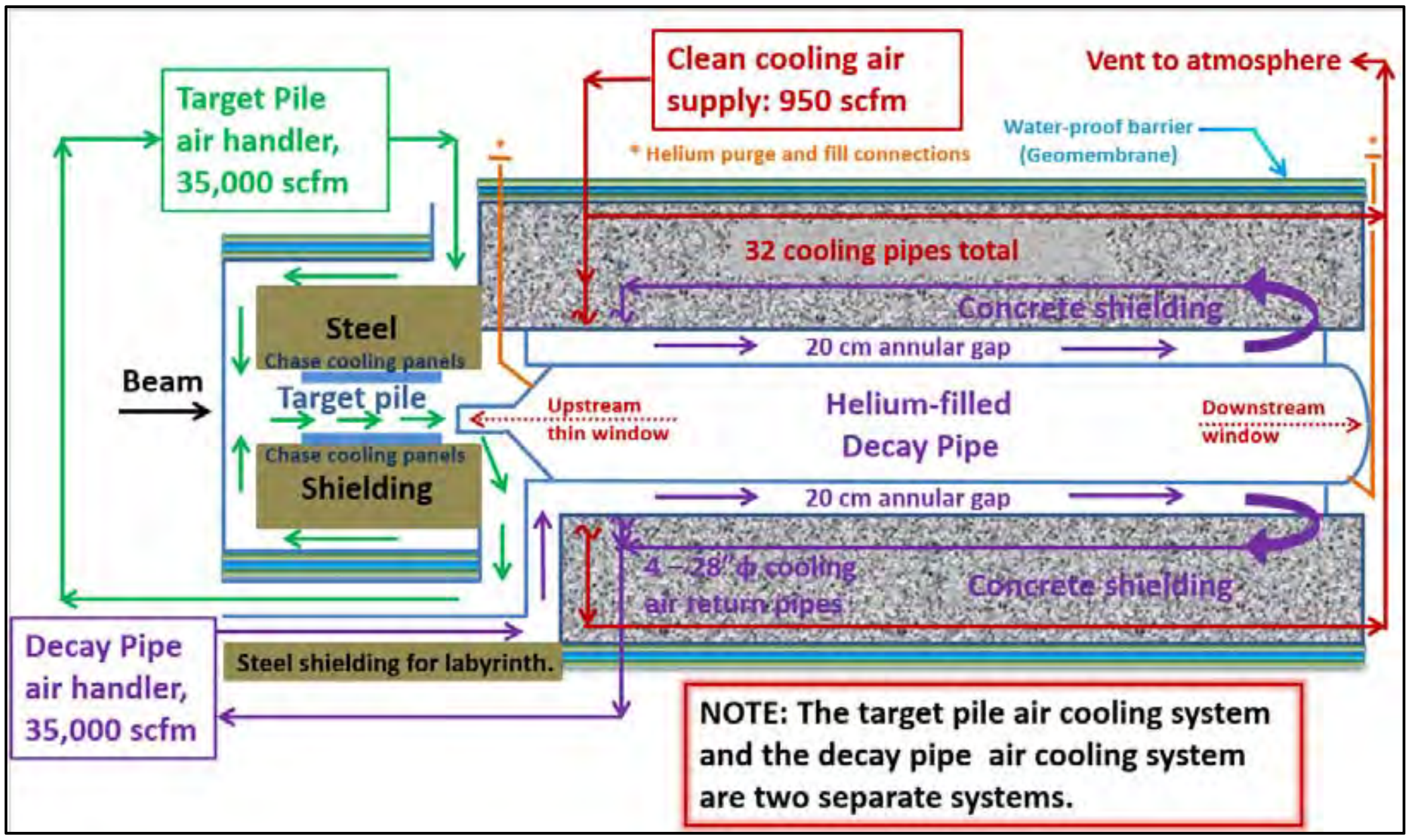

Figure 6-14: Schematic of the target pile and decay pipe air-cooling systems

## 6.4.6.2 Downstream Decay Pipe Window

The decay pipe ends with a window located in the upstream wall of the Absorber Hall. The downstream window is a 6-mm thick aluminum, dished cylindrical plate, 1-meter in diameter. It is centrally set in a hot-rolled steel pressure vessel head. The head is welded to the inner pipe of the concentric decay pipe. The head and aluminum window are cooled by natural convection with air on the exterior, and by natural convection with helium on the interior. For all of the cases, the temperatures and stresses are below the allowable values. All of the temperatures vary little per pulse so fatigue can be neglected. All of the temperatures are also below the creep temperature limit so creep can be neglected.

The concentric decay pipe has a design pressure of 5 psig as required by the helium system to purge, fill and operate the decay pipe, as discussed in 6.4.7.3.

## 6.4.7 Beam Windows

This WBS includes two objects: the primary beam window (the window through which the primary proton beam exits the vacuum beam pipe before hitting the target) and the decay pipe upstream window (the large diameter window that the secondary beam passes through after the second horn and before entering the decay pipe).

Both windows will be exposed to very high levels of beam power and are expected to have lifetimes significantly shorter than the anticipated operational lifetime of the facility. Therefore, both windows will need to be replaceable. Because both windows will become significantly activated as a result of beam operation, means to replace the windows with minimal human exposure to activated components is part of this design effort.

### 6.4.7.1 Primary Beam Window

The upstream beamline enclosures are separated from the target chase by a 3.9 m thick concrete shielding wall to isolate the upstream beamline components from high radiation dose rates. The primary protons enter the target chase through a window in the wall; it is a beryllium foil that seals the evacuated primary beam pipe. The NuMI 708 kW primary beam window is an air cooled, 0.25 mm thick, 25.4 mm diameter beryllium grade PF-60 foil. Experience from NuMI shows that the primary-beam window has an estimated lifetime of three years at 708 kW.

The LBNF primary proton window has been designed as a 50 mm diameter, 0.2 mm thick, partial-hemispherical beryllium window. At 1.2 MW the window is able to withstand the stress waves and also pressure and thermal loading given a beam spot size of 1.7 mm sigma while air-cooled. At 2.4 MW the window may require water cooling at its periphery. Optimization of a peripherally air-cooled, hemispherical tapered beryllium window shape with a thin center and gradually thicker outer crown which allows greater conduction is worthy of further investigation.

### 6.4.7.2 Upstream Decay Pipe Window

The upstream decay pipe window (see Figure 6-15**)** is comprised of a 1.25 mm thick beryllium or a beryllium-aluminum alloy foil welded to a heavier aluminum ring. This heavier ring includes a seal groove for an all metal seal. The center portion of the decay pipe window through which the proton beam will pass will be made of either beryllium or a beryllium-aluminum alloy. The annular disc outside of the center region will be made of a beryllium-aluminum alloy or aluminum.

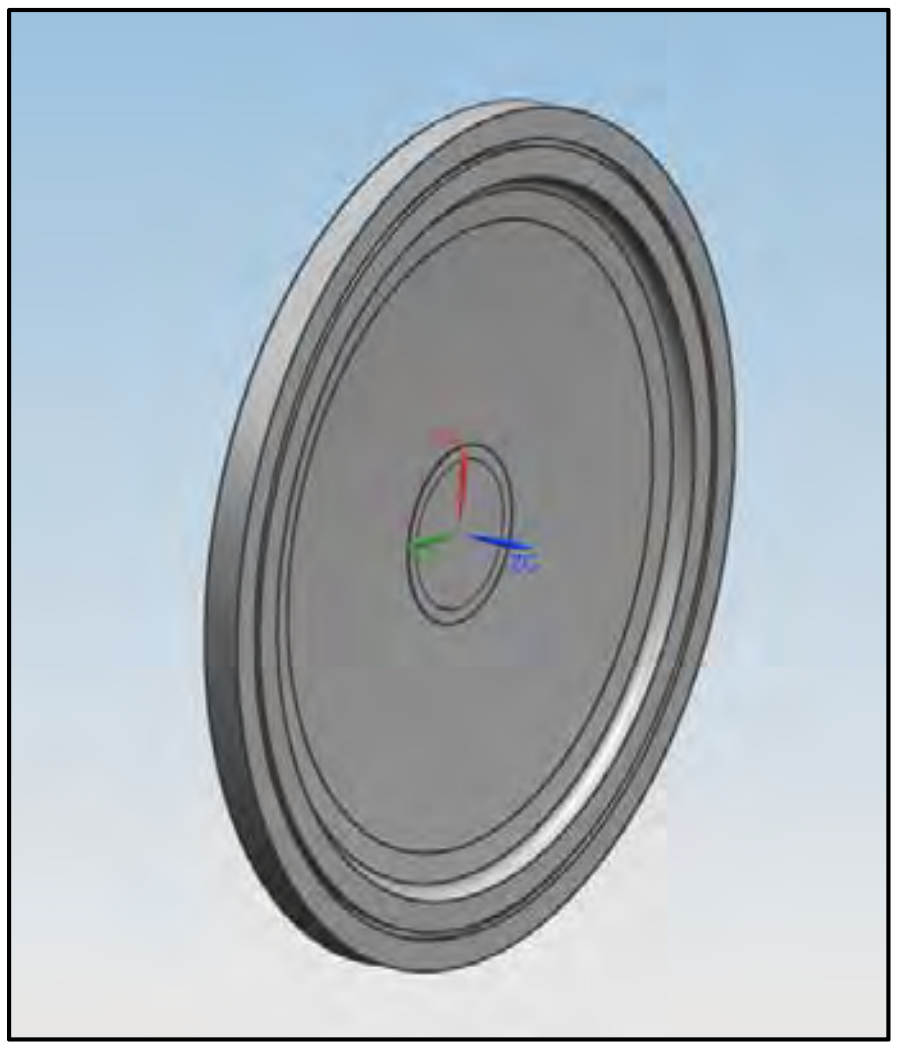

Figure 6-15: Upstream decay pipe window. The central part is a beryllium or a beryllium-aluminum alloy section

The metal seal will be a commercially produced metal seal. Seal requirements include: 1) 5 psig (0.3 bar) maximum internal decay pipe helium pressure; 2) 1.5 psig normal working internal decay pipe pressure; 3) survive a high radiation and a corrosive atmosphere; 4) remote actuation (area will become too radioactive for a person to access); 5) an allowable leak rate of approximately 10 cubic centimeters per minute.

There are several candidate designs for applying the seal load and include a wedge system, a four-bar mechanism, and a compressed gas system. Evaluation of the candidate seal loading designs and selection of an optimal design are part of the advanced conceptual design phase.

Removal of heat deposited by the beam in the window has been considered. Convection heat transfer has been applied to both surfaces using forced convection values. Actively cooling the heavier outer ring which houses the seal has been considered and will likely be incorporated into the final design. The provisions for this cooling will be included in the advanced conceptual design.

### 6.4.7.3 Decay Pipe Helium Fill

The decay pipe will be filled with air during construction and at the completion of the construction activities. The design prohibits evacuation and backfilling with helium gas. Therefore, a method of purging the air out of the vessel with inexpensive, relatively heavy, carbon dioxide and then displacing the carbon dioxide with helium has been developed. This system will allow 99% helium concentration to be achieved and offers the possibility of increasing the helium purity by absorbing the carbon dioxide.

## 6.4.8 Hadron Absorber

The hadron-absorber structure (also called simply the "absorber") is located directly downstream of the decay pipe. The absorber, an assembly of aluminum, steel, and concrete is designed to absorb the residual energy from protons and the secondary particles (hadrons) which have not decayed. Approximately 750 kW of the 2.4 MW beam power is deposited into the absorber and must be properly contained to prevent activation of soil and groundwater. The absorber is designed for the worst case condition: 2.4 MW operation, helium-filled decay pipe (194 m long) and the shortest target envisioned at 2 interaction lengths.

The absorber consists of two major sections, as shown in the left image of Figure 6-16. The core, a section consisting of replaceable water-cooled blocks, is shown inside the green box. It is enlarged in the right image of Figure 6-16 . The core consists of an aluminum spoiler block to initiate the particle shower, aluminum mask blocks with air space in the center to allow the shower to spread, a sculpted aluminum region of reduced central density to further distribute the heat load, solid aluminum blocks, and solid steel blocks. The beam power deposited into the core is 519 kW, which is the majority of the incoming beam power into the absorber. Outside of the core is forced-air cooled steel and concrete shielding.

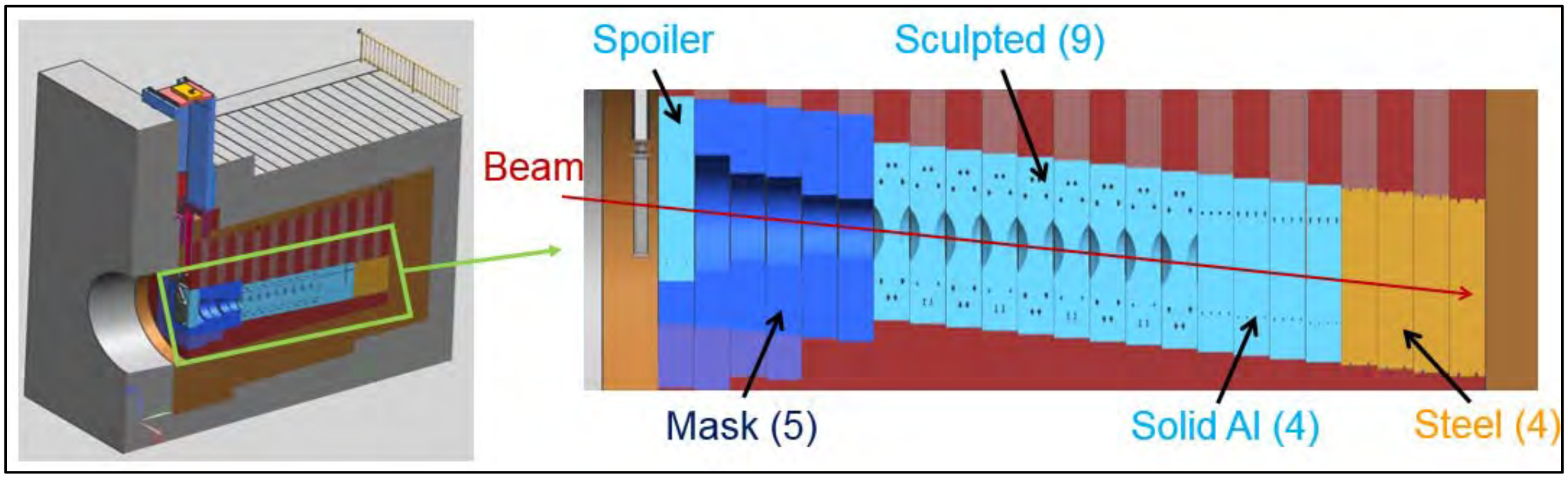

Figure 6-16: Left: Cross section of absorber through beam axis. Right: Cross section of absorber core.

## 6.4.8.1 Steady State Normal Operation

The energy deposition calculations were performed using a unified computer model that includes the Target Hall, target chase and Absorber Hall. A distribution of total power deposited in elements of the central region of the absorber is shown in Figure 6-17.

Using the MARS energy deposition results as a basis for heat load on the absorber and its core blocks, many iterative simulations between MARS and ANSYS have been carried out to determine the final configuration of the absorber. The main driver of this optimization is reduction of temperature and stress to acceptable levels for the materials in both normal operation and accident scenarios. Creep and fatigue effects have been considered when applicable.

Aluminum core blocks are all water cooled via four 1" diameter gun-drilled channels in the aluminum with 20 gallons per minute (gpm) volumetric flow rate through each channel. The water will be cooled to 10°C. Steel blocks are cooled via two 1" diameter stainless steel lines along the perimeter of the block with 20 gpm flow rate each.

As a result of an FEA analysis, temperatures, stresses and safety factors can be calculated for each location. These values are shown in Table 6-6 and are not compensated for any additional load or uncertainties. The lowest safety factors are for fatigue to the center of the sculpted block, and to yield for the steel block. The fatigue limit set for the center of the sculpted block is very conservative from the actual case. It uses nearly 3 times the number of cycles to failure (1e5) and the test temperature is 60°C higher (150°C). Also, this area is in compression, so concerns about opening a crack are small. The steel block does have a low safety factor to yield, although yielding the material does not necessarily indicate failure.

### 6.4.8.2 Accident Conditions

The absorber must be able to handle, without loss of function or damage, an accident condition where two pulses of the full proton beam do not hit the baffle or target and travels down the decay pipe. Two accident scenarios were considered. First, an on-axis accident, in which the beam travels down the center of the absorber and strikes the region that already has the highest temperature and largest stress from normal operation. Second, an off-axis accident where the beam strikes the absorber offset from the on-axis accident and passes directly through the water lines, where the water-line geometry might induce stress-risers and where one does not have the shower-spreading advantage of the central sculpting region. The FEA analysis has shown that the absorber is capable to handle a maximum of 2 accident pulses at full beam intensity at 2.4 MW.

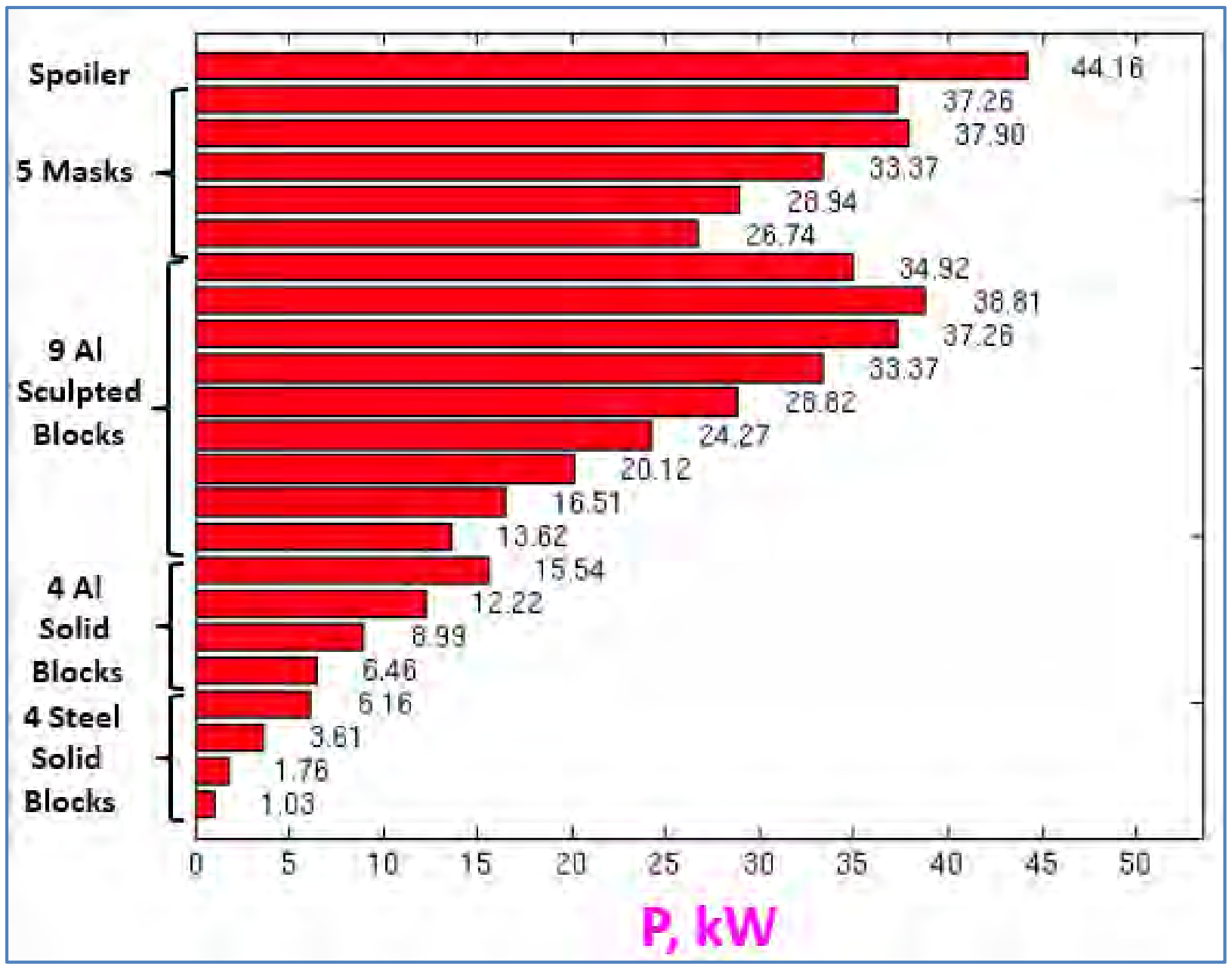

Figure 6-17: Distribution of total power deposited in the central part of the absorber (See Figure 6-16).

Table 6-6: Steady state normal operation safety factors

| | Steady State Maximum | | Safety Factor to Yield | Safety Factor to Fatigue | Satisfies Creep Criteria? |
|---|---|---|---|---|---|
| | Temp (°C) | Stress (MPa) | | | |
| Spoiler | 60 | 34 | 6.8 | 7.6 | Y |
| Sculpted, Center | 88 | 103 | 2.3 | 1.6 | Y |
| Sculpted, Water Line | 25 | 74 | 3.7 | 3.5 | Y |
| Solid Al | 84 | 48 | 4.8 | 3.4 | Y |
| Steel | 225 | 199 | 1.4 | - | - |

### 6.4.8.3 Steel Shielding Air Cooling

While the bulk of the beam energy reaching the absorber is deposited in the core, the outer steel shielding receives the remaining 30% of the energy deposited in the absorber. However, this energy deposition is not as concentrated as the core, and lends itself well to air cooling. Air from the air handling room flows over the top of the absorber, then flows downward through 5mm gaps in the steel shielding. After passing through the 5mm gaps, the air flows through a duct on the bottom of the absorber formed by a gap in the shielding and then is fed back to the air handling room. The NuMI target chase air cooling system operates at 25,000 cfm and removes approximately the same heat load, so this flow rate was selected for calculations in the absorber air cooling model. The calculations show that 130°C is the maximum temperature the steel could achieve. This is well within limits for steel and the paint applied.

## 6.4.9 Remote Handling Equipment

Technical components installed in the Target Hall enclosure are subjected to intense radiation from the primary or secondary beam. The level of irradiation in some LBNF environments will reach levels that are unprecedented at Fermilab. These radiation levels will be too high for workers near such components. The failure of some of the technical components (such as target or horns) is likely over the lifetime of the DUNE experiment. Therefore remotely operated removal and handling systems are an integral part of the Target Complex design (Target Hall enclosure and neighboring service areas). Because the remote handling systems are integrated into the infrastructure of the Target Complex and cannot be upgraded after irradiating the Target Complex areas, they must be designed to be sufficient for 2.4-MW beam power.

The LBNF remote-systems reference design includes equipment and systems in two functional locations. These are the surface Target Complex and the underground Absorber Hall. Along with shielded, remote-capable work areas, each of these locations will have the variety of equipment, lifting fixtures and vision systems required to carry out needed operations.

### 6.4.9.1 Target Complex Remote-Handling Facilities

The Target Hall enclosure contains the components for generating neutrinos and focusing them toward the near and far detectors. The layout of the Target Complex is shown in Figure 6-18.

The main hallway for transport of equipment shielding, and components is located at the upstream portion of the Target Hall enclosure and connects to the Morgue/Maintenance areas (floors of both Target Hall and Morgue/Maintenance areas are at the same elevation). It is through this hallway that radioactive components must pass to get from the Target Hall to the maintenance and morgue (short-term storage) areas. Since most of the service areas are planned to be occupiable with beam on, a shield door must be provided to shield the service areas from the Target Hall. The shield door will incorporate an air seal to separate the radioactive air volume of the Target Hall enclosure from the air volume of the neighboring service areas.

The Target Complex remote operations plan incorporates one hot storage rack in the Target Hall enclosure, designed to provide short- or long-term storage for Horn Module “T-blocks” during component replacement activities, and a work cell is located at the upstream end of the Target Hall enclosure. The work cell is primarily used to remotely remove a horn, target/baffle, or decay pipe window that has reached its end of life from an activated module and attach a new replacement component.

### 6.4.9.2 Absorber Hall Remote Handling Facilities

The Absorber Hall remote handling facilities are similar in concept to those for the Target Hall in that they will include a bridge crane, cask system and long-reach tools to enable the replacement of the hadron monitor upstream of the hadron absorber. However, unlike in the Target Hall, replacement of components will not require a work cell. Although the absorber components are designed to last the lifetime of the facility and will include redundant water-cooling lines, the consequences of complete failure are significant. Therefore, provisions will be made in the design of the Absorber Hall components and shielding to allow future replacement. However, because of the low probability of complete failure, final design and construction of remote handling equipment for absorber modules and water-cooled shielding will not be included in the LBNF project. If complete failure occurs during operation, a long downtime (6 months to 1 year) would then be required to final design, build, develop procedures and safely replace the failed component(s).

## 6.4.10 Radioactive Water Systems

Many components in Target Hall as well as the core of the absorber are water-cooled. Since these elements are operated in an environment with a high flux of energetic particles from the beam interacting with the target, the cooling water itself will be activated and cannot be allowed to mix with un-activated water. Therefore, these components are cooled using a closed-circuit water system; the heat being moved by conduction and convection to secondary water heat-exchanger/chiller system connected to the outside world.

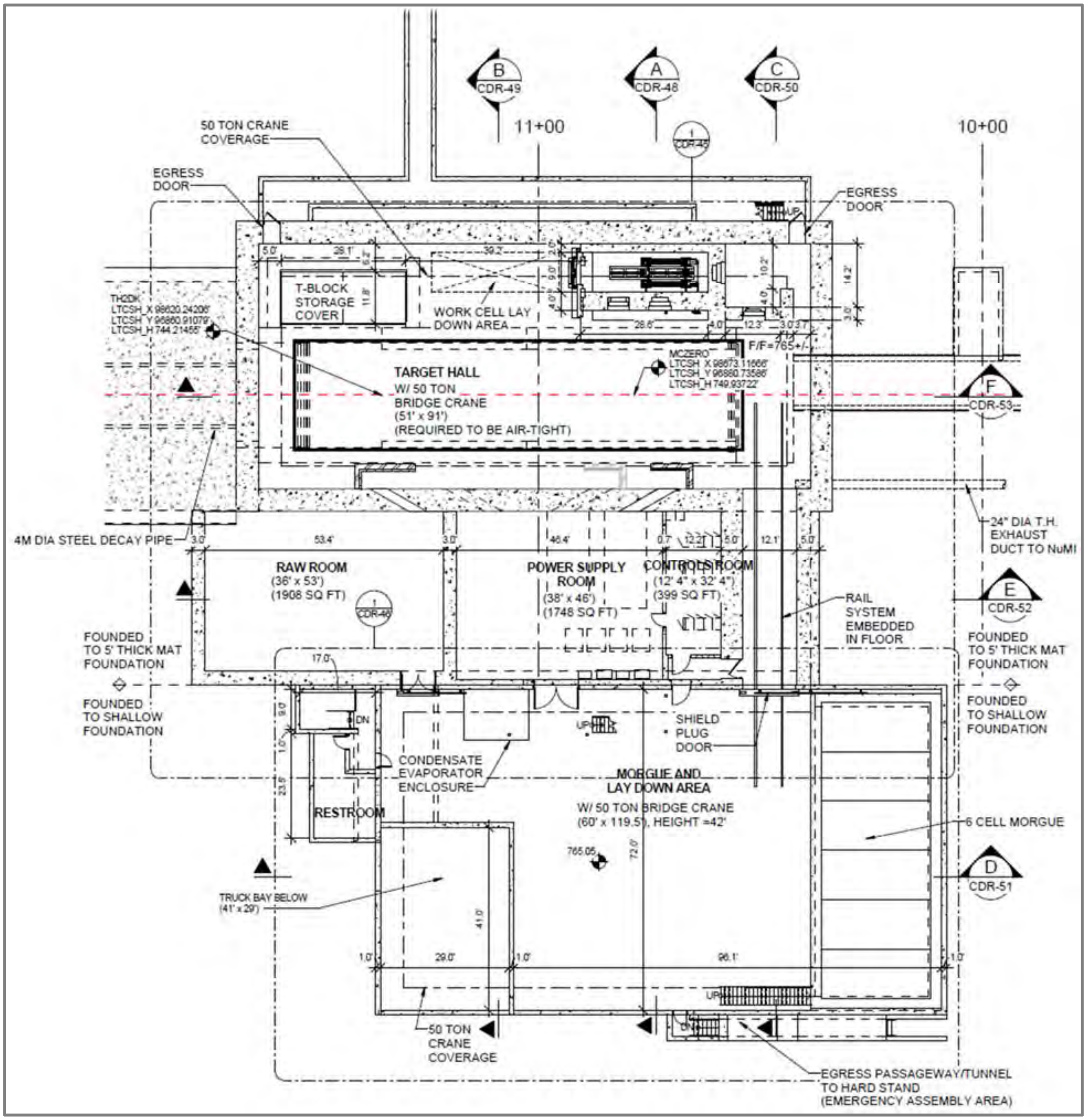

Figure 6-18: Target Complex plan view

Table 6-7: Summary of RAW skids heat loads for the Target Hall

| System | Heat Loads @ 1.2MW (kW) | Heat Loads @ 2.4MW (kW) |
|---|---|---|
| Target and Baffle RAW skid | 28 | 56 |
| Horn 1 RAW Skid | 64 | 128 |
| Horn 2 RAW Skid | 49 | 98 |
| Target Chase Shielding / Window Skid | 683 | 1166 |
| RAW Exchange System | none | none |
| Intermediate Cooling System (RAW total + pump + 10%) | 958 | 1669 |

Table 6-8: Summary of RAW skids heat loads for the Absorber Hall

| System | Heat Load @ 1.2MW (kW) | Heat Load @ 2.4MW (kW) |
|---|---|---|
| Absorber RAW System | 375 | 660 |
| RAW Exchange System | none | none |
| Intermediate Cooling System (RAW total + pump + 10%) | 425 | 740 |

### 6.4.10.1 Target Hall Systems

Located outside the Target Hall will be a Radioactive Water (RAW) equipment room, which will hold the majority of the equipment for RAW skids, for cooling of the Target, Horns 1 and 2, and the target chase shielding panels.

The estimated overall beam heat load in the different components is around 367 kW (at 1.2 MW-beam power). Anticipated heat loads for the various systems are as shown in Table 6-7, and include added heat from pumps, and heat exchanger efficiencies:

### 6.4.10.2 Absorber Hall Systems

Located outside the main Absorber Hall will be a RAW equipment room, which will hold the majority of the equipment for RAW skids for cooling of the absorber. The estimated total beam heat load for the Absorber Hall RAW systems is approximately 260 kW (at 1.2 MW beam power).

Anticipated heat loads for the absorber systems are as shown in Table 6-8 and include added heat from pumps, and heat exchanger efficiencies:

# 6.5 Radiological Considerations

## 6.5.1 Overview

In the 2012 version of this CDR, it was envisioned that the beam line would start operating with a 700 kW beam and after the accelerator complex upgrades were completed, about five years later, the beam line would be running a 2.3 MW proton beam. Since then the accelerator upgrade plan has changed. Currently, it is planned to start the operation of the beam line at 1.2 MW for five years followed by fifteen years at 2.4 MW, when the accelerator complex upgrade is ready. It is expected that this 20 years of running beam to be accomplished over the span of thirty years of accelerators operations. Change in radiological quantities, in going from 2.3 MW to 2.4 MW, is only a few percent. In this section, some of the radiological calculations presented have not yet been updated to 2.4 MW beam power. This actual beam power used will be noted where needed.

Only radiological issues for operation of the LBNF beam line with 2.4 MW beam power are considered, since retrofitting the LBNF facility for 2.4 MW after years of operation at 1.2 MW is very costly and not practical. The scope of radiological issues includes the primary transport line, Target Hall, decay pipe and the Absorber Hall. The analyses contained in this section are based on current requirements of the Fermilab Radiological Control Manual [36]. Other measurements and verification data available are also used where applicable.

The posting and entry control requirements for access to areas outside of beam enclosures where prompt radiation exposure may exist for normal and accident conditions are given in the Fermilab Radiological Control Manual. All results presented in the following subsections are based on the MARS modeling of the LBNF.

In the NuMI (700kW) primary beam line, fractional beam losses are controlled to better than $10^{-5}$. To maintain the same radiation level when scaling to 2.4 MW (desirable to keep residual radiation at a level where maintenance is not hindered), corresponds to controlling the losses at $3\times10^{-6}$ for LBNF. Control of the LBNF beam average operational losses is assumed to be $10^{-5}$ for shielding purposes, which gives a sensitivity/safety factor of more than 3 larger. While accidental beam losses are difficult to estimate from first principles, again the NuMI beam can be used as the analog to LBNF for this estimation. During the six years of NuMI primary beam operation, more than 50 million beam pulses were transported to the NuMI target with a total of more than $1.2\times10^{21}$ protons on target at 120 GeV. A total of 6 beam pulses have experienced primary beam loss at the 1% level, all due to Main Injector radio frequency (RF) problems. Since 2005 only one full intensity pulse has been lost. Therefore, it is assumed that control of LBNF primary beam losses to less than 2 pulses/hour is possible by using a controls system similar to that developed for the NuMI beam.

A computer model for the entire beamline has been built in the framework of the MARS15 Monte-Carlo code [NM1]. This model includes all the essential components of the target chase, decay channel, hadron absorber and the steel and concrete shielding in the present design. The MARS simulations are used specifically in the calculation of: (1) beam-induced energy deposition in components for engineering design, (2) prompt dose rates in halls and outside shielding, (3) residual dose rates from activated components, (4) radionuclide production in components, shield and rock, (5) horn focusing design and optimization of neutrino flux.

The radiological requirements outlined in this section are applied to the designs of the technical systems and equipment for the Beamline discussed in this volume, as well as to the associated conventional facilities.

## 6.5.2 Shielding

## 6.5.3 Primary Beamline

The Conceptual Design for the LBNF has been developed with external primary beam soil shielding of 25 feet (7.6 m). This provides an additional safety factor beyond the calculated LBNF required shielding for both the normal and accidental losses. The calculated soil shielding required for 2.4 MW beam, for unlimited occupancy classification, is 22.5 feet (6.9 m) for continuous fractional beam loss of $10^{-5}$ level and 25 feet (7.6 m) for 2 localized full beam pulses lost per hour.

To reduce the contribution of accidental dose from muons at the site boundary to less than 1 mrem, MARS simulations show that an additional 54 feet (16.5 m) of soil in the path of the muons, downstream of the Target Hall, is required to shield against losses at the apex of the embankment. The width of the embankment should be no less than 6.5 ft beam-left and 10.5 ft beam-right. It is also possible to accidentally lose the beam on one of the beam line elements, on the uphill side of the embankment. The worst case is when the beam is lost near the top. In this case the upward going muon plume will have the minimal soil shielding in its path, before exiting the transport-line shielding. MARS simulation shows a maximum dose rate of about 35 mrem on a small area on the embankment is possible. Based on the transverse dimension of the plume a fenced area is designated that encloses the area such that the accidental dose at the fence boundary will be less than 1 mrem.

### 6.5.3.1 Target Hall/Target Pile

The Target Hall and target pile shielding is designed to contain prompt radiation, residual radiation, activated air, accidental spills of radioactivated water, and to control a twenty-year buildup of the radionuclides in the soil outside the shielding to below requirements. Another goal of the design is to have an average dose rate of less than 100 mrem/hr in the Target Hall above the target pile during the normal beam operations, to minimize radiation damage to lights, crane, etc. A combination of steel, marble and borated polyethylene is used for shielding on top of the target pile. Because of sky shine considerations, the walls and the roof of the Target Hall are required to be 5 feet (1.5 m) and 7 feet (2.1 m) of concrete, respectively for 2.4 MW operation. Since the roof can be easily upgraded for higher beam power, in the reference, costed design the roof of the Target Hall is expected to be 6 feet (1.8 m) thick which is appropriate for 1.2 MW operation. For the sides and the bottom of the target pile, combinations of steel and concrete shielding are used. Details of the Target Hall and the target pile shielding are given in the corresponding portions of Section 3 of *Annex 3A* [30].

### 6.5.3.2 Decay Pipe

If the decay pipe was constructed horizontally at the elevation of the Main Injector with enough shielding, such that the production of radionuclides in the soil will have concentrations below surface water limits, and given the geology in the region near the Main Injector, no additional mitigation would be required. However, the decay pipe will be partly underground with the downstream end close to the aquifer. Under these conditions, it is prudent (ALARA) to reduce the radionuclide concentrations by two orders of magnitude; e.g. tritium concentration to be 0.3 pCi/ml, which is below the standard level of detection. Based on the requirement set by the project [37], the decay pipe will use 5.6 m of concrete shielding.

Figure 6-19 shows the sum of the ratio of radionuclide concentration in ground water over the Federal drinking water limit as a function of decay pipe shield thickness. To achieve undetectable levels of tritium and $^{22}$Na, the sum of concentration ratio should be 0.1 or less. Additionally, to protect against tritium leaking out of the shielding and being released to the environment, the concrete is surrounded by multiple protective outer layers of water impermeable material. For water drainage, the decay pipe shielding will be surrounded by a large volume of sand, with a sump pump at the downstream end. This sand layer will also be encased in a system of water impermeable layers to further isolate the decay pipe from water from the outside. Additional information about the geomembrane can be found in the Near Site Conventional Facilities, Section 5.5.2.1.

### 6.5.3.3 Absorber Hall Complex

As by-product of neutrino production, a flux of primary protons non-interacted in the target and non-decayed secondary hadrons (mostly $\pi$ and K-mesons) and leptons must be absorbed to prevent them from entering the surrounding rock of the excavation and inducing radioactivity. This is accomplished with an absorber structure which is located directly after the decay pipe. It is a assembly of aluminum, steel and concrete blocks, water- and air-cooled, which must contain the energy of the particles after the decay pipe. The vast majority of these secondary hadrons and primary protons, essentially, all of them are stopped in the absorber. The fluxes of secondary particles (mainly neutrons) escaping the system must be attenuated by the absorber and shielding to the tolerable levels. According to the MARS15 simulation, in total, 31% of the total beam power is deposited in the absorber. The shielding configuration, composition and dimensions have been thoroughly optimized in massive MARS15 simulations [38] to keep the calculated radiation quantities safely below the regulatory limits in the following four areas:

1. Prompt dose (beam-on): Service Building < 0.05 mrem/hr for general public, elevators < 0.25 mrem/hr for unlimited access of radiation workers.

2. Residual dose (beam off) in Absorber Hall, Raw and Sump Rooms: < 5 mrem/hr.

3. Groundwater activation in soil/rock immediately outside the concrete shielding: below the limit for $^{3}$H and $^{22}$Na that corresponds to the hadron flux of 400 cm$^{-2}$ s$^{-1}$ with energy > 30 MeV.

4. The hadron fluxes with energy > 30 MeV in air pockets inside the absorber and in various regions of the Absorber Hall and Muon Alcove below the air release limits (see below).

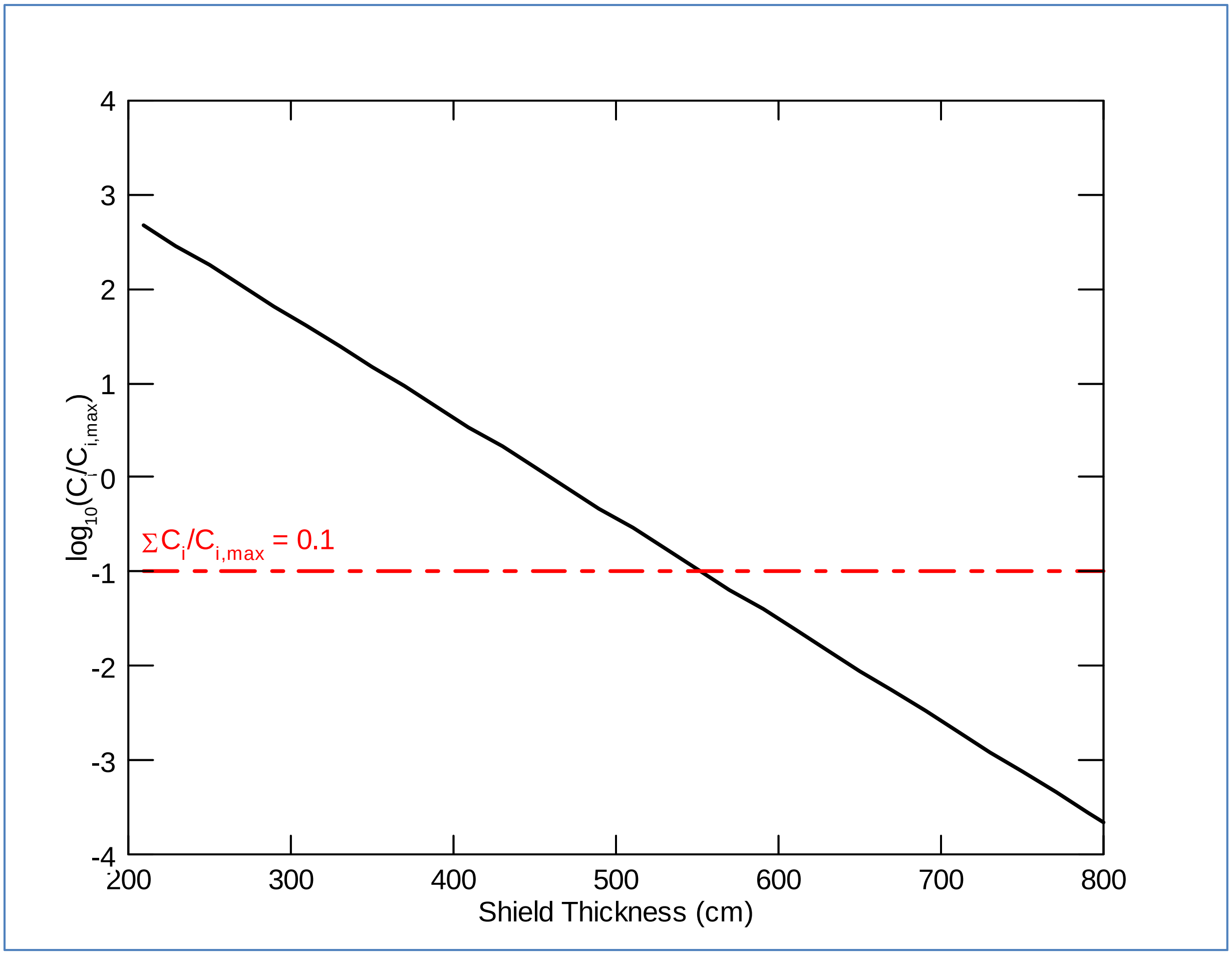

Figure 6-19: Sum of the ratio of radionuclide concentrations ($C_i$) over the Federal drinking water limits ($C_{i,max}$) for tritium and $^{22}$Na as a function of decay pipe shield thickness. The dashed red line shows the self-imposed limit on the ratio sum which will result in the radionuclide concentrations in the ground water to be below the detection limits

The absorber (Section 6.4.8) and the Absorber Hall Complex shielding is based on the ground water management requirement [37] set by the project using the latest MARS model [38]. The shielding is designed to keep a twenty-year buildup of the radionuclides in the soil outside the shielding to below the standard detection levels. Another goal is to reduce the residual dose rates outside the absorber pile to well below 100 mrem/hr to allow for maintenance activities, and to preclude significant activation of the equipment in the absorber RAW room. During proton beam operations, the following areas will be designated as “limited occupancy” for the radiation worker: the service building, personnel elevator and stairway shaft, elevator lobbies at the three stops, the three sump pump rooms and the Instrumentation room. The rest of the complex (Absorber Hall, muon monitor area, RAW room, air handling room) is considered a high-radiation area and will not be accessible during beam-on periods.

The sophisticated MARS15 calculations have shown that a muon plume 170-m long and ~10 m in diameter is steadily generated in the rock with a central region (80-m long and up to 7 m in diameter) where hadron fluxes exceed the above-mentioned limit of 400 $cm^{-2}$ $s^{-1}$ for groundwater activation. It means that such a region needs to be protected from groundwater penetration through it, which is nontrivial and expensive from the civil construction point of view. It has been found in optimization MARS15 studies [38] that the dimensions and cost of such a construction can be substantially reduced by using a steel conical kern 30-m long (7-m maximal and 3-m minimal diameters) immediately downstream the Absorber Hall concrete wall with the excessive hadron fluxes contained in the kern. See Figure 6-20 and refer to Section 5 *Annex 3A* [30] for more information.

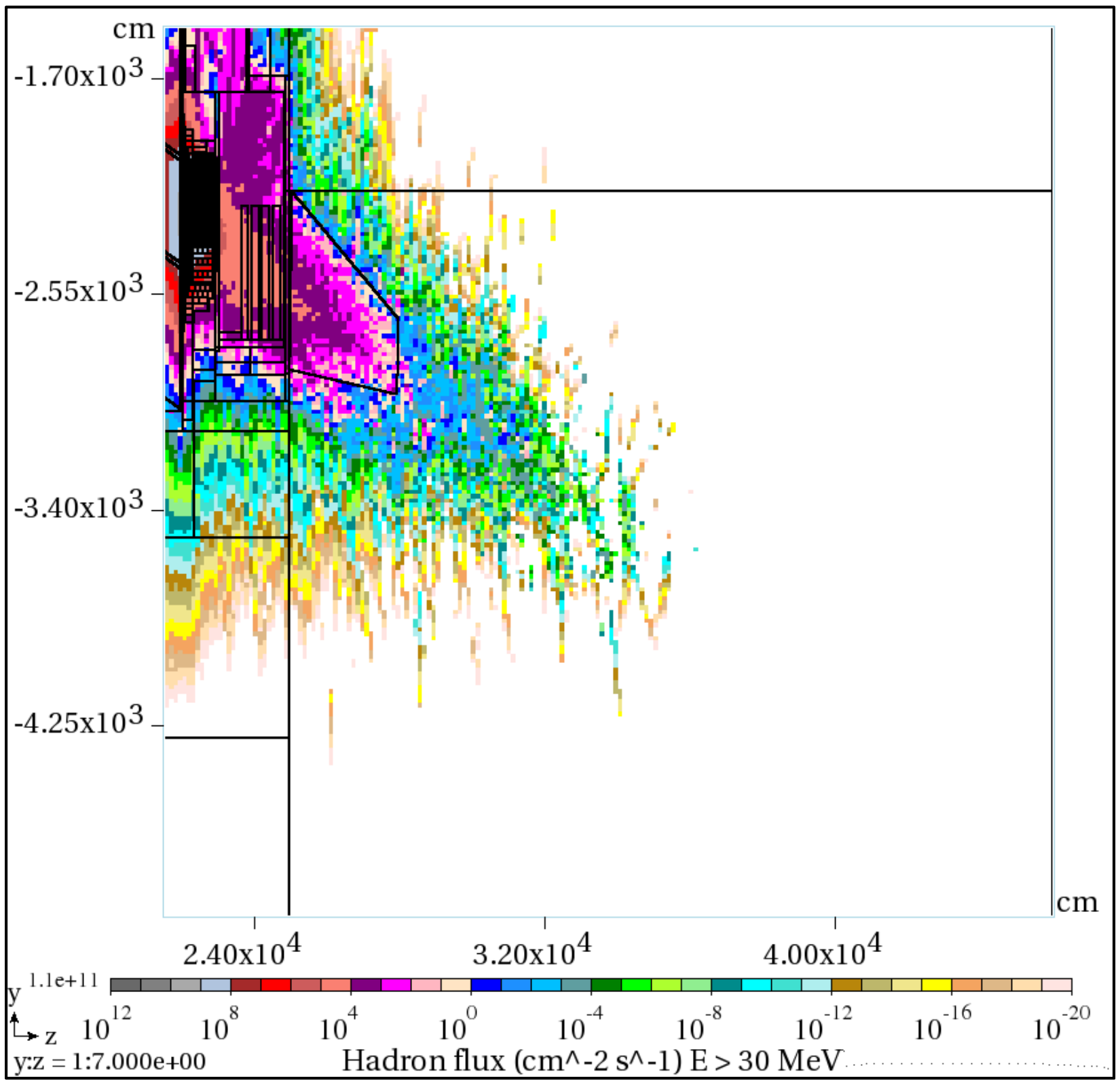

Figure 6-20: Hadron (E > 30 MeV) isoflux contours (cm-2 s-1) in the rock and steel kern downstream the absorber; hadron fluxes outside the steel kern are below the accepted limit of 400 cm-2 s-1 for groundwater activation

### 6.5.4 Other Radiological Design Issues

#### 6.5.4.1 Groundwater and Surface Water Protection

The production of potentially mobile isotopes such as tritium ($^{3}H$) and sodium-22 ($^{22}Na$) is an unavoidable consequence of high-energy particle collisions with nuclei. Since the primary transport line is located in the glacial till, with no direct connection to the aquifer, all radionuclides produced in the soil surrounding the enclosure will have to migrate down through the soil layers to reach the aquifer. These seepage velocities, for the layers in the glacial till, are very small and the concentrations of the radionuclides are reduced by 5 to 7 orders of magnitude, well below detectable values.

The Target Hall and the target chase are also at grade level, located in the glacial till. The shielding of the Target Hall and the target chase is designed such that a 20-year accumulation of radionuclides in the soil immediately outside the shielding, assuming no dispersion, would result in maximum concentrations of 27% of the surface waters limits. Additionally, the target chase and the Target Hall will have a geomembrane barrier system, preventing water from coming in contact with the shielding. For the rest of the beam line, from the decay pipe to the end of Absorber Hall, there will be sufficient shielding and water impermeable layers to render the concentration of the radionuclides of interest, accumulated in the soil over 20 years, to be less than the current standard detection limits [30]. The current accepted detection limits are 1 pCi/ml for tritium and 0.04 pCi/ml for sodium-22.

#### 6.5.4.2 Tritium Mitigation

Tritiated water molecules (e.g. HTOs) are highly mobile, especially in humid air, and can create significant concentrations in drain waters that are then collected in the pumping processes that keep the beam line areas dry. These basic processes, namely tritium production and migration, show that strategies to avoid unnecessary effluent will rely on isolating the materials in which the tritium is produced from water, and in the dehumidification of air in contact with these materials, together with subsequent collection and evaporation of the tritiated condensate. Additionally,

- There will be a geomembrane barrier system installed between the decay pipe concrete and the soil that is largely impervious to water. In this way the decay pipe concrete (in which tritium is created during operations) will be held at a low saturation. Numerical studies using the NuMI system indicate that if shielding concrete is unsaturated, the mobility of created tritium is low [39].

- The operational design of a sampling and monitoring program is straightforward, and allows for maintenance of the drainage system.

#### 6.5.4.3 RAW Systems

The cooling water for the baffle, target, horns and the absorber will be highly activated after a short time of operation. The prompt dose rates from the RAW (RadioActive Water) skids belonging to these devices will be high and in addition to the short lived radionuclides, large concentrations of the tritium will build up in these systems. Shielding and cool-down times will be used to reduce the dose from these systems. Remotely controlled drainage and top up with fresh water will be used to keep the tritium concentrations at manageable levels. Alarms and containment systems will be used to prevent spills and contamination of the soil and surface waters. Water from these systems will be disposed of as low level radioactive waste.

### 6.5.4.4 Activated Air

High levels of radioactive air will be produced in the target chase (at the center of the target pile), in the decay pipe air cooling system, and in the Absorber Hall. The air to the chase and decay pipe cooling are closed, isolated loops, that however are expected to leak at some level. The most significant leaks will be at the air handing systems in the air handling room, with smaller leak expected through seals on the covers of the target pile in the Target Hall. Additionally, the air from the Absorber Hall and the air handling room is sent to the Target Hall, where it combines with the small leakage from the target chase. The air handling systems filter, chill and dehumidify the air in the cooling loops. The air handling room structure and the doors are designed to be fairly air-tight, and the air handling room and Target Hall are maintained at lower pressure than the outside of the structures by pulling air to the NuMI Target Hall through a buried pipe, and through the NuMI pre-target volume and then exhausted. The transit time from the LBNF Target Hall to the NuMI exhaust is a sufficient to allow the airborne radionuclides to decay by orders of magnitude. The current Fermilab radioactive air emissions permit allows the annual exposure of a member of public offsite to the radioactive air emissions, from all sources to be less than 0.1 mrem. It is the goal of the LBNF design is to have the air emissions contribute less than 30% of this limit which allows for the emissions from other accelerators and beam lines at Fermilab.

### 6.5.4.5 Outside Prompt Dose

There are three ways where the prompt dose rates may reach outside the facility: (1) direct attenuated radiation outside the shielding, (2) sky shine, which is radiation, primarily neutrons, due to back scattering from air and (3) released radioactive air. Fermilab Radiological Control Manual (FRCM) Article 1104 [36] describes the regulatory requirements/limits regarding the maximum annual allowable dose to the public. The LBNF primary beam transport line, Target Hall and the decay pipe and the Absorber Service Building can contribute to outdoor doses. Based on the latest MARS calculations [40] [41] both the annual direct and sky shine doses are calculated for both offsite and onsite locations. Direct accidental muon dose at the apex of the transport line is also included in the offsite dose. The dose rate from released radioactive air is discussed in the previous paragraph.

### 6.5.4.6 Offsite Dose

To allow operations of other experiments, beam-lines and accelerators, the offsite goal for LBNF is set at 1±1 mrem in a year, from all radiation sources generated by this beam-line. The total offsite dose, at the nearest site boundary, due to both direct and skyshine is estimated to be 1.02 mrem in a year.

### 6.5.4.7 Onsite Dose

Wilson Hall is the nearest publically occupied building to the LBNF beam line. Both the maximum direct and skyshine annual dose to the occupants of the Wilson Hall has been calculated. The total annual dose, at Wilson Hall, due to both direct and skyshine is estimated to be 0.06 mrem. Doses for other locations onsite, further away, will be less.

#### 6.5.4.8 Residual Radiation

Based on the past experience and the difficulty of component replacement with the steep grades (~10%) of the LBNF primary beam enclosures, the beam loss and beam control devices would be employed to keep the residual radiation inside the beam line to no more than 50 mrem/hr on contact. This allows for repair or replacement of the beam line elements with little programmatic impact and keeping the dose to the workers ALARA.

There are other beam line devices, such as targets/horns and their mounting modules, target pile cooling panels and absorber core modules that are exposed to high levels of beam spray and are expected to become highly radioactive. These devices may need to be repaired or replaced. The LBNF design provides for remote handling and shielded storage of these devices. The shielding of the work/repair cell used for targets/horns and modules is designed such that for a 20 kR/hr object, the dose rate outside the cell is less than 1 mrem/hr. The shielding of the containers used for the over the road transport of such devices will be such that the dose rate outside the containers is less than 100 mrem/hr at one foot.

## 6.6 System Integration

### 6.6.1 Introduction

This section covers the System Integration activity of the LBNF Beamline Project. The System Integration team's responsibilities can be broken into two major areas: first, the oversight of systems for Controls, Alignment and Interlocks, and Installation Coordination. Second, there is the task of ensuring that the interfaces between each of the subsystems of the Beamline Project are complete. The Controls, Alignment, Interlocks and Installation Coordination span the entire Beamline project and must therefore be properly supported by all the interfaces in addition to the relevant components. Interface coordination involves both achieving consensus as to the location and nature of each interface and the party responsible for it. The coordination activity must also ensure proper distribution of requirements and specifications so that all the needed components are accounted for, and that they will be constructed such that they will fit together properly during installation and operate successfully.

System Integration thus has the primary responsibility of facilitating good communication throughout the L2 project in order to prevent deficiencies and scope-related problems, and for any that are introduced, to spot them early on and make sure they get corrected.

### 6.6.2 Controls

#### 6.6.2.1 Introduction

Any high-energy external beamline requires a robust control system to ensure proper operation. The control issues for a beamline like LBNF's are well understood. The control system must be able to perform as follows:

- Reliably log data for every beam pulse (this implies a digitization with appropriate throughput).
- Plot both real-time and logged data in strip-chart form and capture all operational information for the beamline devices in a database.
- Issue alarms for off-nominal conditions and provide power-supply controllers with ramping capability.
- Handle the so-called slow-control subsystems: water, vacuum and temperature.
- Provide environmental monitoring.
- Display information from the position and loss monitors along the beamline and provide an auto-tuning facility to keep the beam centered over its length without significant human intervention.

These functions will be provided for via the existing Accelerator Controls NETwork (ACNET) Control System of the Fermilab Accelerator Complex.

## 6.6.3 Radiation-Safety Interlock Systems

Radiation-Safety Interlock Systems include Electrical Safety interlock System (ESS), Radiation Safety Interlock Systems (RSS), Radiation Monitors, Radiation Air Monitors, and Radiation Frisker Stations. Underlying all safety-system designs is a commitment to providing the necessary hardware, procedures, and knowledge to personnel to ensure their well-being. Inherent in each of these systems is the concept of redundancy.

## 6.6.4 Alignment

### 6.6.4.1 Overview

This section summarizes the concepts, methodology, implementation and commissioning of the geodetic surveying (global positioning) efforts for determining the absolute positions of the LBNF beamline components at Fermilab and the location for the far detector at SURF. This information is critical to achieving proper aim of the neutrino beam. From this information, the beam orientation parameters are computed, as well as the alignment of the LBNF beamline.

### 6.6.4.2 Design Considerations

Clearly, directing the neutrino beam to intersect the far detector located 1,300 km distant from the source, is of paramount importance. Physics requirements will drive the absolute and relative alignment tolerances.

The divergence (spatial spread orthogonal to the line of travel) of the neutrino beam at this distance is on the order of kilometers. The spectrum of neutrino energies varies with their offset from the beam's center line, higher-energy neutrinos are closer to the center, lower-energy ones are farther out. Based on NuMI's requirement for the energy spread, LBNF will require that the combined effect of all alignment errors must cause less than 2% change in any 1-GeV energy interval in predicting the far detector energy spectrum.

To accomplish this, and prorate from NuMI to SURF, the neutrino-beam center must be within ±133 m from its ideal position at the far detector, corresponding to an angular error of $\pm10^{-4}$ radians. Achieving this tolerance requires precise knowledge of the geometry of the neutrino beam. Table 6-9 lists alignment tolerance requirements for the low-energy beam for NuMI, which will also be established for LBNF, with the exception of the Far detector which was prorated to SURF. A Monte Carlo (PBEAM_WMC) was used to calculate the effect of misalignments of each beamline element for the determination of the far detector spectrum (without oscillations) from the NuMI's measured near-detector spectrum.

The requirement on the relative alignments of the beamline components and the target-station components (target and horns) is that they be within ±0.35 mm (These requirements are based on NuMI design; have been studied by an LBNF simulation, and are the accepted criteria at this stage of the LBNF design). To accomplish this, high-accuracy local geodetic and underground networks will be established to support the installation and alignment of the primary-beam components, neutrino-beam devices and the near detector.

Table 6-9: Alignment tolerance requirements

| Position | Tolerance |
|---|---|
| Beam position at target | ±0.45 mm |
| Target position - each end | ±0.5 mm |
| Horn 1 position - each end | ±0.5 mm |
| Horn 2 position - each end | ±0.5 mm |
| Decay pipe position | ±20 mm |
| Downstream Hadron monitor | ±25 mm |
| Muon monitors | ±25 mm |
| Far Detectors | ±21 m |

## 6.6.5 Installation Coordination

This activity provides the management oversight of the day-to-day activities taking place in the installation areas and the framework for sequencing and scheduling the installation tasks. The scope of this role is driven by the need of balance the resources required in four distinct installation sub-projects. In addition, there is a need to ensure that all activities are conducted with a consistent level of safety and quality assurance throughout the entire project. The role of Installation Coordination is distinct from the actual task of installation. Its role is primarily the coordination of installation activities and will be led by an Installation Coordinator. The responsibility for the design, fabrication and installation of each element of the Beamline Project resides in its appropriate subsystem.

Installation Coordination will draw on the experiences of previous installations such as NuMI, and the lessons learned from more recent installation projects such as ANU. In addition, the team will be organized in a manner that advantageously uses the project management tools being implemented throughout the laboratory. The implementation of Installation Coordination will begin with the managerial role of sequencing and controlling the activities in each of the areas (as illustrated below). Each area (e.g., Main Injector, Primary Beamline, Target Complex, and Absorber Hall) will be under the supervision of either an Operations Specialist or a Floor Manager whose job it is to oversee the overall installation activity taking place in the area and to supervise the daily activities of task managers who are leading the work crews in each area. Floor Managers will report directly to the Installation Coordinator.

## 6.7 Alternative Beamline Options

The LBNF Beamline Facility is being designed for twenty years of operation with thirty years of total lifetime of the facilities. During this time period, the beam power is expected to increase from 1.2 MW to 2.4 MW and the facility must accommodate upgraded targets and horns in different configurations to maximize the neutrino flux in the appropriate energy range and to enable tuneability in the neutrino energy spectrum. To allow for flexibility and for improved capabilities in the future, the LBNF Beamline Team is investigating and considering alternative design options. One of those changes affect physics and one is a technical alternative in case the current default design proves insufficient after more detailed design work. The considered alternatives are:

- A further optimized target-horn system
- A gas different than air in the target chase (e.g., nitrogen or helium)

The reference Beamline design uses a NuMI-like target and NuMI-style horns appropriately modified for 1.2 MW operation. Further optimization of the target-horn system has the potential to substantially increase the neutrino flux at the first and especially second oscillation maxima as well as the area in between the two maxima and reduce wrong-sign neutrino background, thereby increasing the sensitivity to CP violation and mass hierarchy determination; see discussion in *Volume 2: The Physics Program for DUNE at LBNF* [2]. Target R&D and target-horn optimization work is on-going and may yield further improvements beyond those currently achieved. Engineering studies of the proposed target and horn designs and methods of integrating the target into the first horn must be performed to turn these concepts into real buildable and reliable structures. These studies will be carried out between CD-1 and CD-2 to determine the baseline design for the LBNF target-horn system. In addition, since targets and horns are consumables, more advanced ones could very well be designed in the future as 2$^{nd}$ generation components.

The more advanced target and focusing system described in *Volume 2* [2] utilizes two horns that are longer, of larger diameter and that are spaced farther apart than in the reference design. The first horn in particular is ~5.5 m long and ~1.3 m in diameter and functions effectively as two horns in one structure. The second horn, is NuMI-style but wider and longer, and is 7.8 m farther downstream than that of the reference design. The size of the target chase in the reference design is based on these dimensions, and is therefore larger than is required for the current reference target-horn system. It is likely that the dimensions of the target chase will be further modified between CD-1 and CD-2 as the optimization and engineering of the target-horn system advances. It is also necessary to consider that over the multi-decade lifetime of this facility, new target and focusing system designs may emerge or new physics directions may require a different optimization of the beam than the ones currently under consideration for DUNE. Between CD-1 and CD-2, additional conceptual studies will be done of different beam optimizations to guide the development of the baseline design for the target hall complex, to provide adequate flexibility for the future.

As described earlier in this CDR, the target chase is air filled and together with the surrounding target pile it is cooled by air. There are two studies in progress that could eventually determine gas selection for use in the target chase/pile cooling system: (1) the LBNF Corrosion Task Force studies that includes both measurements at NuMI and associated modeling, and (2) LBNF studies of Air Releases to the Atmosphere. The conclusion from either one or both of these studies could require that the oxygen and argon concentrations in the target pile cooling system be minimized to mitigate the possible problems of (1) corrosion due to ozone production, and/or (2) radionuclide emission to the atmosphere. As discussed in the Neutrino Beam section of the CDR, compliance with a requirement to minimize the oxygen concentration will be accomplished by changing the cooling gas from air to nitrogen gas or possibly to helium. The conclusion of the LBNF Beamline Air-Releases Design Review that took place in April 2015 indicates that the reference design scheme to keep air-releases under control is reasonable and with sufficient safety factor [42]. The air-releases calculations will be revisited between CD-1 and CD-2 after the volumes of the target chase and decay pipe are finalized.